# Überprüfung von Integritätsbedingungen in Deduktiven Datenbanken

Diplomarbeit

von

Stefan Decker

November 1994

AG Grundlagen der Programmierung
Fachbereich Informatik
Universität Kaiserslautern

Betreuer: Prof. Dr. Otto Mayer
Dr. Christoph Lingenfelder

## Erklärung

Hiermit erkläre ich, Stefan Decker, daß ich die vorliegende Diplomarbeit selbständig verfaßt und keine anderen als die angegebenen Hilfsmittel verwendet habe.

Kaiserslautern, den 4. November 1994

# Inhaltsverzeichnis

# Abbildungsverzeichnis

# Kapitel 1

# Einleitung

Die Fortschritte in der Informatik und der künstlichen Intelligenz lassen die Entwicklung und Anwendung immer größerer Wissensbasen zu. Selten werden diese Wissensbasen direkt vollständig erstellt und nach ihrer Konstruktion nicht mehr verändert, sondern sie entwickeln sich im Laufe der Zeit, werden größer und damit auch unübersichtlicher. Mit der Größe wächst aber auch die Gefahr, widersprüchliche Aussagen in eine Wissensbasis aufzunehmen, insbesondere wenn mehrere Experten an der Erstellung einer Wissensbasis arbeiten. Ein einzelner Wissensingenieur kann die zu verändernde Wissensbasis dann nicht in allen Einzelheiten kennen und somit die Auswirkungen einer Änderung nicht abschätzen. Aus diesem Grunde werden Verfahren benötigt, die bei Änderungen die Integrität einer Wissensbasis gewährleisten.

Die vorliegende Arbeit beschäftigt sich ausgehend vom Gebiet des *logischen Programmierens* mit dieser Problemstellung. Wir zeigen, daß *Integritätsverletzungen* als spezielle Operationen auf Beweise von *Integritätsbedingungen* (hier im speziellen SLDNF-Beweise) interpretiert werden können. Hierzu geben wir die Definition eines *Beweisbaumes*, einer speziellen Datenstruktur, und zeigen, daß die Existenz eines Beweisbaumes die Existenz eines SLDNF-Beweises impliziert. Ein Beweisbaum ist im Vergleich zu einem SLDNF-Baum eine handlichere Datenstruktur: sie ermöglicht die mengenorientierte Betrachtung eines Beweises, wie sie bei den Bottom-up Auswertungsstrategien im Bereich der deduktiven Datenbanken eingesetzt werden. Weiter gibt ein Beweisbaum die Struktur eines Beweises klarer als ein SLDNF-Baum wieder, was weitere Anwendungsmöglichkeiten in Aussicht stellt.

Mit Hilfe dieser Struktur kommen wir zu einer minimalen Menge von Bedingungen, die angibt, wann eine Änderung der Wissensbasis einen gegebenen Beweis und damit die Gültigkeit der zugehörigen Integritätsbedingung beeinträchtigt. Weiterhin können wir bei der Suche nach einem neuen Beweis große Teile des alten Beweises wiederverwenden, so daß sich die Suche nach einem neuen Beweis i.a. weniger aufwendig gestaltet als mit bisherigen Ansätzen.

## 1.1 Gliederung

- Das Kapitel 2 gibt eine Zusammenfassung der wichtigsten Grundlagen auf dem Gebiet der Logikprogrammierung und der deduktiven Datenbanken und kann von auf dem Gebiet erfahrenen Lesern in großen Teilen überschlagen werden. Lediglich die Abschnitte über deduktive Datenbanken sollten etwas genauer durchgearbeitet werden, da die dort definierten Begriffe in der Literatur nicht immer einheitlich gebraucht werden.
- Das Kapitel 3 enthält eine Zusammenfassung und Überblick über die wichtigsten bekannten Methoden zur inkrementellen Integritätsüberprüfung. Insbesondere zeigen wir Fehler eines bereits in der Literatur bekannten Verfahrens auf.
- Kapitel 4 führt unseren Ansatz zunächst an einem Beispiel vor. Leser, die sich intensiver mit der formalen Methode befassen wollen, sollten zunächst versuchen, die Beispiele nachzuvollziehen, um ein Gefühl für die Methode zu entwickeln.
- Kapitel 5 führt den für die späteren Definitionen und Beweise wichtigen Begriff des *unvollständigen SLDNF-Baumes* ein, der eine echte Erweiterung des üblichen Begriffs darstellt. Unvollständige SLDNF-Bäume können kombiniert und somit zur Zerlegung eines Beweises in Teilbeweise benutzt werden.
- Kapitel 6 führt die Sichtweise von Klauseln als Filter bei der Suche nach einen endlichen gescheiterten SLDNF-Baum ein und gibt darauf aufbauend die Definition der Beweisbaumstruktur. Weiter wird ein Korrektheitsbeweis gegeben, sowie einige später benötigte spezielle Beweisbäume definiert und ihre Eigenschaften geklärt.
- Kapitel 7 gibt ein Vollständigkeitsresultat, indem ein Algorithmus für die Transformation eines SLDNF-Beweises via einer bestimmten Berechnungsregel in einen Beweisbaum angegeben wird.
- Kapitel 8 gibt verschiedenen Arten an, wie die Änderung der Wissensbasis den Beweisbaum beeinflussen kann. Darauf aufbauend werden Algorithmen angege-ben, die durch Modifikation der Beweise die eigentliche Integritätsüberprüfung durchführen.

- Kapitel 9 gibt eine Zusammenfassung und zeigt verschiedene Richtungen auf, in denen eine weitere Untersuchung im Zusammenhang mit der vorliegenden Arbeit sinnvoll erscheint. Es konnten bei weitem nicht alle Zusammenhänge und Erweiterungen untersucht werden, jedoch werden die wichtigsten Möglichkeiten hier aufgezählt.

## 1.2 Rahmen der Arbeit

Diese Diplomarbeit wurde im Rahmen des Projektes ESM (Enterprise Security Manager) bei der IBM Deutschland Entwicklung GmbH in Böblingen erstellt. Ziel dabei war es, eine Methode zu finden, die es ermöglicht, Datenobjekte und ihre Relationen zueinander in einem großen Netzwerk automatisch auf die Einhaltung von Bedingungen zu überprüfen. Hierbei flossen Erfahrungen mit ein, die mit dem Projekt KBSSM (siehe [Lingenfelder und Schmücker-Schend, 1992, Decker und Lingenfelder, 1992]) am Institut für Wissensbasierte Systeme der IBM in Heidelberg gemacht wurden.

# Kapitel 2

# Grundlagen

Die vorliegende Arbeit bewegt sich im formalen Rahmen der Prädikatenlogik erster Stufe, teilweise unsortiert und um hierarchische Sorten erweitert. In der Notation orientieren wir uns an [Lloyd, 1987] und [Walther, 1987]. Umfassendere Einführungen in die Prädikatenlogik 1. Stufe geben z.B. [Ebbinghaus et al., 1986] oder [Shoenfield, 1967]. Einführungen in deduktive Datenbanksysteme finden sich in [Lloyd, 1987] und [Cremers et al., 1994]. Das Gebiet des automatischen Beweisens wird in [Hofbauer und Kutsche, 1991] ausführlich dargestellt. Eine etwas kürzere Einführung findet sich in [Walther, 1993].

## 2.1 Prädikatenlogik erster Stufe

Zunächst führen wir die Syntax und Semantik der Prädikatenlogik erster Stufe ein und erweitern diese dann um den Begriff der Sorten.

### 2.1.1 Syntax von $\mathcal{PL}_1$

**Definition 2.1.1 (Signatur)** *Eine* Signatur $\Sigma = \langle \mathbf{C}, \mathbf{F}, \mathbf{P}, \mathbf{V} \rangle$ *der Sprache* $\mathcal{PL}_1$ *(Prädikatenlogik 1. Stufe) besteht aus*

1. *einer abzählbaren Menge* $\mathbf{C}$ *von Konstanten* $\{a, b, c, \ldots\}$*;*
2. *einer Familie* $\mathbf{F} = (F_n)$ *von abzählbaren Mengen n–stelliger Funktionssymbole* $\{f, g, h, \ldots\}$*;*

3. *einer Familie* $\mathbf{P} = (P_n)$ *von abzählbaren Mengen n-stelliger Prädikatensymbole* $\{p, q, r, \ldots\}$*;*

4. *einer abzählbaren Menge* $\mathbf{V}$ *von Variablen* $\{x, y, z, \ldots\}$*.*

**Definition 2.1.2 (Terme)** *Die Menge der Terme* $T^\Sigma$ *über* $\Sigma$ *ist definiert durch die kleinste Menge, für die gilt:*

1. $v \in T^\Sigma$ *für jede Variable* $v \in \mathbf{V}$*;*

2. $c \in T^\Sigma$ *für jede Konstante* $c \in \mathbf{C}$*;*

3. *falls* $f \in F_n$ *und* $t_1, \ldots, t_n \in T^\Sigma$*, dann ist auch* $f(t_1, \ldots, t_n) \in T^\Sigma$*.*

$t \in T^\Sigma$ *heißt ein* Grundterm*, falls* $t$ *keine Variablen enthält.*

**Definition 2.1.3 (Formeln)** *Die Menge der* atomaren Formeln $At^\Sigma$ *über* $\Sigma$ *ist wie folgt definiert:*

$p(t_1, \ldots, t_n) \in At^\Sigma$*, falls* $p \in P_n$*,* $P_n \in \mathbf{P}$ *und* $t_i \in T^\Sigma$ *für* $1 \leq i \leq n$*.* $p(t_1, \ldots, t_n) \in At^\Sigma$ *heißt* Grundatom*, falls alle* $t_i$*,* $1 \leq i \leq n$*, Grundterme sind. Die Menge der Formeln* $For^\Sigma$ *über* $\Sigma$ *ist definiert durch die kleinste Menge, für die gilt:*

1. $At^\Sigma \subseteq For^\Sigma$

2. $\neg\varphi \in For^\Sigma$*, falls* $\varphi \in For^\Sigma$ *(Negation);*

3. $(\varphi \vee \phi) \in For^\Sigma$*, falls* $\varphi, \phi \in For^\Sigma$ *(Disjunktion);*

4. $(\forall\ x\ \varphi) \in For^\Sigma$*, falls* $x \in \mathbf{V}$ *und* $\varphi \in For^\Sigma$ *(universelle Quantifikation).*

*Es werden die folgenden abkürzenden Schreibweisen verwendet*

- $(\varphi \wedge \phi)$ *für* $\neg(\neg\varphi \vee \neg\phi)$ *(Konjunktion);*
- $(\varphi \rightarrow \phi)$ *für* $(\neg\varphi \vee \phi)$ *(Implikation);*
- $(\varphi \leftarrow \phi)$ *für* $(\phi \rightarrow \varphi)$*;*
- $(\varphi \leftrightarrow \phi)$ *für* $((\varphi \rightarrow \phi) \wedge (\varphi \leftarrow \phi))$*;*
- $(\exists\ x\ \varphi)$ *für* $\neg(\forall\ x\ \neg\varphi)$ *(existentielle Quantifikation).*

*Die Symbole* $\forall$ *und* $\exists$ *bezeichnen wir als* Quantoren*, die Symbole* $\neg,\vee,\wedge,\rightarrow,\leftarrow$ *und* $\leftrightarrow$ *bezeichen wir als* Junktoren*. Enthält eine Formel keine Quantoren, so heißt sie* quantorenfrei.

*Zur Vereinfachung der Schreibweise dürfen nachfolgende Klammern weggelassen werden. Beispielsweise steht* $\varphi_1 \wedge \varphi_2 \vee \varphi_3$ *für* $((\varphi_1 \wedge \varphi_2) \vee \varphi_3)$, *wobei die üblichen Bindungsprioritäten* $\neg; \wedge; \forall$ *vorausgesetzt werden. Außerdem steht* $(\forall\ x\ y \ldots)$ *abkürzend für eine Folge von* Allquantoren $(\forall\ x\ (\forall\ y \ldots)$.

*Eine Variable x kommt* gebunden *in einer Formel* $\varphi$ *vor, falls in* $\varphi$ *ein Ausdruck* $(\forall\ x\ \phi)$ *enthalten ist. Eine Variable x kommt* frei *in einer Formel* $\varphi$ *vor, falls sie in* $\varphi$ *vorkommt und nicht gebunden ist. Eine Formel heißt* geschlossen, *falls sie keine freien Variablen enthält. Eine atomare Formel oder eine negierte atomare Formel heißt* Literal. *Eine atomare Formel heißt auch* positives Literal, *eine negierte atomare Formel heißt auch* negatives Literal. *Ein Literal L heißt* Grundliteral, *wenn die in L vorkommende atomare Formel ein Grundatom ist. Die Menge aller Literale heißt* $L^{\Sigma}$.

Wir machen bei Formeln noch folgende Einschränkungen:

1. In einer Formel dürfen Variablen nicht gleichzeitig frei und gebunden auftreten.
2. In einer Formel darf keine Variable mehrfach quantifiziert werden, d.h. jedes Vorkommen eines Quantors besitzt eine eigene Variable.
3. Wir betrachten nur solche Mengen von Formeln $\Phi$, in denen für alle Formeln $\varphi_1, \varphi_2 \in \Phi$ mit $\varphi_1 \not\equiv$ [1]$\varphi_2$ die Menge der gebundenen Variablen von $\varphi_1$ und die Menge aller Variablen von $\varphi_2$ disjunkt sind.

Formeln, die den genannten Eigenschaften genügen, werden als *rektifizierte Formeln* bezeichnet. Die gestellten Forderungen können stets erfüllt werden, indem gebundene Variablen umbenannt werden.

**Definition 2.1.4 (Allabschluß, Existenzabschluß)**
*Sei* $\varphi$ *eine Formel und* $V = \{x_1, \ldots, x_n\}$ *die Menge der freien Variablen von* $\varphi$. *Der* Allabschluß *von* $\varphi$ *ist die Formel*

$$\forall\ x_1, \ldots, x_n\ \varphi.$$

*Wir schreiben den Allabschluß auch als* $\forall\varphi$. *Der* Existenzabschluß *von* $\varphi$ *ist die Formel*

$$\exists\ x_1, \ldots, x_n\ \varphi.$$

*Wir schreiben den Existenzabschluß auch als* $\exists\varphi$.

[1] Das Symbol $\equiv$ werden wir in dieser Arbeit als Symbol für die Gleichheit von Formeln verwenden.

### 2.1.2 Semantik von $\mathcal{PL}_1$

Nachdem die Syntax der klassischen Prädikatenlogik beschrieben wurde, benötigen wir nun noch die Definition der Semantik.

**Definition 2.1.5 (Struktur)** *Eine* Struktur $M = (D, I)$ *der Sprache* $\mathcal{PL}_1$ *besteht aus einem nichtleeren* Grundbereich $D$ (Domäne) *und einer Interpretationsfunktion* $I$*, die*

1. *jeder Konstanten* $c \in \mathbf{C}$ *ein Element* $d \in D$ *aus dem Grundbereich zuordnet; (Schreibweise:* $I(c) = d$*)*
2. *jedem Funktionssymbol* $f \in F_n$ *eine* $n$*-stellige Funktion* $F : D^n \mapsto D$ *zuordnet; (Schreibweise:* $I(f) = F$*)*
3. *jedem Prädikatensymbol* $p \in P_n$ *eine* $n$*-stellige Relation* $R \subseteq D^n$ *zuordnet; (Schreibweise:* $I(p) = R$*)*

Um allgemein einem Term mit Variablen ein Objekt der Domäne zuweisen zu können, benötigen wir den Begriff der *Variablenbelegung.*

**Definition 2.1.6 (Variablenbelegung)** *Eine* Variablenbelegung $B : \mathbf{V} \mapsto D$ *zu einer Struktur* $M = (D, I)$ *ist eine Funktion, die jeder Variablen* $v \in \mathbf{V}$ *ein Element* $d \in D$ *der Domäne zuordnet.*

Nachdem die Bedeutung von Konstanten, Funktionen, Variablen und Prädikaten definiert ist, kann die Bedeutung von Termen und Formeln erklärt werden.

**Definition 2.1.7 (Termdenotation)** *Die* Denotation eines Terms $t^{M,B}$*,* $t \in T^{\Sigma}$ *in der Struktur* $M = (D, I)$ *unter der Variablenbelegung* $B$ *ist wie folgt definiert:*

1. $v^{M,B} = B(v)$*, für alle* $v \in \mathbf{V}$
2. $c^{M,B} = I(c)$*, für alle* $c \in \mathbf{C}$
3. $f(t_1, \ldots, t_n)^{M,B} = I(f)(t_1^{M,B}, \ldots, t_n^{M,B})$*, für alle* $f \in F_n, t_i \in T^{\Sigma}$

**Definition 2.1.8 (Formeldenotation)** *Eine Struktur* $M = (D, I)$ *erfüllt eine Formel* $\varphi \in For^{\Sigma}$ *unter der Variablenbelegung* $B$ *(in Zeichen:* $M, B \models \varphi$*) unter den folgenden Bedingungen:*

1. $M, B \models p(t_1, \ldots, t_n)$ *gdw.* $\langle t_1^{M,B}, \ldots, t_n^{M,B} \rangle \in I(p)$*;*
2. $M, B \models \neg\varphi$ *gdw.* $M, B \not\models \varphi$

3. *$M, B \models \varphi \vee \psi$ gdw. $M, B \models \varphi$ oder $M, B \models \psi$*

4. *$M, B \models \forall x\ \varphi$ gdw. für jedes $d \in D$ gilt $M, B_x^d \models \varphi$, wobei*

$$B_x^d(y) := \begin{cases} d & : x = y \\ B(y) & : sonst \end{cases}$$

*$M$ ist* Modell einer Formel *$\varphi \in For^\Sigma$ ($M \models \varphi$) gdw. für alle Belegungen $B$ gilt: $M, B \models \varphi$. $M$ ist* Modell einer Formelmenge *$\Phi \subseteq For^\Sigma$ ($M \models \Phi$) gdw. für alle Belegungen $B$ und alle $\varphi \in \Phi$ gilt: $M, B \models \varphi$.*

Wir benötigen noch den Begriff der *semantischen Folgerung.*

**Definition 2.1.9 (Semantische Folgerung)**

- *Eine Formel $\varphi$ folgt (semantisch) aus einer Formelmenge $\Phi$ ($\Phi \models \varphi$)*[2] *gdw. $\varphi$ in allen Modellen von $\Phi$ erfüllt ist, d.h. $M \models \varphi$ für alle $M$ mit $M \models \Phi$. Analog folgt eine Formelmenge $\Psi$ aus einer Formelmenge $\Phi$, ($\Phi \models \Psi$), wenn alle $\varphi \in \Psi$ aus $\Phi$ folgen, d.h. $\Phi \models \varphi$ für alle $\varphi \in \Psi$.*

- *Folgen zwei Formelmengen wechselseitig auseinander, d.h. $\Phi \models \Psi$ und $\Psi \models \Phi$, so heißen sie (semantisch) äquivalent.*

**Definition 2.1.10** *Eine Formelmenge $\Phi$ (eine Formel $\varphi$) heißt:*

- erfüllbar*, wenn es ein Modell $M$ von $\Phi$ (von $\varphi$) gibt;*

- allgemeingültig *(oder* Tautologie*), wenn für jede Struktur $M$ und jede Belegung $B$ gilt $M, B \models \Phi$ ($M, B \models \varphi$);*

- widersprüchlich *(oder* unerfüllbar*), wenn es kein Modell $M$ von $\Phi$ (von $\varphi$) gibt.*

## 2.2 Normalformen

Die semantische Äquivalenz von Formelmengen kann ausgenutzt werden, um eine Formelmenge in eine andere Menge von Formeln zu überführen, deren Elemente in irgendeiner Hinsicht einfacher strukturiert sind.

**Definition 2.2.1 (disjunktive, konjunktive Normalform)**

[2] Das Zeichen $\models$ wird in zwei verschiedenen Bedeutungen gebraucht. Die jeweils gemeinte wird aus dem Zusammenhang deutlich.

1. *Eine Formel $\varphi$ ist in* Pränex-Normalform *gdw. $\varphi$ folgenden Aufbau hat:*

$$\varphi \equiv Q_1\ x_1 \ldots, Q_m\ x_m \varphi',$$

   *wobei $Q_i \in \{\forall, \exists\}$,$1 \leq i \leq m$ und $\varphi'$ quantorenfrei ist. Der innere, quantorenfreie Teil $\varphi'$ von $\varphi$ heißt* Matrix, *die Folge der Quantoren bezeichnen wir als* Präfix *der Formel.*

2. *Eine Formel $\varphi$ ist in* konjunktiver Normalform (KNF) *gdw. $\varphi$ in Pränex-Normalform ist und die Matrix von $\varphi$ eine Konjunktion von Disjunktionen von Literalen $L_{i,j} \in L^{\Sigma}$ ist, d.h.*

$$\varphi \equiv Q_1\ x_1 \ldots Q_m\ x_m (L_{1,1} \vee \ldots \vee L_{1,n_1}) \wedge \ldots \wedge (L_{k,1} \vee \ldots \vee L_{k,n_k})$$

3. *Eine Formel ist in* Skolem-Normalform, *wenn sie in Pränex-Normalform ist und das Präfix keine Existenzquantoren enthält.*

**Lemma 2.2.1**

- *Jede Formel läßt sich in eine äquivalente Formel überführen, die in konjunktiver Normalform ist.*
- *Jede geschlossene Formel $\varphi$ in Pränex-Normalform kann in eine Formel $\varphi'$ in Skolem-Normalform abgebildet werden, wobei $\varphi$ genau dann erfüllbar ist, wenn auch $\varphi'$ erfüllbar ist.*

**Beweis** Siehe [Cremers et al., 1994, S. 35-36].

Den Vorgang, bei dem eine geschlossene Formel in Skolem-Normalform umgewandelt wird, nennt man *Skolemisierung*. Beide Formeln, das Original und seinen skolemisierten Gegenpart, nennt man *erfüllbarkeitsgleich*.

**Definition 2.2.2** *Sei $\Sigma$ eine Signatur. Eine* Klausel *ist der Allabschluß einer Disjunktion von Literalen, d.h.*

$$\forall(L_1 \vee \ldots \vee L_n),$$

*wobei die $L_i \in L^{\Sigma}$. Eine Formel ist in* Klauselnormalform, *wenn sie eine Konjunktion von Klauseln ist.*

Da die Variablen in Klauseln immer allquantifiziert sind, werden die Quantoren i.a. weggelassen.

Betrachtet man nur eine Klausel, so kann man diese noch strukturierter aufschreiben: alle positiven Literale können durch einfaches Umsortieren auf die linke Seite, alle negativen auf die rechte Seite der Klausel gebracht werden. Es ergibt sich folgender Ausdruck:

$$A_1 \vee \ldots \vee A_k \vee \neg A_{k+1} \vee \ldots \vee \neg A_n.$$

Durch eine einfache Umformung ergibt sich:

$$A_1 \vee \ldots \vee A_k \leftarrow A_{k+1} \wedge \ldots \wedge A_n.$$

**Definition 2.2.3** *Sei $\Sigma$ eine Signatur. Eine Klausel der Form*

$$P_1 \vee \ldots \vee P_n \leftarrow Q_1 \wedge \ldots \wedge Q_m$$

*heißt* Gentzenformel[3]*. Dabei zeichnen wir insbesondere den Spezialfall $n \leq 1$ aus. Formeln dieser Art heißen* Hornformeln*. Hier unterscheiden wir folgende Fälle:*

| | | |
|---|---|---|
| $n = 1, m \geq 1$ | $P_1 \leftarrow Q_1 \wedge \ldots \wedge Q_m$ | definite Programmklausel |
| $n = 1, m = 0$ | $P_1 \leftarrow$ | (positive) Einheitsklausel |
| $n = 0, m \geq 1$ | $\leftarrow Q_1 \wedge \ldots \wedge Q_m$ | definite Zielklausel |
| $n = 0, m = 0$ | $\leftarrow$ | Widerspruch (wird auch mit $\Box$ bezeichnet) |

**Lemma 2.2.2** *Eine geschlossene Formel ist erfüllbar gdw. ihre Klausel-Normalform erfüllbar ist.*

**Beweis** Folgt direkt aus 2.2.1.

## 2.3 Ordnungssortierte Prädikatenlogik erster Stufe

Der Grundbereich bei unsortierter Prädikatenlogik ist homogen, d.h. er weist keine Struktur auf. In Modellen der realen Welt ist dies häufig anders, so kann z.B. die Menge der Tiere aufgeteilt werden in Säugetiere, Reptilien und Insekten. Säugetiere lassen sich weiter in Pflanzenfresser, Fleischfresser o.ä. aufteilen. Ähnliche Strukturierungen finden sich auch in der Mathematik wieder, so z.B. bei der Definition der Zahlen.

Will man einen Ausschnitt der realen Welt modellieren, so liegt es nahe, den Grundbereich ähnlich zu strukturieren. Dies legt die Verwendung einer ordnungssortierten Logik nahe [Oberschelp, 1962]. Aus Effizienzgründen sind sortierte Logiken im Bereich des automatischen Beweisens eingeführt worden [Walther, 1987]. Die Signatur einer ordnungssortierten Sprache sieht neben den Symbolen und Prädikaten auch eine partiell geordnete Menge von Sortensymbolen vor.

[3] Dies entspricht nicht genau der üblichen Definition, da dort $\vee$ und $\wedge$ vertauscht und statt $\leftarrow$ ein $\rightarrow$ verwendet wird.

### 2.3.1 Syntax Ordnungssortierter Prädikatenlogik

**Definition 2.3.1 (Ordnungssortierte Signatur)**
*Die Signatur* $\Sigma_{SORT} = \langle \mathbf{S}, \mathbf{C}, \mathbf{F}, \mathbf{P}, \mathbf{V} \rangle$ *der Sprache* $\mathcal{PL}_{SORT}$ *(Ordnungssortierte Prädikatenlogik 1. Stufe) besteht aus*

1. *einer endlichen, partiell geordneten Menge* $(\mathbf{S}, \leq)$ *von* Sortensymbolen, *der sogenannten* Sortenhierarchie. $\mathbf{S}$ *enthält als ausgezeichnete Sorten die Elemente:*
   - $\top$ *oder TOP als größtes Element,*
   - $\bot$ *oder BOTTOM als kleinstes Element;*
2. *einer Familie abzählbarer* $\mathbf{S}$*-indizierter Mengen von* Konstanten $\mathbf{C} = (C_S), S \in \mathbf{S}, S \neq \bot$
3. *einer Familie abzählbarer* $\mathbf{S}^+ \times \mathbf{S}$*-indizierter Mengen von* Funktionssymbolen $\mathbf{F} = (F_{w,S}), w \in \mathbf{S}^+, S \in \mathbf{S}$
4. *einer Familie abzählbarer* $\mathbf{S}^*$*-indizierter Mengen von* Prädikatensymbolen $\mathbf{P} = (P_w), w \in \mathbf{S}^*, \bot \notin w$
5. *einer Familie abzählbarer* $\mathbf{S}$*-indizierter Mengen von Variablen* $\mathbf{V} = (V_S), S \in \mathbf{S}$, *wobei* $V_\bot = \emptyset$

Da $\mathbf{S}$ eine endliche partiell geordnete Menge ist, gibt es insbesondere nur endliche absteigende Ketten in $(\mathbf{S}, \leq)$.

**Definition 2.3.2 (Wohlsortierte Terme)**
*Die* wohlsortierten Terme $T^{\Sigma_{SORT}} = (T_S)$ *über* $\Sigma_{SORT}$ *sind definiert durch die kleinste Familie* $\mathbf{S}$*-indizierter Mengen, für die gilt*

1. $v \in T_S$ *für jede Variable* $v \in V_S$*;*
2. $c \in T_S$ *für jede Konstante* $c \in C_S$*;*
3. $f(t_1, \ldots, t_n) \in T_S$, *falls* $f \in F_{S_1 \ldots S_n, S}$ *und für alle* $1 \leq i \leq n$ *gilt:* $t_i \in T_{S'_i}$ *und* $S'_i \leq S_i$.

$t \in T_S$ *ist ein* Grundterm der Sorte $S$, *falls* $t$ *keine Variablen enthält.*

**Definition 2.3.3 (Wohlsortierte Formeln)** *Die Menge der* wohlsortierten atomaren Formeln $At^{\Sigma_{SORT}}$ *über* $\Sigma_{SORT}$ *ist wie folgt definiert:* $p(t_1,\ldots,t_n) \in At^{\Sigma_{SORT}}$, *falls* $p \in P_{S_1\ldots S_n}$ *und für alle* $1 \leq i \leq n$ *gilt:* $t_i \in T_{S'_i}$ *und* $S'_i \leq S_i$. *Die Menge der wohlsortierten Formeln* $For^{\Sigma_{SORT}}$ *wird mit Ausnahme der quantifizierten Formeln entsprechend Definition 2.1.3 gebildet. Für den Allquantor gilt:*

$$(\forall x{:}S\ \varphi) \in For^{\Sigma_{SORT}}, \quad \textit{falls} \quad \varphi \in For^{\Sigma_{SORT}}, x \in V_S \quad \textit{und} \quad S \in \mathbf{S}$$

*Die Begriffe* Grundatom, Literal *und* Grundterm *für den sortierten Fall werden analog zu den unsortierten Begriffen definiert.*

### 2.3.2 Semantik Ordnungssortierter Prädikatenlogik

Die Semantik von Sorten und ihrer partiellen Ordnung ist die Menge und die Teilmengenbeziehung bezüglich der Domäne.

**Definition 2.3.4 (Ordnungssortierte Struktur)**
*Eine* ordnungssortierte Struktur $M = (D,I)$ *über einer Signatur* $\Sigma_{SORT} = \langle \mathbf{S},\mathbf{C},\mathbf{F},\mathbf{P},\mathbf{V}\rangle$ *besteht aus*

1. *einer Familie* $\mathbf{S}$*-indizierter Mengen* $D = (D_S)$ *mit*
   (a) $D_\perp = \emptyset$
   (b) $D_\top = \bigcup_{S\in\mathbf{S}} D_S$ *und*
   (c) $D_S \subseteq D'_S$ *für alle* $S, S' \in \mathbf{S}$ *mit* $S \leq S'$;
2. *aus einer Interpretationsfunktion* $I$, *die*
   (a) *jeder Konstante* $c \in C_S$ *ein Element* $I(c) \in D_S$ *zuordnet;*
   (b) *jedem Funktionssymbol* $f \in F_{S_1\ldots S_n,S}$ *eine Funktion* $I(f) : D_{S_1} \times \ldots \times D_{S_n} \mapsto D_S$ *zuordnet;*
   (c) *jedem Prädikatensymbol* $p \in P_{S_1\ldots S_n}$ *eine Relation* $I(p) \subseteq D_{S_1} \times \ldots \times D_{S_n}$ *zuordnet.*

**Definition 2.3.5 (Ordnungssortierte Variablenbelegung)** *Eine ordnungssortierte Variablenbelegung* $B$ *zu einer Struktur* $M = (D,I)$ *ist eine Familie von* $F$*-indizierten Funktionen*

$$B = \{B_S : V_S \mapsto D_S\},$$

*wobei die $B_S$ bezüglich $V_S$ Variablenbelegungen im Sinne von Definition 2.1.6 sind.*

Die Definition der Term- und Formeldenotation kann im wesentlichen aus den Definitionen 2.1.7 und 2.1.8 entnommen werden, es muß nur auf die Einschränkung der quantifizierten Variablen auf eine bestimmte Sorte und damit auf eine Teilmenge der Domäne geachtet werden.

### Übersetzung sortierter in unsortierte Logik

**Definition 2.3.6** *Die Funktion $'$ : $STRINGS \mapsto STRINGS$ ist eine Funktion auf Zeichenketten mit der Eigenschaft, daß eine Zeichenketten bijektiv auf eine neue Zeichenkette abgebildet wird*[4]. *Statt $'(str)$ schreiben wir $str'$ (Postfixnotation).*

**Definition 2.3.7 (Signaturumwandlung)**
*Gegeben ist die sortierte Signatur $\Sigma_{SORT} = \langle \mathbf{S}, \mathbf{C}, \mathbf{F}, \mathbf{P}, \mathbf{V} \rangle$. Die* korrespondierende unsortierte Signatur $\Sigma = \langle \mathbf{C}', \mathbf{F}', \mathbf{P}', \mathbf{V}' \rangle$ *wird folgendermaßen gebildet:*

- $\mathbf{C}' = \bigcup_{S \in \mathbf{S}} \{c' \mid c \in C_S\}$
- $\mathbf{F}' = (F'_n)$ *mit*
  - $F'_n = \bigcup_{S_1 \ldots S_n, S_{n+1}} \{f'_{S_1 \ldots S_n, S_{n+1}} \mid f_{S_1 \ldots S_n, S_{n+1}} \in F_{S_1 \ldots S_n, S_{n+1}}\}$ *mit* $S_i \in \mathbf{S}$ *für alle $i$.*
- $\mathbf{P}' = (P'_n)$ *mit*
  - $P'_1 = \bigcup_{S \in \mathbf{S}}(\{q'_S \mid q_S \in P_S\}) \cup \{p_S\})$. *Dabei ist jedes $p_S$ ein neues, bisher nicht verwendetes Prädikatensymbol.*
  - $P'_n = \bigcup_{S_1 \ldots S_n} \{p'_{S_1 \ldots S_n} \mid p_{S_1 \ldots S_n} \in P_{S_1 \ldots S_n}\}$ *für $n \geq 2$ und $S_i \in \mathbf{S}$ für alle $i$.*
- $\mathbf{V}' = \bigcup_{S \in \mathbf{S}} \{v' \mid v \in V_S\}$

**Definition 2.3.8 (Relativierung)** *Sei $\Sigma_{SORT} = \langle \mathbf{S}, \mathbf{C}, \mathbf{F}, \mathbf{P}, \mathbf{V} \rangle$ eine sortierte Signatur und $\Sigma$ eine unsortierte Signatur nach Definition 2.3.7. Sei $\varphi_{SORT} \in For^{\Sigma_{SORT}}$. Die zu $\varphi_{SORT}$ bezüglich $\Sigma$* korrespondierende Formel $\varphi \in For^{\Sigma}$ *erhält man durch folgende Operationen:*

1. *Jedes Symbol in $\varphi_{SORT}$ aus $\Sigma^{SORT}$ wird gegen das korrespondierende Symbol aus $\Sigma$ ausgetauscht.*

[4] Die Funktion ist einfach realisierbar: Es reicht aus, wenn sie an eine Zeichenkette ein beliebiges Zeichen anhängt.

2. *Jeder Allquantor* $(\forall\ x{:}S\ \phi)$ *wird gegen* $(\forall\ x\ p_S(x) \rightarrow \phi)$ *ausgetauscht.*

*Die* korrespondierende Menge $\Phi$ *zu einer Menge von sortierten Formeln* $\Phi_{SORT}$ *ist die kleinste Menge, die für jede Formel aus* $\Phi_{SORT}$ *die korrespondierende Formel enthält.*

Die Übersetzung wie oben angegeben berücksichtigt noch nicht die Sortenstruktur. Diese ist in einen Hornklauselteil zu übersetzen.

## 2.4 Domänenunabhängigkeit

Im folgenden schlagen wir eine Definition von *Domänenunabhängigkeit* aufbauend auf der bereits eingeführten Terminologie vor.

**Definition 2.4.1 (Domänenunabhängige Formel)** *Sei* $\Sigma$ *eine Signatur. Eine Formel* $\varphi$ *bezüglich* $\Sigma$ *heißt* domänenunabhängig *(eng.: domain independent), wenn für alle Strukturen* $M_1 = (D_1, I_1)$ *und* $M_2 = (D_2, I_2)$ *und Variablenbelegungen* $B_1$ *zu* $M_1$*, die die folgenden Eigenschaften erfüllen:*

- $I_1$ *und* $I_2$ *bilden alle Konstantensymbole in* $\varphi$ *auf die gleichen Domänenelemente ab und*
- $I_1$ *und* $I_2$ *bilden alle Prädikatensymbole in* $\varphi$ *auf die gleichen Relationen ab,*

*folgendes gilt:*

- *falls* $M_1, B_1 \models \varphi$*, dann existiert eine Variablenbelegung* $B_2$ *zu* $M_2$*, die die freien Variablen von* $\varphi$ *auf die gleichen Domänenelemente abbildet wie* $B_1$ *und für die gilt:* $M_2, B_2 \models \varphi$.
- *falls* $M_1, B_1 \not\models \varphi$*, dann existiert eine Variablenbelegung* $B_2$ *zu* $M_2$*, die die freien Variablen von* $\varphi$ *auf die gleichen Domänenelemente abbildet wie* $B_1$ *und für die gilt:* $M_2, B_2 \not\models \varphi$.

Domänenunabhängigkeit bedeutet also, daß sich die Domäne einer Interpretation ändern kann, ohne die Gültigkeit einer Formel zu beeinflussen (soweit die Interpretation der Konstantensymbole und Prädikatensymbole nicht geändert wird).

**Beispiel 2.4.1** *Wir betrachten die folgenden Strukturen* $M_1$ *und* $M_2$ *sowie die Formel* $\exists\ x\ \neg p(x)$*:*

- $M_1 = (D_1, I_1)$ *mit* $D_1 = \{a\}$*,* $I_1(p) = \{\langle a\rangle\}$ *und*

- $M_2 = (D_2, I_2)$ *mit* $D_2 = \{a, b\}$, $I_2(p) = \{\langle a \rangle\}$

*$M_1$ und $M_2$ unterscheiden sich nur in der Domäne. Da keine freien Variablen vorhanden sind gilt mit beliebigen Variablenbelegungen $B_1$ zu $M_1$ und $B_2$ zu $M_2$:*

- $M_1, B_1 \not\models \exists\ x\ \neg p(x)$
- $M_2, B_2 \models \exists\ x\ \neg p(x)$.

*Obwohl die Interpretationsfunktionen $I_1$ und $I_2$ dem Prädikatensymbol $p$ die gleiche Relation zuweisen, erfüllt $M_1$ die Formel und $M_2$ nicht. Damit ist die Formel nicht domänenunabhängig.*

Domänenunabhängigkeit ist für Datenbanken sehr wichtig, da sich die Domäne ändern kann, ohne daß sich die Interpretationen ändern. Leider gilt aber folgendes:

**Lemma 2.4.1** *Die Klasse der domänenunabhängigen Formeln ist nicht rekursiv aufzählbar.*

**Beweis** [DiPaola, 1969, Vardi, 1981]

Jedoch lassen sich leicht einige syntaktische Einschränkungen identifizieren, die eine Formel domänenunabhängig machen. Die Relativierung reicht dazu aus.

**Lemma 2.4.2** *Sei $\Sigma_{SORT}$ eine sortierte Signatur und $\Sigma$ die korrespondierende unsortierte Signatur. Weiterhin sei $\varphi$ eine geschlossene Formel bezüglich $\Sigma_{SORT}$ und $\varphi'$ die korrespondierende relativierte Formel zu $\varphi$ bezüglich $\Sigma$. Dann ist $\varphi'$ domänenunabhängig.*

**Beweis** Folgt aus der Definition der Relativierung und [Topor und Sonnenberg, 1988, S. 227].

Weitere allgemeinere syntaktische Einschränkungen sind die der *sicheren Programme* bzw. *range-restricted formulas* [Decker, 1986].

Domänenunabhängigkeit ist eine modelltheoretische Eigenschaft. Deshalb bleibt bei Umformungen der Formel, die die semantische Äquivalenz erhalten, die Eigenschaft bestehen.

## 2.5 Herbrandstrukturen

**Definition 2.5.1 (Herbrand-Universum, Herbrand-Basis)**
*Sei $\Sigma$ eine Signatur. Dann definieren wir:*

1. *das* Herbrand-Universum *als die Menge aller Grundterme* $T_G^\Sigma \subseteq T^\Sigma$*, und*

2. *die* Herbrand-Basis *als die Menge aller Grundatome* $At_G^\Sigma \subseteq At^\Sigma$*.*

Ist $T_G^\Sigma = \emptyset$, so erweitern wir die Signatur $\Sigma$ um ein Konstantensymbol.

**Definition 2.5.2 ($\Sigma$-Struktur)**
*Sei* $\Sigma = \langle \mathbf{C}, \mathbf{F}, \mathbf{P}, \mathbf{V} \rangle$ *eine Signatur. Eine Struktur* $M = (D, I)$ *heißt* $\Sigma$*-Struktur, wenn* $D = T_G^\Sigma$ *und jedes Funktions- und Konstantensymbol durch sich selbst interpretiert wird:*

- $I(c) = c$ *für alle* $c \in \mathbf{C}$*.*
- $I(f(t_1, \ldots, t_n)) = f(I(t_1, \ldots, t_n)$ *für alle* $f \in F_n, F_n \in \mathbf{F}$*.*

**Definition 2.5.3 (Herbrand-Modell)** *Ein* Herbrand-Modell *für eine Menge von geschlossenen Formeln* $\Phi$ *ist ein* $\Sigma$*-Struktur* $M$*, das Modell für* $\Phi$ *ist.*

**Lemma 2.5.1** *Sei* $\Phi$ *eine Menge von Klauseln.*

1. *Wenn* $\Phi$ *ein Modell hat, dann hat* $\Phi$ *auch ein Herbrand-Modell.*

2. $\Phi$ *ist unerfüllbar gdw.* $\Phi$ *kein Herbrand-Modell hat.*

**Beweis** Siehe [Cremers et al., 1994, S. 40].

## 2.6 Kalkül und Resolution

### 2.6.1 Kalkül

**Definition 2.6.1 (Kalkül)**
*Ein* Kalkül $\mathcal{D} = (\mathcal{L}, \mathcal{R})$ *besteht aus:*

- *einer rekursiven Menge von Formeln* $\mathcal{L}$ *und*
- *einer rekursiven Menge von* Deduktionsregeln $\mathcal{R} \subseteq \mathcal{L}^* \times \mathcal{L}$ *der Form*

$$\frac{\varphi_1 \cdots \varphi_m}{\varphi}$$

  *mit* $\varphi_i, \varphi \in \mathcal{L}$*,*$1 \leq i \leq m$*.*

**Definition 2.6.2 (Ableitung)** *Sei* $D = (\mathcal{L}, \mathcal{R})$ *ein Kalkül,* $\Phi \subseteq \mathcal{L}$ *eine rekursive Menge von Formeln und* $\varphi \in \mathcal{L}$ *eine Formel.*

1. *Die* Ableitung *der Formel $\varphi$ aus $\Phi$ in $\mathcal{D}$ ist eine endliche Folge $\varphi_1, \ldots, \varphi_n$ mit $\varphi_n \equiv \varphi$, wobei die $\varphi_i \in \mathcal{L}$, $1 \leq i \leq n$, Formeln sind und für alle $\varphi_i$ gilt:*

   *(a) $\varphi_i \in \Phi$, oder*

   *(b) es gibt natürliche Zahlen $i_1, \ldots, i_k \leq i$ und es existiert eine Regel*

   $$\frac{\varphi_{i_1} \ldots \varphi_{i_k}}{\varphi_i} \in \mathcal{R}.$$

   *(c) Eine Formel $\varphi$ heißt* ableitbar *aus $\Phi$ (in Zeichen: $\Phi \vdash \varphi$), wenn es eine Ableitung von $\varphi$ aus $\Phi$ in $\mathcal{D}$ gibt.*

**Definition 2.6.3 (Adäquatheit)**
*Sei $\mathcal{D} = (\mathcal{L}, \mathcal{R})$ ein Kalkül, $\Phi \subseteq \mathcal{L}$ eine Formelmenge und $\varphi \in \mathcal{L}$ eine Formel.*

1. *$\mathcal{D}$ heißt* korrekt *gdw. für alle $\Phi$ und $\varphi$ gilt: Wenn $\Phi \vdash \varphi$ dann $\Phi \models \varphi$.*

2. *$\mathcal{D}$ heißt* vollständig *gdw. für alle $\Phi$ und $\varphi$ gilt: Wenn $\Phi \models \varphi$ dann $\Phi \vdash \varphi$.*

3. *$\mathcal{D}$ heißt* widerlegungsvollständig *gdw. für alle $\Phi$ gilt: Wenn $\Phi \models \Box$ dann $\Phi \vdash \Box$.*

4. *$\mathcal{D}$ heißt* adäquat *(*widerlegungsadäquat*) gdw. $\mathcal{D}$ korrekt und vollständig (widerlegungsvollständig) ist.*

### 2.6.2 Substitutionen

**Definition 2.6.4** *Sei $\Sigma = \langle \mathbf{C}, \mathbf{F}, \mathbf{P}, \mathbf{V} \rangle$ eine Signatur. Eine* Substitution *$\sigma$ ist eine Abbildung von $\mathbf{V}$ nach $T^\Sigma$, so daß die Menge*

$$dom(\sigma) = \{x \in \mathbf{V} \mid \sigma(x) \neq x\}$$

*endlich ist. Die Menge $dom(\sigma)$ heißt auch* Domain *von $\sigma$. Wir schreiben i.a. $x\sigma$ statt $\sigma(x)$. Eine Substitution wird üblicherweise als eine Menge $\{x_i \backslash t_i \mid x_i \in dom(\sigma), t_i = x_i\sigma\}$ dargestellt.*

*$\sigma$ heißt* Grundsubstitution, *falls für alle $x \in dom(\sigma)$ der Term $t = x\sigma$ ein Grundterm ist.*

*Die Substitution, deren Domain die leere Menge ist, heißt* identische Substitution *und wird mit $\epsilon$ bezeichnet.*

**Definition 2.6.5** *Sei $\Sigma = \langle \mathbf{C}, \mathbf{F}, \mathbf{P}, \mathbf{V} \rangle$ eine Signatur und $\sigma$ eine Substitution. Wir definieren induktiv eine Abbildung $\sigma' : T^\Sigma \mapsto T^\Sigma$ durch*

$$\begin{array}{lcl} x\sigma' & = & x\sigma \quad \textit{für alle } x \in \mathbf{V} \\ c\sigma' & = & c \quad \textit{für alle } c \in \mathbf{C} \\ f(t_1, \ldots, t_n)\sigma' & = & f(t_1\sigma', \ldots, t_n\sigma') \quad \textit{für alle } f \in F_n, F_n \in \mathbf{F} \end{array}$$

*und eine Abbildung $\sigma^* : For^\Sigma \mapsto For^\Sigma$ durch*

$$\begin{array}{lcl} P(t_1, \ldots, t_n)\sigma^* & = & P(t_1\sigma', \ldots, t_n\sigma') \quad \textit{für alle } P(t_1, \ldots, t_n) \in At^\Sigma \\ (\neg\varphi)\sigma^* & = & (\neg\varphi\sigma^*) \quad \textit{für alle } (\neg\varphi) \in For^\Sigma \\ (\varphi \vee \phi)\sigma^* & = & (\varphi\sigma^* \vee \phi\sigma^*) \quad \textit{für alle } (\varphi \vee \phi) \in For^\Sigma \\ (\forall\, x\; \varphi)\sigma^* & = & \varphi\sigma^* \quad \textit{für alle } (\forall\, x\; \varphi) \in For^\Sigma \textit{ falls } x \in dom(\sigma) \\ (\forall\, x\; \varphi)\sigma^* & = & (\forall\, x\; \varphi\sigma^*) \quad \textit{für alle } (\forall\, x\; \varphi) \in For^\Sigma \textit{ falls } x \notin dom(\sigma) \end{array}$$

*Wir werden im Folgenden $\sigma'$ und $\sigma^*$ mit $\sigma$ identifizieren.*

*Ist $t$ ein Term und $\sigma$ eine Substitution, so nennen wir den Term $t\sigma$ eine Instanz von $t$ (Wir sagen auch „$t$ wird zu $t\sigma$ instanziiert“). Ist $t\sigma$ ein Grundterm, so heißt $t\sigma$* Grundinstanz *von $t$.*

**Definition 2.6.6 (Komposition von Substitutionen)** *Seien $\sigma$ und $\theta$ Substitutionen. Wir definieren eine Substitution $\sigma\theta$ durch $x(\sigma\theta) = (x\sigma)\theta$ für alle $x \in \mathbf{V}$. Die Substitution $\sigma\theta$ heißt* Komposition *von $\sigma$ und $\theta$.*

**Definition 2.6.7**

- *Eine Substitution $\sigma$ heißt* idempotent *gdw. $\sigma = \sigma\sigma$.*
- *Eine Substitution $\sigma$ heißt* Umbenennung*, falls es eine Substitution $\sigma'$ gibt, so daß $\sigma\sigma' = \sigma'\sigma = \epsilon$ gilt.*
- *Seien $s$ und $t$ Terme. Dann heißt $t$ eine* Variante *von $s$, falls es eine Umbenennung $\sigma$ gibt, so daß $s = t\sigma$.*

**Definition 2.6.8** *Seien $s$ und $t$ Terme (oder Atome [5]).*

1. *Ein* Unifikator *von $(s, t)$ ist eine Substitution $\sigma$, so daß $s\sigma = t\sigma$ gilt. Wir sagen $\sigma$ unifiziert das Paar $(s, t)$. $\sigma$ heißt* allgemeinster Unifikator *von $(s, t)$, wenn es für jeden Unifikator $\theta$ von $(s, t)$ eine Substitution $\tau$ gibt, so daß $\theta = \sigma\tau$. Wir schreiben*

[5] Die Unifikation wird hier zwar nur für Terme formuliert, sie läßt sich aber direkt auf Atome übertragen.

*für einen allgemeinsten Unifikator $\sigma$ von $(s,t)$ auch $\sigma = mgu(s,t)$ (mgu = most general unifier).*

**Definition 2.6.9 (Einschränkung einer Substitution)** *Die Einschränkung einer Substitution $\sigma$ auf eine Variablenmenge $V$, geschrieben $\sigma|_V$, ist eine Substitution $\sigma'$ mit $\sigma'(x) = t$ mit $t \neq x$ gdw. $x \in V$, $\sigma(x) = t$ und $x \neq t$.*

### 2.6.3 Resolution

**Satz 2.6.1 (Herbrand)** *Eine Menge $\Phi$ von Klauseln ist unerfüllbar gdw. es eine endliche, unerfüllbare Menge von Grundinstanzen der Klauseln in $\Phi$ gibt.*

**Definition 2.6.10 (Resolution)**
*Der Resolutionskalkül für die Prädikatenlogik besteht aus der Resolutionsregel (1) für zwei Gentzenformeln und der Faktorisierungsregel, die in einer positiven (2) und einer negativen (3) Ausprägung vorhanden ist:*

1. $$\frac{P_1 \vee ... \vee R \vee ... \vee P_n \leftarrow Q_1 \wedge ... \wedge Q_m \qquad P'_1 \vee ... \vee P'_{n'} \leftarrow Q'_1 \wedge ... \wedge R' \wedge ... \wedge Q'_{m'}}{(P_1 \vee ... \vee P_n \vee P'_1 \vee ... \vee P'_{n'} \leftarrow Q_1 \wedge ... \wedge Q_m \wedge Q'_1 \wedge ... \wedge Q'_{m'})mgu(R,R')}$$
2. $$\frac{P_1 \vee ... \vee F_1 \vee ... \vee F_2 \vee ... \vee P_n \leftarrow Q_1 \wedge ... \wedge Q_m}{(P_1 \vee ... \vee F_1 \vee ... \vee P_n \leftarrow Q_1 \wedge ... \wedge Q_m)mgu(F_1,F_2)}$$
3. $$\frac{P_1 \vee ... \vee P_n \leftarrow Q_1 \wedge ... \wedge F_1 \wedge ... \wedge F_2 \wedge ... \wedge Q_m}{(P_1 \vee ... \vee P_n \leftarrow Q_1 \wedge ... \wedge F_1 \wedge ... \wedge Q_m)mgu(F_1,F_2)}$$

Im Folgenden meint das Symbol $\vdash$ das Resolutionskalkül.

**Satz 2.6.2 (Korrektheit der Resolution)**
*Sei $\Phi$ eine beliebige Klauselmenge und $\varphi$ eine beliebige Klausel. Dann gilt:*

$$\textit{Wenn} \quad \Phi \vdash \varphi \quad \textit{dann} \quad \Phi \models \varphi$$

**Beweis** Siehe [Chang und Lee, 1973, S. 72], dabei sind die Definitionen anzupassen.

**Satz 2.6.3 (Widerlegungsvollständigkeit der Resolution)**
*Sei $\Phi$ eine beliebige Klauselmenge und $\varphi$ eine beliebige Klausel. Dann gilt:*

$$\textit{Wenn} \quad \Phi \models \Box \quad \textit{dann} \quad \Phi \vdash \Box$$

**Beweis** Siehe [Chang und Lee, 1973, S. 85]

Daraus folgt direkt:

$$\Phi \models \neg\varphi \quad \text{gdw.} \quad \Phi \cup \{\varphi\} \vdash \Box$$

## 2.7 Logikprogramme und Deduktive Datenbanken

Bei der Definition von *Gentzenformeln* haben wir bereits einige Spezialfälle dieser Formeln identifiziert, die im logischen Programmieren eine wichtige Rolle spielen.

**Definition 2.7.1 (definites Programm)** *Sei $\Sigma$ eine Signatur. Ein* definites Programm *$\Phi$ ist eine endliche Menge von definiten Programmklauseln bezüglich $\Sigma$. Eine* definite Anfrage *an ein definites Programm ist eine definite Zielklausel bezüglich $\Sigma$.*

Definite Programme haben die angenehme Eigenschaft, daß ihnen in Bezug auf eine definite Anfrage eine prozedurale Semantik verliehen werden kann. Der Kopf einer definiten Programmklausel kann als Prozedureinsprung aufgefaßt werden, der Rumpf der Klausel realisiert dann die Prozedur. Allerdings haben sie eine Einschränkung — im Rumpf einer definiten Programmklausel können definitionsgemäß keine negativen Literale enthalten sein. Dies stellt aber eine starke Restriktion der Ausdruckskraft dar. Um diese Restriktion zumindest teilweise zu umgehen, führen wir einen speziellen Negationsjunktor $\neg^*$ ein, mit dem wir zunächst die Syntax von *normalen Programmen* definieren.

### 2.7.1 Syntax normaler Programme

**Definition 2.7.2** *Sei $\Sigma$ eine Signatur.*

1. *Eine* normale Programmklausel *bezüglich $\Sigma$ ist eine Formel der Form $A \leftarrow L_1 \wedge \ldots \wedge L_n$, wobei $A \in At^\Sigma$ ein Atom ist und die $L_i$ entweder atomare Formeln $A_i \in At^\Sigma$ oder negierte atomare Formeln $\neg^* A_i$, $A_i \in At^\Sigma$ für $1 \leq i \leq n$ sind. Für die $L_i$ definieren wir auf analoge Weise die Begriffe* Literal, positives Literal, negatives Literal *und* Grundliteral*. $A$ heißt* Kopf*, $L_1 \wedge \ldots \wedge L_n$ heißt* Rumpf *der Programmklausel.*

2. *Eine* normale Zielklausel *bezüglich $\Sigma$ ist eine Formel der Form $\leftarrow L_1 \wedge \ldots \wedge L_n$, wobei die $L_i$, $1 \leq i \leq n$, Literale im Sinne von Punkt 1 sind.*

3. *Ein* normales Programm *bezüglich $\Sigma$ ist eine endliche Menge von normalen Programmklauseln bezüglich $\Sigma$. Die Menge der Programmklauseln, deren Kopf das Prädikatensymbol p enthält, bezeichnen wir als* Definition *von p.*

4. *Die Begriffe* Einheitsklausel *und* Programmklausel *sind analog zum definiten Fall definiert.*

### 2.7.2 Semantik normaler Programme

Während die Semantik bei definiten Programmen durch die Semantik von $\mathcal{PL}_1$ gegeben ist, ist dies bei normalen Programmen, bedingt durch die Einführung des Junktors $\neg^*$, nicht mehr der Fall. Aus diesem Grunde ist die Semantik (insbesondere von $\neg^*$) separat zu klären. Hierzu benötigen wir den Begriff der *Vervollständigung* (engl. completion) eines Logikprogrammes, der sich aus der *Vervollständigung* von Prädikatendefinitionen und einer Gleichheitstheorie zusammensetzt. Dabei nehmen wir die Existenz eines bisher unbenutzten zweistelligen Prädikatensymbols $=$ an. $\neq$ ist dabei die übliche Abkürzung für $\neg =$.

**Definition 2.7.3 (Gleichheitstheorie)** *Sei* $\Sigma = \langle \mathbf{C}, \mathbf{F}, \mathbf{P}, \mathbf{V} \rangle$ *eine Signatur und D ein normales Programm. Eine* Gleichheitstheorie $EQT^\Sigma$ *besteht aus folgenden Formeln:*

1. *Gleichheitsaxiome:*

   (a) $\forall\, x\; x = x$

   (b) $\forall\, x_1 \ldots x_n\; y_1 \ldots y_n$
   $$((x_1 = y_1 \wedge \ldots \wedge x_n = y_n) \rightarrow f(x_1, \ldots, x_n) = f(y_1, \ldots, y_n))$$
   *für alle* $f \in F_n, F_n \in \mathbf{F}$.

   (c) $\forall\, x_1 \ldots x_n\; y_1 \ldots y_n$
   $$((x_1 = y_1 \wedge \ldots \wedge x_n = y_n) \rightarrow (p(x_1, \ldots, x_n) \rightarrow p(y_1, \ldots, y_n)))$$
   *für alle* $p \in P_n, P_n \in \mathbf{P}$.

2. *Freie Gleichheitsaxiome:*

   (a) $\forall\, x_1 \ldots x_n\; y_1 \ldots y_n\; f(x_1, \ldots, x_n) \neq g(y_1, \ldots, y_n)$ *für alle* $f, g \in F_n$, $F_n \in \mathbf{F}$ *und* $f \not\equiv g$.

   (b) $\forall\, x_1 \ldots x_n\; y_1 \ldots y_n$
   $$f(x_1, \ldots, x_n) = f(y_1, \ldots, y_n) \leftarrow (x_1 = y_1 \wedge \ldots \wedge x_n = y_n)$$
   *für alle* $f \in F_n, F_n \in \mathbf{F}$.

   (c) *Vorkommensaxiom:*
   $\forall\, x\; (t \neq x)$ *für alle Terme* $t \in T^\Sigma$ *mit* $t \neq x$ *und* $x$ *Element der Menge der Variablen von* $t$.

**Definition 2.7.4 (Vervollständigung)** *Sei* $\Sigma = \langle \mathbf{C}, \mathbf{F}, \mathbf{P}, \mathbf{V} \rangle$ *eine Signatur und D ein normales Programm.*

1. *Sei $p \in P_n$, $P_n \in \mathbf{P}$ ein n-stelliges Prädikatensymbol, das durch die normalen Programmklauseln*

$$\begin{array}{rcl} p(t_{1,1},\ldots,t_{1,n}) & \leftarrow & W_1 \\ & \vdots & \\ p(t_{m,1},\ldots,t_{m,n}) & \leftarrow & W_m \end{array}$$

*in D definiert wird. Die* Vervollständigung *comp(p) des Prädikatensymbols p ist die Formel*

$$\begin{array}{l} \forall\, x_1 \ldots x_n (p(x_1,\ldots,x_n) \leftrightarrow \\ \qquad \exists_{Y_1}(x_1 = t_{1,1} \wedge \ldots \wedge x_n = t_{1,n} \wedge W_1) \vee \ldots \vee \\ \qquad \exists_{Y_m}(x_1 = t_{m,1} \wedge \ldots \wedge x_n = t_{m,n} \wedge W_m)), \end{array}$$

*wobei $Y_i$ die Tupel der Variablen aus $W_i$ für $1 \leq i \leq m$ sind und $x_1,\ldots,x_n$ neue Variablen, die in keinem $W_i$ vorhanden sind*[6]*. Jedes $\neg^*$ in einem $W_i$ wird dabei zu einem $\neg$.*

2. *Die* Vervollständigung *comp(D) des Programms D ist die Formelmenge*

$$comp(D) = \{comp(p) \mid p \in P_n, P_n \in \mathbf{P}\} \cup EQT^{\Sigma}$$

Für die vervollständigte Formelmenge kann die Standardsemantik von $\mathcal{PL}_1$ angewandt werden. Falls eine Verwechslung ausgeschlossen ist, werden wir auch in normalen Programmen statt $\neg^*$ das Symbol $\neg$ verwenden.

**Definition 2.7.5 (Korrekte Antwortsubstitution)** *Sei D ein normales Programm und $\leftarrow W$ eine Anfrage. Eine* Antwortsubstitution *für D und $\leftarrow W$ ist eine Substitution für die freien Variablen von W, für die $comp(D) \models \forall(W\sigma)$.*

### 2.7.3 SLDNF-Resolution

Die Resolution für die volle Prädikatenlogik haben wir bereits kennengelernt. Die SLDNF-Resolution bildet das operationale Gegenstück zur Vervollständigung.

**Definition 2.7.6 (Berechnungsregel)** *Eine* Berechnungsregel *R ist eine Abbildung aus der Menge der Zielklauseln in die Menge der Literale, d.h. R wählt aus einer Zielklausel $\leftarrow L_1 \wedge \ldots \wedge L_n$ ein Literal $L_i$ aus.*

[6] Formal müßte man hier die Signatur erweitern. Wir behelfen uns mit der Annahme, daß die Variablen schon in der Signatur waren, nur nicht verwendet wurden.

Die Berechnungsregel $R$ mit $R(\leftarrow L_1 \wedge \ldots \wedge L_n) = L_1$, die das am weitesten links stehende Literal auswählt, heißt *Standardberechnungsregel.* Die Berechnungsregeln, die in der SLDNF-Resolution verwendet werden, dürfen keine variablenbehafteten negativen Literale selektieren. Diese Einschränkung liegt in der Behandlung negativer Literale begründet und ist notwendig, um die Korrektheit der SLDNF-Resolution zu garantieren.

**Definition 2.7.7 (Sichere Berechnungsregel)** *Ein Literal heißt* ausführbar *gdw. es ein positives Literal oder ein negatives Grundliteral ist. Eine Zielklausel heißt* ausführbar *gdw. sie mindestens ein ausführbares Literal enthält. Eine* sichere Berechnungsregel *ist eine Abbildung aus der Menge der ausführbaren Zielklauseln in die Menge der ausführbaren Literale, d.h. $R$ wählt aus einer ausführbaren Zielklausel $\leftarrow L_1 \wedge \ldots L_n$ ein ausführbares Literal $L_i$ aus.*

**Definition 2.7.8 (SLDNF-Baum)** *Sei $D$ ein normales Programm, $\leftarrow G$ eine Zielklausel und $R$ eine Berechnungsregel. Ein* SLDNF-Baum *für $D \cup \{\leftarrow G\}$ via $R$ ist ein Baum mit folgenden Eigenschaften:*

1. *Jeder Knoten ist mit einer Zielklausel beschriftet.*[7]

2. *Die Wurzel des Baumes ist $\leftarrow G$, die* angewendete Substitution *ist $\epsilon$.*

3. *Sei $G' \equiv \leftarrow L_1 \wedge \ldots \wedge L_i \wedge \ldots \wedge L_n$ $(n \geq 1)$ ein Knoten mit Nachfolgern und $L_i$ das von $R$ selektierte Literal. Wir unterscheiden zwei Fälle:*

   (a) *$L_i$ ist ein positives Literal:*
   *Für jede Programmklausel $(A \leftarrow W) \in D$, so daß $\theta = mgu(A\sigma, L_i)$ existiert, hat $G'$ einen Nachfolgerknoten*

   $$\leftarrow (L_1 \wedge \ldots \wedge L_{i-1} \wedge W\sigma \wedge L_{i+1} \wedge \ldots \wedge L_n)\theta,$$

   *wobei $\sigma$ eine Umbenennung ist, die alle Variablen in $A \leftarrow W$ durch neue Variablen ersetzt. Der Unifikator $\theta$ ist die in diesem Schritt* verwendete Substitution.
   *$G'$ ist ein Blatt und zählt zu den* Mißerfolgsknoten*, wenn keine derartige Programmklausel existiert.*

   (b) *$L_i \equiv \neg A$ ist ein negatives Grundliteral:*
   *Der Knoten $G'$ hat einen Nachfolgerknoten*

   $$\leftarrow L_1 \wedge \ldots \wedge L_{i-1} \wedge L_{i+1} \wedge \ldots \wedge L_n,$$

[7] Der Kürze halber identifizieren wir im folgenden die Knoten mit ihren Beschriftungen.

*wenn es einen endlichen SLDNF-Mißerfolgsbaum für $D$ und $\leftarrow A$ via einer beliebigen Berechnungsregel gibt. Die in diesem Schritt* verwendete Substitution *ist die identische Substitution $\epsilon$.*

4. *Sei $G' \equiv \leftarrow L_1 \wedge \ldots \wedge L_n$ mit $n > 0$ nicht ausführbar.*
   *In diesem Fall ist $G'$ ein Blatt.*

5. *Die leere Klausel ist stets ein Blatt und heißt* Erfolgsknoten.

*Ein SLDNF-Baum, der endlich ist und dessen Blätter Mißerfolgsknoten sind, heißt* endlicher Mißerfolgsbaum. *Alle SLDNF-Bäume, die mindestens einen Erfolgsknoten haben, werden als* SLDNF-Erfolgsbäume *bezeichnet.*

Wie in [Cremers et al., 1994, S. 78] erwähnt, ist die vorstehende Definition streng genommen keine mathematisch korrekte Definition, da in Teil 3b zirkulär auf das zu definierende Konzept zurückgegriffen wird. Eine mathematisch korrekte, aber wesentlich kompliziertere Definition von SLDNF-Bäumen findet sich in [Lloyd, 1987, S. 84-88].

**Definition 2.7.9 (SLDNF-Ableitung, SLDNF-Widerlegung)**
*Sei $D$ ein normales Programm und $G$ eine normale Zielklausel. Jeder in der Wurzel beginnende Pfad des SLDNF-Baumes für $D$ und $G$ heißt* SLDNF-Ableitung von $D$ und $G$. *Eine SLDNF-Ableitung heißt* erfolgreich, *wenn sie in einem Erfolgsknoten endet. Eine erfolgreiche SLDNF-Ableitung heißt auch* SLDNF-Widerlegung. *Eine SLDNF-Ableitung heißt* gescheitert, *falls sie mit einem Mißerfolgsknoten endet. Eine SLDNF-Ableitung heißt* blockiert, *falls sie in einer nicht ausführbaren Zielklausel endet.*

*Sei $R$ eine Berechnungsregel. Man spricht von einer SLDNF-Ableitung von $D$ und $G$ via $R$, wenn die Berechnungsregel $R$ für den Aufbau des SLDNF-Baumes verwendet wird.*

Wir betrachten eine SLDNF-Ableitung als Folge von Zielklauseln. Eine SLDNF-Widerlegung ist in diesem Sinne immer eine endliche Folge von Zielklauseln.

**Definition 2.7.10 (Faire Berechnungsregel)** *Eine SLDNF-Ableitung heißt* fair, *wenn sie entweder gescheitert ist oder jedes vorkommende Literal bzw. eine Instanz des Literals nach endlich vielen Schritten selektiert wird. Eine sichere Berechnungsregel $R$ heißt* fair, *wenn jede SLDNF-Ableitung via $R$ fair ist.*

**Definition 2.7.11 (Berechnete Antwortsubstitution)** *Sei $D$ ein normales Programm, $G$ eine normale Zielklausel und $\theta_1, \ldots, \theta_n$ die Folge der in einer SLDNF-Widerlegung von $D$ und $G$ verwendeten Substitutionen. Eine* berechnete Antwortsub-stitution $\theta$ *für D und G ist die Komposition der Substitutionen* $\theta_1 \ldots \theta_n$, *eingeschränkt auf die Variablen von G.*

### Korrektheit der SLDNF-Resolution

**Satz 2.7.1 (Korrektheit der „Negation as Failure"-Regel)** *Sei D ein normales Programm und* $G \equiv\leftarrow L_1 \wedge \ldots \wedge L_n$ *eine normale Zielklausel. Ist der SLDNF-Baum für D und G ein endlicher Mißerfolgsbaum, dann gilt:*

$$comp(D) \models \forall(\neg(L_1 \wedge \ldots \wedge L_n)).$$

**Beweis** Siehe [Cremers et al., 1994, S. 82-83].

**Satz 2.7.2 (Korrektheit der SLDNF-Resolution)** *Sei D ein normales Programm und G eine normale Zielklausel. Jede berechnete Antwortsubstitution* $\theta$ *für D und G ist eine korrekte Antwortsubstitution:*

$$comp(D) \models \forall(G\theta).$$

**Beweis** Siehe [Cremers et al., 1994, S. 82-83].

**Vollständigkeitsergebnisse zur SLDNF-Resolution** SLDNF-Resolution ist im allgemeinen unvollständig. Man kann allerdings Programmklassen identifizieren, für die die SLDNF-Resolution vollständig ist. Diese Programmklassen lassen sich oft mit einfachen syntaktischen Merkmalen definieren.

**Definition 2.7.12 (Prädikat-Abhängigkeiten)** *Sei D ein normales Programm und p und q seien Prädikatensymbole, die in D auftreten.*

1. $p$ hängt direkt positiv von $q$ ab *(in Zeichen:* $p \sqsubset_{+1} q$*) gdw. eine Programmklausel* $A \leftarrow W \in D$ *existiert, so daß p das Prädikatensymbol von A ist und q ein Prädikatensymbol, das in einem positiven Literal in W vorkommt.*

2. $p$ hängt direkt negativ von $q$ ab *(in Zeichen:* $p \sqsubset_{-1} q$*) gdw. eine Programmklausel* $A \leftarrow W \in D$ *existiert, so daß p das Prädikatensymbol von A ist und q ein Prädikatensymbol, das in einem negativen Literal in W vorkommt.*

3. $p$ hängt positiv von $q$ ab *(in Zeichen:* $p >_{+1} q$*) gdw.* $p \sqsupset_{+1} q$ *oder es gibt ein Prädikatensymbol r in D, so daß* $p \sqsupset_{+1} r$ *und* $r >_{+1} q$, *oder* $p \sqsupset_{-1} r$ *und* $r >_{-1} q$.

4. $p$ hängt negativ von $q$ ab *(in Zeichen:* $p >_{-1} q$*) gdw.* $p \sqsupset_{-1} q$ *oder es gibt ein Prädikatensymbol* $r$ *in* $D$*, so daß* $p \sqsupset_{+1} r$ *und* $r >_{-1} q$*, oder* $q \sqsupset_{-1} r$ *und* $r >_{+1} q$*.*
5. $p$ hängt von $q$ ab *(in Zeichen:* $p > q$*) gdw.* $p >_{+1} q$ *oder* $p >_{-1} q$*.*
6. *$p$ und $q$ sind gegenseitig abhängig (in Zeichen:* $p \approx q$*) gdw.* $p > q$ *und* $q > p$*.*

**Definition 2.7.13 (Hierarchisches Programm)** *Ein normales Programm* $D$ *heißt* hierarchisch *gdw. keine Prädikatensymbole gegenseitig voneinander abhängen, d.h. es gibt keine Prädikatensymbole* $p$ *und* $q$ *aus* $D$ *mit* $p \approx q$*.*

**Definition 2.7.14 (Stratifiziertes Programm)** [8] *Ein normales Programm* $D$ *heißt* stratifiziert *gdw. keine Prädikatensymbole gleichzeitig gegenseitig und negativ voneinander abhängen, d.h. es gibt keine Prädikatensymbole* $p$ *und* $q$ *aus* $D$ *mit* $p \approx q$ *und* $p >_{-1} q$*.*

**Definition 2.7.15 (Striktes Programm)** *Ein normales Programm* $D$ *heißt* strikt *gdw. keine Prädikatensymbole gleichzeitig positiv und negativ voneinander abhängen, d.h. es gibt keine Prädikatensymbole* $p$ *und* $q$ *aus* $D$ *mit* $p >_{+1} q$ *und* $p >_{-1} q$*.*

*Ein normales Programm heißt* strikt bezüglich der normalen Zielklausel $\leftarrow W$*, wenn* $D \cup \{q \leftarrow W\}$ *strikt ist, wobei* $q$ *ein neues Prädikatensymbol ist, das in* $D$ *nicht auftritt.*

**Definition 2.7.16 (Semi-striktes Programm)** *Ein normales Programm heißt* semi-strikt *gdw. kein Prädikatensymbol negativ von sich selbst abhängt, d.h. es gibt kein Prädikatensymbol* $p$ *aus* $D$ *mit* $p >_{-1} p$*.*

### Sichere und Quasi-sichere Programme

**Definition 2.7.17 (Sicherheit)**

1. *Eine normale Programmklausel* $C$ *ist* bedingt sicher *(engl. admissible) gdw. jede Variable* $v \in vars(C)$ *im Kopf von* $C$ *oder in einem positiven Literal im Rumpf von* $C$ *vorkommt.*
2. *Eine normale Programmklausel* $G$ *ist* sicher *gdw. jede Variable* $v \in vars(G)$ *in einem positiven Literal von* $G$ *vorkommt.*

[8] Zum Sprachgebrauch: Das Substantiv zu „stratifiziert“ sollte nach Duden „Stratifikation“ lauten. Da in [Cremers et al., 1994] jedoch der nicht im Duden aufgeführte Begriff „Stratifizierung“ verwendet wird, werden wir diesen Begriff im folgenden ebenfalls verwenden, um eine einheitliche Terminologie beizubehalten.

3. *Ein normales Programm $D$ ist* sicher *gdw. alle Klauseln aus $D$ sicher sind.*

4. *Ein normales Programm $D$ heißt* sicher bezüglich einer normalen Zielklausel $G$ *gdw.*

   (a) *jede Klausel in $D$ bedingt sicher ist,*

   (b) *jede Klausel in der Definition eines Prädikatensymbols, das positiv in $G$ oder im Rumpf einer Klausel aus $D$ vorkommt, sicher ist und*

   (c) *die Zielklausel $G$ sicher ist.*

**Lemma 2.7.1** *Sei $D$ ein normales Programm und $G$ eine normale Zielklausel, so daß $D$ bezüglich $G$ sicher ist.*

1. *Keine SLDNF-Ableitung von $D$ und $G$ ist blockiert.*

2. *Jede berechnete Antwortsubstitution für $D$ und $G$ ist eine Grundsubstitution bezüglich der Variablen aus $G$.*

**Lemma 2.7.2** *Sei $D$ ein normales Programm und $G$ eine normale Zielklausel, so daß $D$ bezüglich $G$ sicher ist. Wenn comp($D$) erfüllbar ist, dann ist jede korrekte Antwortsubstitution für $D$ und $G$ eine Grundsubstitution bezüglich der Variablen aus $G$.*

**Beweis** Siehe [Shepherdson, 1991].

**Satz 2.7.3 (Vollständigkeit der fairen SLDNF-Resolution)** *Sei $D$ ein stratifiziertes normales Programm und $\leftarrow W$ eine normale Zielklausel, so daß $D$ bezüglich $\leftarrow W$ sicher und strikt ist. Sei $R$ eine faire Berechnungsregel.*

1. *Gilt comp($D$) $\models (\neg W)$, dann existiert der SLDNF-Baum via $R$ und ist ein endlicher Mißerfolgsbaum für $D$ und $\leftarrow W$.*

2. *Jede korrekte Antwortsubstitution für $D$ und $\leftarrow W$ ist eine via $R$ berechnete Antwortsubstitution für $D$ und $\leftarrow W$.*

**Beweis** Siehe [Cavedon und Lloyd, 1989].

Faire Berechnungsregeln sind wegen des Breitensuchansatzes in der Praxis kaum im Einsatz. Deswegen werden wir uns in späteren Kapiteln auf die Standardberechnungsregel (wie sie in PROLOG zum Einsatz kommt) beschränken.

### 2.7.4 Deduktive Datenbanken

**Definition 2.7.18 (Intensionale, extensionale Prädikate)** *Sei* $\Sigma = \langle \mathbf{C}, \mathbf{F}, \mathbf{P}, \mathbf{V} \rangle$ *eine Signatur und D ein normales Programm.*

- *Die Menge der* extensionalen Prädikatensymbole $\Psi^{\Sigma}_{EDB}$ *ist definiert als:* $\Psi^{\Sigma}_{EDB} = \{p \in P_n \mid P_n \in \mathbf{P}$ *und es existiert eine Einheitsklausel in D, deren Kopf das Prädikatensymbol p enthält* $\} \cup Q$*, dabei ist* $Q \subseteq \{p \in P_n \mid P_n \in \mathbf{P}$ *und p wird in D nicht verwendet*$\}$.

- *Die Menge der* intensionalen Prädikatensymbole $\Psi^{\Sigma}_{IDB}$ *ist definiert als:* $\Psi^{\Sigma}_{IDB} = \{p \in P_n \mid P_n \in \mathbf{P}$ *und es existiert eine Klausel mit nichtleerem Rumpf in D, deren Kopf das Prädikatensymbol p enthält*$\} \cup Q'$*, dabei ist* $Q' \subseteq \{p \in P_n \mid P_n \in \mathbf{P}$ *und p wird in D nicht verwendet*$\}$.

**Definition 2.7.19 (Integritätsbedingung)** *Sei* $\Sigma_{SORT}$ *eine ordnungssortierte Signatur. Eine* Integritätsbedingung *ist eine geschlossene Formel aus* $For^{\Sigma_{SORT}}$.

**Definition 2.7.20 (Deduktive Datenbank)** *Sei* $\Sigma_{SORT}$ *eine ordnungssortierte Signatur und* $\Sigma$ *die korrespondierende unsortierte Signatur. Eine* deduktive Datenbank $D$ *ist ein normales, sicheres, stratifiziertes, striktes Programm, so daß die beiden Mengen* $\Psi^{\Sigma}_{EDB}$ *und* $\Psi^{\Sigma}_{IDB}$ *disjunkt sind, d.h.* $\Psi^{\Sigma}_{EDB} \cap \Psi^{\Sigma}_{IDB} = \emptyset$*. Die Menge* $I^{SORT}_{D}$ *ist eine zu D gehörende Menge von Integritätsbedingungen zu* $\Sigma_{SORT}$ *und* $I^{D}$ *die zu* $I^{SORT}_{D}$ *korrespondierende Menge von Integritätsbedingungen zu* $\Sigma$*. Weiterhin ist für alle Programmklauseln* $A \leftarrow L_1 \wedge \ldots \wedge L_n$, $n > 0$*, aus D A ein Atom, bei dem alle Argumente Variablen sind.*

Die letzte Eigenschaft stellt keine Einschränkung dar, da eine Programmklausel $p(t_1, \ldots, t_n) \leftarrow W$ immer in eine äquivalente Programmklausel $p(x_1, \ldots, x_n) \leftarrow x_1 = t_1 \wedge \ldots \wedge x_n = t_n \wedge W$ transformiert werden kann. Die Einschränkung vereinfacht aber die nötigen Definitionen, Verfahren und Beweise. Die Eigenschaft der Sicherheit bedingt z.B., daß alle Einheitsklauseln aus einer Datenbank nur ein Grundatom enthalten.

**Definition 2.7.21 (Datenbankänderung)** *Sei D eine deduktive Datenbank zur Signatur* $\Sigma$*. Eine* Datenbankänderung *ist ein Paar* $\langle Del, Add \rangle$*. Del und Add sind hierbei zwei Mengen von normalen Programmklauseln zur Signatur* $\Sigma$*, für die gilt:*

1. *für alle* $W \in Del$ *gilt auch* $W \in D$.

2. *für alle $W \in Add$ gilt: falls $W$ eine Einheitsklausel $A \leftarrow$ ist, dann ist das Prädikatensymbol von $A \in \Psi^{\Sigma}_{EDB}$ und $A$ ist eine Grundatom. Falls $W$ eine Programmklausel $A \leftarrow L_1 \wedge \ldots \wedge L_n$ mit $n > 0$ ist, dann ist das Prädikatensymbol von $A \in \Psi^{\Sigma}_{IDB}$.*

3. *es gibt keine Klausel, die gleichzeitig in Add und Del ist.*

*Die* resultierende deduktive Datenbank $D'$ *bezüglich einer deduktiven Datenbank $D$ und einer Datenbankänderung $d = \langle Del, Add\rangle$ ist die deduktive Datenbank $D' = (D \setminus Del) \cup Add$.*

**Definition 2.7.22** *Sei $D$ eine deduktive Datenbank und $I_D$ die zugehörige Menge von Integritätsbedingungen. Sei weiter $\varphi \in I_D$ eine relativierte Integritätsbedingung. $D$* genügt *$\varphi$ gdw. $comp(D) \models \varphi$.*

Diese Definition beruht auf der *Folgerungssichtweise*. Sie ist in der Literatur zu deduktiven Datenbanken nicht unumstritten. Eine konkurrierende Definition besagt, daß eine deduktive Datenbank $D$ einer Integritätsbedingung $IC$ dann genügt, wenn $comp(D) \cup \{IC\}$ erfüllbar ist (*Erfüllbarkeitssichtweise*). Definitionen dieser Art finden sich in [Sadri und Kowalski, 1988, Asirelli et al., 1989]. Obwohl die Entscheidung für eine von diesen Sichtweisen letztlich Geschmackssache ist, sind beide nicht äquivalent. Das folgende Beispiel hierzu stammt aus [Das, 1992, S. 283].

**Beispiel 2.7.1** *Sei $D = \{p(a) \leftarrow p(a)\}$ eine deduktive Datenbank und $IC = p(a)$ eine Integritätsbedingung. Die Vervollständigung $comp(D)$ von $D$ ist $p(x) \leftrightarrow x = a \wedge p(a)$ (zuzüglich der Gleichheitstheorie). Bezüglich der ersten Definition ist $D$ inkonsistent, da $comp(D) \not\models p(a)$. Bezüglich der zweiten Definition ist $D$ konsistent, da das Modell $\{p(a)\}$ sowohl Modell von $D$ als auch von $IC$ ist.*

Praktisch wirkt sich der Unterschied folgendermaßen aus:

**Beispiel 2.7.2** *Wir betrachten im folgenden die existenzquantifizierte Integritätsbedingung $\exists\, x : S\ p(x)$.*

1. *Unter der Erfüllbarkeitssichtweise wird versucht zu zeigen, daß für alle $x : S$ gilt $\neg p(x)$. Eine Menge von Programmklauseln, die dann diese Integritätsbedingung realisieren, sieht folgendermaßen aus:*
   - $ic_E \leftarrow \neg r$
   - $r \leftarrow s(x) \wedge p(x)$

*Die Datenbank $D$ genügt dabei der Integritätsbedingung, wenn es einen endlichen gescheiterten SLDNF-Baum für $D \cup \{\leftarrow ic_E\}$ gibt, also die Anfrage $\leftarrow ic_E$ fehlschlägt.*

2. *Eine Klauselmenge, die die gegebene Integritätsbedingung in der Folgerungssichtweise überprüft, ist die folgende:*

   - $ic_F \leftarrow s(x) \wedge p(x)$

   *Falls die Datenbank die Integritätsbedingung erfüllt, würde die Anfrage $D \cup \{\leftarrow ic_F\}$ erfolgreich sein.*

### 2.7.5 Relativierte Integritätsbedingungen

Eine deduktive Datenbank $D$ genügt ihren Integritätsbedingungen $I_D$, wenn für jedes $\varphi \in I_D$ gilt $comp(D) \models \varphi$. Dies legt ein einfaches Verfahren der Überprüfung der Integritätsbedingungen nahe. Hierbei werden die Integritätsbedingungen direkt in normale Programme übersetzt[9] und durch eine Inferenzmaschine (z.B. einen PROLOG-Interpretierer) evaluiert. Hierzu ist ein Verfahren bekannt, das in [Lloyd und Topor, 1984, Lloyd, 1987] veröffentlicht wurde. Dieses Verfahren hat aber im allgemeinen Effizienzprobleme, was hauptsächlich an der Übersetzung von Disjunktionen liegt. Die Behandlung von Disjunktionen im Lloydschen Verfahren ist wie folgt:

Aus $A \leftarrow W_1 \wedge \ldots \wedge W_{i-1} \wedge (V \vee W) \wedge W_{i+1} \wedge \ldots \wedge W_m$
werden $A \leftarrow W_1 \wedge \ldots \wedge W_{i-1} \wedge V \wedge W_{i+1} \wedge \ldots \wedge W_m$
und $A \leftarrow W_1 \wedge \ldots \wedge W_{i-1} \wedge W \wedge W_{i+1} \wedge \ldots \wedge W_m$

Werden diese Programme tatsächlich von einem PROLOG-System ausgeführt, dann führt dies zu Effizienzproblemen. Falls die Beweissuche für $V$ scheitern sollte, müssen $W_1 \wedge \ldots \wedge W_{i-1}$ für den Beweisversuch von $W$ noch einmal gezeigt werden, was unnötig ist.

Besser wäre ein Verfahren, das bei Disjunktionen nicht die zu kodierende Klausel komplett kopiert, sondern die Disjunktion durch die Einführung neuer Prädikatensymbole auflöst. Eine entsprechende Regel wäre z.B. die folgende:

Aus $A \leftarrow W_1 \wedge \ldots \wedge W_{i-1} \wedge (V \vee W) \wedge W_{i+1} \wedge \ldots \wedge W_m$
werden $A \leftarrow W_1 \wedge \ldots \wedge W_{i-1} \wedge p \wedge W_{i+1} \wedge \ldots \wedge W_m$

[9] Dabei wird $\neg^*$ als $\neg$ interpretiert. Da wir die vervollständigte deduktive Datenbank betrachten, ist diese Interpretation erlaubt.

und $p \leftarrow V$
sowie $p \leftarrow W$

Semantisch sind diese beiden Umformungen äquivalent, von einem Prolog-System ausgeführt ist die zweite i.a. erheblich effizienter. Im Anhang geben wir einen Algorithmus, der normale Programme daraufhin optimiert und leicht implementierbar ist, was für zukünftige Anwendungen besonders wichtig ist.

Bisher wurden die Begriffe *deduktive Datenbank* und *Integritätsbedingungen* dieser Datenbank getrennt betrachtet. Diese Trennung wird nun aufgegeben und die Menge der Programmklauseln, die durch die Übersetzung einer Integritätsbedingung erzeugt wurden, zu der deduktiven Datenbank hinzufügt. Dies hat den Nachteil, daß einige Eigenschaften der Datenbank verloren gehen, insbesondere die Eigenschaft der Sicherheit. Für unsere Zwecke ist das nicht weiter relevant, da die Datenbank dann immer noch quasisicher ist und somit keine SLDNF-Ableitung blockiert [Cremers et al., 1994, S.99], weil die Domänenunabhängigkeit der Integritätsbedingung für die Existenz einer nicht blockierten SLDNF-Ableitung ausreicht (zumindest für die Anfrage der Integritätsbedingungen und die Anfragen, die nur die ursprüngliche Datenbank betreffen). Aus diesem Grund werden wir eine deduktive Datenbank im folgenden immer noch als „sicher“ bezeichnen, auch wenn dies formal (für den neu hinzugekommenen Teil) nicht korrekt ist.

# Kapitel 3

# Vorhandene Ansätze zur Integritätsüberprüfung

Das einfachste Verfahren zur Überprüfung der Integritätsbedingungen nach einer Datenbankänderung ist es, alle Bedingungen erneut vollständig zu überprüfen. Dies ist aber aus Effizienzgründen nicht wünschenswert. Deshalb werden Verfahren benötigt, die abhängig von der speziellen Datenbankänderung Integritätsbedingungen identifizieren, deren Gültigkeit verletzt bzw. nicht verletzt sein kann. Hierbei nutzt man aus, daß die Datenbank die Integritätsbedingungen vor der Datenbankänderung erfüllt hat. Im folgenden werden wir einige Ansätze, die diese Idee verfolgen, genauer betrachten. Einen Überblick über das Gebiet bis ca. 1989 gibt [Bry et al., 1992, Das und Williams, 1989a, Das, 1992].

Viele der veröffentlichten Verfahren lassen eine gemeinsame Idee erkennen: Ein $\Delta$ von Konsequenzen einer Datenbankänderung wird durch Vorwärtsberechnung bestimmt, die Integritätsbedingungen werden damit vereinfacht. Diese vereinfachten Bedingungen werden dann überprüft. Vielfach unterscheiden sich diese Verfahren dann in der Granularität der Vereinfachung und diversen Optimierungen. Verfahren, die hauptsächlich diesem Ansatz folgen, nennen wir *konventionelle Verfahren*, Verfahren, deren Ansatz noch andere Elemente aufweisen, nennen wir dementsprechend *unkonventionelle Verfahren*[1].

[1] Tatsächlich ist mit diesem Kriterium keine genaue Unterscheidung der Verfahren möglich, da *alle* Verfahren Elemente der beschriebenen Vorgehensweise enthalten, die mehr oder weniger stark ausgeprägt sind. Die gewählte Differenzierung sollte man deswegen nicht ganz so genau nehmen.

## 3.1 Konventionelle Verfahren

Grundlegend für die weiteren Ansätze ist die im folgenden beschriebene Arbeit, die nur relationale Datenbanken betrachtet.

### 3.1.1 Logic for Improving Integrity Checking in Relational Databases

*Jean-Marie Nicolas*

Die Beispiele und die Beschreibung des Algorithmus wurden [Griefahn, 1989] entnommen. Der von Jean-Marie Nicolas in [Nicolas, 1982] veröffentlichte Ansatz ist für viele andere Ansätze fundamental. Ursprünglich nur für relationale Datenbanken entwickelt, wurde dieser Ansatz von vielen aufgegriffen und für deduktive Datenbanken weiterentwickelt. Die Hauptidee des Ansatzes ist es, die Integritätsbedingungen im Hinblick auf erfolgte Datenbankänderungen zu vereinfachen und nur diese vereinfachten Integritätsbedingungen erneut zu überprüfen. Hierbei geht man von der Annahme aus, daß eine relationale Datenbank vor eventuellen Änderungen ihren Integritätsbedingungen genügt hat. Einfügungen (bzw. Löschungen) können eine Integritätsbedingung nur dann betreffen, wenn das entsprechende Literal in der Bedingung negativ (bzw. positiv) enthalten ist. Auf der Basis dieser Beobachtungen wird ein Algorithmus entwickelt, der ausgehend von einer Änderung eine Integritätsbedingung $W$ vereinfacht. Wir geben ein Beispiel für diese Vereinfachung ausgehend von der Einfügung eines Faktes $F = p(t_1, \ldots, t_n)$ in die Datenbank $D$ sowie einer Integritätsbedingung $W$ in konjunktiver Pränexnormalform. Die vereinfachte Integritätsbedingung notieren wie mit $S^+(W)$.

**Algorithmus 3.1.1 (Vereinfachung von Integritätsbedingungen)**

Schritt 1: *Ausgehend von der Annahme, daß eine Integritätsbedingung vor einer Änderung der Datenbank erfüllt war, reicht es aus, nur bestimmte Instanzen der Integritätsbedingung in Abhängigkeit der konkreten Änderung zu überprüfen. Aus diesem Grund wird eine Menge von Substitutionen $\Gamma^+$ ausgehend von $p(t_1, \ldots, t_n)$ und möglichen Vorkommen von $\neg p(t_1, \ldots, t_n)$ in $W$ berechnet. Diese Substitutionen werden dann eingeschränkt auf alle universell quantifizierten Variablen von $W$, deren Quantor vor allen Existenzquantoren steht d.h.*

$$\Gamma^+ = \{\sigma|_V \mid \neg p(g_1, \ldots, g_n) \in W, \sigma = mgu(p(g_1, \ldots, g_n), p(t_1, \ldots, t_n))\}$$

*mit $W \equiv \forall x_1, \ldots, x_m W'$ und $V = \{x_1, \ldots, x_m\}$.*

*Das folgende Beispiel erklärt diese Einschränkung:*

**Beispiel 3.1.1**

– *Sei* $W \equiv \forall\ x\ \exists\ y\ (\neg p(x) \vee q(x,y))$ *eine Integritätsbedingung, die in einer Datenbank* $D$ *gültig ist. Das Einfügen des Fakts* $p(a)$ *kann die Gültigkeit von* $W$ *möglicherweise verletzen. Offensichtlich reicht es aber aus, die Gültigkeit folgender Instanz von* $W$ *zu überprüfen:*

$$\exists\ y\ (\neg p(a) \vee q(a,y))$$

– *Sei* $W \equiv \exists\ y\ \forall\ x\ (\neg p(x) \vee q(x,y))$*. Fügen wir hier das Fakt* $p(a)$ *hinzu und instanziieren* $W$ *entsprechend, so ergibt sich wiederum:*

$$\exists\ y\ (\neg p(a) \vee q(a,y))$$

*Die Überprüfung dieser Instanz kann aber nicht die Gültigkeit der Integritätsbedingung sicherstellen.*

*Schritt 2:* *In diesem Schritt wird die Menge* $\Gamma^+$ *weiter vereinfacht: es können redundante Substitutionen in* $\Gamma^+$ *vorhanden sein, so z.B. bei der Substitutionsmenge* $\Gamma^+ = \{\{x/a\},\{x/a,y/b\}\}$*. Die erste Substitution ist eine Teilmenge der zweiten und in diesem Sinne allgemeiner. Daher werden alle Substitutionen* $\theta \in \Gamma^+$*, die eine andere Substitution* $\sigma \in \Gamma^+$ *enthalten*[2] *aus* $\Gamma^+$ *gelöscht. Wir erhalten die neue Menge* $\Gamma_R^+$*. Damit definieren wir die Menge* $T_R^+(W)$ *als die Konjunktion aller* $W\sigma$ *mit* $\sigma \in \Gamma_R^+$*, d.h. als die Konjunktion aller instanziierten Formeln.*

*Schritt 3:* *Die neue Formel* $T_R^+(W)$ *kann ebenfalls noch weiter vereinfacht werden:*

– *Von einigen Grundliteralen ist der Wahrheitswert aus den Änderungsoperationen bekannt (die gelöschten Grundfakten haben den Wahrheitswert* $false$*, die hinzugefügten den Wahrheitswert* $true$*). Diese instanziierten Literale innerhalb der Integritätsbedingungen können durch ihren Wahrheitswert ersetzt werden.*

– *Durch die Anwendung der Absorptionsregeln kann die Formel dann vereinfacht werden. Diese Regeln sind im einzelnen:*

$$\begin{array}{lcrlcr} \neg true & \Rightarrow & false & \neg false & \Rightarrow & true \\ true \vee A & \Rightarrow & true & false \vee A & \Rightarrow & A \\ true \wedge A & \Rightarrow & A & false \wedge A & \Rightarrow & false \end{array}$$

[2] Falls man die Substitutionen als Menge von Variable – Term-Paaren betrachtet.

*Schritt 4: In einem letzten Schritt können noch alle $W\theta$ aus der Formel entfernt werden, für die eine anderes $W\sigma$ in der Formel enthalten ist, mit $W\theta \equiv W\sigma$ bis auf Permutationen und Variablenumbenennung. Die resultierende Formel ist dann die vereinfachte Integritätsbedingung $S^+(W)$.*

Um zu zeigen, daß die Integritätsbedingung $W$ nach der Änderung der Datenbank noch gilt, reicht es aus, die Gültigkeit von $S^+(W)$ zu zeigen.

### 3.1.2 Integrity Constraint Checking in Stratified Databases

*J.W. Lloyd, E.A. Sonenberg, and R.W. Topor*

Dieser Ansatz, den wir im folgenden als „LST-Ansatz“ bezeichnen und der ausführlich in [Lloyd et al., 1987, Lloyd, 1987] beschrieben ist, ist eine Weiterentwicklung der oben beschriebenen Simplifikationsmethode für deduktive Datenbanken. Die Grundidee dieses Ansatzes besteht in der Berechnung der Modelldifferenzen von zwei verschiedenen Datenbankzuständen $D$ und $D'$, wobei $D'$ aus $D$ durch die Anwendung einer Datenbankänderung $d$ hervorgeht.

#### Modelldifferenzen

Sei $D$ eine deduktive Datenbank, $d = \langle Del, Add \rangle$ eine Datenbankänderung und $D'$ die resultierende Datenbank. Sei $D'' = D \setminus Del$. Es gilt: $D'' \subseteq D$ sowie $D'' \subseteq D'$. Bei relationalen Datenbanken besteht der Unterschied zwischen den Modellen von $D$ und $D'$ gerade aus der Interpretation der in $Del$ und $Add$ enthaltenen Fakten.[3] Im Fall von deduktiven Datenbanken können aber bedingt durch Regeln auch größere Differenzen entstehen. Will man also den Unterschied zwischen diesen $D$ und $D'$ berechnen, so muß dieser Umstand mit berücksichtigt werden. Dies verlangt die Berechnung von vier Mengen von Atomen und zwei Substitutionsmengen. Diese werden im folgenden definiert:

**Definition 3.1.1** *Seien $D$ und $D'$ zwei Datenbanken mit $D \subseteq D'$.[4] Die Mengen $pos_{D,D'}$ und $neg_{D,D'}$ werden induktiv wie folgt definiert:*

- $pos^0_{D,D'} = \{A \mid A \leftarrow W \in D' \setminus D\}$
- $neg^0_{D,D'} = \{\}$

[3] Wir identifizieren im folgenden die Datenbankrelationen und deren Interpretationen.

[4] Man beachte, daß $D$ und $D'$ *nicht* mit den oben erwähnten $D$ und $D'$ übereinstimmen.

- $pos^{n+1}_{D,D'} = \{A\theta \mid A \leftarrow W \in D,\ B$ *ist positives Literal in* $W, C \in pos^n_{D,D'}$ *und* $\theta$ *ist der mgu von* $B$ *und* $C\}$
- $neg^{n+1}_{D,D'} = \{A\theta \mid A \leftarrow W \in D,\ B$ *ist positives Literal in* $W,\ C \in neg^n_{D,D'}$ *und* $\theta$ *ist der mgu von* $B$ *und* $C\}$
- $pos_{D,D'} = \bigcup_{n\geq 0} pos^n_{D,D'}$
- $neg_{D,D'} = \bigcup_{n\geq 0} neg^n_{D,D'}$

Intuitiv erfaßt $pos_{D,D'}$ den Teil, der bei der Änderung von $D$ zu $D'$ zu dem Modell von $comp(D)$ hinzugefügt wird. $neg_{D,D'}$ erfaßt den Teil, der verloren geht. Wichtig ist, daß zur Berechnung dieser Mengen nur die Regeln benutzt werden und kein Zugriff auf die Fakten (die möglicherweise extern gespeichert sind) erfolgt. Die Berechnung dieser Mengen wird im allgemeinen nicht terminieren, wenn rekursive Regeln in der Datenbank vorhanden sind. Daher ist es wichtig, Kriterien zu haben, wann die Berechnung aufhören kann. Zu diesen Zweck wird der Begriff der *Stopp-Regeln* (eng. stopping rules) eingeführt, die Kriterien liefern, wann man mit der Berechnung der Mengen aufhören kann: eine mögliche Stopp-Regel führt zur Berechnung von zwei Mengen $P$ und $N$ (anstatt $pos_{D,D'}$ und $neg_{D,D'}$). Die Berechnung dieser Mengen ist ähnlich wie die bereits beschriebenen, nur werden im Laufe der Berechnung von $P^n$ alle Atome, die Instanzen eines Elementes aus $P^k$ für $0 \leq k \leq n$ sind, weggelassen. Wird ein $P^n$ dabei leer, kann man mit der Berechnung aufhören. $P$ ist dann wieder die Vereinigung aller $P^n$.

Bei der Nichtbeachtung der Struktur des aktuellen Beweises einer Integritätsbedingung, ist das Problem wegen der Berechenbarkeitsproblematik nicht lösbar, da nicht bestimmt werden kann, ob ein beliebiges Programm terminiert oder nicht.

Aus diesen Mengen werden dann zwei Mengen von Substitutionen berechnet, die mit denen analog sind, die beim Ansatz von Nicolas (siehe oben) verwendet wurden. Damit können dann die Integritätsbedingungen vereinfacht werden.

### 3.1.3 Improving Integrity Constraint Checking in Deductive Databases

*P. Asirelli, P. Inverardi, and A. Mustaro*

Dieser Ansatz [Asirelli et al., 1988] ist dem LST-Ansatz sehr ähnlich. Die Verfeinerung liegt in der Vermeidung von unnötigen Substitutionen bei der Berechnung der Differenzmengen der Datenbankzustände. Allerdings ist die Darstellung im Artikel an mehreren

Stellen fehlerhaft, und der im Artikel angegebene Beweis paßt nicht zum Verfahren. Deshalb geben wir zunächst einige Gegenbeispiele und zeigen dann auf, aus welchem Grunde der im Artikel beschriebene Algorithmus fehlerhaft ist.

**Beispiel 3.1.2**

*Gegeben sei folgende Datenbank:*

*Integritätsbedingungen:* $\neg p(a)$

*Regeln:* $p(x) \leftarrow q(x,y) \wedge r(y)$

*Fakten:* $q(a,b)$

*Datenbankänderung* $d = \langle Del, Add \rangle$ *mit* $Del = \emptyset$ *und* $Add = \{r(b)\}$

*Im Ausgangszustand ist die Integritätsbedingung erfüllt (*$p(a)$ *kann nicht abgeleitet werden), nach der Anwendung der Datenbankänderung ist die Integritätsbedingung jedoch verletzt (mit Hilfe der Regel kann* $p(a)$ *abgeleitet werden). Wir betrachten nun die Berechnung der Änderungsmengen nach dem im Artikel angegebenen Verfahren:*

- $pos^0_{ins} = \{\{r(b), \#\}\}$
- $pos^1_{ins} = \{\{p(x), r(b), q(x,b)\}\}$
- $pos^2_{ins} = \{\}$
- $neg^0_{ins} = \{\}$
- $neg^1_{ins} = \{\}$
- $neg^2_{ins} = \{\}$
- $pos^0_{del} = \{\}$
- $neg^0_{del} = \{\}$

*Dies führt zu* $P_{ins} = \{\{r(b), \#\}, p(x), r(b), q(x,b)\}\}$, *was wiederum zu* $P^*_{ins} = \{\}$ *führt. Dieses Ergebnis hat aber zur Folge, daß die Integritätsbedingung* nicht *überprüft wird, obwohl sie nach der Datenbankänderung verletzt ist!*

**Beispiel 3.1.3** *Ein Großteil des in dem Artikel angegebenen Beweises beruht auf der Behauptung, daß die Berechnung der Änderungsmengen zu den Berechnungen im LST-Verfahren äquivalent sind. Dies ist jedoch nicht der Fall, wie das folgende Beispiel zeigt:*

*Integritätsbedingung:* $\neg s(b)$

*Regeln:* $p(x,y) \leftarrow q(x,y) \wedge r(y)$

$s(y) \leftarrow p(a,y)$

*Fakten:* $q(a, b)$

*Datenbankänderung* $d = \langle Del, Add \rangle$ *mit* $Del = \emptyset$ *und* $Add = \{r(b)\}$

| *Berechnung nach Asirelli* | | | | | *Berechnung nach LST* | | |
|---|---|---|---|---|---|---|---|
| $pos^0_{D,D'}$ | $=$ | $\{r(b)\}$ | $=$ | $P^0$ | $pos^0_{D,D'}$ | $=$ | $\{r(b)\}$ |
| $neg^0_{D,D'}$ | $=$ | $\{\}$ | $=$ | $N^0$ | $neg^0_{D,D'}$ | $=$ | $\{\}$ |
| $pos^1_{D,D'}$ | $=$ | $\{p(x, b)\}$ | $=$ | $P^1$ | $pos^1_{D,D'}$ | $=$ | $\{p(x, b)\}$ |
| $neg^1_{D,D'}$ | $=$ | $\{\}$ | $=$ | $N^1$ | $neg^1_{D,D'}$ | $=$ | $\{\}$ |
| $pos^2_{D,D'}$ | $=$ | $\{r(b)\}$ | $=$ | $P^2$ | $pos^2_{D,D'}$ | $=$ | $\{\}$ |
| $neg^2_{D,D'}$ | $=$ | $\{\}$ | $=$ | $N^2$ | $neg^2_{D,D'}$ | $=$ | $\{\}$ |
| $pos^3_{D,D'}$ | $=$ | $\{r(b)\}$ | $=$ | $P^3$ | | | |
| $neg^3_{D,D'}$ | $=$ | $\{\}$ | $=$ | $N^3$ | | | |

Das Verfahren zeigt folgende Fehler:

1. Die Definition der Änderungsmengen ist inkorrekt: Auf S. 74 lauten die Definitionen 11 und 12 der Mengen $pos^n_{ins}$ und $neg^n_{ins}$:

Def. 11 $pos^{n+1}_{ins} := \{[B\sigma, A', L_1\sigma, \ldots, L_{i-1}\sigma, L_{i+1}\sigma, \ldots, L_n\sigma] \mid \exists E \in P^n_{ins}$ with
$-head(E) = A'$
$-\exists$a rule $B \leftarrow L_1, \ldots, L_{i-1}, A^\sim, L_{i+1}, \ldots, L_n \in D$ **s.t.** $\mathbf{A}' = \mathbf{A}^\sim\sigma$
$\cup\{[B\sigma, A', L_1\sigma, \ldots, L_{i-1}\sigma, L_{i+1}\sigma, \ldots, L_n\sigma] \mid \exists E \in N^n_{ins}$with
$-\exists head(E) = A'$
$-\exists$a rule $B \leftarrow L_1, \ldots, L_{i-1}, \neg A^\sim, L_{i+1}, \ldots, L_n \in D$ **s.t.** $\mathbf{A}' = \mathbf{A}^\sim\sigma$

Def. 12 $neg^{n+1}_{ins} := \{[B\sigma, A', L_1\sigma, \ldots, L_{i-1}\sigma, L_{i+1}\sigma, \ldots, L_n\sigma] \mid \exists E \in P^n_{ins}$ with
$-head(E) = A'$
$-\exists$a rule $B \leftarrow L_1, \ldots, L_{i-1}, \neg A^\sim, L_{i+1}, \ldots, L_n \in D$ **s.t.** $\mathbf{A}' = \mathbf{A}^\sim\sigma$
$\cup\{[B\sigma, A', L_1\sigma, \ldots, L_{i-1}\sigma, L_{i+1}\sigma, \ldots, L_n\sigma] \mid \exists E \in N^n_{ins}$with
$-\exists head(E) = A'$
$-\exists$a rule $B \leftarrow L_1, \ldots, L_{i-1}, A^\sim, L_{i+1}, \ldots, L_n \in D$ **s.t.** $\mathbf{A}' = \mathbf{A}^\sim\sigma$

Die Mengen $pos^{n+1}_{ins}$ und $neg^{n+1}_{ins}$ werden nur dann erweitert, wenn der Regelkopf, der in einer vorangegangenen Berechnung bestimmt wurde, eine *Instanz* eines positiven bzw. negativen Literals ist, welches in der Datenbank $D$ enthalten ist (wegen der Bedingung $\mathbf{A}' = \mathbf{A}^\sim\sigma$). Tatsächlich ist diese Bedingung unzureichend, denn es genügt, wenn der Regelkopf und das Literal *unifizierbar* sind. Dies wird in Beispiel 3.1.3 verdeutlicht.

2. Die Definitionen 15 und 16 auf S. 75 lauten:

Def. 15 $P^*_{ins} = \{E \mid E \in P_{ins}$ and $head(E)$ is an instance of an atom which occurs negatively in $W\}$

Def. 16 $N^*_{ins} = \{E \mid E \in N_{ins}$ s.t. $head(E)$ is an instance of an atom which occurs positively in $W\}$

Wie Beispiel 3.1.2 zeigt, genügt die Bedingung „Instanz“ (instance) nicht, da $p(x) \in P_{ins}$ keine Instanz von $p(a)$ (aus der Integritätsbedingung) ist. Daher werden Situationen, in denen die Integritätsbedingung möglicherweise verletzt wurde, nicht erkannt.

### 3.1.4 Integrity Enforcement on Deductive Databases

*Hendrik Decker*

Dieser in [Decker, 1986] veröffentlichte Ansatz behandelt ebenfalls die Simplifizierung von Integritätsbedingungen. Im Gegensatz zu anderen Ansätzen werden aber mögliche Vereinfachungen schon vor den eigentlichen Datenbankänderungen gefunden, d.h. die Integritätsbedingungen werden im voraus analysiert und bei der eigentlichen Überprüfung als eine Menge von *Änderungsbedingungen* betrachtet. Zu jedem negativen (bzw. positiven) Vorkommen eines Atomes $A$ in einer Integritätsbedingung $W$ wird eine *Einfüge- bzw. Löschbedingung* konstruiert:

$$\begin{array}{l} insert\ A\ only_if\ C_1 \\ delete\ A\ only_if\ C_2 \end{array}$$

Dabei erhalten wir die Bedingung $C_1$ (bzw. $C_2$) aus $W$ durch Umbenennen aller freien Variablen und Weglassen aller negativen (bzw. positiven) Vorkommen von $A$.

Sei $D$ eine deduktive Datenbank, $d$ eine Datenbankänderung und $D'$ die resultierende Datenbank. Für eine Programmklausel $C \equiv H \leftarrow B$ sind die beiden Faktenmengen $db^C$ und $db_C$ wie folgt definiert:

$$\begin{array}{rcl} db^C & = & \{H\theta \mid H\theta \text{ Grundatom}, D' \models B\theta \text{ und } D \not\models H\theta\} \\ db_C & = & \{H\theta \mid H\theta \text{ Grundatom}, D \models B\theta \text{ und } D' \not\models H\theta\} \end{array}$$

Die Mengen enthalten genau die Fakten, die aus der Datenbank mittels der Datenbankänderung gelöscht bzw. zu ihr hinzugefügt wurden. Die vollständig Modelldifferenz $db^\Delta$ (bzw. $db_\Delta$) ist die Vereinigung aller $db^C$ (bzw. $db_C$).

- $db^{\Delta}$ ist definiert als die Menge aller Fakten, die aus $D'$ ableitbar sind, aber nicht aus $D$.
- $db_{\Delta}$ ist definiert als die Menge aller Fakten, die aus $D$ ableitbar sind, aber nicht mehr aus $D'$.

Die eigentliche Integritätsüberprüfung findet nun mit einem einfachen Algorithmus statt, der für jede hinzuzufügende (bzw. zu löschende) Klausel $C$ aus der Datenbankänderung aufgerufen wird:

Eingabe: $add(C)$ oder $remove(C)$

1. Für jedes Fakt $A'$ in $db^C$ (bzw. $db_C$) und für jede Änderungsbedingung *insert A only_if C'*, falls die Eingabe $add(C)$ war (bzw. *delete A only_if C'*, falls die Eingabe $remove(C)$ war), berechne $\theta = mgu(A', A)$. Damit wird dann $C'\theta$ in $D'$ neu evaluiert. Wenn die Evaluation scheitert, ist die Integritätsbedingung verletzt.
2. Für jedes Fakt $A \in db^C$ (bzw. $db_C$) und für jede Regel $R$ aus $D$ sei $A'$ ein positives Literal aus dem Rumpf von $R$ und $\theta = mgu(A, A')$. Der Algorithmus wird dann rekursiv mit $add(R\theta)$ (bzw. $remove(R\theta)$) aufgerufen.
3. Für jedes Fakt $A \in db^C$ (bzw. $db_C$) und für jede Regel $R$ aus $D$ sei $\neg A'$ ein negatives Literal aus dem Rumpf von $R$ und $\theta = mgu(A, A')$. Der Algorithmus wird dann rekursiv mit $remove(R\theta)$ (bzw. $add(R\theta)$) aufgerufen.

Der LST-Ansatz und Deckers Ansatz sind sich sehr ähnlich: die Vereinigung aller $db^C$ (bzw. $db_C$) ergibt gerade die relevanten Modellunterschiede. Eine Weiterentwicklung dieses Ansatzes im Hinblick auf eine datenbankadäquatere Darstellung findet sich in [Martens und Bruynooghe, 1988].

### 3.1.5 A uniform Approach to Constraint Satisfaction and Constraint Satisfiability in Deductive Databases

*François Bry, Hendrik Decker and Rainer Manthey*

Dieser in [Bry et al., 1988] veröffentlichte Ansatz beschäftigt sich neben der Frage der inkrementellen Integritätsüberprüfung auch mit der Erfüllbarkeit einer Menge von Integritätsbedingungen. Während der Ansatz zur inkrementellen Integritätsüberprüfung hauptsächlich eine Weiterentwicklung des Ansatzes von Decker ist, ist der Ansatz zum zweiten Problem im Bereich der deduktiven Datenbanken neu.

Da das Problem der Erfüllbarkeit einer Menge von prädikatenlogischen Formeln i.a. nicht entscheidbar ist, beschränkt man sich hier auf genau den Teil, der mindestens semientscheidbar ist: mit Hilfe eines modifizierten Tableauverfahrens [Smullyan, 1968, Fitting, 1990] wird ein Modell für eine gegebene Formelmenge konstruiert. Gibt es kein oder ein endliches Modell, so terminiert der Algorithmus. Gibt es ein unendliches Modell, so terminiert der Algorithmus i.a. nicht. Eine Formelmenge mit einem unendlichen Modell ist z.B. die Formalisierung folgender Aussagen (aus [Bry und Manthey, 1986]):

- Jeder arbeitet für jemanden.
- Keiner arbeitet für sich selbst.
- Wenn $x$ für $y$ arbeitet, und $y$ arbeitet für $z$, dann arbeitet $x$ für $z$.

Eine solche Formelmenge ist prinzipiell als Menge von Integritätsbedingungen für eine deduktive Datenbank nicht akzeptabel, da sie

- keiner wirklich speicherbaren Datenbank entsprechen und
- nicht in endlicher Zeit überprüfbar sind.

### 3.1.6 A Path Finding Method for Constraint Checking in Deductive Databases

*S.K. Das and M.H. Williams*

Das in [Das und Williams, 1989b, Das, 1992] veröffentlichte Verfahren beruht auf der Idee, daß eine Integritätsbedingung nur dann verletzt sein kann, wenn es gelingt, einen *Pfad* von einem Literal aus einer Datenbankänderung zu einem Literal aus der Integritätsbedingung zu konstruieren. Ein *Pfad* zu einer Menge $\mathbf{D}$ von Programmklauseln ist definiert als eine Folge von Literalen

$$L_0 \stackrel{R_1}{\rightarrow} L_1 \stackrel{R_2}{\rightarrow} \ldots \stackrel{R_n}{\rightarrow} L_n,$$

dabei ist $L_0$ der *Start* des Pfades, $L_n$ das *Ziel*, $n$ die *Länge* und $R_1, \ldots, R_n$ sind Programmklauseln aus $\mathbf{D}$, die zur Bildung des Pfades benutzt werden.

Ausgehend von einem $L_i$ erhält man den Nachfolger $L_{i+1}$ auf folgende Weise:

1. Wenn $L_i$ positiv und $\sigma = mgu(L_i, L)$ existiert, wobei $L$ ein positives Literal aus dem Rumpf einer Regel $R_{i+1} \equiv H \leftarrow B$, $G$ der Rest von $B$ ohne das Literal $L$ und $\theta$ die berechnete Antwort von $D \cup \leftarrow G\sigma$ ist, dann ist $L_{i+1} = H\sigma\theta$.

2. Wenn $L_i$ positiv ist und $\sigma = mgu(L_i, A)$ existiert, wobei $\neg A$ ein negatives Literal aus dem Rumpf einer Regel $R_{i+1} \equiv H \leftarrow B$ sowie $\neg H\sigma$ nicht Instanz einer $L_j$, $0 \leq j \leq i$, ist, dann ist $L_{i+1} = \neg H\sigma$.

3. Wenn $L_i$ negativ und $\sigma = mgu(L_i, L)$ existiert, wobei $L$ ein negatives Literal aus dem Rumpf einer Regel $R_{i+1} \equiv H \leftarrow B$ und $\theta$ die berechnete Antwort von $D \cup \leftarrow B\sigma$ ist, dann ist $L_{i+1} = H\sigma\theta$.

4. Wenn $L_i$ negatives Literal $\neg A$ ist und $\sigma = mgu(A, L)$ existiert, wobei $L$ ein positives Literal aus dem Rumpf einer Regel $R_{i+1} \equiv H \leftarrow B$ ist, $\neg H\sigma$ ist nicht Instanz eines $L_j$, $0 \leq j \leq i$, dann ist $L_{i+1} = \neg H\sigma$.

Obwohl die vorstehende Aufzählung anders aussieht, ist hier auch zu erkennen, daß von der Datenbankänderung ausgehend eine Vorwärtsberechnung durchgeführt wird.

Sei $D$ deduktive Datenbank, $d = \langle Del, Add \rangle$ eine Datenbankänderung und $D'$ die daraus resultierende Datenbank.

Zur Integritätsüberprüfung wird als Start ein *Änderungsliteral* benutzt. Ein Änderungsliteral ist

1. ein Atom $A \in Add$[5],

2. ein Literal $\neg A$, mit $A \in Del$,

3. $H\sigma$, wenn eine Regel $H \leftarrow B \in Add$ existiert und $\sigma$ die berechnete Antwort von $D \cup \{\leftarrow B\}$ ist,

4. $\neg H$, wenn eine Regel $H \leftarrow B \in Del$ existiert.

Eine Integritätsbedingung $W$, die als normales Programm $ic \leftarrow W'$ vorliegt[6], ist genau dann verletzt, wenn es einen Pfad von einem Änderungsliteral zum Literal $ic$ gibt.

Die Vorwärtsberechnung, die in den bereits zitierten Ansätzen enthalten ist, findet sich auch hier wieder.

---

[5] Eigentlich haben wir nur Regeln $H \leftarrow B \in Add$ und $Del$, wobei der Rumpf $B$ möglicherweise leer ist. Legt man diese Terminologie zugrunde, so kann diese Aufzählung auf zwei Punkte beschränkt werden. Wir zitieren hier jedoch die Original-Literatur.

[6] $ic$ ist in diesem Sinne ein Startatom, d.h. bei der Anfrage $D \cup \{\leftarrow ic\}$ wird die Integritätsbedingung $W$ überprüft.

## 3.2 Unkonventionelle Ansätze

### 3.2.1 A Theorem-Proving Approach for Database Integrity

*Fariba Sadri and Robert Kowalski*

Diesem in [Sadri und Kowalski, 1988] veröffentlichten Ansatz liegt eine andere als die bisher benutzte Sichtweise zugrunde: Wir gingen bisher davon aus, daß eine deduktive Datenbank $D$ eine Integritätsbedingung $W$ erfüllt, wenn gilt: $comp(D) \models W$. In diesem Ansatz wird angenommen, daß $D$ die Integritätsbedingung $W$ erfüllt, wenn gilt: $comp(D) \cup \{W\}$ ist erfüllbar (siehe Abschnitt 2.7.4 und [Reiter, 1992, S. 133] für einen Überblick über verschiedene Konsistenzbegriffe).

Diese Methode versucht einen *Widerlegungsbaum* zu finden, d.h. einen Baum, bei dem *ein* Blatt die leere Klausel ist. Dabei wird als Wurzel jede Klausel, die in einer Datenbankänderung $d$ enthalten ist, benutzt. Wenn jeder mögliche Konstruktionsversuch scheitert, ist die Integrität garantiert. Zur Konstruktion wird eine erweiterte SLDNF-Strategie benutzt, wobei der Unterschied in der Wahl der Klauseln besteht. Bei der SLDNF-Resolution ist die Menge der Klauseln, die in einer Ableitung vorkommen können, beschränkt auf Zielklauseln und die leere Klausel. Um Vorwärtsableitungen zu erlauben, läßt die hier erweiterte SLDNF-Strategie als Wurzelknoten jede mögliche Klausel zu. Eine Klausel, die in einer Ableitung vorkommt, ist entweder die leere Klausel oder eine Klausel der Art

$$\neg A \leftarrow L_1 \wedge \ldots \wedge L_n \qquad n \geq 1,$$

wobei $A$ ein Atom ist und die $L_i$, $1 \leq i \leq n$, Literale. Ein Literal wird durch eine sichere Berechnungsregel selektiert. Weiterentwicklungen des Verfahrens finden sich in [Decker und Celma, 1994].

### 3.2.2 Integrity constraint checking in deductive databases using the Prolog not-predicate

*Tok-Wang Ling*

Ling verwendet in seinem Artikel [Ling, 1987] andere Definitionen, als sie in allen anderen Ansätzen üblich sind, dies auch bei bei der Definition von Integritätsbedingungen und

Datenbankänderungen. Unter einer Datenbankänderung versteht er eine Transaktion bestehend aus

- Einfügungen von Atomen
- Löschen von Atomen
- Modifikationen von Atomen

Integegritätsbedingungen sind spezielle, eingeschränkte Formeln der Prädikatenlogik erster Stufe, die im folgenden definiert werden:[7]

Zunächst jedoch die Definition der Normalform einer negativen Formel:

**Definition 3.2.1 (NF-negative Formel)** *Eine Formel $F = not(P)$ ist eine* negative Formel in Normalform (NF-negative Formel)*, wenn*

1. *$P$ ein positives Literal ist oder*
2. *$P$ die folgende Form hat:*

$$A_1 \wedge \ldots \wedge A_n \wedge not(B_1) \wedge \ldots \wedge not(B_m)$$

*mit positiven Literalen $A_i$, $1 \leq i \leq n$, und NF-negativen Formeln $not(B_j)$, $1 \leq j \leq m$.*

Integritätsbedingungen heißen bei Ling *IC-Formeln* und sind folgendermaßen definiert:

**Definition 3.2.2 (IC-Formel)** *Eine* IC-Formel *ist eine geschlossene Formel der Prädikatenlogik erster Stufe in der folgenden Form:*

$$A_1, \ldots, A_n, not(B_1), \ldots, not(B_m) \rightarrow C_1 \vee \ldots \vee C_p$$

*mit*

1. $n \geq 0, m \geq 0, p \geq 0$
2. *Jedes $A_i$, $1 \leq i \leq n$, ist ein positives Literal*
3. *Jedes $not(B_j)$, $1 \leq j \leq m$, ist eine NF-negative Formel*
4. *Jedes $C_k$, $1 \leq k \leq p$, hat eine ähnliche Form wie die $B_j$, d.h. jedes $C_j$ ist eine Konjunktion von positiven Literalen und/oder NF-negativen Formeln.*

[7] Es werden alle Definitionen und Verweise aus der Originalarbeit, die mit von PROLOG ausführbaren Prädikaten zu tun haben, weggelassen, da sie für das Verständnis nicht wichtig sind.

Eine IC-Formel wird wie folgt interpretiert:

- Das „$\leftarrow$“-Zeichen steht für die Implikation, das „ , “-Zeichen für die Konjunktion;
- Variablen in einem positiven Literal $A_i$ werden als universell quantifiziert angenommen
- Variablen, die in einem positiven Literal von einem $C_k$ vorkommen aber in keinem positiven Literal $A_i$, werden als existenzquantifiziert mit Gültigkeitsbereich $C_k$ angenommen
- Variablen, die in einem $B_j$ vorkommen aber in keinem $A_i$, werden als lokale Variablen von $not(B_j)$ betrachtet
- Variablen, die in NF-negativen Formeln von $C_k$, z.B. $not(E)$, vorkommen, aber in keinem positiven Literal von $C_k$ oder $A_i$, werden als lokale Variablen von $not(E)$ betrachtet.

**Beispiel 3.2.1** *Die Formel*

$$a(x), b(x,y) \rightarrow c(x) \vee d(x,y,z)$$

*wird interpretiert als*

$$\forall\, x\, \forall\, y\, ((a(x) \wedge b(x,y)) \rightarrow (c(x) \vee \exists\, z\, d(x,y,z)))$$

Diese IC-Formeln können direkt in NF-negative Formeln übersetzt werden. Diese wiederum haben die Eigenschaft, direkt von einem PROLOG-System interpretierbar zu sein. Die Übersetzung geschieht durch folgende Äquivalenz: jede IC-Formel

$$A_1, \ldots, A_m, not(B_1), \ldots, not(B_m) \rightarrow C_1 \vee \ldots \vee C_p$$

ist logisch äquivalent zu einer Formel $W'$ mit

$$W' \equiv not(A_1, \ldots, A_n, not(B_1), \ldots, not(B_m), not(C_1), \ldots, not(C_p)),$$

welche direkt als Anfrage an PROLOG stellbar ist. Falls PROLOG eine SLDNF-Widerlegung von $\leftarrow W'$ findet, erfüllt die Datenbank die IC-Formel.

Die Polarität eines Literals in einer IC-Formel definieren wir über die Anzahl der *not*-Prädikate, die das Literal direkt oder indirekt negieren. Ist die Anzahl der not-Prädikate ungerade, dann ist die Polarität negativ, ist die Anzahl der not-Prädikate gerade, ist die Polarität positiv.[8]

Abschließend stellen wir nun die Möglichkeiten für inkrementelles Überprüfen vor, beschränken uns hier aber auf das Einfügen von Literalen und relationale Datenbanken.

- Das Einfügen eines Atoms $p(t_1, \ldots, t_n)$ kann nur dann eine Integritätsbedingung verletzen, wenn es mit einem Atom negativer Polarität unifiziert.
- Modifikationen eines Attributes $X$ einer Relation $R$ können auf den Wahrheitwert einer IC-Formel $F$ keinen Einfluß haben, wenn folgende drei Bedingungen erfüllt sind:
  - Jede Variable, die für das Attribut $X$ steht, kommt nur einmal in $F$ vor.
  - Der alte Wert von $X$ ist nicht gleich dem Wert von $X$ in einem Vorkommen der Relation $R$ in einem Literal positiver Polarität.
  - Der neue Wert von $X$ ist nicht gleich dem Wert von $X$ in einem Vorkommen der Relation $R$ in einem Literal negativer Polarität.

Der eigentliche Vorgang des Wiederüberprüfens nach eventuell die IC-Formel verletzenden Änderungen, erfolgt weitestgehend nach der Vorgehensweise von Nicolas:

Mit Hilfe der möglicherweise die Integrität verletzenden Änderungen wird eine neue IC-Formel generiert (mit *ic_insert* und *ic_delete* bezeichnet), die den vereinfachten Integritätsbedingungen von Nicolas entsprechen. Es werden gerade alle in Frage kommenden allquantifizierten Variablen, die vor allen existenzquantifizierten stehen, instanziiert. Dies entspricht (s.o.) gerade den Variablen, die in den Atomen $A_1, \ldots, A_m$ vorkommen. Das entsprechende Atom kann durch *true* ersetzt werden und die Formel wird gemäß den Absorptionsregeln vereinfacht, um dann neu evaluiert zu werden.

Falls in der Datenbank Regeln vorliegen, wird auf den IC-Formeln eine Abflachungsoperation durchgeführt: Die Regeln werden anstelle der durch sie definierten Prädikate sukzessive eingesetzt.

Zusammenfassend gibt dieser Ansatz zu folgenden Kritikpunkten anlaß:

[8] Im Originalpapier wird statt dieser Definition eine „Level"-Definition benutzt (ohne explizit auf Negationen einzugehen), die aber im Grunde nichts anderes aussagt. Die Polarität eines Literales bestimmt dessen Relevanz für Änderungen und wird auch in anderen Ansätzen benutzt.

- Die Wahl der Syntax für die Integritätsbedingungen (die IC-Formeln) wirkt wenig problemangemessen. Von der mangelnden Vertrautheit abgesehen wirken IC-Formeln künstlich und es ist nicht zu sehen, wie eine gegebene (natürlichsprachliche) Integritätsbedingung als IC-Formel formuliert werden kann.
- IC-Formeln besitzen nur eine eingeschränkte Ausdrucksstärke: selbst einfache Formeln wie $\exists\, x\; q(x) \vee p(x)$ sind nicht als IC-Formeln formulierbar.

Allerdings ist mit diesem Artikel der einzige uns bekannte Ansatz vertreten, der auch Modifikationen betrachtet. Damit sind Situationen feststellbar, in denen eine Integritätsbedingung trotz scheinbarer Verletzung ihre Gültigkeit behält.

### 3.2.3 Integrity Constraints Checking in Deductive Databases

*Antoni Olivé*

Dieser in [Olivé, 1991] veröffentlichte Ansatz übersetzt in einem vorbereitenden Schritt die gesamte Datenbank. Dabei wird die deduktive Datenbank als konsistent angesehen, wenn eine (bereits in Klauselform vorliegende) Integritätsbedingung *IC nicht* aus der Datenbank per SLDNF-Resolution ableitbar ist. Sei $D$ die deduktive Datenbank, $d$ eine Datenbankänderung und $D'$ die resultierende Datenbank. Hierzu werden zunächst zu jedem Prädikatensymbol $p$ aus $D$ drei neue Prädikatensymbole $p'$, $tp$ und $\delta p$ eingeführt:

1. $p'$ entspricht dem Prädikatensymbol $p$ in der deduktiven Datenbank $D'$,
2. Die Prädikatensymbole $tp$ und $\delta p$ werden durch folgende Äquivalenzen eingeführt:
   (a) $\forall\, x\; (tp(x) \leftrightarrow p'(x) \wedge \neg p(x))$
   (b) $\forall\, x\; (\delta p(x) \leftrightarrow p(x) \wedge \neg p'(x)$

Jedes $x$ ist dabei als Vektor von Variablen gemäß der Stelligkeit von $p$ bzw. $p'$ aufzufassen. Die Prädikatensymbole $tp$ und $\delta p$ erfassen also einen Gültigkeitswechsel von $p$ in $D$ und $D'$. Damit kommt man durch einfache logische Umformungen zu folgenden Formeln:

1. $\forall\, x\; (p'(x) \leftrightarrow (p(x) \wedge \neg\delta p(x)) \vee tp(x))$
2. $\forall\, x\; (\neg p'(x) \leftrightarrow (\neg p(x) \wedge \neg tp(x)) \vee \delta p(x))$

Die Datenbank $D$ wird nun folgendermaßen verändert: In jeder Programmklausel wird jedes Literal abhängig von der Polarität durch eine der beiden obigen Formeln ersetzt. Falls z.B. das Literal $\neg q$ in einer Programmklausel auftaucht, so wird $\neg q$ durch $(\neg q(x) \wedge \neg tq(x)) \vee \delta p(x)$ ersetzt. Zusätzlich werden für jedes Prädikatensymbol $p$ in $D$ die Programmklauseln $tp(x) \leftarrow p'(x) \wedge \neg p(x)$, sowie $\delta p(x) \leftarrow p(x) \wedge \neg p'(x)$ hinzugefügt.

Faßt man nun eine Datenbankänderung $d$ als Hinzufügen von $tp$ und Löschen von $\delta p$ auf, so kann mit einer SLDNF-Anfrage von $D \cup \{\leftarrow tIC\}$ nach der Gültigkeit der Integritätsbedingung $IC$ gefragt werden[9].

Mit der angegebenen Methode ist es weiterhin möglich, dynamische Integritätsbedingungen zu formulieren, d.h. Integritätsbedingungen, die Aussagen über Zustandswechsel von Datenbanken machen.

### 3.2.4 Selective Refutation of Integrity Constraints in Deductive Databases

*P. Asirelli, C. Billi, and P. Inverardi*

Dieser Ansatz aus [Asirelli et al., 1989] behandelt nur Hornklauselprogramme zusammen mit normalen Zielklauseln in der Erfüllbarkeitssichtweise, d.h. die Datenbank erfüllt eine Integritätsbedingung $IC$ genau dann, wenn $comp(D) \cup \{IC\}$ erfüllbar ist. Dazu muß gezeigt werden, daß $comp(D) \cup \{\neg IC\}$ einen endlich gescheiterten SLDNF-Baum besitzt.

Die Grundidee hierbei ist, daß für die Gültigkeit einer definiten Zielklausel nach einer Datenbankänderung nur solche SLDNF-Pfade relevant sind (d.h. Einfluß auf die Gültigkeit haben können), in denen eine Klausel aus der Datenbankänderung für einen Beweisschritt benutzt wird. Zu diesem Zweck wird die Berechnungsregel, die aus einer Zielklausel ein Literal auswählt, so modifiziert, daß solche Berechnungen bevorzugt werden. Falls alle Berechnungen, die nun neu möglich sind, fehlschlagen, so ist die Gültigkeit der Integritätsbedingung gewährleistet.

Der Ansatz ist etwas problematisch: Durch die Festlegung auf definite Programme ist man in der Ausdrucksstärke der Integritätsbedingungen sehr festgelegt. Beispielsweise sind keine ausdrucksstärkeren allquantifizierten Integritätsbedingungen realisierbar, da in diesem Fall praktisch immer ein nichtdefinites (normales) Programm aus der Integritätsbedingung generiert wird.

[9] Natürlich vorausgesetzt, die Integritätsbedingung ist auch in die bereits übersetzten Programmklauseln der Datenbank transformiert worden.

### 3.2.5 Integrity Constraint Checking for Deductive Databases

*Ulrike Griefahn und Stefan Lüttringhaus*

Beschreibungen des folgenden Verfahrens sind zu finden in [Griefahn, 1989, Griefahn und Lüttringhaus, 1990, Cremers et al., 1994]. Statt wie die meisten anderen Verfahren von den Änderungen ausgehend Modelldifferenzen zu berechnen, wird hier (zumindest teilweise) die Struktur des Beweises selbst betrachtet. Die zentrale Datenstruktur dieses Verfahrens ist ein UND-ODER-Baum (siehe [Nilsson, 1980]). Der UND-ODER-Baum wird hier zur Darstellung für die Teile eines SLDNF-Beweises benutzt, die unabhängig voneinander ausgewertet werden können. Die Abschnitte werden als *unabhängige Segmente* bezeichnet. Hierzu zunächst einige Definitionen:

**Definition 3.2.3** **($\Delta^+, \Delta^-$)** *Sei $\Sigma = \langle \mathbf{C}, \mathbf{F}, \mathbf{P}, \mathbf{V} \rangle$ eine Signatur, D ein normales Programm bezüglich $\Sigma$ und $d = \langle Del, Add \rangle$ eine Datenbankänderung. Für eine Klauselmenge R bezeichnet pred(R) die Menge aller Prädikatensymbole, die in den Köpfen der Klauseln aus R auftreten. Die Mengen $\Delta^+ \subseteq (\bigcup_{P_n \in \mathbf{P}} P_n)$ und $\Delta^- \subseteq (\bigcup_{P_n \in \mathbf{P}} P_n)$ enthalten alle Prädikatensymbole, die von der Datenbankänderung d betroffen sein können.*

$$\begin{aligned}\Delta^+ &= pred(Add) \cup \{p \mid q \in pred(Add) \wedge p >_{+1} q\} \\ &\qquad \cup \{p \mid q \in pred(Del) \wedge p >_{-1} q\} \\[1em] \Delta^- &= pred(Del) \cup \{p \mid q \in pred(Del) \wedge p >_{+1} q\} \\ &\qquad \cup \{p \mid q \in pred(Add) \wedge p >_{-1} q\}\end{aligned}$$

*Die Abhängigkeiten zwischen den Prädikaten beziehen sich dabei auf die Datenbank D, nicht auf die resultierende Datenbank D'.*

**Definition 3.2.4** *Sei A ein Atom mit $pred(A) = p$ und $L_1 \wedge \ldots \wedge L_n$ eine Konjunktion von Literalen und $d = \langle Del, Add \rangle$ eine Datenbankänderung zu einer deduktiven Datenbank D. Für $\pi \in \{+, -\}$ ist $\pi^{-1}$ invers zu $\pi$, d.h. $(+)^{-1} = -$ und $(-)^{-1} = +$. Weiter sei $d^- = Del$ und $d^+ = Add$. Wir definieren:*

1. *$A \in \Delta^\pi$ gdw. $p \in \Delta^\pi$*
2. *$\neg A \in \Delta^\pi$ gdw. $A \in \Delta^{\pi^{-1}}$*
3. *$(L_1 \wedge \ldots \wedge L_n) \in \Delta^\pi$ gdw. es ein $i \in \{1, \ldots, n\}$ gibt, so daß $L_i \in \Delta^\pi$*

**Lemma 3.2.1** *Sei D ein normales Programm und $d = \langle Del, Add \rangle$ eine Datenbankänderung zu D mit resultierender Datenbank D'. Sei W eine Konjunktion von Literalen und*

*S bzw. S′ die Menge der Antwortsubstitutionen für D bzw D′ und ← W. Dann gilt:*

$$\begin{aligned} S' \setminus S \neq \emptyset &\implies W \in \Delta^+ \\ S \setminus S' \neq \emptyset &\implies W \in \Delta^- \end{aligned}$$

**Definition 3.2.5 ($\delta^+, \delta^-$)** *Sei D ein normales Programm und $d = \langle Del, Add \rangle$ eine Datenbankänderung von D. Für Konjunktionen von Literalen W definieren wir die Mengen $\delta^\pi(W)$ mit $\pi \in \{+, -\}$ der möglichen neuen bzw. fehlenden Antwortsubstitutionen. Dabei bezeichnen wir Del mit $d^-$ und Add mit $d^+$.*

$$\begin{aligned} \delta^\pi(A) &= \{\sigma \mid (B \leftarrow W) \in d^\pi \quad \text{und} \quad \sigma = mgu(A, B)\} \\ &\cup \{\sigma_1\sigma_2 \mid (B \leftarrow W) \in D, \sigma_1 = mgu(A, B) \quad \text{und} \\ &\qquad \sigma_2 \in \delta^\pi(W)\} \\ \\ \delta^\pi(\neg A) &= \{\sigma \mid \sigma \in \delta^{\pi^{-1}}\} \\ \\ \delta^\pi(L_1 \wedge \ldots \wedge L_n) &= \{\sigma \mid \quad \text{es gibt ein } L_i \text{ mit } \sigma \in \delta^\pi(L_i)\} \end{aligned}$$

**Lemma 3.2.2** *Sei D ein normales Programm und $d = \langle Del, Add \rangle$ eine Datenbankänderung zu D mit resultierender Datenbank D′. Sei W eine Konjunktion von Literalen und S bzw. S′ die Menge der korrekten Antwortsubstitutionen für D bzw. D′ und ← W. Dann gilt:*

$$\begin{aligned} \sigma \in (S' \setminus S) &\implies \text{es gibt ein } \theta \in \delta^+(W) \text{ und eine Subsitution } \theta' \text{ mit } \theta\theta' = \sigma \\ \sigma \in (S \setminus S') &\implies \text{es gibt ein } \theta \in \delta^-(W) \text{ und eine Subsitution } \theta' \text{ mit } \theta\theta' = \sigma \end{aligned}$$

Lemma 3.2.2 zeigt, wie die Substitutionsmengen bei der Integritätsprüfung eingesetzt werden können. Wenn für eine Konjunktion von Literalen neue Antworten hinzukommen, kann $\delta^+$ nicht leer sein. Wenn $\delta^+$ leer ist, sind keine neuen Antwortsubstitutionen hinzugekommen. Analoges gilt für $\delta^-$.

**Definition 3.2.6 (Minimales unabhängiges Segment)** *Sei $G \equiv \leftarrow L_1 \wedge \ldots \wedge L_n$ eine Zielklausel. Wir verwenden in dieser Definition die Mengenschreibweise für Klauseln und fassen G daher als Menge $\{L_1, \ldots, L_n\}$ auf.*

1. *Ein* Segment *G′ von G ist eine nichtleere Teilmenge von G.*

2. *Ein Segment G′ von G heißt* unabhängig, *wenn es mit den restlichen Literalen von G keine gemeinsamen Variablen hat:*

$$vars(G') \cap vars(G \setminus G') = \emptyset$$

3. *Ein Segment $G'$ von $G$ heißt* minimal*, wenn es nicht in zwei unabhängige Segmente zerlegt werden kann.*

Minimale unabhängige Segmente sind offensichtlich Teile einer Klausel, die unabhängig vom Rest der Klausel ausgewertet werden können. Somit sind wir nun in der Lage, die zentrale Datenstruktur des Verfahrens zu definieren. Um die minimalen unabhängigen Segmente in einer Klausel voneinander abzugrenzen, verwenden wir das Zeichen „&".

**Definition 3.2.7 (UND-ODER-Baum)** *Sei $D$ ein normales Programm und $G$ eine Zielklausel. Ein* UND-ODER-Baum *für $D$ und $G$ ist ein Baum mit zwei unterschiedlichen Knotenarten, UND-Knoten sowie ODER-Knoten, mit den folgenden Eigenschaften:*

1. *Jeder UND-Knoten ist eine Zielklausel.* [10]

   (a) *Die Wurzel des UND-ODER-Baumes ist der UND-Knoten $G$.*

   (b) *Sei $W' \equiv S_1 \& \ldots \& S_n$ mit $n > 0$ ein UND-Knoten, wobei die $S_i$ die minimalen unabhängigen Segmente von $W'$ sind. $W'$ hat $n$ Söhne, die ODER-Knoten sind.*

   *Ist $S_i$ ein positives Literal oder eine Konjunktion von mindestens zwei Literalen, dann ist $\leftarrow S_i$ ein Sohn von $W'$.*

   *Ist $S_i \equiv \neg A$ ein negatives Literal, dann ist $\leftarrow A$ ein Sohn von $W'$. Die Kante von $W'$ nach $\leftarrow A$ wird mit $\neg$ gekennzeichnet.*

   *Die Söhne eines UND-Knotens werden von links nach rechts in der Reihenfolge der unabhängigen Segmente des UND-Knotens angeordnet.*

   (c) *Der UND-Knoten $\Box$ ist stets ein Blatt.*

2. *Jeder ODER-Knoten ist eine Zielklausel, die ein minimales Segment ist.*

   (a) *Sei $G' \equiv \leftarrow A$ ein ODER-Knoten, wobei $A$ ein Atom ist, in dem keine Variable mehrfach auftritt.*

   *Dann hat $G'$ Söhne, die UND-Knoten sind. Der UND-Knoten $\leftarrow W\sigma$ ist ein Sohn von $G'$ gdw. eine normale Klausel $(B \leftarrow W) \in D$ existiert und $\sigma = mgu(A, B)$ ist.*

---

[10] Wir identifizieren wieder Knoten und Zielklauseln.

*Die Söhne eines ODER-Knotens werden von links nach rechts in Abhängigkeit von der zugrundeliegenden SLDNF-Widerlegungsprozedur angeordnet.*

(b) *Jeder ODER-Knoten $G \equiv \leftarrow S$, wobei $S$ ein minimales Segment ist, in dem mindestens eine Variable mehrfach auftritt, ist ein SLDNF-Knoten und somit ein Blatt.*

**Beispiel 3.2.2** *Sei $D$ folgendes normale Programm:*

$$\begin{array}{rcl} p(x) & \leftarrow & q(x,x) \\ p(x) & \leftarrow & \neg s(a) \\ r(x) & \leftarrow & u(x) \wedge \neg v(x) \\ t(a) & \leftarrow & \\ q(a,a) & \leftarrow & \end{array}$$

*Der UND-ODER-Baum für $D$ und die Zielklausel $\leftarrow p(x)\&t(y)\&r(z)$ ist der folgende:*

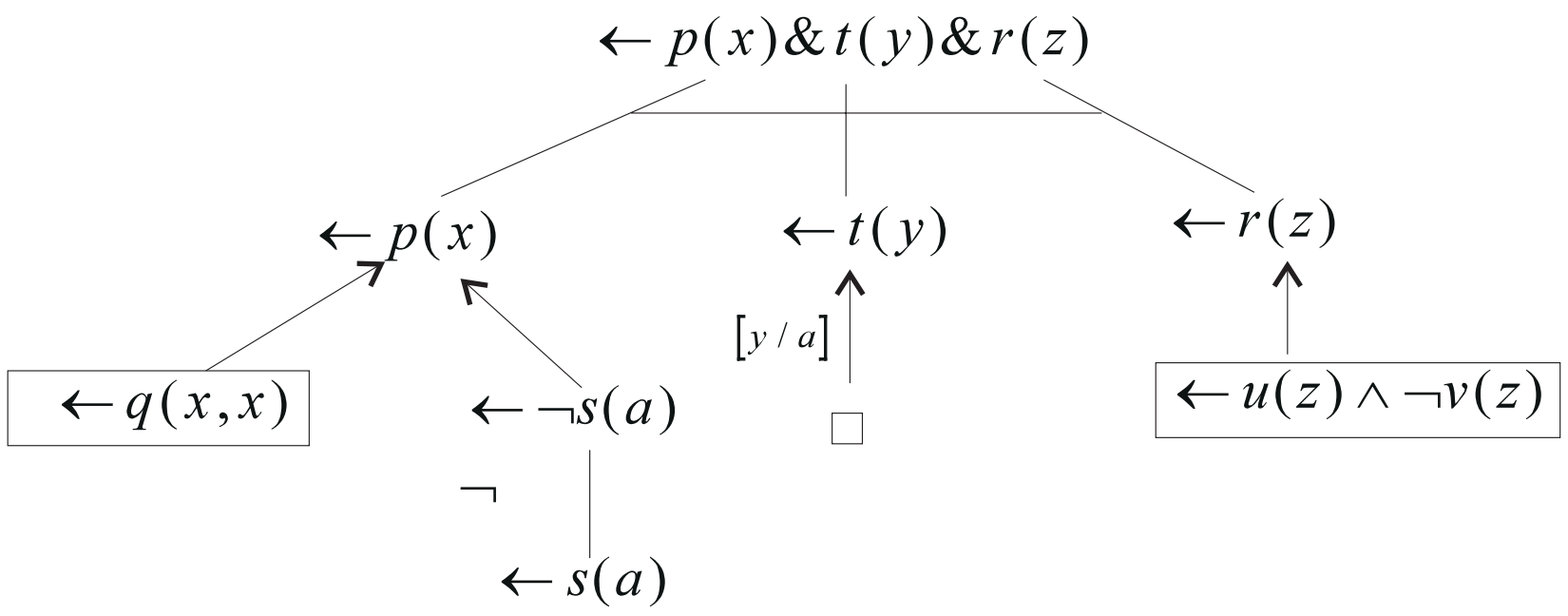

Abbildung 3.1: UND-ODER-Baum

Knoten nach Definition 3.2.7 Punkt 2b heißen auch *SLDNF-Knoten*, da deren Auswertung durch SLDNF-Resolution geschieht. Ein wichtiger Unterschied, der uns auch später noch beschäftigen wird, ist die Behandlung der Negation: die Negation ist hier direkt im Baum integriert, ganz im Gegensatz zu SLDNF-Bäumen, wo mit jeder Negation ein neuer SLDNF-Baum aufgebaut wird. Um diesen UND-ODER-Baum zu einem *Beweis* zu machen, benötigen wir noch die Definition der *Markierung*.

**Definition 3.2.8 (Markierung von UND-ODER-Bäumen)** *Sei $D$ ein normales Programm, $G$ eine Zielklausel und $T$ der UND-ODER-Baum für $D$ und $G$. Eine* Markierung *für $T$ ist eine partielle Abbildung aus der Menge der Knoten von $T$ in die Menge $\{s, f\}$. Eine Markierung $\Gamma$ für $T$ wird* endlich *genannt gdw. $\Gamma$ eine endliche Menge von Knoten markiert.*

Unter der Annahme, daß „$s$" für „success " und „$f$" für „failure" steht und wir die Knoten eines UND-ODER-Baumes so markieren, wie die unterliegende SLDNF-Inferenzmaschine die Knoten auswertet, so können wir aus dem UND-ODER-Baum einen Beweis machen. Die formalen Bedingungen hierzu lauten:

**Definition 3.2.9 (Korrekte Markierung)** *Sei $D$ ein normales Programm und $G$ eine Zielklausel. Der UND-ODER-Baum für $D$ und $G$ hat eine korrekte Markierung, wenn diese endlich ist und die folgenden Eigenschaften erfüllt:*

1. *Die Wurzel von $T$ ist markiert.*
2. *Alle Söhne $W'$ eines $s$–markierten UND-Knotens $W$ haben die Markierung $s$, wenn die Kante von $W$ nach $W'$ nicht mit $\neg$ versehen ist, und $f$, wenn sie mit einem $\neg$ versehen ist.*
3. *Jeder markierte UND-Knoten $\Box$ ist mit $s$ markiert.*
4. *Mindestens ein Sohn $W'$ eines $f$–markierten UND-Knotens $W$ hat die Markierung $f$, wenn die Kante von $W$ nach $W'$ nicht mit $\neg$ versehen ist, oder $s$, wenn die Kante mit einem $\neg$ versehen ist.*
5. *Ein SLDNF-Knoten $G$ hat die Markierung $s$ (falls er überhaupt markiert ist), wenn es eine SLDNF-Widerlegung von $D$ und $G$ gibt. Er ist mit $f$ markiert, wenn ein endlicher SLDNF-Mißerfolgsbaum für $D$ und $G$ existiert.*
6. *Alle Söhne eines $f$-markierten ODER-Knotens haben die Markierung $f$.*
7. *Mindestens ein Sohn eines mit $s$ markierten ODER-Knotens hat die Markierung $s$.*

Damit kommen wir zu folgendem Lemma:

**Lemma 3.2.3** *Sei $D$ ein normales Programm und $G$ eine Zielklausel. Es existiert genau dann eine korrekte Markierung für den UND-ODER-Baum von $D$ und $G$, bei der die Wurzel mit $s$ (bzw. $f$) markiert ist, wenn es eine SLDNF-Widerlegung (bzw. einen endlichen SLDNF-Mißerfolgsbaum) für $D$ und $G$ gibt.*

Damit ist klar, daß ein UND-ODER-Baum mit korrekter Markierung im Griefahnschen Sinne äquivalent zu der Existenz einer SLDNF-Widerlegung bzw. eines endlich gescheiterten SLDNF-Baumes ist. Damit sind integritätsüberprüfende Algorithmen auf UND-ODER-Bäumen definierbar.

Da wir nur unabhängige Segmente betrachten, gibt es für jeden Knoten eine eindeutige Antwortsubstitution. Dies wird ausgenutzt, um die Algorithmen zur Überprüfung der Integrität zu optimieren. Dabei nehmen wir an, daß die Antwortsubstitutionen wie in Abbildung 3.1 an den Knoten gespeichert werden.

Den Vorgang der Integritätsprüfung machen wir an einem Beispiel deutlich:

**Beispiel 3.2.3** *Sei D das folgende normale Programm:*

$$\begin{array}{rcl} p(x) & \leftarrow & \neg s(b) \\ p(x) & \leftarrow & r(x,y) \wedge q(x,y) \\ r(a,c) & \leftarrow & \\ q(a,c) & \leftarrow & \\ s(b) & \leftarrow & \end{array}$$

*sowie ic die Integritätsbedingung* $p(a)$*. Ein korrekt markierter UND-ODER-Baum zu D und* $\leftarrow p(a)$ *ist der folgende:*

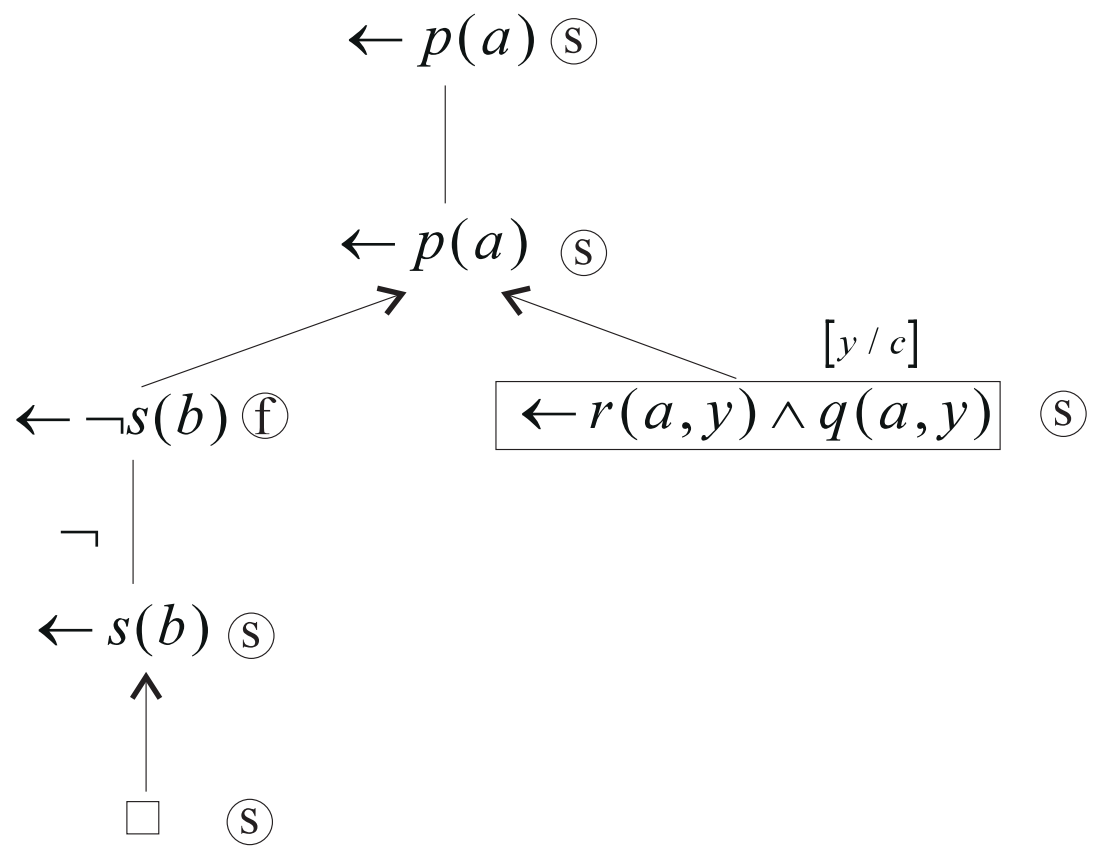

Abbildung 3.2: UND-ODER-Baum zu $D \cup \{\leftarrow p(a)\}$

*Sei* $d = \langle Del, Add \rangle$ *eine Datenbankänderung auf D mit*

- $Del = \{r(a,c), r(a,c), s(b)\}$
- $Add = \{r(a,e), q(a,f)\}$

*und resultierender Datenbank* $D'$*. Die Markierungen der Knoten des UND-ODER-Baumes stimmen nun nicht mehr, sie werden neu markiert. Dazu werden für alle Knoten die notwendigen* $\Delta^+$ *und* $\Delta^-$*-Werte bezüglich der Markierung berechnet. So ist* $p(x)$ *sowohl Element von* $\Delta^+$ *als auch Element von* $\Delta^-$*, d.h. die Integrität ist möglicherweise verletzt. Beginnen wir zunächst mit dem SLDNF-Knoten* $\leftarrow r(a,y) \wedge q(a,y)$*: dieser Knoten ist Element von* $\Delta^-$*. Die Berechung von* $\delta^-$ *ergibt, daß die bisherige Lösungssubstitution nicht mehr gültig ist. Da der SLDNF-Knoten aber auch Element von* $\Delta^+$ *ist, gibt es möglicherweise neue Lösungen. Dies wird per SLDNF-Resolution überprüft. Da es keine neuen Lösungen gibt, wird der Knoten mit einem* $f$ *markiert.*

*Eine andere Situation ergibt sich bei dem Knoten* $\neg s(b)$*: dieser ist Element von* $\Delta^+$*, d.h.* $s(b)$ *ist Element von* $\Delta^-$*. Da* $s(b) \in \delta^-$ *wird der Knoten* $s(b)$ *mit einem* $f$ *markiert und die Markierungen entsprechend nach oben propagiert. Damit ergibt sich folgender neuer korrekt markierter UND-ODER-Baum:*

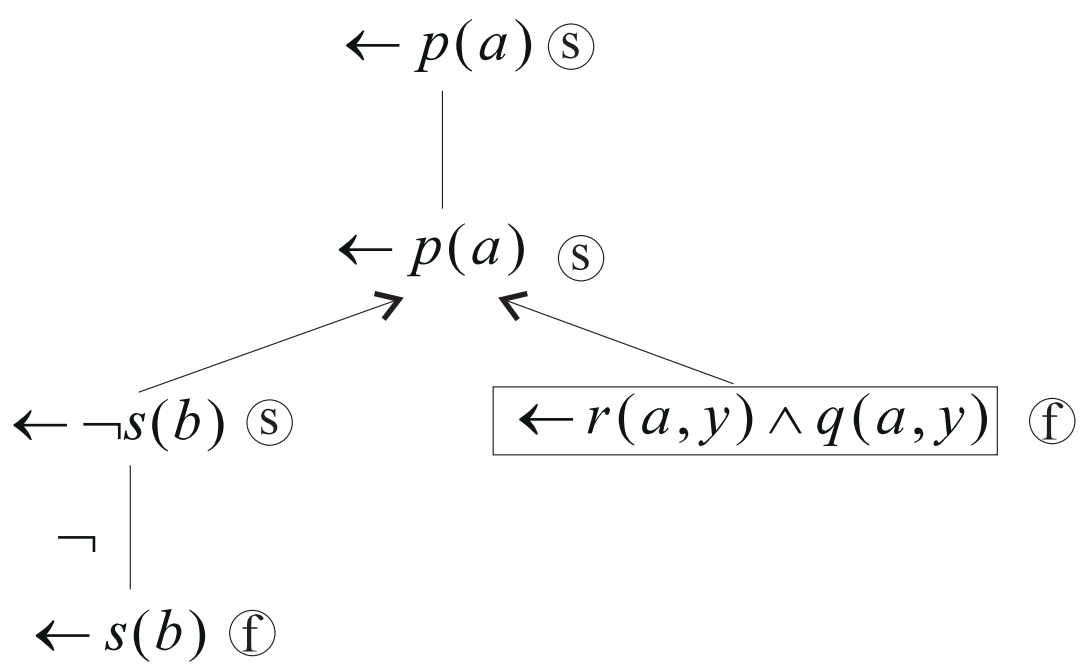

Abbildung 3.3: Neuer UND-ODER-Baum zu $D' \cup \{\leftarrow p(a)\}$

Die bei den Knoten gespeicherten Substitutionen vereinfachen im Normalfall die Berechnung von $\delta^+$ und $\delta^-$.

## 3.3 Diskussion

Die meisten der bisherigen Verfahren beruhen auf der Simplifikationsmethode von Nicolas, verbunden mit einer Art von Vorwärtsberechnung, um von den Datenbankänderungen zu den Integritätsbedingungen zu kommen. Die dabei verwendete Information ist die *Existenz* eines Beweises, eventuell vorliegende Informationen über die *Struktur* (z.B. benutzte Klauseln etc.) eines Beweises werden i.a. nicht genutzt. Zu dieser Klasse gehören die Verfahren von Lloyd, Bry et al, Asirelli et al, Williams et al, Griefahn und Lüttringhaus. Die Ansätze haben den Nachteil, daß

1. durch Vorwärtsberechnung im allgemeinen zu viel berechnet wird,

2. durch die unzureichende Ausnutzung von Instanziierungen viel zu allgemeine Bedingungen überprüft werden,
3. nicht auf die Struktur des Beweises einer Integritätsbedingung Rücksicht genommen wird.

Der Griefahn-Ansatz nutzt noch am ehesten die Struktur eines Beweises durch die Generierung eines UND-ODER-Baumes aus. Dies bringt in der Praxis aber im allgemeinen nicht viel. So ist in [Cremers et al., 1994, S. 273] zu lesen:

> *In der Praxis verwendete Integritätsbedingungen enthalten vielfach Allquantoren. Durch die Normalisierung dieser Bedingungen erhält man Programme, die in den Und-Oder-Bäumen sehr früh SLDNF-Knoten erzeugen* [...]. *Da SLDNF-Knoten nicht expandiert werden, sind die Und-Oder-Bäume in der Regel nicht sehr groß.*

Dies bedeutet, daß auch in diesem Verfahren vielfach vorhandene Informationen bezüglich der Beweisstruktur einer Integritätsbedingung nicht ausgenutzt werden, da die Bäume meist sehr klein sind und somit die Beweisstruktur kaum wiederspiegeln. Verwendet man relativierte Integritätsbedingungen, dann bestehen die im Griefahnschen Verfahren verwendeten Bäume sogar nur aus einem Knoten. Im folgenden stellen wir einen Ansatz vor, der versucht, einen SLDNF-Beweis in geeigneter Weise zu strukturieren und diese Struktur zur Integritätsüberprüfung einzusetzen.

# Kapitel 4

# Unser Ansatz - ein Beispiel

Dieses Kapitel dient dem Zweck, beim Leser ein intuitives Verständnis für unseren Ansatz zu wecken. Aus diesem Grunde wird hier keine formale Definition des Verfahrens oder der verwendeten Datenstrukturen gegeben, sondern es wird zunächst versucht, die Konstruktion und die prinzipielle Vorgehensweise anhand eines Beispieles zu verdeutlichen.

## 4.1 UND-ODER-Bäume

Gegeben sei folgendes logisches Programm $D$, in dem die extensionalen Relationen $employee$, $manager$, $owner$, $classification$ und $clearance$ sowie eine intensionale Relation $access$ definiert sind. Die $access$-Relation ist definiert durch drei Programmklauseln und bestimmt, welcher Arbeitnehmer ($employee$) Zugriff auf Dateien hat (die einzige Datei ist hier der Speiseplan ($menu$)). Das gesamte Programm beschreibt somit die Zugriffsrelationen auf ein Dateisystem in einer Rechnerumgebung.

$$
\begin{array}{lcl}
access(E,F) & \leftarrow & owner(E,F) \\
access(E,F) & \leftarrow & manager(E,E_2) \wedge owner(E_2,F) \\
access(E,F) & \leftarrow & classification(F,C_1) \wedge clearance(E,C_2) \wedge C_1 \leq C_2 \\
employee(hans) & \leftarrow & \\
employee(peter) & \leftarrow & \\
owner(hans, menu) & \leftarrow & \\
manager(peter, hans) & \leftarrow & \\
clearance(hans, 1) & \leftarrow & \\
clearance(peter, 2) & \leftarrow &
\end{array}
$$

$D$ soll folgende Integritätsbedingung erfüllen:

$$\forall E{:}employee\ access(E, menu)$$

Das vorstehende Logikprogramm (bzw. dessen Vervollständigung) erfüllt diese Integritätsbedingung genau dann, wenn jeder Arbeitnehmer Zugriff auf den Speiseplan hat. Diese Integritätsbedingung wird durch die in Abschnitt 2.7.5 angesprochene Methode in folgende Menge von Programmklauseln übersetzt.

$$
\begin{array}{lcl}
ic & \leftarrow & \neg p_1 \\
p_1 & \leftarrow & employee(E) \wedge \neg access(E, menu)
\end{array}
$$

Diese Programmklauseln werden dann zu dem Logikprogramm $D$ hinzugefügt. Um die Integritätsbedingung zu überprüfen, reicht es aus, die Anfrage $\leftarrow ic$ an ein deduktives Datenbanksystem zu stellen. In der vorgestellten Datenbank ist die Integritätsbedingung erfüllt, da Hans Eigentümer des Speiseplans ist (und somit Zugriff hat) und Peter der Manager von Hans ist und deswegen Zugriff auf die Datei hat. Die dritte *access*-Regel wird für eine Widerlegung von $D \cup \{\leftarrow ic\}$ nicht benötigt.

Zu der Zielklausel $\leftarrow ic$ und der Datenbank $D$ kann auf einfache Weise ein Baum wie in Abbildung 4.1 generiert werden. Die Knoten des Baumes bestehen abwechselnd aus UND-Knoten, die durch die Rümpfe von Programmklauseln bzw. der Anfrage gebildet werden und aus ODER-Knoten, die von einzelnen Atomen aus einem Literal des Vorgängerknotens gebildet werden. Ist das zugehörige Literal in dem Vorgängerknoten negiert, so wird die Kante vom UND-Knoten zum ODER-Knoten mit einem $\neg$ markiert.

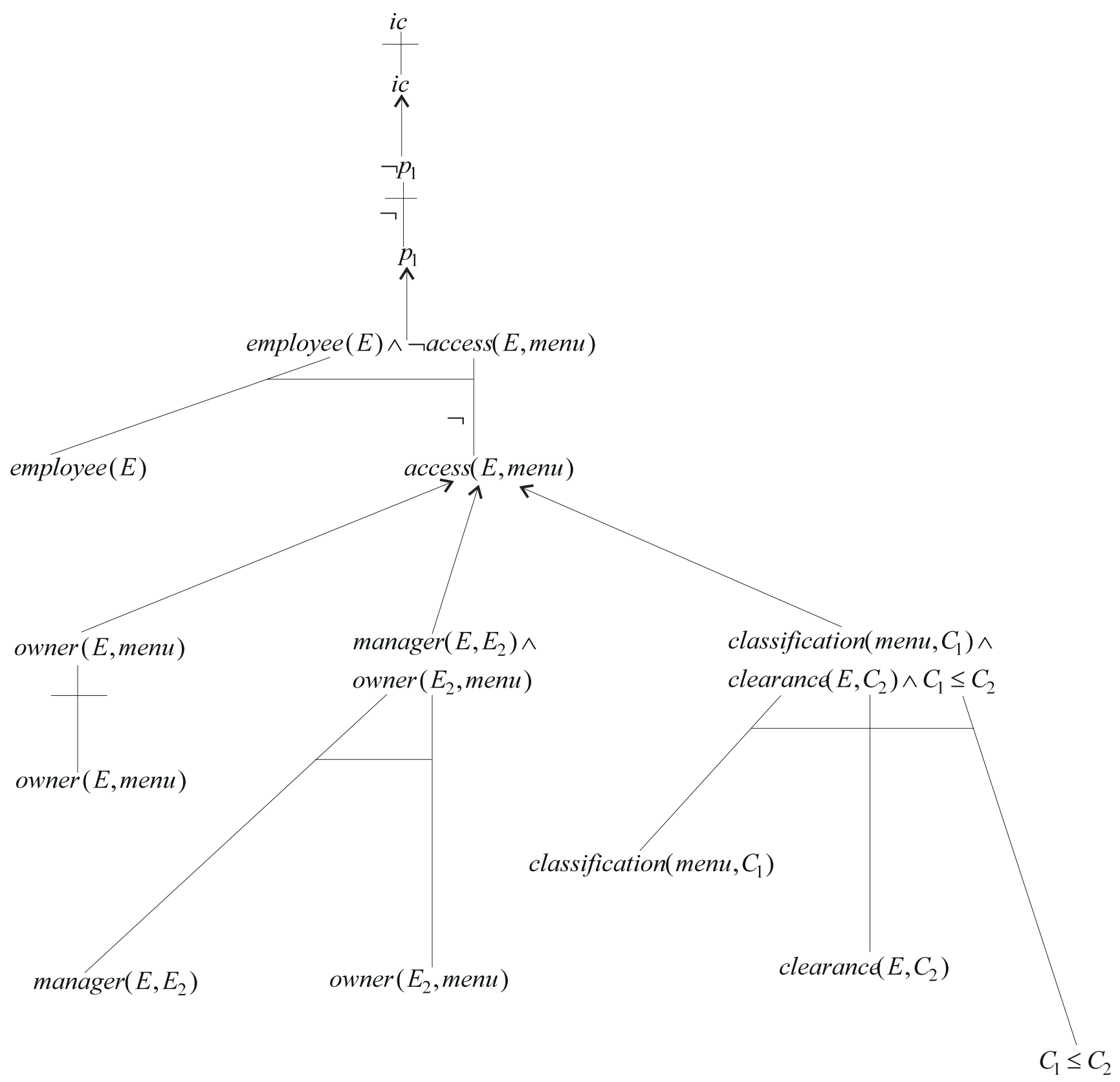

Abbildung 4.1: UND-ODER-Baum zu $D \cup \{\leftarrow ic\}$

In diesem UND-ODER-Baum ist der gesamte für die Anfrage relevante Teil der Datenbank berücksichtigt. Ein Beweissystem (z.B. Prolog) bewegt sich nur innerhalb dieses Baumes. Bei Programmen in denen rekursive Programmklauseln vorkommen, kann der so gebildete UND-ODER-Baum prinzipiell unendlich groß sein. Der LST-Ansatz (siehe Abschnitt 3.1.2) zur Integritätsüberprüfung läßt sich mit Hilfe eines derartig gebildeten UND-ODER-Baumes formulieren: es werden ausgehend von Datenbankänderungen und den Stellen im UND-ODER-Baum, wo diese geänderten Datenbankrelationen vorkommen, Mengen von Substitutionen berechnet. Diese werden von unten nach oben durch den Baum propagiert. Da der Baum prinzipiell unendlich groß ist, kann es unter Umständen unendlich viele Ansatzpunkte im Baum geben, bei der mit der Propagierung zu beginnen ist. Daraus resultiert das Problem, daß die Berechnung im LST-Ansatz nicht notwendigerweise terminiert.

## 4.2 Beweis-Schablonen

Dieses Vorgehen berücksichtigt nicht die Tatsache, daß ein Beweis, der von einem Beweissystem gefunden wurde, endlich sein muß. Es ist immer möglich, einen endlichen Teilbaum aus einem UND-ODER-Baum zu identifizieren, der für einen endlichen Beweis relevant ist. Dieser Teilbaum, den wir im folgenden *Beweisschablone* nennen, ist nicht nur endlich, sondern im allgemeinen auch weniger verzweigt als der vollständige UND-ODER-Baum.

In Abbildung 4.2 wird eine derartige Beweisschablone gezeigt. Die dritte Regel zum *access*-Prädikat ist nicht vorhanden, da sie für den Beweis nicht benötigt wurde.

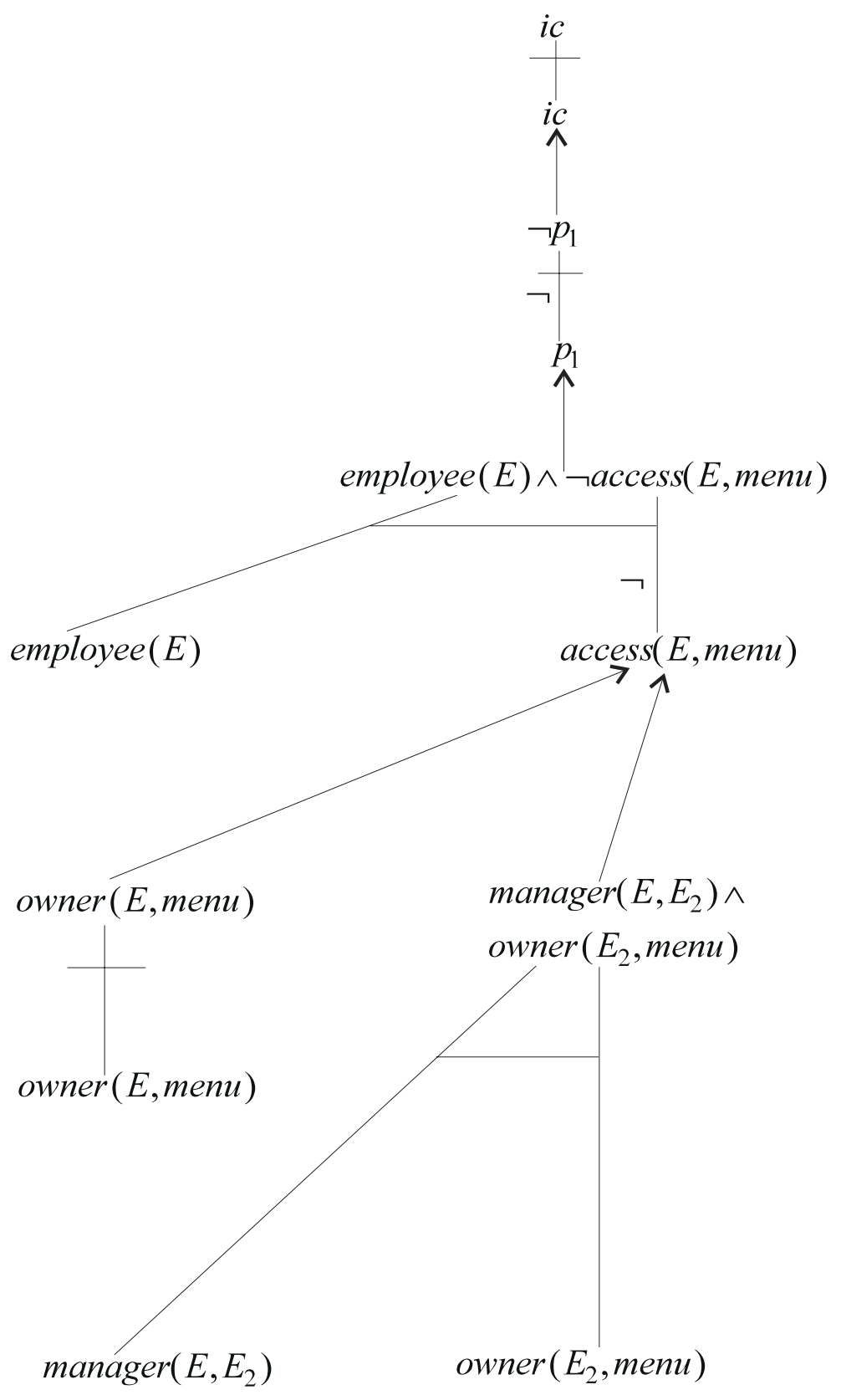

Abbildung 4.2: Relevanter UND-ODER-Baum zu $D \cup \{\leftarrow ic\}$

Ausgehend von dieser Überlegung läßt sich jetzt ein Verfahren analog zum LST-Ansatz definieren, das Datenbankänderungen nur durch die Beweisschablone propagiert und wegen der Endlichkeit der Beweisschablone immer terminiert. Diesen Ansatz verfolgen wir hier nicht weiter, sondern betrachten die Änderungen, die noch nötig sind, um aus einer Beweisschablone einen Beweis zu machen.

## 4.3 Beweisbäume

Abbildung 4.2 ist nur eine Schablone für einen Beweis, da die Substitutionen, die während eines Beweiserlaufes (von z.B. PROLOG) relevant sind, nicht im Baum vorhanden sind. Versehen wir alle Knoten der Beweisschablone noch mit geeigneten Substitutionen, so erhalten wir einen vollständigen Beweis und können bei Datenbankänderungen erkennen, ob diese Änderung den Beweis ungültig macht. Diese ausgefüllte Beweisschablone nennen wir im folgenden einen *Beweisbaum*. Die Substitutionsmengen lassen sich aus einem SLDNF-Baum mittels eines später vorgestellten Verfahrens gewinnen. Der Beweisbaum zu dem eingangs vorgestellten Beispiel ist der folgende:

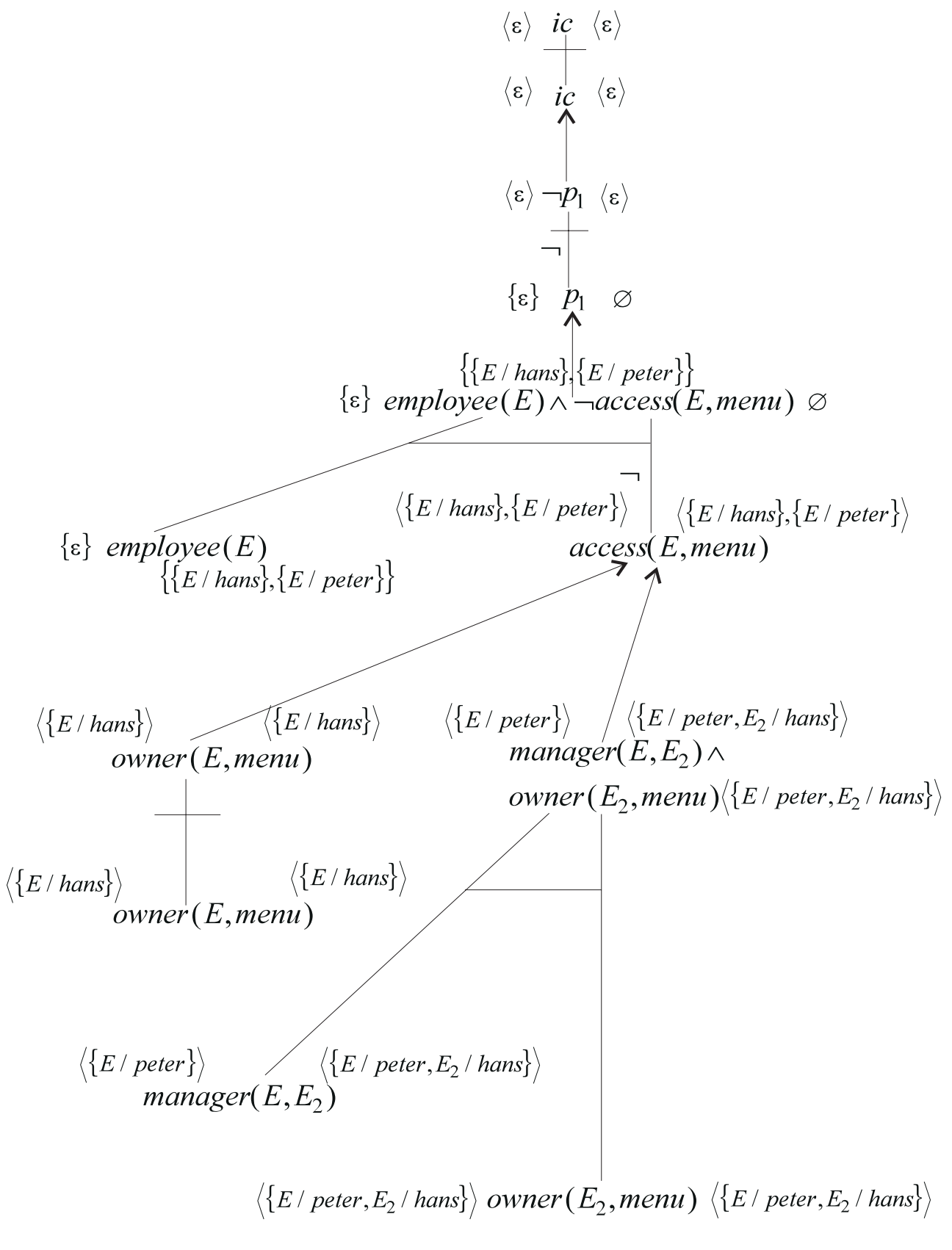

Abbildung 4.3: Beweisbaum für $D \cup \{\leftarrow ic\}$

## 4.4 Änderungen in Datenbanken

Wir machen die Möglichkeiten, die sich durch eine solche Beweisdarstellung bei der Integritätsüberprüfung ergeben, an mehreren Beispielen deutlich. Diese nachfolgenden Beispiele gehen jeweils von der oben angegebenen Datenbank $D$ aus.

1. In der Datenbank $D$ wird das Fakt $clearance(hans, 1)$ gelöscht. Es gibt im Baum keinen ODER-Knoten, der mit $clearance(hans, 1)$ unifiziert d.h. in dem Beweis wird dieses Fakt nicht verwendet. Somit kann der Beweisbaum (und damit der Beweis) nicht ungültig werden.

2. In der Datenbank $D$ werden die beiden Fakten $manager(peter, hans)$ sowie $employee(peter)$ gelöscht. Während nach dem Löschen des ersten Fakts $manager(peter, hans)$ die Integritätsbedingung verletzt ist, wird nach dem Löschen des Fakts $employee(peter)$ die Integritätsbedingung wieder gültig. Dies kann durch das Löschen von Substitutionen, die im Baum vom Fakt $employee(peter)$ abhängig waren, ohne einen Neubeweis erkannt werden. Der daraus resultierende Baum ist in Abbildung 4.4 zu sehen.

3. In der Datenbank $D$ wird das Fakt $manager(peter, hans)$ gelöscht und ein Fakt $classification(menuplan, 1)$ hinzugefügt. Der Beweisbaum wird ungültig, es muß nach einer neuen Begründung für die Integritätsbedingung gesucht werden. Glücklicherweise muß der Beweisbaum nicht vollständig neu aufgebaut werden, sondern kann in großen Bereichen wiederverwendet werden. In diesen Fall wird ein neuer Ast für die dritte $access - Regel$ zu dem Baum hinzugefügt, so daß sich wieder ein gültiger Beweisbaum ergibt. Der nun konstruierte Beweisbaum ist in Abbildung 4.5 zu sehen.

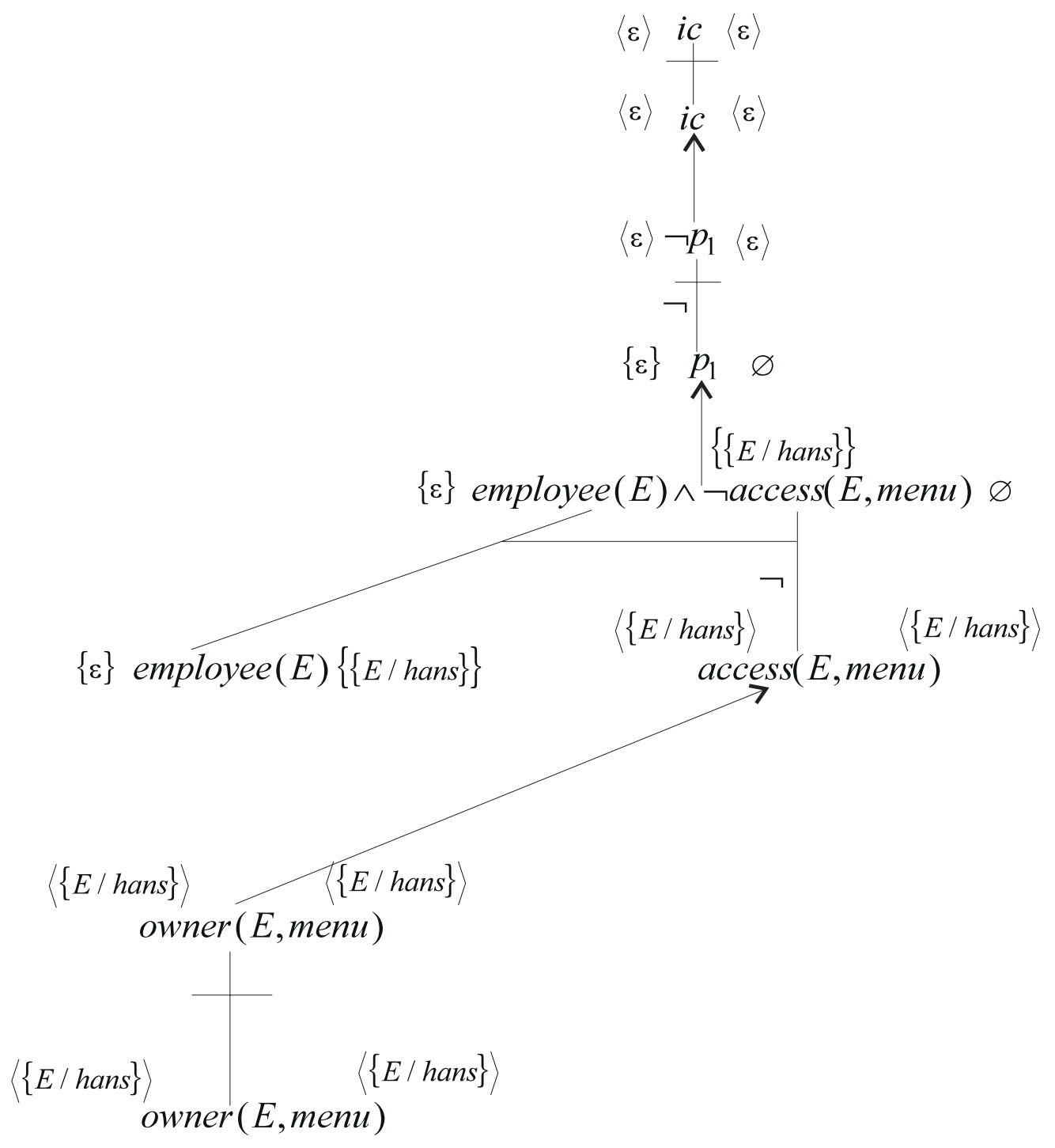

Abbildung 4.4: Beweisbaum zur Anfrage $\leftarrow ic$ in modifizierter Datenbank von Punkt 2

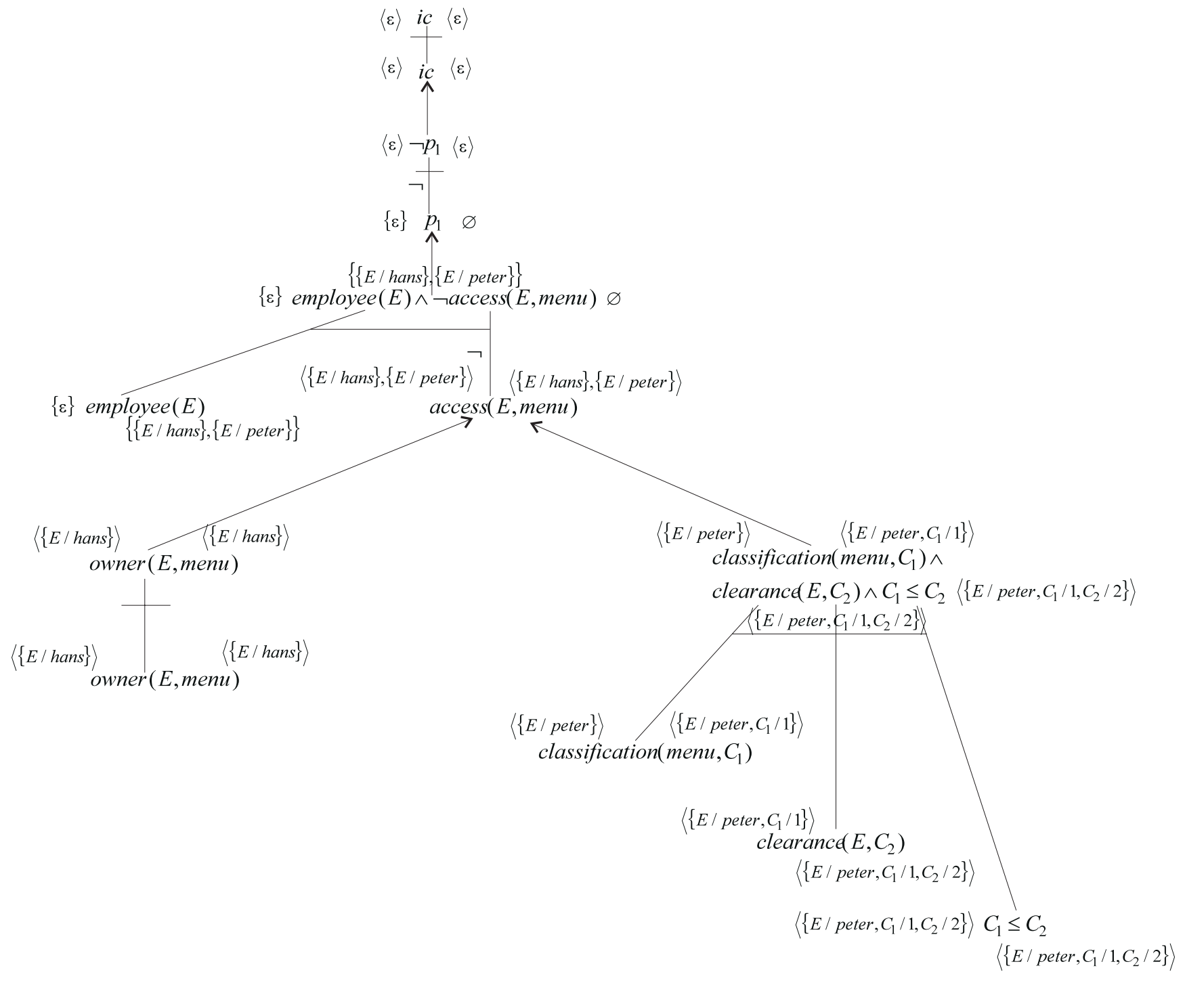

Abbildung 4.5: Beweisbaum zur Anfrage $\leftarrow ic$ in modifizierter Datenbank von Punkt 3

# Kapitel 5

# Komposition von SLDNF-Bäumen

Ein Logikprogrammier-System wie Prolog führt $\leftarrow L_1 \wedge L_2$ aus, indem zunächst $\leftarrow L_1$ ausgeführt und eine Lösungssubstitution $\sigma$ berechnet wird. Darauffolgend wird $\leftarrow L_2\sigma$ ausgeführt und auch hierfür eine Lösungssubstitution berechnet. Bei der Untersuchung der Bedeutung der *Hintereinanderausführung* von zwei Zielen von z.B. Prolog in einem SLDNF-Baum sind folgende Punkte zu betrachten:

1. die Suche nach einer Widerlegung — hierfür benötigt man den Begriff der Komposition von Widerlegungen.

2. die Suche nach einem endlich gescheiterten Mißerfolgsbaum — hierfür benötigt man den Begriff der Komposition von SLDNF-Bäumen. Um dies sinnvoll bewerkstelligen zu können, muß man den Begriff des SLDNF-Baumes erweitern zu dem Begriff des *unvollständigen SLDNF-Baumes*, der eine echte Verallgemeinerung des üblichen Begriffs (siehe Def. 2.7.8) darstellt.

Bemerkenswert ist noch, daß für die folgenden Aussagen nur angenommen wird, daß $\leftarrow L_1$ *vor* $\leftarrow L_2$ ausgewertet wird. Über die verwendete Berechungsregel bei der Auswertung von $\leftarrow L_1$ und $\leftarrow L_2$, also die Berechnungsregel der einzelnen Anfragen, wird nichts vorausgesetzt.

## 5.1 SLDNF-Widerlegungen

Die Komposition von SLDNF-Widerlegungen besteht im wesentlichen aus der Verkettung von zwei einzelnen Widerlegungen.

**Definition 5.1.1 (Komposition von SLDNF-Widerlegungen)** *Sei $D$ ein normales Programm, $\leftarrow G_1, \leftarrow G_2$ (nicht notwendigerweise variablendisjunkte) Zielklauseln, $R_1$ eine SLDNF-Widerlegung von $G_1$ mit berechneter Antwort $\sigma_1$ und $R_2$ eine SLDNF-Widerlegung von $G_2\sigma_1$. Die* Komposition *$R_3$ von $R_1$ und $R_2$ wird folgendermaßen konstruiert:*

- *Jede Zielklausel $\leftarrow G$ von $R_1$ wird zu $\leftarrow G \wedge G_2\theta'$ modifiziert. $\theta'$ ist die relevante Substitution von $\leftarrow G$ in $R_1$. Die selektierten Literale bleiben gleich.*
- *Anstelle der letzten Klausel (= leere Klausel) von $R_1$ wird die Widerlegung $R_2$ angehängt.*

**Beispiel 5.1.1** *Sei $D$ folgendes normales Programm:*

$$\begin{array}{rcl}
p(x) & \leftarrow & t(x) \wedge \neg s(x) \\
r(x,y) & \leftarrow & u(y) \wedge \neg v(x) \\
q(x,y) & \leftarrow & t(y) \wedge u(x) \\
t(x) & \leftarrow & u(x) \\
t(a) & \leftarrow & \\
u(b) & \leftarrow & \\
s(b) & \leftarrow &
\end{array}$$

*und $G_1 \equiv \leftarrow p(x)$ sowie $G_2 \equiv \leftarrow r(x,y)$.*

*Dann ist*

$$R_1 \equiv \leftarrow p(x), \leftarrow t(x) \wedge \neg s(x), \leftarrow \neg s(a), \Box$$

*eine SLDNF-Widerlegung von $D \cup \{G_1\}$ mit berechneter Antwortsubstitution $\sigma_1 = \{x/a\}$ und*

$$R_2 \equiv \leftarrow r(a,y), \leftarrow u(y) \wedge \neg v(a), \leftarrow \neg v(a), \Box$$

*eine SLDNF-Widerlegung von $D \cup \{G_2\sigma_1\}$ mit berechneter Antwortsubstitution $\sigma_2 = \{y/b\}$.*

*Die Komposition $R_3$ von $R_1$ und $R_2$ ergibt sich dann durch folgende Schritte:*

- *Jede Zielklausel* $\leftarrow G$ *von* $R_1$ *wird zu* $\leftarrow G \wedge r(x,y)\sigma$ *modifiziert. Dabei ist* $\sigma$ *die relevante Substitution von* $\leftarrow G$ *in* $R_1$.
- *Anstelle der leeren Klausel von* $R_1$ *wird nun* $R_2$ *angehängt.*

*Es ergibt sich die Komposition* $R_3$ *zu*

$$R_3 \equiv \quad \leftarrow p(x) \wedge r(x,y), \leftarrow t(x) \wedge \neg s(x) \wedge r(x,y), \leftarrow \neg s(a) \wedge r(a,y),$$
$$\leftarrow r(a,y), \leftarrow u(y) \wedge \neg v(a), \leftarrow \neg v(a), \Box$$

$R_3$ *ist eine Widerlegung von* $D \cup \{\leftarrow p(x) \wedge r(x,y)\}$ *mit berechneter Antwortsubstitution* $\sigma_3 = \sigma_1\sigma_2 = \{x/a, y/b\}$.

**Lemma 5.1.1** *Sei* $D$ *ein normales Programm,* $\leftarrow G_1, \leftarrow G_2$ *Zielklauseln,* $R_1$ *eine SLDNF-Widerlegung von* $G_1$ *mit berechneter Antwort* $\sigma_1$ *und* $R_2$ *eine SLDNF-Widerlegung von* $G_2\sigma_1$ *mit berechneter Antwort* $\sigma_2$. *Die Komposition* $R_3$ *von* $R_1$ *und* $R_2$ *ist eine SLDNF-Widerlegung von* $\leftarrow G_1 \wedge G_2$ *mit Antwort* $\sigma_1\sigma_2$.

**Beweis** Einfache Folgerung aus den Definitionen.

## 5.2 SLDNF-Bäume

Im folgenden erweitern wir den Begriff des SLDNF-Baumes zu dem Begriff des *unvollständigen SLDNF-Baumes*. Im Gegensatz zu einem normalen SLDNF-Baum ist es bei einem unvollständigen SLDNF-Baum erlaubt, daß in einem nichtleeren Knoten kein Literal selektiert ist. Dieser Knoten ist dann ein Blattknoten. Da über den Status des Knotens nicht entschieden ist (Erfolgsknoten oder Mißerfolgsknoten) heißt er *potentieller Erfolgsknoten.*

**Definition 5.2.1 (unvollständiger SLDNF-Baum)** *Sei* $D$ *ein normales Programm und* $\leftarrow G$ *eine Zielklausel. Ein* unvollständiger SLDNF-Baum *für* $D \cup \{\leftarrow G\}$ *ist ein Baum mit folgenden Eigenschaften:*

1. *Jeder Knoten ist mit einer Zielklausel beschriftet.*[1]
2. *Die Wurzel des Baumes ist* $\leftarrow G$, *die* angewandte Substitution *ist* $\epsilon$.
3. *Sei* $\leftarrow L_1 \wedge \ldots \wedge L_i \wedge \ldots \wedge L_n$ $(n \geq 1)$ *ein Knoten mit Nachfolgern und* $L_i$ *das selektierte Literal. Dann gilt genau einer der folgenden Punkte:*

[1] Der Kürze halber identifizieren wir im folgenden die Knoten mit ihren Beschriftungen.

(a) $L_i$ *ist ein Atom und für jede Programmklausel* $A \leftarrow W$ *in* $D$*, so daß* $\theta = mgu(L_i, A\sigma)$ *existiert, hat der Knoten einen Nachfolger*

$$\leftarrow (L_1 \wedge \ldots \wedge L_{i-1} \wedge W\sigma \wedge L_{i+1} \wedge \ldots \wedge L_n)\theta,$$

*wobei* $\sigma$ *eine Variablenumbenennung ist, die alle Variablen in* $A \leftarrow W$ *durch neue Variablen ersetzt. Der Unifikator* $\theta$ *ist die in diesem Schritt* verwendete Substitution.

(b) $L_i$ *ist ein negatives Grundliteral* $\neg A_i$*. Dann gibt es einen endlichen SLDNF-Mißerfolgsbaum für* $D \cup \{\leftarrow A_i\}$*. Der Knoten hat den einzigen Nachfolger*

$$\leftarrow L_1 \wedge \ldots \wedge L_{i-1} \wedge L_{i+1} \wedge \ldots \wedge L_n$$

*Die in diesem Schritt* verwendete Substitution *ist die identische Substitution* $\epsilon$.

4. *Sei* $\leftarrow L_1 \wedge \ldots \wedge L_i \wedge \ldots \wedge L_n$ $(n \geq 1)$ *ein Blatt mit selektiertem Literal* $L_i$*. Dann gilt genau einer der folgenden Punkte:*

   (a) $L_i$ *ist ein Atom und es gibt keine Programmklausel* $A \leftarrow W$ *in* $D$*, so daß* $L_i$ *und* $A\sigma$ *unifizierbar sind, wobei* $\sigma$ *eine Variablenumbenennung von* $A \leftarrow W$ *ist, die alle Variablen durch neue Variablen ersetzt. Dann zählt der Knoten zu den* Mißerfolgsknoten.

   (b) $L_i$ *ist ein negatives Grundliteral* $\neg A_i$*, und gibt es eine endliche SLDNF-Widerlegung von* $D \cup \{\leftarrow A_i\}$*. Dann zählt der Knoten zu den* Mißerfolgsknoten.

5. *Sei* $\leftarrow L_1 \wedge \ldots \wedge L_n$ $(n \geq 1)$ *ein Blatt-Knoten und* kein *Literal selektiert. Dann gehört der Knoten zu den* potentiellen Erfolgsknoten.

6. *Die leere Klausel ist stets ein Blatt und zählt zu den* potentiellen Erfolgsknoten.

*Sei* $B$ *ein Knoten des SLDNF-Teilbaumes* $T$ *für* $D \cup \{\leftarrow G\}$*. Die Komposition der auf dem Pfad von der Wurzel* $G$ *zu* $B$ *verwendeten Substitutionen heißt* für $B$ relevante Substitution. *Die Abbildung „überdeckung“ ist eine Funktion von endlichen unvollständigen SLDNF-Bäumen auf Substitutionsmengen. Ein endlicher unvollständiger SLDNF-Baum* $T$ *für* $D \cup \{\leftarrow G\}$ *wird dabei auf die Menge der relevanten Substitutionen aller potentiellen Erfolgsknoten von* $T$*, eingeschränkt auf die Variablen von* $G$*, abgebildet.*

Mit Hilfe der unvollständigen SLDNF-Bäume können wir beschreiben, was bei der Suche nach einem endlichen Mißerfolgsbaum die Hintereinanderausführung von zwei Literalen bedeutet.

**Definition 5.2.2 (Komposition von unvollständigen SLDNF-Bäumen)** *Sei $D$ ein normales Programm, $\leftarrow G_1$ und $\leftarrow G_2$ zwei (nicht notwendigerweise variablendisjunkte) Zielklauseln. Sei $T$ ein endlicher unvollständiger SLDNF-Baum für $D \cup \{\leftarrow G_1\}$, $S$ die Überdeckung der berechneten Antworten von $T$. $S' \subseteq S$ ist eine Indexmenge. $\mathbf{K}$ sei eine Familie von $\sigma$- indizierten Mengen $K_\sigma$, $\sigma \in S'$, von potentiellen Erfolgsknoten aus $T$, deren relevante Substitution eingeschränkt auf die Variablen von $G_1$ gleich $\sigma$ ist. $\mathbf{T}$ ist eine Menge von $T_\sigma$, $\sigma \in S'$ von endlichen unvollständigen SLDNF-Beweisbäumen für $D \cup \{\leftarrow G_2\sigma\}$. Ein Baum $T'$ heißt die* Komposition *von $T$ und $\mathbf{T}$ bzgl $S'$ und $\mathbf{K}$, wenn er auf folgende Weise konstruiert wird:*

1. *Zunächst wird ein Baum $T'_1$ konstruiert, den aus $T$ auf folgende Art hervorgeht: Jeder Knoten $\leftarrow k$ aus $T$ wird durch $\leftarrow k \wedge G_2\theta$ ersetzt. $\theta$ ist die relevante Substitution von $k$ in $T$. Ein eventuell selektiertes Literal bleibt in jedem Knoten gleich.*
2. *Danach wird für alle $\sigma \in S'$ und jeden Knoten $\leftarrow k \in K_\sigma$ aus $T_\sigma \in \mathbf{T}$ ein Baum $T_k^\sigma$ auf folgende Art und Weise konstruiert: Jeder Knoten $\leftarrow k'$ aus $T_\sigma$ wird durch $\leftarrow k\theta \wedge k'$ ersetzt. $\theta$ ist dabei die relevante Substitution von $\leftarrow k'$ in $T_\sigma$. Ein eventuell selektiertes Literal bleibt in jedem Knoten gleich.*
3. *$T'$ wird durch folgende Operation konstruiert: Für alle $\sigma \in S'$ wird für alle Knoten $\leftarrow k \in K_\sigma$ der Knoten $\leftarrow k \wedge G_2\theta'$ aus $T'_1$ (mit $\theta'$ ist relevante Substitution für $\leftarrow k$ in $T$) durch die Wurzel von $T_k^\sigma$ ersetzt. Damit wird der Baum $T_k^\sigma$ angehängt.*

Die Komposition von unvollständigen SLDNF-Bäumen ist wieder ein unvollständiger SLDNF-Baum, diesmal jedoch für $D \cup \{\leftarrow G_1 \wedge G_2\}$, wobei das selektierte Literal in jedem Schritt immer aus dem jeweiligen Baum stammt, der ergänzt wurde. Ein Knoten $k \in K_\sigma$ wird in $T'_1$ zu $k \wedge G_2\theta'$, wobei $\theta'$ die relevante Substitution von $k$ in $T$ ist, ergänzt. Dieser Knoten ist identisch (bis auf das selektierte Literal) zu der Wurzel von $T_k^\sigma$, d.h. zu der Wurzel von $T_\sigma \in \mathbf{T}$ ergänzt um $k\theta$ (wobei $\theta = \epsilon$). $T$ und $T_\sigma$ können höchstens Variable gemeinsam haben, die auch $G_1$ und $G_2$ gemeinsam haben. Der Kompositionsprozeß terminiert immer, da wir nur endlich viele Operationen auf endlichen Strukturen ausführen. Das folgende Beispiel soll die Definition verdeutlichen.

**Beispiel 5.2.1** *Sei $D$ das normale Programm aus Beispiel 5.1.1. Sei $G_1 \equiv \leftarrow p(x)$ und $G_2 \equiv \leftarrow q(x,y)$. In den folgenden Abbildungen sind die in einem Knoten eines SLDNF-*

*Baumes selektierten Literale mit einem Pfeil und die Mißerfolgsknoten mit einem Haken markiert.*

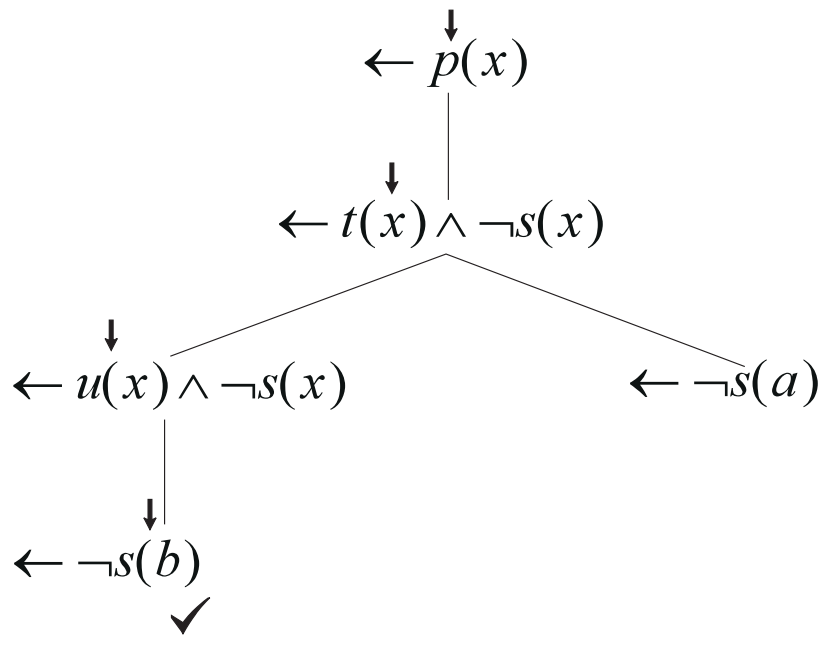

Abbildung 5.1: Unvollständiger SLDNF-Baum für $D \cup \{\leftarrow p(x)\}$

*Sei $T$ der in Abbildung 5.1 gezeigte unvollständige SLDNF-Baum für $D \cup \{\leftarrow p(x)\}$. Dann ist die Überdeckung der Antworten von $T$ gleich $S = \{\{x/a\}\}$, da nur ein potentieller Erfolgsknoten (in dem kein Literal selektiert ist) im Baum vorhanden ist. Dann wählen wir die Indexmenge $S' = S = \{\{x/a\}\}$ und somit $\mathbf{K} = \{K_{\{x/a\}}\}$ mit $K_{\{x/a\}} = \{\leftarrow \neg s(a)\}$. Die Menge $\mathbf{T} = \{T_{\{x/a\}}\}$ enthält nur einen endlichen unvollständigen SLDNF-Baum $T_{\{x/a\}}$ für $D \cup \{\leftarrow G_2\sigma\}$ mit $\sigma = \{x/a\}$. Der Baum $T_{\{x/a\}}$ ist in Abbildung 5.2 zu sehen.*

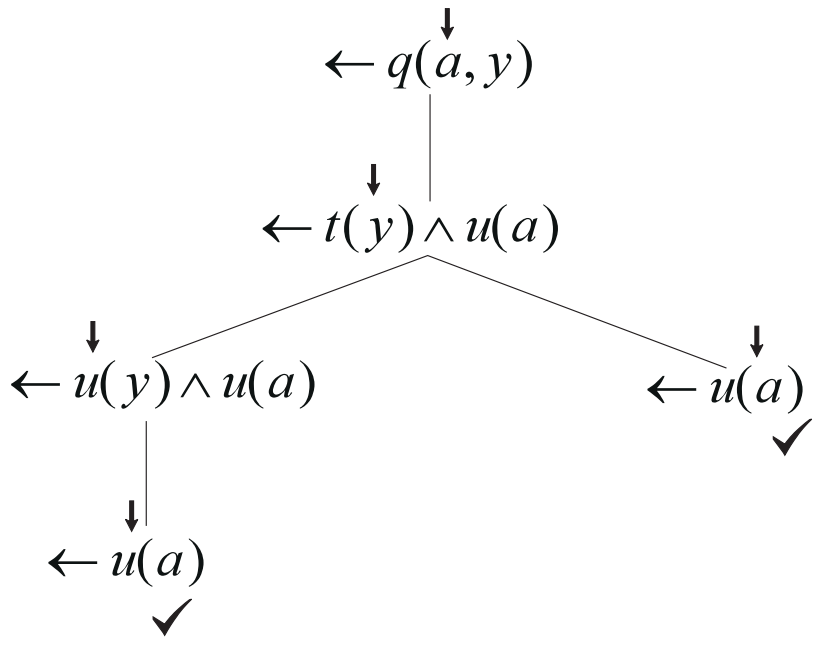

Abbildung 5.2: Unvollständiger SLDNF-Baum für $D \cup \{\leftarrow q(a, y)\}$

*Die Komposition $T'$ von $T$ und $\mathbf{T}$ bzgl. $S'$ und $\mathbf{K}$ erhält man nun durch die angegebenen Schritte:*

- *Zunächst wird der Baum $T'_1$ konstruiert, indem jeder Knoten $\leftarrow k$ aus $T$ durch $\leftarrow k \wedge q(x, y)\theta$ ersetzt wird. Dabei ist $\theta$ die relevante Substitution von $\leftarrow k$ in $T$. $T'_1$ ist in Abbildung 5.3 zu sehen.*

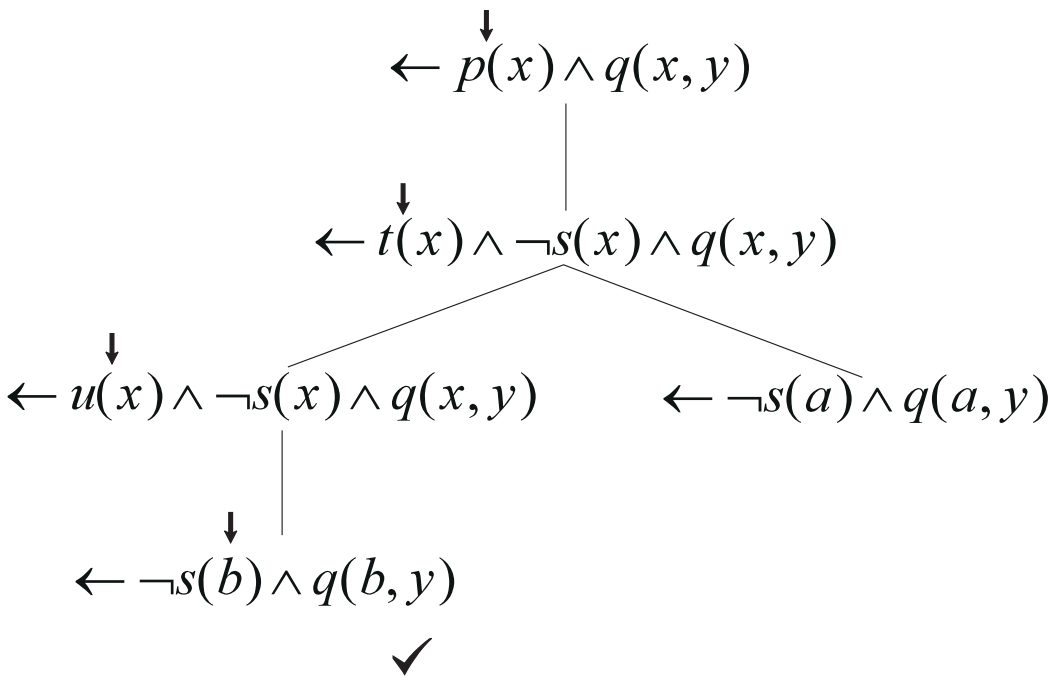

Abbildung 5.3: Unvollständiger SLDNF-Baum $T_1'$

- *Nun wird für alle Knoten $\leftarrow k \in K_\sigma$, $K_\sigma \in \mathbf{K}$ ein Baum $T_k^\sigma$ aus $T_\sigma \in \mathbf{T}$ konstruiert, indem jeder Knoten $\leftarrow k'$ aus $T_\sigma$ durch $\leftarrow k\theta \wedge k'$ ersetzt wird. Dabei ist $\theta$ die relevante Substitution von $\leftarrow k' \in T_\sigma$. In diesem Beispiel ist $\mathbf{K} = \{K_{\{x/a\}}\}$ mit $K_{\{x/a\}} = \{\neg s(a)\}$. Somit ist nur der Baum $T_{\{x/a\}}$ zu überführen. Dies ergibt den resultierenden Baum $T_{\neg s(a)}^{\{x/a\}}$, der in Abbildung 5.4 zu sehen ist.*

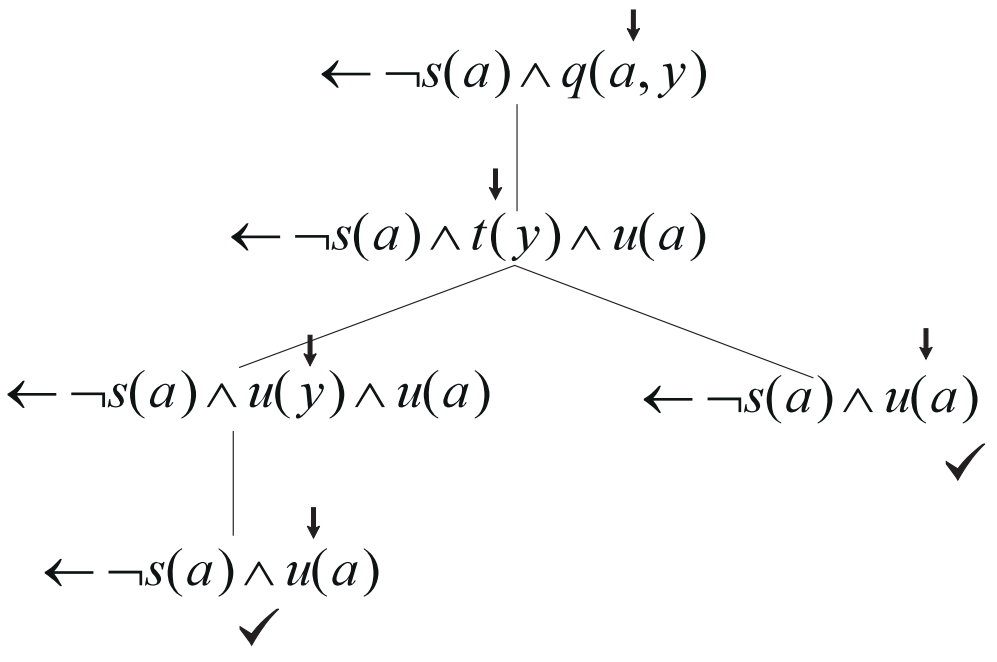

Abbildung 5.4: Unvollständiger SLDNF-Baum $T_{\neg s(a)}^{\{x/a\}}$

- *Im letzten Schritt wird der Baum $T_{\neg s(a)}^{\{x/a\}}$ an den Baum $T_1'$ beim Knoten $\neg s(a) \wedge q(a, y)$ angehängt (genauer: in $T_1'$ wird der Knoten $\leftarrow \neg s(a)$ durch den Baum $T_{\neg s(a)}^{\{x/a\}}$ ersetzt). Dies ergibt nun den Baum $T'$, der in Abbildung 5.5 dargestellt wird. Der Teilbaum, der aus $T_{\neg s(a)}^{\{x/a\}}$ stammt, ist in der Abbildung gestrichelt umrandet.*

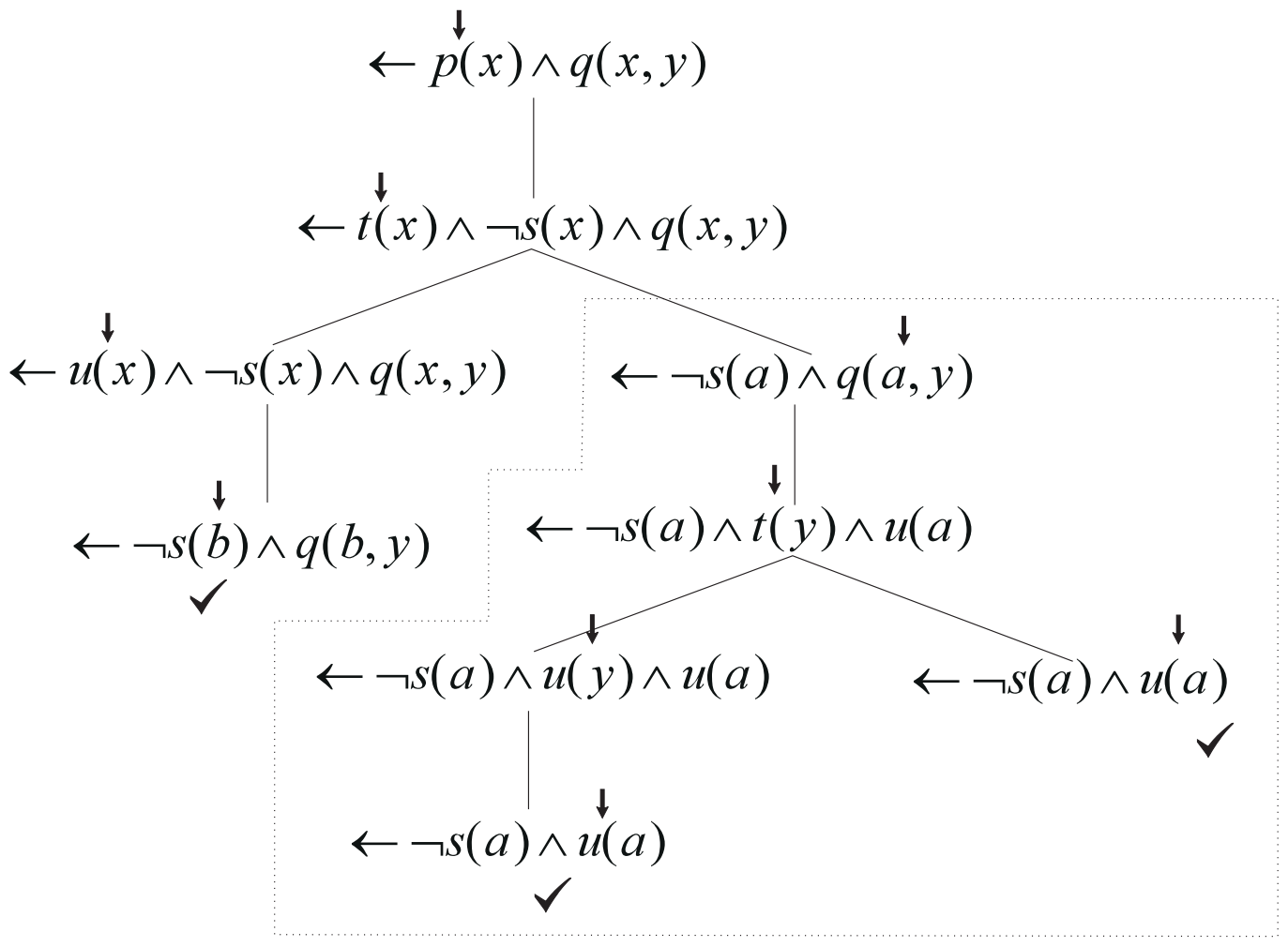

Abbildung 5.5: Resultierender Baum $T'$

*Der Baum $T'$ ist in diesem Fall sogar ein endlicher Mißerfolgsbaum, was allgemein nicht so sein muß. Außerdem wurde bei der Konstruktion von $T$ und $T_\sigma$ die Standardberechnungregel verwendet. Dies ist ebenfalls nicht unbedingt nötig. Interessant ist aber, daß in diesem Fall die Komposition einen SLDNF-Baum ergibt, der nicht mit der Standardberechnungsregel hergeleitet werden kann.*

Die Aussagen im Beispiel werden auch hier wieder formalisiert dargestellt:

**Lemma 5.2.1** *Sei $D$ ein normales Programm und $\leftarrow G_1$ und $\leftarrow G_2$ zwei Zielklauseln. Sei $T$ ein endlicher unvollständiger SLDNF-Baum für $D \cup \{\leftarrow G_1\}$, $S$ die Überdeckung der Antworten von $T$. $S' \subseteq S$ ist eine Indexmenge. Sei $\mathbf{K}$ eine Familie von $\sigma$- indizierten Mengen $K_\sigma$, $\sigma \in S'$, von potentiellen Erfolgsknoten aus $T$, deren relevante Substitution eingeschränkt auf die Variablen von $G_1$ gleich $\sigma$ ist. $\mathbf{T}$ ist eine Menge von $T_\sigma$, $\sigma \in S'$ von endlichen unvollständigen SLDNF-Beweisbäumen für $D \cup \{\leftarrow G_2\sigma\}$. Sei $T'$ die* Komposition *von $T$ und $\mathbf{T}$ bzgl. $S'$ und $\mathbf{K}$. Dann ist $T'$ ein endlicher unvollständiger SLDNF-Baum für $D \cup \{\leftarrow G_1 \wedge G_2\}$.*

**Beweis**

Es ist folgendes zu zeigen:

- Die Wurzel des Baumes $T'$ ist die Zielklausel $\leftarrow G_1 \wedge G_2$: Nach Definition 5.2.2 Punkt 1 besteht die Wurzel aus der Zielklausel $\leftarrow G_1 \wedge G_2\epsilon$.
- Jeder Schritt von einer Klausel auf eventuell vorhandene Nachfolger ist ein im Sinne von Definition 5.2.1 gültiger Schritt. Für einen Knoten $k'$ aus $T'$ gibt es

folgende Möglichkeiten:

1. der Knoten $k'$ ist ein modifizierter Knoten $k$ aus $T$ und hat nur Nachfolger, die auch modifizierte Knoten aus $T$ sind.
2. der Knoten $k'$ ist ein modifizierter Knoten $k$ aus $T$ und hat keinen Nachfolger.
3. der Knoten $k'$ ist ein modifizierter Knoten $k$ aus $T$ und hat auch Nachfolger, die modifizierte Knoten aus einem der $T_\sigma$ sind.
4. der Knoten ist ein modifizierter Knoten aus einem der $T_\sigma$ und hat nur Nachfolger, die ebenfalls modifzierte Knoten aus $T_\sigma$ sind.
5. der Knoten ist ein modifizierter Knoten aus einem der $T_\sigma$ und hat keinen Nachfolger.

Wir zeigen, daß für jede dieser Möglichkeiten der Übergang von einem Knoten zu seinen Nachfolgern ein korrekter SLDNF-Schritt ist.

Zu 1: Sei zu $k'$ aus $T'$ der dazugehörige Knoten $k = L_1 \wedge \ldots \wedge L_i \wedge \ldots \wedge L_n$ aus $T$, $L_i$ das selektierte Literal und $\theta$ die relevante Substitution von $k$. Dann ist $k' = L_1 \wedge \ldots L_i \wedge \ldots \wedge L_n \wedge G_2\theta$. Für $L_i$ gibt es zwei Möglichkeiten:

* Sei $L_i$ ein Atom. Dann hat $k$ für jede Programmklausel $A \leftarrow W$, für die $\theta' = mgu(L_i, A\sigma)^2$ existiert, einen Nachfolger in $T$. Sei $k_{(A\leftarrow W)\sigma} = (L_1 \wedge \ldots \wedge L_{i-1} \wedge W\sigma \wedge L_{i+1} \wedge \ldots \wedge L_n)\theta'$ ein beliebiger dieser Nachfolger. Nach Konstruktion hat aber $k' \in T'$ für jeden Nachfolger $k_{(A\leftarrow W)\sigma}$ von $k \in T$ einen Nachfolger $k'_{(A\leftarrow W)\sigma}$. Nach Konstruktion ist $k'_{(A\leftarrow W)\sigma} = (L_1 \wedge \ldots \wedge L_{i-1} \wedge W\sigma \wedge L_{i+1} \wedge \ldots \wedge L_n)\theta' \wedge G_2\theta''$. Dabei ist $\theta''$ die relevante Substitution von $k_{(A\leftarrow W)\sigma} \in T$. Nach Definition ist $\theta'' = \theta\theta'$. Dies ist aber gerade die Zielklausel, die sich durch Resolution von $k'$ mit $(A \leftarrow W)\sigma$ ergibt. Da $k$ für jede auf das selektierte Literal anwendbare Regel einen Nachfolger besitzt, besitzt also auch $k'$ einen solchen Nachfolger. Dann ist aber der Schritt von $k'$ auf die Nachfolger von $k'$ ein gültiger SLDNF-Schritt mit selektiertem Literal $L_i$.
* Sei $L_i$ ein negiertes Literal $\neg A$. Da $k$ einen Nachfolger in $T$ hat gibt es einen endlich gescheiterten SLDNF-Baum für $D \cup \{\leftarrow A\}$. Dann hat $k$ aber nach Definition einen Nachfolger $k_\neg = L_1 \wedge \ldots \wedge L_{i-1} \wedge L_{i+1} \ldots \wedge L_n$.

[2] $\sigma$ ist wieder die hier nötige Variablenumbenennung

Daraus folgt, daß dann nach Konstruktion auch $k'$ einen Nachfolger $k'_{\neg} = L_1 \wedge \ldots \wedge L_{i-1} \wedge L_{i+1} \wedge \ldots \wedge L_n \wedge G_2\theta'$ hat. Dabei ist $\theta'$ relevante Substitution von $k_{\neg}$. Da aber $\theta' = \theta$, ist dies ein gültiger SLDNF-Schritt in $T'$.

Zu 2: Es ergeben sich zwei Fälle:

- $k$ ist ein potentieller Erfolgsknoten. Dann ist auch $k'$ ein potentieller Erfolgsknoten.
- $k$ ist ein Mißerfolgsknoten. Dann ist auch $k'$ ein Mißerfolgsknoten, da das selektierte Literal erhalten bleibt und es keine Programmklausel gibt, die mit dem selektierten Literal unifiziert.

Zu 3: Es reicht zu zeigen, daß die Wurzel des angehängten Beweisbaumes $T_\sigma$ gleich dem Knoten ist, der durch die Wurzel ersetzt wurde. Dann folgt der Rest aus Punkt 1.

Sei der Knoten $k = L_1 \wedge \ldots \mathrm{L}_n$, $n \geq 0$. Der Knoten $k$ wird zu $k' = L_1 \wedge \ldots \wedge L_n \wedge G_2\theta$ modifiziert, dabei ist $\theta$ die relevante Substitution von $k$ in $T$. Die Wurzel von $T_\sigma^k$ ist aber $k\epsilon \wedge G_2\sigma$. Da $\sigma = \theta|_{var(G_1)}$ und $k$ und $G_2$ nur Variablen gemeinsam haben können, die in $G_1$ vorhanden sind, müssen beide Knoten identisch sei.

Zu 4: Der Beweis ist analog zum Beweis von Punkt 1.

Zu 5: Der Beweis ist analog zum Beweis von Punkt 1.

**Definition 5.2.3 (Vollständige Komposition)** *Sei $D$ ein normales Programm, $\leftarrow G_1$ und $\leftarrow G_2$ zwei Zielklauseln. Sei $T$ ein endlicher unvollständiger SLDNF-Baum für $D \cup \{\leftarrow G_1\}$, $S$ die Überdeckung der Antworten von $T$. Sei* $\mathbf{K}$ *eine Familie von $\sigma$-indizierten Mengen $K_\sigma$, $\sigma \in S$,* **aller** *potentiellen Erfolgsknoten aus $T$, deren relevante Substitution eingeschränkt auf die Variablen von $G_1$ gleich $\sigma$ ist und* $\mathbf{T}$ *eine Menge von $T_\sigma$, $\sigma \in S'$, von endlichen unvollständigen SLDNF-Beweisbäumen für $D \cup \{\leftarrow G_2\sigma\}$. Ein Baum $T'$ heißt die* vollständige Komposition *von $T$ und* $\mathbf{T}$*, wenn $T'$ die Komposition von $T$ und* $\mathbf{T}$ *bzgl. $S$ und* $\mathbf{K}$ *ist.*

Mit Hilfe des Begriffs der *Komposition* können nun Aussagen über die Überdeckung der Antworten des resultierenden Baumes gewonnen werden.

**Lemma 5.2.2** *Sei $D$ ein normales Programm, $\leftarrow G_1$ und $\leftarrow G_2$ zwei Zielklauseln. Sei $T$ ein unvollständiger SLDNF-Baum für $D \cup \{\leftarrow G_1\}$, $S$ die Überdeckung der Antworten* *von $T$, $\mathbf{T}$ ist eine Menge von endlichen unvollständigen SLDNF-Bäumen $T_\sigma$, $\sigma \in S$. $T_\sigma$ ist hierbei ein endlicher unvollständiger SLDNF-Beweisbaum für $D \cup \{\leftarrow G_2\sigma\}$. $T'$ sei die vollständige Komposition von $T$ und $\mathbf{T}$. Dann ist*

1. *$T'$ ein endlicher unvollständiger SLDNF-Baum für $D \cup \{\leftarrow G_1 \wedge G_2\}$ und*
2. *$überdeckung(T') = \{\sigma\theta \mid \sigma \in S, \theta \in überdeckung(T_\sigma)\}$*

**Beweis** Die erste Behauptung folgt aus Lemma 5.2.1.

Die zweite Behauptung zeigen wir auf folgende Weise: Sei $R$ ein Pfad aus $T'$ zu einem potentiellen Erfolgsknoten und $\theta_1, \ldots, \theta_n, \theta'_1, \ldots, \theta'_m$ die Folge der in $R$ verwendeten Substitutionen, wobei die $\theta_i$ aus dem Baum $T$ stammen und die $\theta'_j$ aus einem Baum $T_\sigma$. Damit ist also die Komposition der Substitutionen $(\theta_1 \ldots \theta_n \theta'_1 \ldots \theta'_m)|_{var(G_1 \wedge G_2)} \in$ *$überdeckung(T')$*. Da in der Komposition der Substitutionen $\theta_1 \ldots \theta_n$ keine Variablen gebunden werden, die nur in $G_2$ enthalten sind und in der Komposition $\theta'_1 \ldots \theta'_m$ keine Variablen gebunden werden, die nur in $G_1$ enthalten sind, gilt also

$$(\theta_1 \ldots \theta_n)|_{var(G_1)}(\theta'_1, \ldots, \theta'_m)|_{var(G_2)} = (\theta_1 \ldots \theta_n \theta'_1 \ldots \theta'_m)|_{var(G_1 \wedge G_2)}.$$

Da $(\theta_1 \ldots \theta_n)|_{var(G_1)} = \sigma$ und $(\theta'_1 \ldots \theta'_m)|_{var(G_2)} = \theta$ mit $\theta \in überdeckung(T_\sigma)$ gilt $\sigma\theta \in überdeckung(T')$.

Sei jetzt $\sigma'\theta' \in \{\sigma\theta \mid \sigma \in S, \theta \in überdeckung(T_\sigma)\}$. Im Konstruktionsprozeß wurde $T_\sigma$ an den Knoten, dessen relevante Substitution eingeschränkt auf die Variablen von $G_1$ $\sigma$ ergibt, angehängt (mit den entsprechenden Modifikationen). Also gibt es einen Pfad von der Wurzel von $T'$ zu einem potentiellen Erfolgsknoten, dessen verwendete Substitutionen eingeschränkt auf die Zielklausel des jeweiligen Baumes, aus dem die Substitution stammt, die Komposition $\sigma\theta$ ergibt. Daraus folgt dann die Behauptung.

# Kapitel 6

# Beweisdarstellung durch Beweisbäume

## 6.1 Einführung

Wie in Kapitel 3 bereits gezeigt, verwenden die meisten der bekannten Ansätze zur Integritätsüberprüfung von deduktiven Datenbanken zwar die Information, daß zu einer Integritätsbedingung ein Beweis vorhanden ist, aber nicht welche Struktur er hatte. Bei der Untersuchung von SLDNF-Beweisen ergeben sich mehrere Probleme:

- ein SLDNF-Baum ist zur Beweisanalyse sehr ungeeignet: Er wächst mit der Anzahl der in der Datenbank vorhandenen Regeln, und, was bei deduktiven Datenbanken sehr viel mehr ins Gewicht fällt, mit der Anzahl der vorhanden Fakten sehr in die Breite. Dies macht die Analyse eines Beweises sehr aufwendig.

- Die Definition eines SLDNF-Baumes verlangt im Fall der Selektion eines negierten Literales den Aufbau eines neuen Baumes. Dies macht die Behandlung von Negationen unübersichtlich.

- Ein SLDNF-Baum reflektiert nicht die intuitive Aufrufhierarchie, die ein Datenbankdesigner modellieren will, wenn er eine Menge von Programmklauseln niederschreibt.

Aus diesen Gründen benötigen wir eine alternative Beweisdarstellung. Als Grundbaustein dieser Beweisdarstellung definieren wir einen UND-ODER-Baum, aus dem wir einen Teilbaum identifizieren und mit Substitutionsmengen versehen. Dabei werden folgende Strukturen betrachtet:

1. Ein *UND-ODER-Baum*, dieser dient zunächst nur als Darstellung der Abhängigkeiten der Programmklauseln innerhalb einer Datenbank von einer bestimmten Anfrage.
2. Eine *Beweisschablone*, diese dient dazu, die *Struktur* eines Beweises zu identifizieren.
3. Ein *Beweisbaum*: dieser ist eine Beweisdarstellung, die äquivalent ist zu einem SLDNF-Beweis mit bestimmten Berechnungsregeln.

Dabei sollte man aus der Reihenfolge der Aufzählung keine logischen Abhängigkeiten herleiten; Sie gibt nur den Informationsgehalt der einzelnen Darstellungsmöglichkeiten wieder. So ist die Beweisschablone erst identifizierbar, wenn ein Beweisbaum vorliegt.

### 6.1.1 UND-ODER-Baum

Zunächst geben wir die Definition eines UND-ODER-Baumes. Ein UND-ODER-Baum gibt zunächst nur die Aufrufhierarchie eines Logik-Progammes wieder.

**Definition 6.1.1 (UND-ODER-Baum)** *Sei $D$ ein normales Programm und $W$ eine Konjunktion von Literalen. Ein* UND-ODER-Baum *aus $D$ und der Zielklausel $\leftarrow W$ ist ein Baum mit UND- sowie ODER-Knoten mit den folgenden Eigenschaften:*

- *Die Wurzel des Baumes ist der UND-Knoten $W$.*
- *Jeder UND-Knoten bestehend aus einer Konjunktion von Literalen $W \equiv L_1 \wedge \ldots \wedge L_n$ hat $n$ ODER-Knoten als Söhne, die sich folgendermaßen aus den einzelnen Literalen $L_i, 1 \leq i \leq n$ ergeben: Ist $L_i$ ein negiertes Atom $A_i$, so besteht der $i-$te Nachfolgerknoten aus dem Atom $A_i$ und die Kante von $W$ zu $A_i$ wird mit einem $\neg$ markiert. Ist $L_i$ ein Atom, so besteht der i-te Nachfolgerknoten aus $L_i$.*
- *Jeder ODER-Knoten bestehend aus einem Atom $A$ besitzt für jede Regel $A' \leftarrow W' \in D$ mit nichtleerem Rumpf $W'$ einen Nachfolger UND-Knoten $W'\sigma\theta$. Dabei ist $\theta = mgu(A, A'\sigma)$ und $\sigma$ eine Variablenumbenennung, die alle Variablen von $A' \leftarrow W'$ durch neue Variablen ersetzt.*[1]

[1] Der Grund für die Einführung einer Variablenumbenennung ist hier der Gleiche, wie bei der Definition des SLDNF-Baumes.

- *Die Blätter des UND-ODER-Baumes bestehen aus ODER-Knoten, deren Atome ein Prädikatensymbol* $\in \Psi^{\Sigma}_{EDB}$ *besitzen.*

**Beispiel 6.1.1** *Sei D folgendes normales Programm:*

$$\begin{array}{rcl} p(x) & \leftarrow & q(x,x) \\ p(x) & \leftarrow & t(x) \wedge \neg s(x) \\ r(x) & \leftarrow & u(x) \wedge \neg v(x) \\ t(a) & \leftarrow & \\ q(a,a) & \leftarrow & \\ u(a) & \leftarrow & \end{array}$$

*Der UND-ODER-Baum für D und die Zielklausel* $W \equiv \leftarrow p(x) \wedge t(y) \wedge r(z)$ *ist der folgende:*

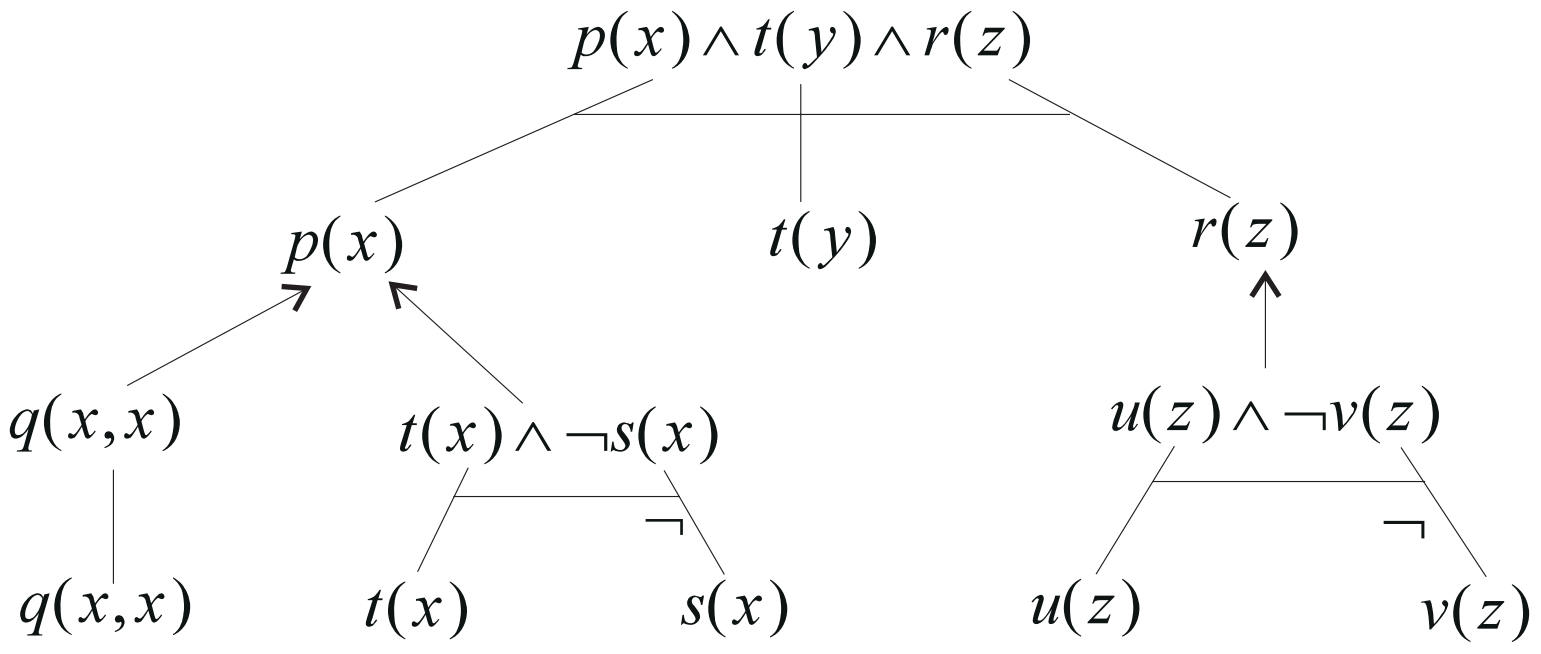

Abbildung 6.1: UND-ODER-Baum

Ein UND-ODER Baum kann bei Existenz rekursiver Regeln prinzipiell unendlich groß sein. Dies ist aber hier kein Problem, da falls ein Beweis der Wurzel des Baumes existiert, immer ein endlicher Teilbaum identifiziert werden kann, der für den Beweis relevant ist. Ein UND-ODER-Baum ist also nur eine Hilfskonstruktion, welche wir zur Bestimmung des Begriffs eines Beweisbaumes benötigen: er enthält immer die vollständige Aufrufhierarchie eines Logikprogrammes. Für einen Beweis ist aber meistens nur ein Teilbaum nötig und dieser Teilbaum ist immer endlich. Unsere weitere Aufgabe ist es also, aus dem vollständigen UND-ODER-Baum einen endlichen Teilbaum zu identifizieren, der einen Beweis ermöglicht (wobei ein Beweis gleichbedeutend ist mit der Existenz einer SLDNF-Widerlegung für $\leftarrow W$, falls $W$ die Wurzel eines UND-ODER-Baumes ist).

Für die weitere Vorgehensweise benötigen wir den Begriff der *Polarität eines Knotens* in einem UND-ODER-Baum. Dies hängt unter anderem mit der schon in anderen Ansätzen gemachten Beobachtung zusammen, daß Datenbankänderungen nur Integritätsbedingungen beeinflussen können, in denen die Änderung mit umgekehrter Polarität vorhanden ist (siehe Abschnitt 3.1.1).

**Definition 6.1.2 (Polarität eines Knotens)** *Die* Polarität *eines Knotens $k$ in einem UND-ODER-Baum mit Wurzel $W$ heißt:*

- positiv, *wenn auf dem Pfad im Baum ausgehend von $W$ zu $k$ eine gerade Anzahl von mit $\neg$ markierten Kanten liegen.*
- negativ, *wenn auf dem Pfad im Baum ausgehend von $W$ zu $k$ eine ungerade Anzahl von mit $\neg$ markierten Kanten liegen.*

**Definition 6.1.3 (Gleiche Polaritätsstufe)** *Sei $T$ ein UND-ODER-Baum. Zwei Knoten $k$ und $k'$ liegen auf der* gleichen Polaritätsstufe, *wenn es einen Pfad ausgehend von der Wurzel von $T$ gibt, auf dem sowohl $k$ als auch $k'$ liegen und auf dem Pfad von $k$ zu $k'$ keine mit $\neg$ markierte Kante liegt.*

### 6.1.2 Klauseln als Filter

Unser Ziel ist es, aus einem vollständigen UND-ODER-Baum einen für einen Beweis relevanten Teilbaum zu identifizieren. Probleme tauchen dabei insbesondere bei der Behandlung der Negation auf.

**Beispiel 6.1.2** *Sei $D$ folgendes normales Programm:*

$$\begin{array}{rcl} p(x) & \leftarrow & q_1(x) \wedge q_2(x) \wedge q_3(x) \wedge q_4(x) \wedge q_5(x) \\ q_1(a) & \leftarrow & \\ q_2(a) & \leftarrow & \\ q_3(a) & \leftarrow & \\ q_5(a) & \leftarrow & \end{array}$$

*Der UND-ODER-Baum für $D$ und die Zielklausel $W \equiv \leftarrow \neg p(a)$ ist der folgende:*

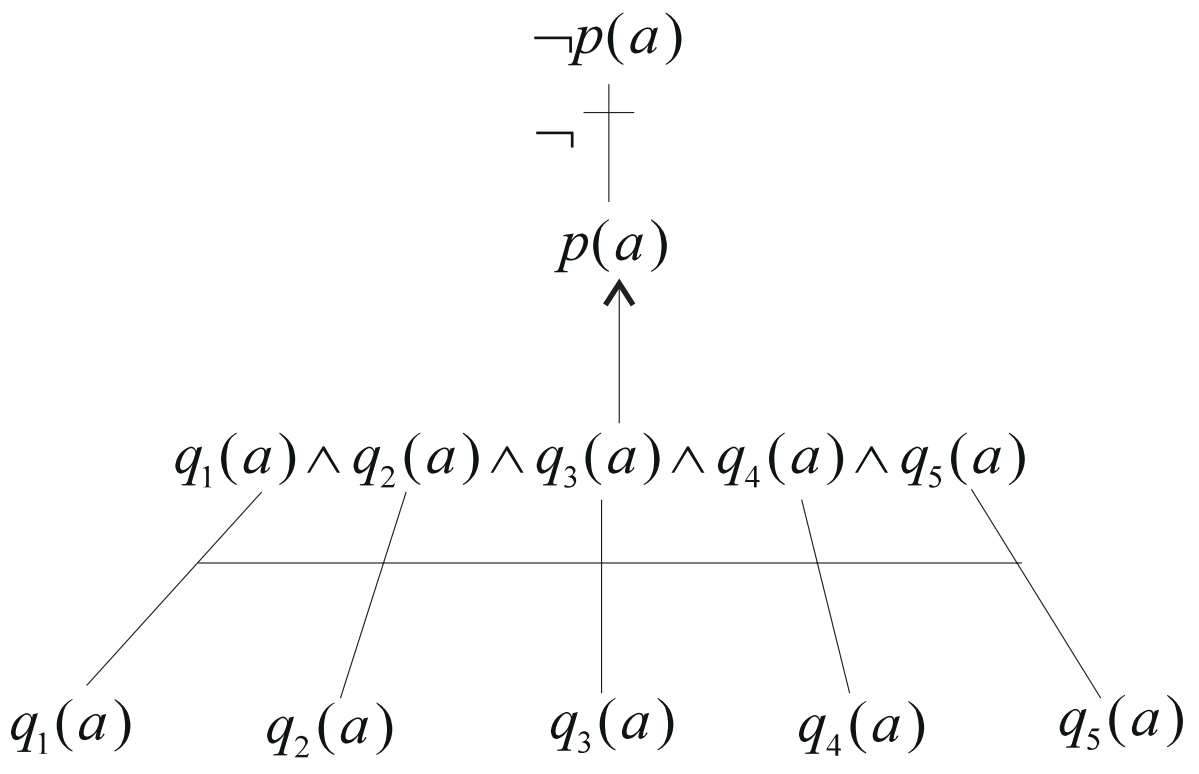

Abbildung 6.2: UND-ODER-Baum für $D \cup \{\leftarrow \neg p(a)\}$

*Tatsächlich reicht aber für einen SLDNF-Beweis von* $D \cup \{\leftarrow \neg p(a)\}$ *aus, daß kein Fakt* $q_4(a)$ *in der Datenbank (d.h. in D) enthalten ist. Ein Baum, der nur das unbedingt Nötige für einen Beweis enthält, sieht also so aus:*

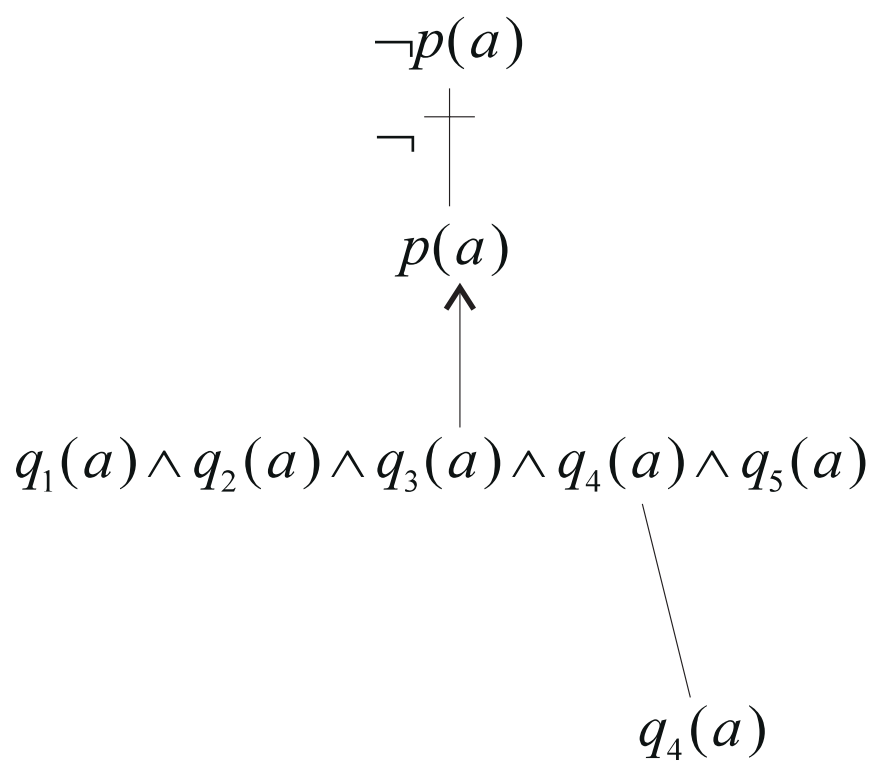

Abbildung 6.3: Beweisschablone für $D \cup \{\leftarrow \neg p(a)\}$

Eine solche Darstellung nennen wir im folgenden eine *Beweisschablone*. Bei der Identifikation von überflüssigen Teilen eines UND-ODER-Baumes hilft eine neue Sichtweise: wir betrachten Klauseln als Filter.

Sei $W \equiv\leftarrow L_1 \wedge \ldots \wedge L_n$ eine normale Zielklausel, für die ein endlich gescheiterter SLDNF-Baum gesucht wird. Die dabei verwendete Berechnungsregel sei die Standardberechnungsregel, d.h. die Literale in einer Zielklausel werden von links nach rechts selektiert. Die SLDNF-Resolution versucht nun, einen endlichen gescheiterten SLDNF-Baum für $W$ zu bestimmen. Dabei werden zunächst *alle* berechneten Antwortsubstitutionen von $\leftarrow L_1$ bestimmt. Sei $\Phi_1$ die Menge dieser Antwortsubstitutionen von $L_1$.

Jedes $\sigma \in \Phi_1$ kann als *Ausgangssubstitution* von $L_1$ und *Eingangssubstitution* von $L_2$ aufgefaßt werden, da $L_2$ das nächste durch die Berechnungsregel selektierte Literal ist. Die Menge der *Ausgangssubstitutionen* von $L_2$ ist demnach $\Phi_2 = \{\sigma\sigma' \mid \sigma'$ ist berechnete Antwortsubstitution von $\leftarrow L_2\sigma$, $\sigma \in \Phi_1$ $\}$. Die Folge dieser Mengen für $0 \leq i \leq n$ läßt sich leicht induktiv definieren:

1. $\Phi_0 = \{\epsilon\}$
2. $\Phi_i = \{\sigma'\sigma \mid \sigma'$ ist berechnete Antwortsubstitution von $\leftarrow L_i\sigma$, $\sigma \in \Phi_{i-1}\}$ für $1 \leq i \leq n$.

Gibt es zu einem $\sigma \in \Phi_{i-1}$, $1 \leq i \leq n$, keine berechnete Antwortsubstitution für $\leftarrow L_i\sigma$, so sagen wir: „$L_i$ hat die Substitution $\sigma$ *herausgefiltert*“. Ist $\Phi_n = \emptyset$, und alle SLDNF-Anfragen $\leftarrow L_i\sigma$, $\sigma \in \Phi_{i-1}$, $1 \leq i \leq n$ terminieren, so gibt es einen endlich gescheiterten SLDNF-Baum für $\leftarrow L_1 \wedge \ldots \wedge L_n$.

Damit kann nun eine Bedingung angegeben werden, unter der bei der Suche nach einem endlich gescheiterten SLDNF-Baum ein Literal $L_i$ in einer Zielklausel als unwichtig identifiziert werden kann: Ist $\Phi_{i-1} \subseteq \Phi_i$, so hat $L_i$ keine Substitution herausgefiltert und trägt nicht zu der Suche nach einem endlich gescheiterten SLDNF-Baum bei.

## 6.2 Beweisbäume

Bevor wir zu der Definition der Beweisbäume kommen, benötigen wir noch einige Definitionen über Substitutionsmengen und Tupel.

### 6.2.1 Substitutionsmengen und -tupel

Im folgenden lassen wir nur idempotente Substitutionen zu, d.h. Substitutionen $\sigma$ mit $\sigma = \sigma\sigma$.

**Definition 6.2.1**

- *Eine Substitution $\sigma$ heißt* allgemeiner *als eine Substitution $\sigma'$, geschrieben $\sigma \geq \sigma'$, wenn eine Substitution $\theta$ existiert mit $\sigma\theta = \sigma'$.*
- *Ein $n$-Tupel von Substitutionen $U = \langle\sigma_1, \ldots, \sigma_n\rangle$ heißt* allgemeiner *als ein $n$-Tupel von Substitutionen $U' = \langle\sigma'_1, \ldots, \sigma'_n\rangle$ ($U \geq U'$), wenn $\sigma_i \geq \sigma'_i$ für alle $1 \leq i \leq n$.*

- *Eine Menge von Substitutionen $S$ heißt* allgemeiner *als eine Menge von Substitutionen $S'$ ($S \geq S'$), wenn für alle $\sigma' \in S'$ ein $\sigma \in S$ existiert, so daß $\sigma \geq \sigma'$.*
- *Eine Menge von Substitutionen $S$ heißt* echt allgemeiner *als eine Menge von Substitutionen $S'$, geschrieben $S > S'$, wenn $S \geq S'$ und $S \not\leq S'$.*
- *Die* Differenz *zweier Substitutionsmengen $S$ und $S'$, geschrieben $S - S'$, ist die Menge $G \equiv \{\sigma \in S \mid$ für kein $\sigma' \in S'$ ist $\sigma \geq \sigma'\}$.*
- *Eine Folge von Mengen von Substitutionen $(S_i)$, $0 \leq i \leq n$, heißt* monoton fallend, *wenn $S_{i-1} \geq S_i$ für alle $i$ mit $1 \leq i \leq n$.*
- *Eine Folge von Tupeln von Substitutionen $(U_i)$, $0 \leq i \leq n$, heißt* monoton fallend, *wenn $U_{i-1} \geq U_i$ für alle $i$ mit $1 \leq i \leq n$.*
- *Die* Domain *eines Tupels von $U$ von Substitutionen, $dom(U)$, ist die Vereinigung der Domains aller Elemente von $U$.*
- *Die* Domain *einer Substitutionsmenge $S$, $dom(S)$, ist die Vereinigung der Domains aller Elemente von $S$.*
- *Eine Folge von Mengen von Substitutionen $(S_i)$, $0 \leq i \leq n$, heißt* streng monoton fallend, *wenn $S_{i-1} > S_i$ für alle $i$ mit $1 \leq i \leq n$.*

Aus den vorangestellten Definitionen ergeben sich einige elementare Folgerungen:

**Lemma 6.2.1**

1. *Seien $\sigma, \sigma'$ und $\sigma''$ Substitutionen mit $\sigma \geq \sigma'$ und $\sigma' \geq \sigma''$. Dann ist auch $\sigma \geq \sigma''$ (Transitivität der $\geq$-Relation auf Substitutionen).*
2. *Seien $S, S', S''$ Substitutionsmengen mit $S \geq S'$ und $S' \geq S''$. Dann gilt auch $S \geq S''$ (Transitivität der $\geq$-Relation auf Substitutionsmengen).*
3. *Für jede Substitutionsmenge $S$ gilt: $\{\epsilon\} \geq S \geq \emptyset$.*
4. *Seien $\sigma$ und $\sigma'$ Substitutionen und $t$ ein beliebiger Term. Wenn $\sigma \geq \sigma'$, dann ist auch $t\sigma'$ Instanz von $t\sigma$.*

**Beweis**

Zu 1. Sei $\sigma\theta_1 = \sigma'$ und $\sigma'\theta_2 = \sigma''$. Daraus folgt direkt $\sigma\theta_1\theta_2 = \sigma''$. Daraus folgt die Behauptung.

Zu 2. Sei $\sigma''$ beliebige Substitution aus $S''$. Da $S' \geq S''$, existiert eine Substitution $\sigma' \in S'$ mit $\sigma' \geq \sigma''$. Da $S \geq S'$, existiert auch ein $\sigma \in S$ mit $\sigma \geq \sigma'$. Aus Teil 1 folgt $\sigma \geq \sigma''$. Da $\sigma''$ beliebig ist folgt daraus die Behauptung.

Zu 3. Folgt unmittelbar aus der Definition.

Zu 4. Folgt unmittelbar aus der Definition.

**Definition 6.2.2** *Eine monoton fallende Folge von Substitutionsmengen* $S_0, \ldots, S_n$ *heißt eine* Nullfolge, *wenn ein* $S_i = \emptyset$ *für ein* $i$ *mit* $0 \leq i \leq n$.

Daraus folgt $S_j = \emptyset$ für alle $j$ mit $i \leq j \leq n$. Betrachten wir Klauseln als Filter wie bereits erwähnt, dann bildet die zugeordnete Folge von Substitutionsmengen eine monoton fallende Folge.

**Definition 6.2.3 (Einschränkung einer Substitutionsmenge, -tupel)** *Die Einschränkung einer Menge von Substitutionen* $S$ *auf eine Variablenmenge* $V$, *geschrieben* $S|_V$, *ist eine Substitutionsmenge* $S|_V = \{\sigma|_V \mid \sigma \in S\}$.

*Die Einschränkung eines Tupels* $U = \langle \sigma_1, \ldots, \sigma_n \rangle$ *auf eine Variablenmenge* $V$, *geschrieben* $U|_V$, *ist das Tupel* $U|_V = \langle \sigma_1|_V, \ldots, \sigma_n|_V \rangle$.

**Definition 6.2.4** *Die Anzahl von Literalen in einer Konjunktion von Literalen* $L_1 \wedge \ldots \wedge L_n$ *bezeichnen wir als die Länge der Konjunktion.*

**Definition 6.2.5** *Eine* Konkatenation *von Tupeln* $U_1, \ldots, U_n$, *geschrieben* $U_1 + \ldots + U_n$ *ist ein neues Tupel* $U$, *das man durch die Aneinanderreihung der Tupel* $U_1, \ldots, U_n$ *erhält.*

*Unter einer* Permutation *verstehen wir eine Abbildung* $\pi$, *die ein* $n$*-Tupel* $U$ *wieder auf ein* $n$*-Tupel* $U'$ *abbildet, bei dem die Elemente von* $U$ *auf eine durch* $\pi$ *festgelegte Art vertauscht sind.*

### 6.2.2 UND-ODER-Bäume und Substitutionen

Unser Ziel ist es, einen SLDNF-Beweis minimal und strukturerhaltend zu charakterisieren. Dazu muß folgendes geleistet werden:

1. Eine Charakterisierung der Knoten des aus einer Zielklausel und einer Datenbank aufgebauten UND-ODER-Baumes, die für einen Beweis relevant sind.

2. Die Charakterisierung der Variablensubstitutionen der relevanten Knoten, die für den Beweis einer Zielklausel relevant sind.

Um dies später ohne großen Aufwand beschreiben zu können, definieren wir zunächst zwei Klassen von Abbildungen $nachfolger$ und *subst*, die das Gewünschte leisten.

**Definition 6.2.6** *Sei $D$ ein normales Programm, $\leftarrow W$ eine Zielklausel und $T$ ein UND-ODER-Baum aus $W$ und $D$. Die Klasse der Abbildungen $nachfolger$ sind alle partiellen Abbildungen der Knoten von $T$ auf Mengen bzw. Tupel von Knoten von $T$ mit folgenden Eigenschaften:*

- *Jeder positive UND-Knoten $L_1 \wedge \ldots \wedge L_n$ wird auf ein n-Tupel $\langle k_1, \ldots, k_n \rangle$ aller seiner direkten Nachfolger abgebildet. Die Reihenfolge der Knoten im Tupel entspricht nicht notwendigerweise der Reihenfolge der Knoten im Baum.*
- *Jeder negative UND-Knoten $L_1 \wedge \ldots \wedge L_n$ wird auf ein m-Tupel $\langle k_1, \ldots, k_m \rangle$, $m \leq n$ seiner Nachfolger abgebildet. Die Reihenfolge der Knoten im Tupel entspricht nicht notwendigerweise der Reihenfolge der Knoten im Baum.*
- *Jeder positive ODER-Knoten $A$ wird auf eine Teilmenge aller seiner Nachfolgerknoten abgebildet.*
- *Jeder negative ODER-Knoten $A$ wird auf die Menge aller seiner Nachfolgerknoten abgebildet.*

Beginnend mit der Wurzel eines UND-ODER-Baumes $T$ bestimmt die $nachfolger$-Funktion eindeutig einen Teilbaum von $T$ bis auf Knotenvertauschungen.

Als nächstes definieren wir eine Funktion *subst*, die Knoten eines UND-ODER-Baumes auf Tupel bzw. Mengen von Substitutionen abbildet. Die Substitutionen entsprechen später berechneten Antworten von Zielklauseln. Hierbei kommt es wieder auf die Polarität der Knoten im UND-ODER-Baum an.

**Definition 6.2.7** *Sei $T$ ein UND-ODER-Baum aus einem normalem Programm und einer Zielklausel. Die Klasse der Abbildungen subst sind alle partiellen Abbildungen der Knoten von $T$ auf Mengen bzw. Tupel von Substitutionen mit folgenden Eigenschaften:*

- *Jeder positive UND-Knoten $L_1 \wedge \ldots \wedge L_n$ wird auf eine Folge von m-Tupeln $(U_i)$, $0 \leq i \leq n$ von Substitutionen abgebildet. $m$ ist für alle Folgenglieder gleich.*
- *Jeder negative UND-Knoten $L_1 \wedge \ldots \wedge L_n$ wird auf eine Folge von Substitutionsmengen $(S_i)$, $0 \leq i \leq m$ mit $m \leq n$ abgebildet.*
- *Jeder positive ODER-Knoten $L$ wird auf ein Paar von m-Tupeln von Substitutionen $\langle U_B, U_E \rangle$ abgebildet. $m$ ist für beide Tupel gleich.*

- *Jeder negative ODER-Knoten $L$ wird auf ein Paar von Substitutionsmengen $\langle S_B, S_E\rangle$ abgebildet.*

### 6.2.3 Syntaktische Definition eines Beweisbaumes

Nun haben wir alle Mittel zur Verfügung, um auf der Basis eines UND-ODER-Baumes einen Beweisbaum zu definieren.

**Definition 6.2.8 (Beweisbaum)** *Sei $D$ ein normales Programm, $\leftarrow W$ eine Zielklausel, $\sigma$ eine beliebige Substitution und $T$ ein UND-ODER-Baum aus $W$ und $D$. Ein durch $W$ und eine Funktion $nachfolger$ gegebener endlicher Teilbaum $T'$ von $T$ heißt zusammen mit einer Funktion $subst$ ein* Beweisbaum *mit Anfangssubstitution $\sigma$ für $D \cup \{\leftarrow W\sigma\}$, wenn folgende Bedingungen erfüllt sind:*

1. *Die Wurzel $W$ mit Länge $n$ von $T$ ist auch Wurzel von $T'$ mit $subst(W) = (U_i)$, $0 \leq i \leq n$, wobei $U_0 = \langle\sigma\rangle$*

2. *Für alle positiven UND-Knoten $W$ aus $T'$ sei $n$ die Länge von $W$, $subst(W) = (U_i)$, $0 \leq i \leq n$ eine Folge von Substitutionstupeln und $K = nachfolger(W)$. Dann ist für alle Knoten mit $k_i \in K$*

   - *$subst(k_i) = \langle U_{i-1}, U_i\rangle$, falls der Knoten $k_i$ positiv ist*

   - *$subst(k_i) = \langle S_B, \emptyset\rangle$ mit $S_B = \{\sigma \mid \sigma \in U_{i-1}\}$, und $U_i = U_{i-1}$, falls $k_i$ negativ.*[2]

3. *Für alle negativen UND-Knoten $W$ aus $T'$ ist $nachfolger(W) = K = \langle k_1, \ldots, k_m\rangle$ ein nicht leeres Tupel von ODER-Knoten und es gilt (mit n ist die Länge von W):*

   - *$subst(W) = (S_i)$, $0 \leq i \leq m$ mit $m \leq n$ und $m > 1$ und $S_{i-1} \not\subseteq S_i$ für $1 \leq i \leq m$.*

   *Für das Tupel $K$ gilt folgendes:*

   - *Für jeden positiven ODER-Knoten $k_i \in K$ ist $subst(k_i) = \langle U_B, U_E\rangle$ mit $U_B = \langle\sigma_1, \ldots \sigma_l\rangle$ mit $\{\sigma_1, \ldots, \sigma_l\} \subseteq S_{i-1}$ und $S_i = S_{i-1} \setminus \{\sigma_1, \ldots, \sigma_l\}$ für ein $l > 0$.*

   - *Für jeden negativen ODER-Knoten $k_i \in K$ ist $subst(k_i) = \langle S_{i-1}, S_i\rangle$ mit $S_{i-1} \not\subseteq S_i$.*

[2] Zur Erinnerung: $S$ bezeichnet Mengen, $U$ bezeichnet Tupel!

4. *Für alle positiven ODER-Knoten A aus $T'$, für die die Menge $K = nachfolger(A)$ nicht leer ist, sei für alle $k \in K$ $subst(k) = (U_j^k)$ die durch die Funktion subst gegebene Folge von Substitutionstupeln und $m_k$ die Länge von k. Weiterhin seien alle Knoten $k \in K$ von 1 bis h durchnumeriert, d.h. wir werden sie mit einem Index versehen. Dann gilt:*
   - *Für alle $k \in K$ ist $U_0^k \neq \langle\,\rangle$*
   - *Es gibt eine Permutation $\pi$ von $TU_1 = U_0^{k_1} + \ldots + U_0^{k_h}$ und $TU_2 = U_{m_{k_1}}^{k_1} + \ldots + U_{m_{k_h}}^{k_h}$ so daß $\pi(TU_1) = U_B$ und die Einschränkung von $\pi(TU_2)$ auf $dom(TU_1) \cup var(A) = U_E$*
5. *Für alle negativen ODER-Knoten A aus $T'$ mit $K = nachfolger(A) \neq \emptyset$ (d.h. der Knoten ist kein Blatt), $m_k$ ist die Länge von $k \in K$, $subst(A) = \langle S_B, S_E\rangle$ und $subst(k) = (S_j^k)$ für $0 \leq j \leq m_k$ gilt:*
   - $S_B \not\subseteq S_E$.
   - *Für alle $k \in K$ gilt $S_0^k = S_B$*
   - $\bigcup_{k\in K}(S_{m_k}^k|_{dom(S_0^k)\cup var(A)}) = S_E$
6. *Für alle negativen ODER-Blätter A aus $T'$ (d.h. $nachfolger(A) = \emptyset$) sei $subst(A) = \langle S_B, S_E\rangle$ und es gilt:*
   - *für alle $\sigma \in S_B$ gibt es einen endlichen unvollständigen SLDNF-Baum $T_\sigma^A$ für $D \cup \{\leftarrow A\sigma\}$ derart, daß*
     - $S_E = \bigcup_{\sigma\in S_B}\{\sigma\theta \mid \theta \in überdeckung(T_\sigma^A)\}$.
   - $S_B \not\subseteq S_E$.
7. *Für alle positiven ODER Blätter A mit $subst(L) = \langle U_B, U_E\rangle$ gilt: $U_B = \langle\sigma_1, \ldots, \sigma_n\rangle$, $U_E = \langle\sigma_1\theta_1, \ldots, \sigma_n\theta_n\rangle$ und für $1 \leq i \leq n$ gibt es eine SLDNF-Widerlegung von $D \cup \{\leftarrow A\sigma_i\}$ mit Antwort $\theta_i$.*

Da die Definition auf Anhieb nicht sehr einfach nachzuvollziehen ist, geben wir nun einige Beispiele:

**Beispiel 6.2.1** *Aus einem Beweisbaum T geben wir nun für jeden Knotentyp ein Beispiel. Die Wahl der Knoten sowie der Substitutionsmengen ist willkürlich und soll nur dazu dienen, die grundsätzliche Struktur zu verdeutlichen. Zur Erinnerung: ob ein Knoten positiv oder negativ ist, hängt von der Anzahl der mit $\neg$ markierten Kanten im UND-ODER-Baum ab.*

1. *Positiver UND-Knoten: Sei* $W \equiv p(x,y) \wedge q(x,y) \wedge \neg r(x,y)$ *ein positiver UND-Knoten für den folgendes gilt:*

   - $nachfolger(W) = \langle q(x,y), p(x,y), r(x,y) \rangle$

   - $subst(W) = \langle \{x/a\} \rangle, \langle \{x/a, y/b\} \rangle, \langle \{x/a, y/b\} \rangle, \langle \{x/a, y/b\} \rangle$

   *In einem Beweisbaum sieht dieser Knoten samt Nachfolgern so aus:*

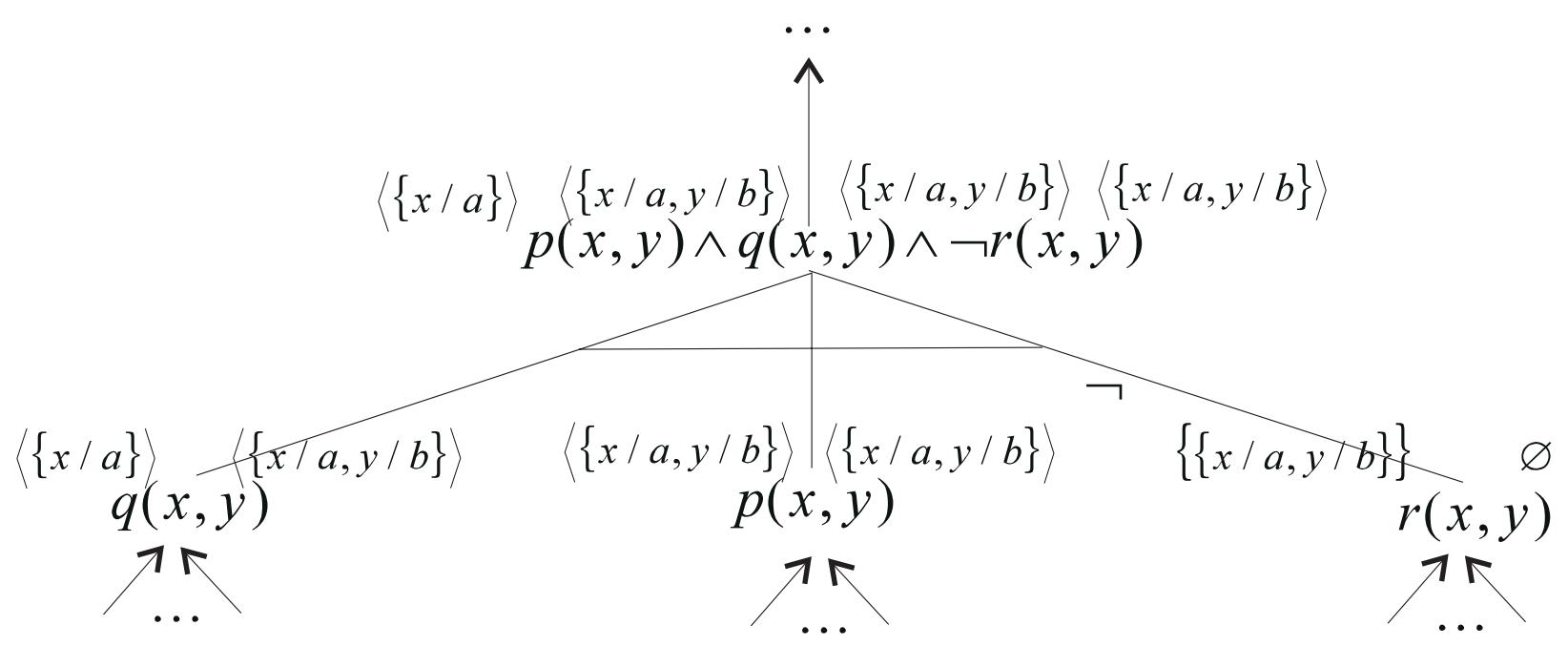

Abbildung 6.4: Positiver UND-Knoten mit Nachfolgern

   *Jedes Literal in einem positiven UND-Knoten erzeugt einen Nachfolgerknoten, da für den Beweis des Knotens jedes Literal gezeigt werden muß. Die Reihenfolge der Nachfolger muß nicht notwendigerweise der Reihenfolge der Literale im UND-Knoten entsprechen. Der Nachfolge-Knoten* $r(x,y)$ *ist in der Abbildung ein negativer ODER-Knoten.*

2. *Negativer UND-Knoten: Sei* $W \equiv p(x,y) \wedge q(x,y) \wedge \neg r(x,y)$ *ein negativer UND-Knoten für den folgendes gilt:*

   - $nachfolger(W) = \langle q(x,y), r(x,y) \rangle$

   - $subst(W) = \langle \{\{x/a\}\}, \{\{x/a, y/b\}, \{x/a, y/c\}, \{x/a, y/d\}\}, \{\{x/a, y/d\}\} \rangle$

   *In einem Beweisbaum kann dieser Knoten samt Nachfolgern wie in Abbildung 6.5 aussehen. Nicht jedes Literal in einem negativen UND-Knoten erzeugt einen Nachfolgerknoten, da ein endlicher gescheiterter SLDNF-Baum gesucht wird und man nur an Literalen interessiert ist, die Lösungen herausfiltern.*

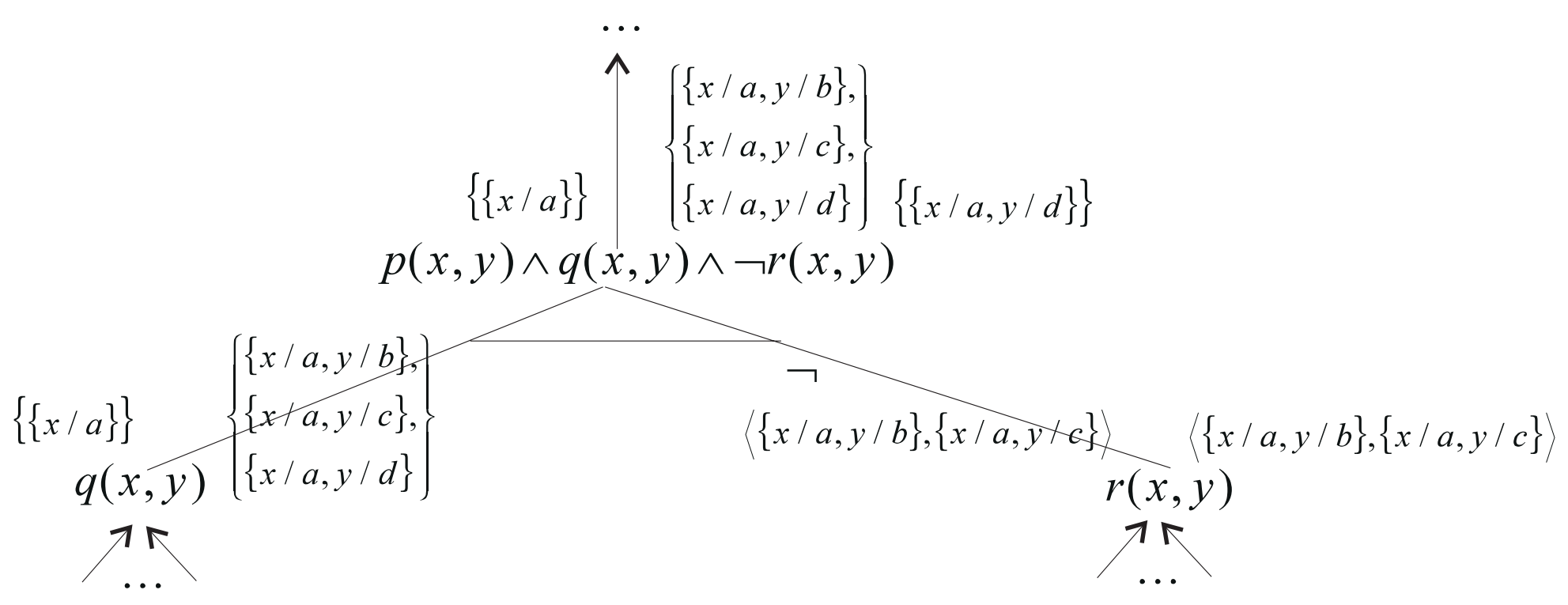

Abbildung 6.5: Negativer UND-Knoten mit Nachfolgern

3. *Positiver ODER-Knoten: Sei* $A \equiv p(x,y)$ *ein positiver ODER-Knoten, für den folgendes gilt:*

   - $nachfolger(A) = \{t(x,y) \wedge g(x,y), l(x,y) \wedge s(x,y)\}$
   - $subst(W) = \langle\langle\{x/a\}\rangle, \langle\{x/a, y/b\}\rangle\rangle$

   *In einem Beweisbaum sieht dieser Knoten samt Nachfolgern wie in Abbildung 6.6 aus.*

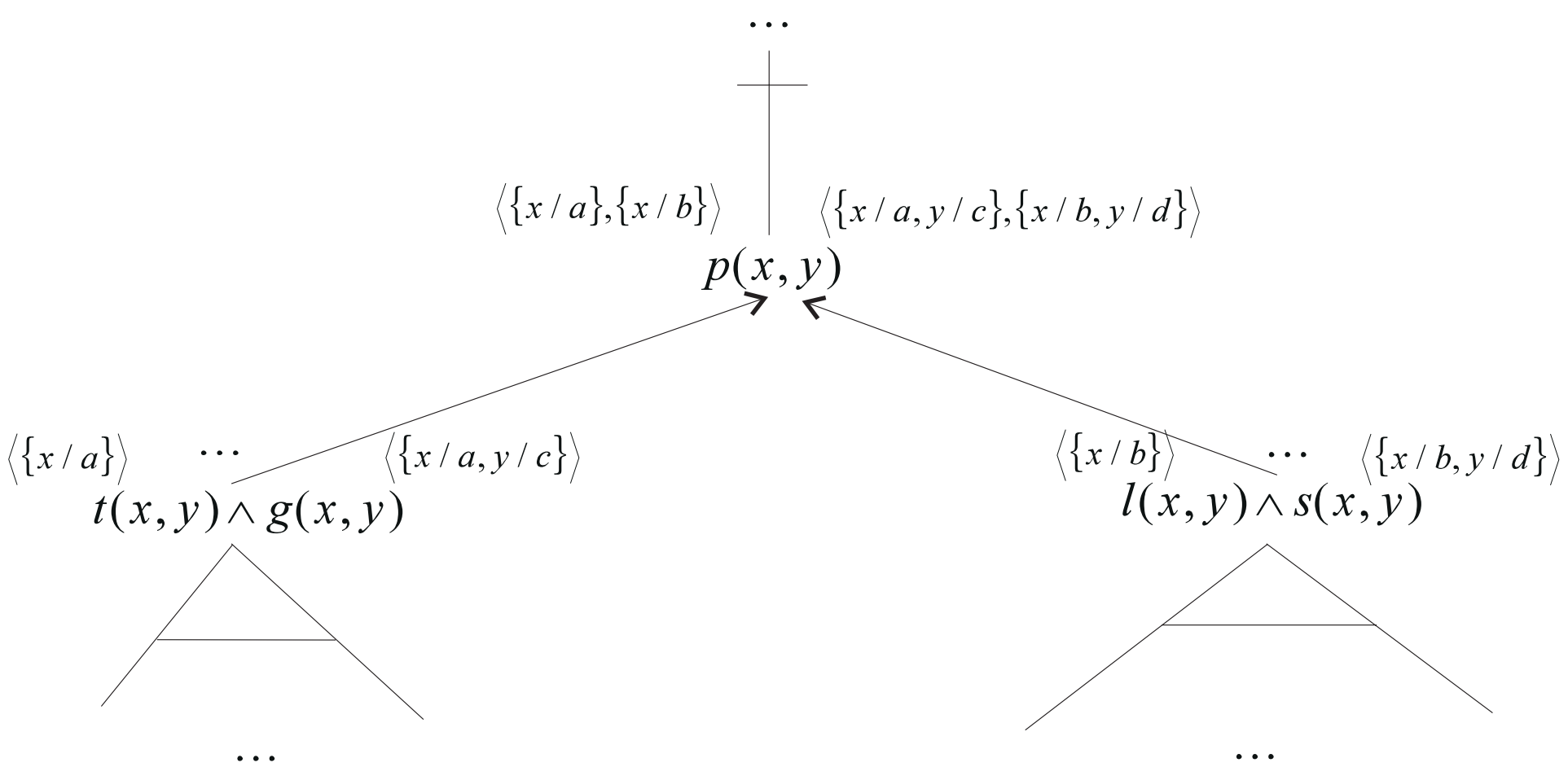

Abbildung 6.6: Positiver ODER-Knoten mit Nachfolgern

*Bei einem positiven ODER-Knoten sind nicht notwendigerweise* alle *Nachfolger im UND-ODER-Baum auch Nachfolger im Beweisbaum, sondern nur die, die für einen Beweis benötigt werden.*

4. *Negativer ODER-Knoten: Sei $A \equiv p(x,y)$ ein negativer ODER-Knoten für den folgendes gilt:*

   - $nachfolger(A) = \{t(x,y) \wedge g(x,y), l(x,y) \wedge s(x,y)\}$
   - $subst(A) = \langle\{\{x/a\}\}, \{\{x/a, y/b\}, \{x/a, y/c\}, \{x/a, y/d\}\}\rangle$

   *In einem Beiweisbaum sieht dieser Knoten samt Nachfolgern so aus:*

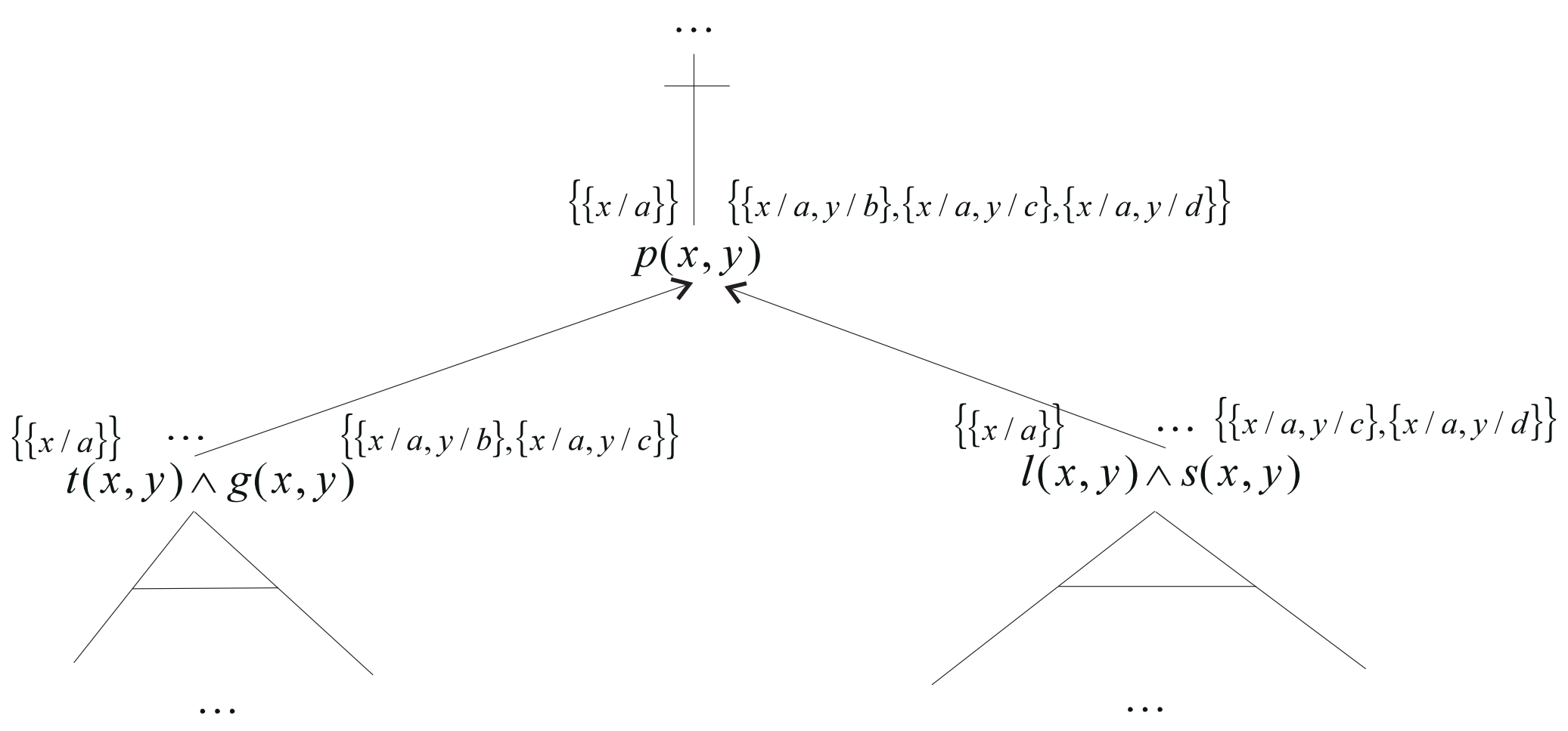

Abbildung 6.7: Negativer ODER-Knoten mit Nachfolgern

   *Bei einem negativen ODER-Knoten sind* alle *Nachfolger im UND-ODER-Baum auch Nachfolger im Beweisbaum, da sichergestellt sein muß, daß sie nicht zusätzliche Lösungen liefern.*

### 6.2.4 Korrektheitsresultate

Nachdem wir die Definition gegeben haben, ist als nächstes zu klären, in welcher Beziehung ein Beweisbaum zu einem SLDNF-Beweis steht. Einen Teil dieser Beziehung verdeutlichen wir durch einen *Korrektheitsbeweis*, in dem wir auf die in dem vorhergehenden Kapitel definierten *unvollständigen SLDNF-Bäume* zurückgreifen werden.

**Definition 6.2.9 (sicherer Beweisbaum)** *Sei $D$ ein normales Programm, $W$ eine Konjunktion von Literalen, $T$ ein UND-ODER-Baum zu $D$ und $W$, sowie $T'$ mit Funktion subst und $nachfolger$ ein Beweisbaum mit Anfangssubstitution $\sigma$. $T'$ heißt* sicherer Beweisbaum, *wenn er folgende Eigenschaften besitzt:*

- *für alle positiven UND-Knoten $W'$ in $T'$ mit $subst(W') = (U_j)$, für die ein negativer ODER-Knoten $k_i \in nachfolger(W')$ ist, ist für alle Substitutionen $\sigma \in U_{i-1}$ $k_i\sigma$ ein Grundatom.*
- *für alle negativen UND-Knoten $W'$ in $T'$ mit $subst(W') = (S_j)$, für die ein positiver ODER-Knoten $k_i \in nachfolger(W')$ ist, ist für alle Substitutionen $\sigma \in (S_{i-1} \setminus S_i)$ $k_i\sigma$ ein Grundatom.*

*Wenn $T'$ nicht sicher ist, dann nennen wir $T'$ einen* unsicheren Beweisbaum.

Ein unsicherer Beweisbaumes korrespondiert mit einer blockierten SLDNF-Ableitung. Wie später gezeigt wird, entspricht der Begriff des *sicheren Beweisbaumes* genau dem Begriff einer SLDNF-Ableitung mit bestimmten Berechnungsregeln, d.h. es gibt zu einer Zielklausel $\leftarrow W$ genau dann eine SLDNF-Ableitung mit bestimmten Berechnungsregeln, wenn es auch einen sicheren Beweisbaum gibt.

**Lemma 6.2.2 (Korrektheit)** *Sei $D$ ein normales Programm, $\leftarrow W$ eine Zielklausel, $T$ ein UND-ODER-Baum aus $W$ und $D$, und $T'$ ein sicherer Beweisbaum von $W$ aus $T$. Dann gilt:*

1. *Für jeden positiven ODER-Knoten $A$ aus $T'$ ist $subst(A) = \langle\langle\sigma_1, \ldots, \sigma_n\rangle$ $,\langle\sigma_1\theta_1, \ldots, \sigma_n\theta_n\rangle\rangle$ und es gilt mit einem beliebigen $n$: Für alle $1 \leq i \leq n$ ist $\theta_i$ eine berechnete Antwort von $D \cup \{\leftarrow A\sigma_i\}$.*
2. *Für jeden negativen ODER-Knoten $A$ aus $T'$ sei $subst(A) = \langle S_B, S_E\rangle$ Dann gilt:*
   (a) *Für alle $\sigma \in S_B$ gibt es gibt es einen endlichen unvollständigen SLDNF-Baum $T^A_\sigma$ für $D \cup \{\leftarrow A\sigma\}$ derart, daß*
      - $S_E = \bigcup_{\sigma \in S_B}\{\sigma\theta \mid \theta$ *ist Element der Überdeckung von* $T^A_\sigma\}$

   (b) $S_B > S_E$

   (c) *Für alle $\sigma \in (S_B - S_E)$ existiert ein endlicher gescheiterter SLDNF-Baum für $D \cup \{\leftarrow A\sigma\}$.*
3. *Für jeden positiven UND-Knoten $W \equiv L_1 \wedge \ldots \wedge L_n$ aus $T'$ mit $subst(W) = \langle\langle\sigma_1, \ldots, \sigma_m\rangle, \langle\sigma_1\theta^1_1, \ldots, \sigma_m\theta^1_m\rangle, \ldots, \langle\sigma_1\theta^1_1 \ldots \theta^n_1, \ldots, \sigma_m\theta^1_m \ldots \theta^n_m\rangle\rangle$ gilt: Für alle $\sigma_j$ ist $\theta = \theta^1_j \ldots \theta^n_j$ eine berechnete Antwort von $D \cup \{\leftarrow W\sigma_j\}$ für alle $1 \leq j \leq m$.*
4. *Für jeden negativen UND-Knoten $W \equiv L_1 \wedge \ldots \wedge L_n$ aus $T'$ mit $subst(W) = (S_i)$,$0 \leq i \leq m$ mit $m \leq n$ gilt:*

*(a)* $S_i = \bigcup_{\sigma \in S_0} \{\sigma\theta \mid \theta$ *ist Element der Überdeckung eines endlichen unvollständigen SLDNF-Baumes für* $D \cup \{\leftarrow (L_1 \wedge \ldots \wedge L_n)\sigma\}\}$ *für alle* $1 \leq i \leq m$.

*(b)* $(S_i)$ *ist eine streng monoton fallende Folge.*

*(c) Für alle* $\sigma \in (S_0 - S_i)$ *gibt es einen endlich gescheiterten SLDNF-Baum für* $D \cup \{\leftarrow W\sigma\}$ *für* $1 \leq i \leq n$

**Beweis** Der Beweis erfolgt über den Aufbau von $T'$, indem zunächst die Behauptungen für alle Blätter gezeigt werden. Dann wird gezeigt, daß, falls für alle Nachfolger eines Knotens die Behauptungen erfüllt sind, die Behauptung auch für den Knoten selbst erfüllt sind. Aus der Endlichkeit des Baumes folgt dann die Korrektheit der Behauptungen.

1. Die Blätter des Beweisbaumes sind nach Definition ODER-Knoten

   - Positive ODER-Blätter: Behauptung folgt direkt aus der Definition 6.2.8 Punkt 7.

   - Negative ODER-Blätter:

   zu 2a. Behauptung folgt aus der Definition 6.2.8 Punkt 6.

   zu 2b. Nach Definition 6.2.8 Punkt 6 ist $S_E = \{\sigma\theta \mid \sigma \in S_B, \theta \in S_\sigma\}$. Dabei ist $S_\sigma$ eine (möglicherweise leere) Menge von Überdeckungen von einem endlichen unvollständigen SLDNF-Baum. Daraus folgt sofort $S_B \geq S_E$. Da $S_E \not\leq S_B$ laut Definition 6.2.8 Punkt 6 folgt sofort $S_E > S_B$.

   zu 2c. Sei $\sigma \in (S_B - S_E)$. Da es kein $\sigma' \in S_E$ mit $\sigma \geq \sigma'$ gibt, muß die Überdeckung von dem endlichen unvollständigen SLDNF-Baum $T_\sigma^A$ leer sein (wäre sie nicht leer, sondern ein $\theta$ in der Überdeckung, so wäre $\sigma\theta \in S_E$ und somit $\sigma \notin (S_B - S_E)$). Wenn die Überdeckung leer ist, so gibt es keinen potentiellen Erfolgsknoten in $T_\sigma^A$. Also sind alle Blätter Mißerfolgsknoten. Ein endlicher unvollständiger SLDNF-Baum, dessen Blätter alle Mißerfolgsknoten sind, ist aber ein endlich gescheiterter SLDNF-Baum.

2. Sei $k$ ein beliebiger Knoten aus $T'$ mit Nachfolgern und es gelten die Behauptungen für alle Nachfolger.

   - $k$ ist negativer UND-Knoten der Länge $n$. Sei $nachfolger(k) = \langle A_1, \ldots, A_m \rangle$ mit $1 \leq m \leq n$ und $subst(k) = (S_j)$, $0 \leq j \leq m$. Sei $L_{i_1} \wedge \ldots \wedge L_{i_m}$ eine Konjunktion, die folgendermaßen konstruiert wird: $L_{i_j} \equiv A_j$, falls $A_j$ negativer ODER-Knoten und $L_{i_j} \equiv \neg A_j$ falls $A_j$ positiver ODER-Knoten ist. Klarerweise ist $L_{i_1} \wedge \ldots \wedge L_{i_m}$ eine Teilkonjunktion von $k$.

zu 4a. Wir zeigen die Behauptung per Induktion nach folgendem Beweisschema:

- Induktionsanfang: Wir zeigen, daß

$$S_i = \bigcup_{\sigma \in S_{i-1}} \{\sigma\theta \mid \theta \in \textit{überdeckung}(T^i_\sigma)\},$$

dabei ist $T^i_\sigma$ ein endlicher unvollständiger SLDNF-Baum für $D \cup \{\leftarrow L_{j_i}\sigma\}$ für $1 \leq i \leq m$.

- Induktionsvoraussetzung

Sei die Behauptung für $S_i$ bzgl. $D \cup \{\leftarrow (L_{j_1} \wedge \ldots \wedge L_{j_i})\sigma$ gezeigt, d.h. zu jedem $\sigma \in S_0$ gibt es einen endlichen unvollständigen SLDNF-Baum $T_\sigma$ für $D \cup \{\leftarrow (L_{j_1} \wedge \ldots \wedge L_{j_i})\sigma\}$, so daß

$$S_i = \bigcup_{\sigma \in S_0} \{\sigma\theta \mid \theta \in \textit{überdeckung}(T_\sigma)\}.$$

- Induktionsschritt:

Wir zeigen, daß die Behauptung dann für $i+1$, $i+1 \leq m$ gilt.

Da $L_{j_1} \wedge \ldots \wedge L_{j_m}$ eine Teilkonjunktion von $k$ ist, gilt somit die Behauptung auch für $k$.

- Induktionsanfang:

Für $L_{j_i}$, $1 \leq i \leq m$, gibt es zwei Möglichkeiten

(a) $L_{j_i}$ ist ein Atom. Dann ist $A_i$ ein negativer ODER-Knoten mit $subst(A_i) = \langle S_{i-1}, S_i \rangle$. Da $A_i$ ein Nachfolger von $k$ ist, folgt die Behauptung aus Behauptung 2a.

(b) $L_{j_i}$ ist ein negatives Literal. Dann ist $A_{j_i}$ ein positiver ODER-Knoten mit $subst(A) = \langle U_B, U_E \rangle$. Nach Definition ist $S_i \subset S_{i-1}$ und $U_B$ ein Tupel genau der $\sigma \in (S_{i-1} \setminus S_i)$. Für ein $\sigma \in S_{i-1}$ gibt es genau zwei Möglichkeiten:

i. $\sigma$ ist aus $U_B$: nach Behauptung 1 ($A_i$ ist ein positiver ODER-Knoten) gibt es für alle $\sigma \in U_B$ eine berechnete Antwort für $D \cup$ $\{\leftarrow A_i\sigma\}$. Also gibt es einen endlichen unvollständigen SLDNF-Baum von $D \cup \{\leftarrow L_{i_j}\sigma\}$, bei dem die Überdeckung leer ist ($L_{i_j}$ ist ein Grundliteral, da $T'$ ein sicherer Beweisbaum ist). Dieser besteht gerade aus der Wurzel $\leftarrow L_{i_j}\sigma$, wobei $L_{i_j}\sigma$ das selektierte Literal der Zielklausel ist.

ii. $\sigma$ ist aus $S_{i-1} \setminus U_B$ (d.h $\sigma \in S_i$). Dann ist $\leftarrow L_{i_j}\sigma$ ein endlicher unvollständiger SLDNF-Baum mit Überdeckung $\{\epsilon\}$ (der nur aus der Wurzel besteht, in der das einzige Literal nicht selektiert ist)

Unter der Verwendung der angegebenen endlichen unvollständigen SLDNF-Bäume ergibt sich die Behauptung.

$*$ Induktionsschritt: Wir zeigen die Behauptung auf folgende Weise:

- $\subseteq$: Zunächst konstruieren wir einen endlichen unvollständigen SLDNF-Baum $T''_\sigma$ für jedes $\sigma \in S_0$ und zeigen, daß $\{\sigma\theta \mid \theta \in \textit{überdeckung}(T''_\sigma)\} \subseteq S_{i+1}$.
- $\supseteq$: Dann zeigen wir, daß es für jedes $\sigma'' \in S_{i+1}$ einen oben konstruierten Baum $T''_\sigma$ gibt, $\sigma'' = \sigma\theta$, $\theta \in \textit{überdeckung}(T''_\sigma)$.

Sei $\sigma \in S_0$ und für jedes $\sigma$ sei $T_\sigma$ der bereits erhaltene endliche unvollständige SLDNF-Baum aus dem letzten Induktionsschritt für $D \cup \{\leftarrow L_{j_1} \wedge \ldots \wedge L_{j_{i-1}})\sigma\}$, so daß

$$S_i = \bigcup_{\sigma \in S_0} \{\sigma\theta \mid \theta \in \textit{überdeckung}(T_\sigma)\}.$$

Aus dem Induktionsanfang folgt für alle $\sigma\theta \in S_i$ die Existenz eines endlichen unvollständigen SLDNF-Baumes $T'_{\sigma\theta}$ für $D \cup \{\leftarrow L_{j_i}\sigma\theta\}$, so daß

$$S_{i+1} = \bigcup_{\sigma\theta \in S_i} \{\sigma\theta\theta' \mid \theta' \in \textit{überdeckung}(T'_{\sigma\theta})\}.$$

Sei $\mathbf{T}'$ die Menge aller $T'_{\sigma\theta}$ mit $\sigma\theta \in S_i$.

Sei für alle $\sigma \in S_0$ $S_\sigma = \{\sigma\theta \mid \theta \in \textit{überdeckung}(T_\sigma)\}$ und $\mathbf{T}_\sigma = \{T'_{\sigma\theta} \mid T'_{\sigma\theta} \in \mathbf{T}'$ und $\sigma\theta \in S_\sigma\}$. Für alle $\sigma \in S_0$ ergibt die vollständige Komposition nach Lemma 5.2.2 von $T_\sigma$ und $\mathbf{T}_\sigma$ einen endlichen unvollständigen SLDNF-Baum $T''_\sigma$, so daß

$$\{\sigma\theta'' \mid \theta'' \in \textit{überdeckung}(T''_\sigma)\} = \bigcup_{T'_{\sigma\theta} \in \mathbf{T}_\sigma} \{\sigma\theta\theta' \mid \theta' \in \textit{überdeckung}(T'_{\sigma\theta})\}.$$

Also ist $\{\sigma\theta'' \mid \theta'' \in \textit{überdeckung}(T''_\sigma)\}$ eine Teilmenge von $S_{i+1}$.

Da jedes $\sigma'' \in S_{i+1}$ Element der Überdeckung eines $T'_{\sigma\theta}$, $\sigma\theta \in S_i$, jedes $\sigma\theta \in S_i$ aber Element der Überdeckung eines $T_\sigma$, $\sigma \in T_\sigma$, ist, ist die Vereinigung aller Mengen $\{\sigma\theta \mid \sigma \in S_0, \theta \in T''_\sigma\}$ gerade $S_{i+1}$.

Zu 4b. Nach 4a ist $S_{i-1} \geq S_i$. Nach Definition 6.2.8 Punkt 3 ist aber $S_i \not\geq S_{i+1}$. Daraus folgt $S_{i-1} > S_i$ für alle $1 \leq i \leq m$.

Zu 4c. Sei $\sigma \in (S_0 - S_i)$ mit $1 \leq i \leq m$. Da es kein $\sigma' \in S_i$ mit $\sigma \geq \sigma'$ gibt, muß die Überdeckung von dem endlichen unvollständigen SLDNF-Baum $T_\sigma$, der in 4b konstruiert wurde, leer sein (wäre sie nicht leer, sondern ein $\theta \in \textit{überdeckung}(T_\sigma)$, so wäre $\sigma\theta \in S_i$ und somit $\sigma \notin (S_0 - S_i)$). Wenn die Überdeckung leer ist, so gibt es keinen potentiellen Erfolgsknoten in $T_\sigma$. Also sind alle Blätter Mißerfolgsknoten. Ein endlicher unvollständiger SLDNF-Baum, dessen Blätter alle Mißerfolgsknoten sind, ist aber ein endlich gescheiterter SLDNF-Baum.

- Sei $k$ ein positiver UND-Knoten der Länge $n$ mit $\textit{subst}(k) = (U_i)$, $0 \leq i \leq n$. Wir zeigen die Behauptung per Induktion über die Nachfolger nach folgendem Beweisschema:

  – Induktionsanfang:

    Sei $A_j$ ein Nachfolger von $k$ und $L \equiv A_j$, falls $A_j$ positiv und $L \equiv \neg A_j$, falls $A_j$ negativ ($A_j$ ist also ein positiver bzw. negativer ODER-Knoten). Sei $\sigma_h$ die $h-$te Substitution aus $U_{j-1}$ und $\sigma_h\theta_h$ die $h-$te Substitution aus $U_j$. Dann ist $\theta_h$ berechnete Antwort für $D \cup \{\leftarrow L\sigma_h\}$. Daraus folgt direkt die Behauptung für $A_1$.

  – Induktionsvoraussetzung:

    Seien $A_1, \ldots, A_i$, $i < n$, die ersten $i$ Nachfolger von $k$ und $L_1, \ldots, L_i$ Literale mit $L_j \equiv A_j$ falls $A_j$ positiver Knoten und $L_j \equiv \neg A_j$, falls $A_j$ negativer Knoten. Für alle $\sigma_h \in U_0$ und $\sigma_h\theta_h^1 \ldots \theta_h^i \in U_i$, wobei $\sigma_h$ und $\sigma_h\theta_h^1 \ldots \theta_h^i$ jeweil an der $h$-ten Stelle im Tupel stehen, ist $\theta = \theta_h^1 \ldots \theta_h^i$ berechnete Antwort für $D \cup \{\leftarrow (L_1 \wedge \ldots \wedge L_i)\sigma_h$.

  – Induktionsschritt:

    Falls für einen Knoten die Behauptungen für die ersten $i$ Nachfolger gilt, dann gilt sie auch für $i+1$ Nachfolger und damit für alle Nachfolger.

Beweis:

- Induktionsanfang: Für den ODER-Knoten $A_j$ gibt es zwei Möglichkeiten:

  (a) $A_j$ ist positiver ODER-Knoten. Dann ist nach Definition $subst(A) = \langle U_{j-1}, U_j \rangle$ und aus der Behauptung für positive ODER-Knoten folgt direkt die Behauptung.

  (b) $A_j$ ist ein negativer ODER-Knoten mit $subst(A_j) = \langle S_B, \emptyset \rangle$. Nach Definition 6.2.8 ist $U_{j-1} = U_j$. Nach Voraussetzung für negative ODER-Knoten gibt es für alle $\sigma \in S_B$ eine endlich gescheiterte SLDNF-Ableitung für $D \cup \{\leftarrow A_j\sigma\}$. Da $A_j\sigma$ ein Grundatom ist, gibt es eine SLDNF-Widerlegung von $D \cup \{\leftarrow \neg A_j\sigma\}$. Da nach Definition in $S_B$ die gleichen Substitutionen enthalten sind, wie in $U_{j-1}$ gibt es also für alle $\sigma \in U_{j-1}$ eine SLDNF-Ableitung von $D \cup \{\leftarrow \neg A_j\sigma\}$ mit $\epsilon$ als berechnete Antwort. Da $U_{j-1} = U_j$ folgt daraus die Behauptung.

- Sei die Behauptung schon für ein $i$, $i < n$ gezeigt, d.h. für ein beliebiges $h$, $1 \leq h \leq m$, existiert eine SLDNF-Widerlegung von $D \cup \{\leftarrow (L_1 \wedge \ldots \wedge L_i)\sigma_h\}$ mit berechneter Antwort $\theta_h$, dabei ist $\sigma_h$ die $h-$te Substitution in $U_0$ und $\sigma_h\theta_h$ die $h-$te Substitution in $U_i$. Aus der Induktionsvoraussetzung folgt die Existenz einer SLDNF-Widerlegung von $D \cup \{\leftarrow L_{i+1}\sigma_h\theta_h\}$ mit berechneter Antwort $\theta'_h$ und die Komposition $\sigma_h\theta_h\theta'_h$ ist die Substitution an der $h-$ten Position in $U_{i+1}$. Die Komposition der SLDNF-Widerlegung für $D \cup \{\leftarrow (L_1 \wedge \ldots \wedge L_i)\sigma_h\}$ mit berechneter Antwort $\theta_h$ und der SLDNF-Widerlegung für $D \cup L_{i+1}\sigma_h\theta_h$ ist nach Lemma 5.1.1 eine SLDNF-Widerlegung von $D \cup \{\leftarrow (L_1 \wedge \ldots \wedge L_{i+1})\sigma_h\}$ mit berechneter Antwort $\theta_h\theta'_h$. Da $h$ beliebig gewählt wurde, gilt dies für alle $h \in \{1, \ldots, m\}$.

Da $L_1 \wedge \ldots \wedge L_n$ bis auf Umordnung der Literale zu $k$ identisch ist, folgt daraus die Behauptung für $k$.

- Sei $k$ ein positiver ODER-Knoten $A$ mit $subst(A) = \langle U_B, U_E \rangle$. Nach Konstruktion sind alle Nachfolger positive UND-Knoten. Nach Definition ist die Menge $M = nachfolger(A)$ nicht leer. Sei für alle $W' \in M$ die Substitutionstupel $subst(W') = (U_i^{W'})$, $0 \leq i \leq m_{W'}$, dabei ist $m_{W'}$ die Länge von $W'$. Nach Punkt 3 gibt es für alle $W' \in M$ und für alle $\sigma \in U_0^{W'}$ eine SLDNF-Ableitung mit berechneter Antwort $\theta$ für $D \cup \{\leftarrow W'\sigma\}$, wobei $\sigma\theta \in U^{W'}_{m_{W'}}$ an der gleichen Position im Tupel steht wie $\sigma$ in $U^{W'}_0$. Nach Definition 6.2.8 Punkt 2 gibt es eine Regel $A' \leftarrow W'' \in D$ mit $\sigma' = mgu(A, A')$ und $W' \equiv W''\sigma'$ bis auf Variablenumbenennungen. Dann gibt es aber auch eine SLDNF-Ableitung von $D \cup \{\leftarrow A\sigma\}$ mit berechneter Antwort $\theta$. Nach Definition 6.2.8 Punkt 2 ist $U_B$ eine Permutation $\pi$ der Konkatenationen aller $U^{W'}_0$ und $U_E$ ist die gleiche Permutation $\pi$ der Konkatenation aller $U^{W'}_{m_{W'}}$ eingeschränkt auf $var(A) \cup dom(U^{W'}_0)$. Daraus folgt die Behauptung.

- Sei $k$ ein negativer ODER-Knoten $A$ mit $subst(A) = \langle S_B, S_E\rangle$ und nichtleerer Menge $M = nachfolger(A)$. Nach Definition 6.2.8 Punkt 5 ist für alle Nachfolger $W' \in M$ ($W'$ ist negativer UND-Knoten) $subst(W') = (S^{W'}_i)$, $0 \leq i \leq m_{W'}$ mit $S^{W'}_0 = S_B$ und

$$S_E = \bigcup_{W' \in M} S^{W'}_{m_{W'}}|_{var(A) \cup dom(S^{W'}_0)}.$$

Dabei ist $m_{W'}$ die Anzahl der Nachfolger von $W'$.

zu 2a. nach Behauptung 4a gibt es für alle $W' \in M$ und alle $\sigma \in S^{W'}_0$ einen endlichen unvollständigen SLDNF-Baum $T^{W'}_\sigma$, so daß

$$S^{W'}_{m_{W'}} = \bigcup_{\sigma \in S^{W'}_0} \{\sigma\theta \mid \theta \in überdeckung(T^{W'}_\sigma)\}.$$

Aus allen $T^{W'}_\sigma$ zusammen mit der Zielklausel $\leftarrow A$ läßt sich ein Baum $T'_\sigma$ auf folgende Art konstruieren:

* $\leftarrow A\sigma$ ist die Wurzel von $T'_\sigma$
* Für alle $W' \in M$ wird der unvollständige SLDNF-Baum $T^{W'}_\sigma$ unter die Wurzel von $T'_\sigma$ gehängt.

$T'_\sigma$ ist wieder ein endlicher unvollständiger SLDNF-Baum da

(a) für jede Regel $A' \leftarrow W \in D$ ein Nachfolgerknoten von $\leftarrow A\sigma$ existiert, so daß mit $\theta = mgu(A\sigma, A')$ gilt $W'\sigma = W\theta$,

(b) jeder $T^{W'}_\sigma$ ein endlicher unvollständiger SLDNF-Baum für $\leftarrow W'\sigma$ ist.

Für $T'_\sigma$ gilt

$$\{\sigma\theta \mid \theta \in überdeckung(T'_\sigma)\} =$$
$$\bigcup_{W' \in M}\{\sigma\theta \mid \theta = \theta'|_{var(A\sigma)} \quad \text{und} \quad \theta' \in überdeckung(T^{W'}_\sigma)\}.$$

Damit gilt

$$S_E = \bigcup_{\sigma \in S_B} \{\sigma\theta \mid \theta \in \textit{überdeckung}(T'_\sigma)\}.$$

zu 2b. Nach 2a gilt: zu jedem $\sigma' \in S_E$ gibt es ein $\sigma \in S_B$ mit $\sigma \geq \sigma'$. Damit ist also $S_B \geq S_E$. Nach Definition 6.2.8 Punkt 5 ist $S_B \not\subseteq S_E$, also $S_B > S_E$.

zu 2c. Da es kein $\sigma' \in S_E$ mit $\sigma \geq \sigma'$ gibt, muß die Überdeckung von dem endlichen unvollständigen SLDNF-Baum $T'_\sigma$ leer sein (wäre sie nicht leer, sondern ein $\theta$ in der Überdeckung, so wäre $\sigma\theta \in S_E$ und somit $\sigma \notin (S_B - S_E)$). Wenn die Überdeckung leer ist, so gibt es keinen potentiellen Erfolgsknoten in $T'_\sigma$. Also sind alle Blätter Mißerfolgsknoten. Ein endlicher unvollständiger SLDNF-Baum, dessen Blätter alle Mißerfolgsknoten sind, ist aber ein endlich gescheiterter SLDNF-Baum.

**Satz 6.2.1 (Korrektheit)** *Sei $D$ ein normales Programm, $\sigma$ eine beliebige Substitution, $W$ eine beliebige Zielklausel der Länge $n$ und $T'$ ein sicherer Beweisbaum für $D \cup \{\leftarrow W\sigma\}$ mit Anfangssubstitution $\sigma$. Dann ist $W$ die Wurzel von $T$, $subst(W) = (U_i)$ und $U_n = \langle\sigma\theta\rangle$ mit $\theta$ ist berechnete Antwort für $D \cup \{\leftarrow W\sigma\}$.*

**Beweis** Nach Definition von $T'$ ist $W$ die Wurzel von $T$ und damit ein positiver UND-Knoten in $T'$ mit $U_0 = \langle\sigma\rangle$. Die Behauptung folgt direkt aus Behauptung 3 aus Lemma 6.2.2.

Die Anfangssubstitution $\sigma$, die auf die Zielklausel $W$ angewendet wird, muß nicht notwendigerweise die Variablen von $W$ belegen. Wird im folgenden im Zusammenhang mit Beweisbäumen keine Anfangssubstitution angegeben, so ist die Anfangssubstitution $\sigma = \epsilon$ (d.h. die identische Substitution).

### 6.2.5 Eigenschaften von Beweisbäumen

In der Definition 6.2.8 wurden noch nicht die Eigenschaften eines Beweisbaumes charakterisiert, die erlaubten Logikprogrammarten entsprechen. Dies holen wir nun nach.

**Definition 6.2.10 (Erlaubter Beweisbaum)** *Sei $D$ normales Programm und $\leftarrow G$ eine Zielklausel. Ein Beweisbaum $T$ für $D \cup \{\leftarrow G\}$ heißt* erlaubter Beweisbaum, *wenn er folgende Eigenschaften hat:*

- *für jeden positiven ODER-Knoten $A$ aus $T$ mit $subst(A) = \langle U_B, U_E\rangle$ gilt: Für jede Substitution $\sigma \in U_E$ ist $A\sigma$ ein Grundatom.*

- *für jeden negativen ODER-Knoten $A$ aus $T$ mit $subst(A) = \langle S_B, S_E \rangle$ gilt: Für jede Substitution $\sigma \in S_E$ ist $A\sigma$ ein Grundatom.*

Ein Beweisbaum ist so definiert, daß seine Konstruktion fast in jeder Höhe abgebrochen werden kann. Für die Blätter sind nur gewisse Eigenschaften verlangt, die mit der SLDNF-Resolution zusammenhängen, aber die Prädikatensymbole der Blätter können sowohl intensional als auch extensional sein. Um einen Beweis vollständig darzustellen, beschränken wir uns auf Beweisbäume, deren Blätter aus der extensionalen Datenbank stammen, d.h. deren Prädikatensymbole werden nur für Fakten verwendet.

**Definition 6.2.11 (vollständiger Beweisbaum)** *Sei $D$ normales Programm und $\leftarrow G$ eine Zielklausel zur Signatur $\Sigma$. Ein Beweisbaum $T$ für $D \cup \{\leftarrow G\}$ heißt* vollständiger Beweisbaum, *wenn für alle Blätter $A$ gilt: $pred(A) \in \Psi^{\Sigma}_{EDB}$.*

**Definition 6.2.12 (Standardbeweisbaum)** *Sei $D$ normales Programm und $\leftarrow G$ eine Zielklausel. Ein Beweisbaum $T$ für $D \cup \{\leftarrow G\}$ heißt* Standardbeweisbaum, *wenn er erlaubt, sicher und vollständig ist.*

Die Substitutionsmengen und -tupel in einem Standardbeweisbaum (bzw. in einem sicheren und erlaubten Beweisbaum, die Vollständigkeit wird nicht benötigt) besitzen bestimmte Eigenschaften, die bei der Integritätsüberprüfung sehr nützlich sind. Diese Eigenschaften werden im folgenden näher geklärt.

**Definition 6.2.13** *Zwei Substitutionen $\sigma$ und $\sigma'$ heißen* unifizierbar, *wenn es zwei Substitutionen $\theta$ und $\theta'$ gibt, so daß $\sigma\theta = \sigma'\theta'$. Gibt es keine zwei solche Substitutionen, so heißen $\sigma$,$\sigma'$ nicht unifizierbar.*

Zwei Substitutionen $\sigma_1$ und $\sigma_2$ sind genau dann unifizierbar, wenn es eine Substitution $\sigma_3$ gibt mit $\sigma_1 \geq \sigma_3$ und $\sigma_2 \geq \sigma_3$.

**Lemma 6.2.3** *Sei $\sigma$ eine beliebige Substitution sowie $\theta$ und $\theta'$ zwei nicht unifizierbare Substitutionen, so daß $dom(\sigma) \cap dom(\theta) = \emptyset$ und $dom(\sigma) \cap dom(\theta') = \emptyset$. Dann sind die Kompositionen $\sigma\theta$ und $\sigma\theta'$ ebenfalls nicht unifizierbar.*

**Beweis** Angenommen $\sigma\theta$ und $\sigma\theta'$ wären unifizierbar mit Substitutionen $\theta_1$ und $\theta_2$. Dann wäre auch $\theta\theta_1 = \theta'\theta_2$.

**Lemma 6.2.4** *Sei $D$ eine deduktive Datenbank, $\leftarrow G$ eine Zielklausel, $T$ ein sicherer erlaubter Beweisbaum. Dann gilt für jeden Knoten $k$ in $T$:*

1. *Wenn $k$ ein positiver UND-Knoten $L_1 \wedge \ldots \wedge L_n$ ist mit $subst(k) = (U_i)$ und $U_i = \langle \sigma_1^i, \ldots, \sigma_m^i \rangle$, dann gilt:*
   - *je zwei Substitutionen $\sigma_j^i$ und $\sigma_p^i$ aus einem beliebigen $U_i$ mit $j \neq p$ sind nicht unifizierbar.*
2. *Wenn $k$ ein positiver ODER-Knoten $A$ ist mit $subst(A) = \langle U_B, U_E \rangle$, dann gilt:*
   - *je zwei Substitutionen $\sigma_j$ und $\sigma_p$ aus $U_B$ (bzw. $U_E$) mit $j \neq p$ sind nicht unifizierbar.*
3. *Wenn $k$ ein negativer UND-Knoten $L_1 \wedge \ldots \wedge L_n$ ist mit $subst(k) = (S_i)$, $0 \leq i \leq m$, dann gilt:*
   - *je zwei verschiedene Substitutionen $\sigma$ und $\sigma'$ aus jedem $S_i$ sind nicht unifizierbar.*
4. *Wenn $k$ ein negativer ODER-Knoten $A$ ist mit $subst(A) = \langle S_B, S_E \rangle$, dann gilt:*
   - *je zwei verschiedene Substitutionen $\sigma$ und $\sigma'$ aus $S_B$ (bzw. $S_E$) sind nicht unifizierbar.*

**Beweis** Wir zeigen die Behauptung ausgehend von der Wurzel für alle Knoten, indem wir die Behauptung für einen Knoten auf die Behauptung für den Vorgänger zurückführen und die Behauptung für die Wurzel direkt zeigen.

- Sei $W$ positiver UND-Knoten. Für $W$ gibt es zwei Möglichkeiten:
  1. $W$ ist die Wurzel von $B$. Dann gilt die Behauptung trivialerweise, da jedes $U_i$ nur eine Substitution enthält.
  2. Sei also $W$ ein Nachfolgerknoten eines positiven ODER-Knotens $A$ und $subst(A) = \langle U_B, U_E \rangle$ und die Behauptung für $A$ sei schon erfüllt. Dann erfüllt auch $U_0$ nach Voraussetzung die Eigenschaft, da $U_B = U_0$. Nach Lemma 6.2.2 gilt für zwei Substitutionen $\sigma_j^i$ und $\sigma_p^i$ aus einem beliebigen $U_i$, daß $\sigma_j^0 \theta_j^i = \sigma_j^i$ und $\sigma_p^0 \theta_p^i = \sigma_p^i$ mit bestimmten Substitutionen $\theta_j^i$ und $\theta_p^i$. Angenommen $\sigma_j^i$ und $\sigma_p^i$ mit $i > 0$ und $p \neq j$ wären unifizierbar . Dann gibt es also auch Substitutionen $\theta$ und $\theta'$ mit $\sigma_j^i \theta = \sigma_p^i \theta'$. Dann wäre aber auch $\sigma_j^0 \theta_j^i \theta = \sigma_p^0 \theta_p^i \theta'$ und somit $\sigma_p^0$ und $\sigma_j^0$ aus $U_0$ unifizierbar im Widerspruch zur Voraussetzung. Also können $\sigma_j^i$ und $\sigma_p^i$ für alle $0 \leq i \leq n$ nicht unifizierbar sein.
- Sei $W$ ein negativer UND-Knoten mit $subst(W) = (S_i)$ mit $n$ Nachfolgern und Vorgänger $A$ mit $subst(A) = \langle S_B, S_E \rangle$, und sei die Behauptung für $A$ schon gezeigt.

Wir zeigen die Behauptung per Induktion über die Anzahl der Nachfolger von W.[3]

1. Induktionsanfang:
   Da $S_0 = S_B$ gilt daher die Behauptung für $S_0$.

2. Induktionsvoraussetzung:
   Sei die Behauptung schon für alle $S_l$, $l \leq i < n$, gezeigt.

3. Induktionsschritt:
   Für den Nachfolger $A_{i+1}$ von $W$ gibt es zwei Möglichkeiten:

   (a) $A_{i+1}$ ist ein negativer ODER-Knoten mit $subst(A_{i+1}) = \langle S'_B, S'_E \rangle$. Seien $\sigma_1$ und $\sigma_2$ zwei beliebige, aber verschiedene Substitutionen aus $S'_E$ (existiert nur eine Substitution in $S'_E$, so ist nichts zu beweisen). Nach Lemma 6.2.2 gibt es für $\sigma_1$ und $\sigma_2$ zwei Möglichkeiten:

      i. $\sigma_1 = \sigma\theta$ mit $\theta \in \mathit{\ddot{u}berdeckung}(T^A_\sigma)$ und $\sigma_2 = \sigma'\theta'$ mit $\theta' \in \mathit{\ddot{u}berdeckung}(T^A_{\sigma'})$. Dabei sind $\sigma$ und $\sigma'$ aus $S'_B$ mit $\sigma \neq \sigma'$ und $T^A_\sigma$ sowie $T'^A_{\sigma'}$ zwei endliche unvollständige SLDNF-Bäume für $D \cup \{\leftarrow A\sigma\}$ bzw. $D \cup \{\leftarrow A\sigma'\}$. Nach Voraussetzung ist aber $S'_B = S_i$ und damit sind auch $\sigma$ und $\sigma'$ nicht unifizierbar. Dann sind aber auch $\sigma\theta$ und $\sigma'\theta'$ nicht unifizierbar.

      ii. $\sigma_1 = \sigma\theta$ und $\sigma_2 = \sigma\theta'$ mit $\theta, \theta' \in \mathit{\ddot{u}berdeckung}(T^A_\sigma)$. Dabei ist $\sigma \in S'_B$ und $T^A_\sigma$ ein endlicher unvollständiger SLDNF-Baum für $D \cup \{\leftarrow A\sigma\}$. Daraus folgt $dom(\theta') \cap dom(\sigma) = \emptyset$. Weiterhin sind $A\sigma\theta'$ und $A\sigma\theta$ Grundatome, da $T$ ein erlaubter Beweisbaum ist. Da $dom(\theta) = dom(\theta') = var(A\sigma)$ ist, gilt also entweder $\theta = \theta'$ oder $\theta$ und $\theta'$ nicht unifizierbar. Wäre $\theta = \theta'$, so wäre aber auch $\sigma_1 = \sigma_2$. Also ist $\theta \neq \theta'$. Da $\theta$ und $\theta'$ Grundsubstitutionen sind, sind sie nicht unifizierbar. Dann ist aber $\sigma_1$ und $\sigma_2$ nicht unifizierbar nach Lemma 6.2.13.

- Sei $A_i$ ein positiver ODER-Knoten mit $subst(A) = \langle U_B, U_E \rangle$ und die Behauptung für den Vorgängerknoten $W$ schon gezeigt. Für den Knoten $W$ gibt es zwei Möglichkeiten:

  1. $W$ ist ein positiver UND-Knoten mit $subst(W) = (U_i)$. Dann ist $U_B = U_{i-1}$ und $U_E = U_i$ und die Behauptung folgt direkt aus der Behauptung für den Vorgängerknoten.

[3] Dies sieht zwar nach einem Zirkelschluß aus, ist es aber nicht, da nicht die zu zeigende Behauptung eingesetzt wird, sondern eine andere Eigenschaft der Nachfolger.

2. $W$ ist ein negativer UND-Knoten mit $subst(W) = (S_i)$. Dann ist $U_B = U_E$ (da $T$ ein sicherer Beweisbaum ist). Da jedes $\sigma \in U_B$ auch Element von $S_{i-1}$ und die Behauptung für den Vorgängerknoten schon gezeigt ist, folgt daraus die Behauptung.

- Sei $A_i$ ein negativer ODER-Knoten mit $subst(A_i) = \langle S_B, S_E \rangle$ und die Behauptung für den Vorgängerknoten $W$ schon gezeigt. Für den Knoten $W$ gibt es wieder zwei Möglichkeiten:

  1. $W$ ist ein negativer UND-Knoten mit $subst(W) = (S_i)$. Dann ist $S_B = S_{i-1}$ und $S_E = S_i$ und die Behauptung folgt aus der Behauptung für den Vorgängerknoten.

  2. $W$ ist ein positiver ODER-Knoten mit $subst(W) = (U_i)$. Dann ist $S_E = \emptyset$ und alle Substitutionen $\sigma \in S_B$ sind auch Element von $U_{i-1}$. Damit folgt die Behauptung aus der Behauptung für den Vorgängerknoten.

Ausgehend von der Wurzel ergibt sich so die Behauptung für alle Blätter des Beweisbaumes.

# Kapitel 7

# SLDNF-Beweise und Beweisbäume

Im vorhergehenden Kapitel haben ist bereits ein Korrektheitsergebnis für Beweisbäume bewiesen worden. Nun interessiert noch ein Vollständigkeitsergebnis, d.h. die Beantwortung folgender Frage: Welche SLDNF-Beweise können als Beweisbäume dargestellt werden? Im folgenden wird ein Verfahren angegeben, mit dem aus einem SLDNF-Beweis, der mit der Standardberechnungsregel durchgeführt wurde, ein Beweisbaum gewonnen werden kann.

## 7.1 SLDNF-Beweise

Der Begriff *SLDNF-Beweis* wurde im Rahmen der Arbeit bereits mehrfach verwendet, ohne daß er formal geklärt wurde. Da zur Transformation eines SLDNF-Beweises eine genaue Begriffsbestimmung notwendig ist, geben wir im folgenden eine Definition.

**Definition 7.1.1 (SLDNF-Beweis)** *Sei $D$ eine deduktive Datenbank, $\leftarrow G$ eine Zielklausel und $\tau$ eine beliebige Substitution. Ein* SLDNF-Beweis *von $D \cup \{\leftarrow G\tau\}$ ist die kleinste endliche Menge* $\mathbf{P}$*, für die gilt:*

- *Eine SLDNF-Widerlegung von $D \cup \{\leftarrow G\tau\}$ ist Element von* $\mathbf{P}$*.*
- *Zu jedem selektierten negativen Literal $\neg A$ in einer SLDNF-Widerlegung aus* $\mathbf{P}$ *existiert genau ein endlich gescheiterter SLDNF-Baum von $D \cup \{\leftarrow A\}$ in* $\mathbf{P}$*.*

- *Zu jedem selektierten negativen Literal $\neg A$ aus einem Mißerfolgsknoten in einem endlich gescheiterten SLDNF-Baum aus* **P** *existiert genau eine SLDNF-Widerlegung von $D \cup \{\leftarrow A\}$ in* **P**.

Wenn jede SLDNF-Ableitung in einem SLDNF-Beweis mit der gleichen Berechnungsregel $R$ durchgeführt wurde, sprechen wir von einem SLDNF-Beweis via $R$. In dieser Definition spielt die Stratifizierung der Datenbank eine Rolle: ist die Datenbank nicht stratifiziert, kann es zu Zyklen im Beweis kommen: zur Widerlegung von $\leftarrow A$ braucht man einen endlich gescheiterten SLDNF-Baum für $\leftarrow B$, für den man wieder eine Widerlegung von $\leftarrow A$ braucht. In solchen Fällen würde das unten angegebene Konstruktionsverfahren für Beweisbäume nicht terminieren — obwohl die Menge **P** selbst endlich ist.

Die obige Definition gibt noch einen weiteren Hinweis auf Optimierung eines SLDNF-Beweises: Offensichtlich reicht es aus, für eine Zielklausel $\leftarrow L$ *eine* SLDNF-Widerlegung für $D \cup \{\leftarrow L\}$ zu finden. Gelangt man während des Beweises zu einem Punkt, an dem $D \cup \{\leftarrow L\}$ noch einmal widerlegt werden muß, so kann die bereits gefundene Widerlegung verwendet werden. Dieses Prinzip findet sich in der OLDT - Resolution (siehe [Tamaki und Sato, 1986, Seki und Itoh, 1988]). Da Beweisbäume ebenfalls eine sparsame Darstellung eines Beweises sind, wäre die Klärung der Verwandtschaft interessant.

Um einen SLDNF-Beweis zu analysieren, werden wir SLDNF-Widerlegungen und endlich gescheiterte SLDNF-Bäume in kleinere Teile zerlegen. Wie dies zu geschehen hat, beschreiben die nachfolgenden Definitionen.

**Definition 7.1.2 (Teilwiderlegung)** *Sei $R$ eine SLDNF-Widerlegung für $D \cup \{\leftarrow L_1 \wedge \ldots \wedge L_n\}$ mit der Standardberechnungsregel. Die* Teilwiderlegung *für $\leftarrow L_1$ ist die kürzeste Teilfolge von $\leftarrow L_1 \wedge \ldots \wedge L_n$ zu $\leftarrow (L_2 \wedge \ldots \wedge L_n)\sigma$, wobei in jeder Zielklausel der Teilfolge die letzten $(n-1)$ Literale gestrichen werden. Die* berechnete Antwort der Teilwiderlegung *ist $\sigma' = \sigma|_{var(L_1)}$. Die Folge $\leftarrow (L_2 \wedge \ldots \wedge L_n)\sigma, \ldots, \Box$ heißt* Restwiderlegung.

Es gilt $(L_2 \wedge \ldots \wedge L_n)\sigma \equiv (L_2 \wedge \ldots \wedge L_n)\sigma'$, da für $L_2 \wedge \ldots \wedge L_n$ nur die Variablen relevant sind, die auch in $L_1$ enthalten sind, da im Ableitungsprozeß immer nur neue Variablen eingeführt werden.

**Beispiel 7.1.1** *Sei $D$ das normale Programm aus Beispiel 5.1.1 und*

$$\begin{aligned} R = \; & \leftarrow p(x) \wedge r(x,y), \leftarrow t(x) \wedge \neg s(x) \wedge r(x,y), \leftarrow \neg s(a) \wedge r(a,y), \\ & \leftarrow r(a,y), \leftarrow u(y) \wedge \neg v(a), \leftarrow \neg v(a), \Box \end{aligned}$$

*eine SLDNF-Widerlegung für* $D \cup \{\leftarrow p(x) \wedge r(x,y)\}$. *Die Teilwiderlegung ergibt sich daraus zu*

$$R' = \leftarrow p(x), \leftarrow t(x) \wedge \neg s(x), \leftarrow \neg s(a), \square.$$

*Die zugehörige Restwiderlegung ist dann:*

$$R'' = \leftarrow r(a,y), \leftarrow u(y) \wedge \neg v(a), \leftarrow \neg v(a), \square.$$

**Lemma 7.1.1** *Sei D deduktive Datenbank und R eine SLDNF-Widerlegung von* $D \cup \{\leftarrow L_1 \wedge \ldots \wedge L_n\}$ *bzgl. der Standardberechnungsregel mit berechneter Antwort* $\sigma$. *Die Teilwiderlegung von R mit berechneter Antwort* $\sigma'$ *ist eine SLDNF-Widerlegung von* $D \cup \{\leftarrow L_1\}$ *bzgl. der Standardberechnungsregel mit berechneter Antwort* $\sigma'$. *Die Restwiderlegung von R ist eine SLDNF-Widerlegung von* $D \cup \{\leftarrow (L_2 \wedge \ldots \wedge L_n)\sigma$ *mit der berechneten Antwort* $\sigma''$. *Es gilt* $\sigma = \sigma'\sigma''$.

**Beweis** Einfache Folgerung aus den Definitionen.

Das Aufspalten einer SLDNF-Widerlegung in eine Teilwiderlegung und eine Restwiderlegung ist die komplementäre Operation zur Komposition von SLDNF-Widerlegungen. Eine ähnliche Aufspaltung benötigen wir noch bei SLDNF-Bäumen.

**Definition 7.1.3 (TeilSLDNF-Baum)** *Sei* $f_\sigma$ *ein endlicher unvollständiger SLDNF-Baum für* $D \cup \{\leftarrow (L_1 \wedge \ldots \wedge L_n)\sigma\}$ *via der Standardberechnungsregel und* $\sigma$ *eine beliebige Substitution. Den TeilSLDNF-Baum* $f'_\sigma$ *von* $f_\sigma$ *erhält man durch folgende Operation:*

- *Ausgehend von der Wurzel werden auf allen Pfaden in allen Knoten die letzten* $n-1$ *Literale* $L'_2 \wedge \ldots \wedge L'_n$ *des Knotens weggelassen solange bis:*
  - *durch das Weglassen in einem Knoten k eine leere Klausel entsteht. Die leere Klausel gehört noch zu* $f'_\sigma$, *die darunterliegenden Äste werden weggelassen. Die Kompositionen der verwendeten Substitutionen, ausgehend von der Wurzel bis zum Knoten k, eingeschränkt auf die Variablen von* $L_1\sigma$, *heißt* berechnete Antwort *auf* $\leftarrow L_1\sigma$. *Der Baum* $f_{\sigma\theta}$, *der sich aus dem (originalen) aktuellen Knoten k und den darunterliegenden (originalen) Ästen zusammensetzt, nennen wir einen* Restbaum *von* $f_\sigma$. $\theta$ *ist dabei die zum Knoten k zugehörige Anwort von* $\leftarrow L_1\sigma$ *in* $f'_\sigma$.
  - *ein Mißerfolgsknoten im SLDNF-Baum erreicht wird. Der Knoten gehört noch zu* $f'_\sigma$.

*Die Menge der Restbäume $F_\sigma$ von $f_\sigma$ ist die Menge aller Restbäume, die bei der Konstruktion von $f'_\sigma$ anfallen.*

**Beispiel 7.1.2** *Sei $D$ das normale Programm aus Beispiel 5.1.1 und $f_\epsilon$ der in Abbildung 7.1 gezeigte SLDNF-Baum für $D \cup \{\leftarrow p(x) \wedge q(x,y)\}$.*

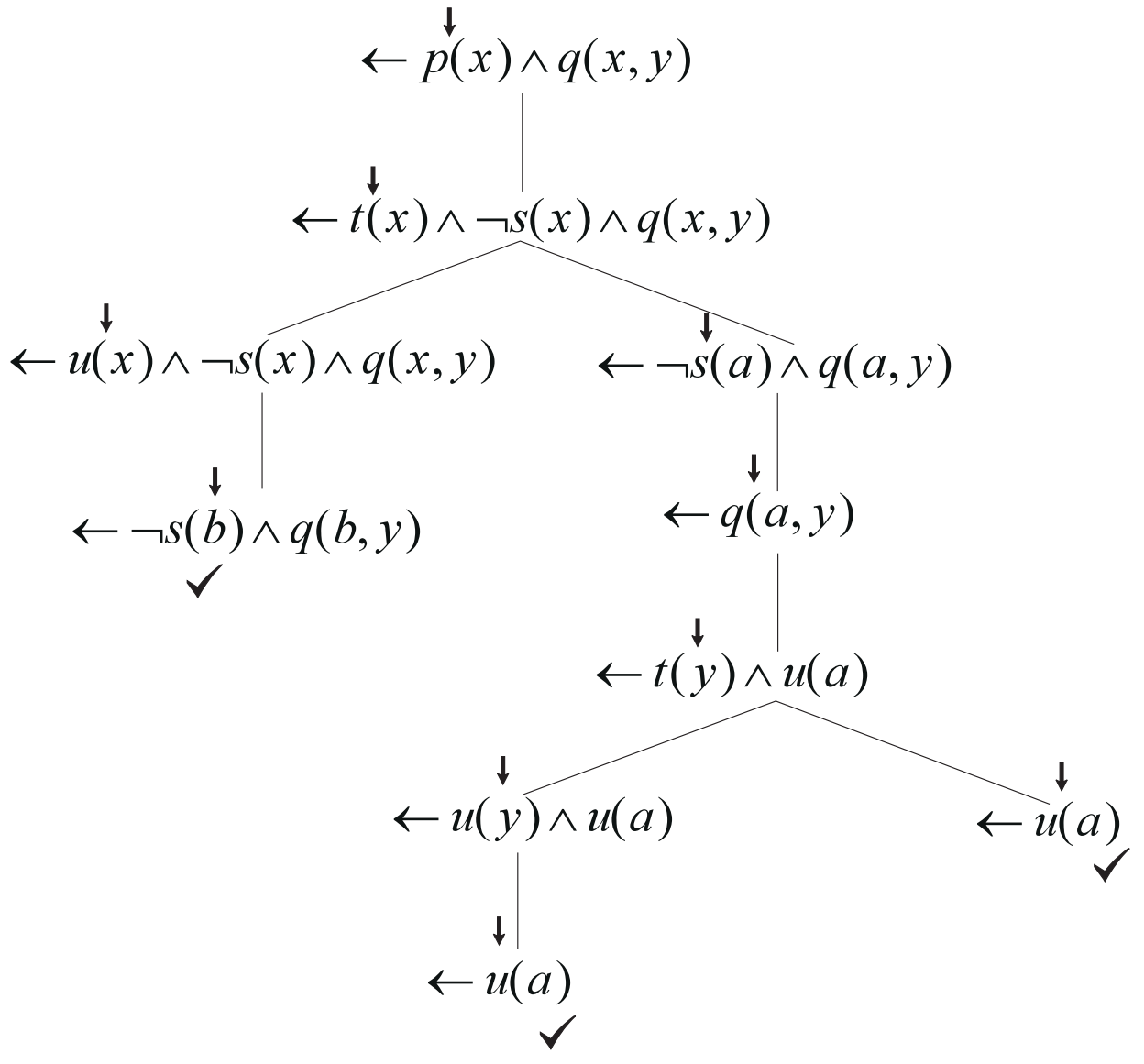

Abbildung 7.1: SLDNF-Baum $f_\epsilon$

*Der TeilSLDNF-Baum $f'_\sigma$ entsteht nun durch das Streichen des letzten Literals in allen Knoten auf allen Pfaden ausgehend von der Wurzel, bis man an einen leeren Knoten gelangt. Der darunterliegende Teilbaum wird weggelassen. Der resultierende Baum $f'_\epsilon$ ist in Abbildung 7.2 zu sehen.*

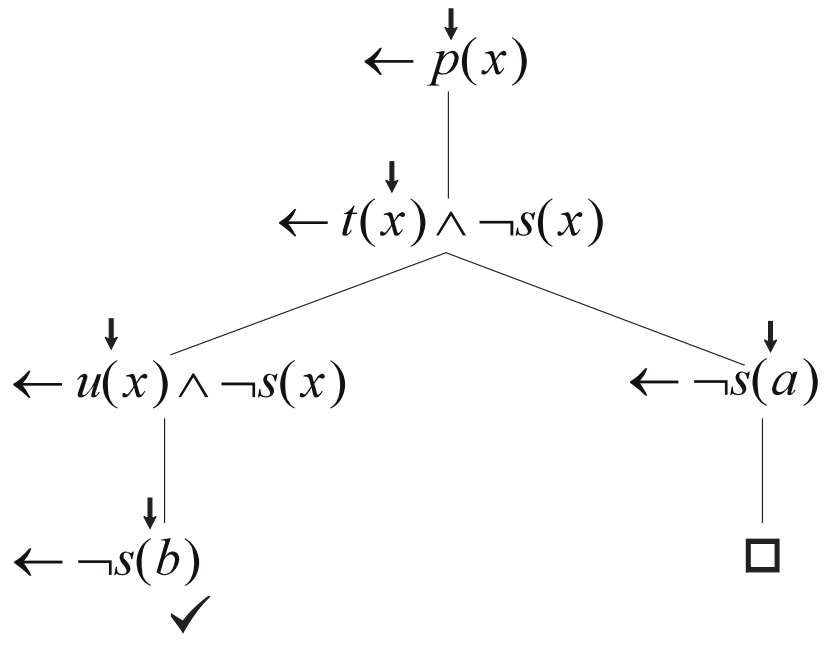

Abbildung 7.2: TeilSLDNF-Baum $f'_\epsilon$

*Es ergibt sich nur ein Restbaum. Dieser Baum ist in Abbildung 7.3 zu sehen.*

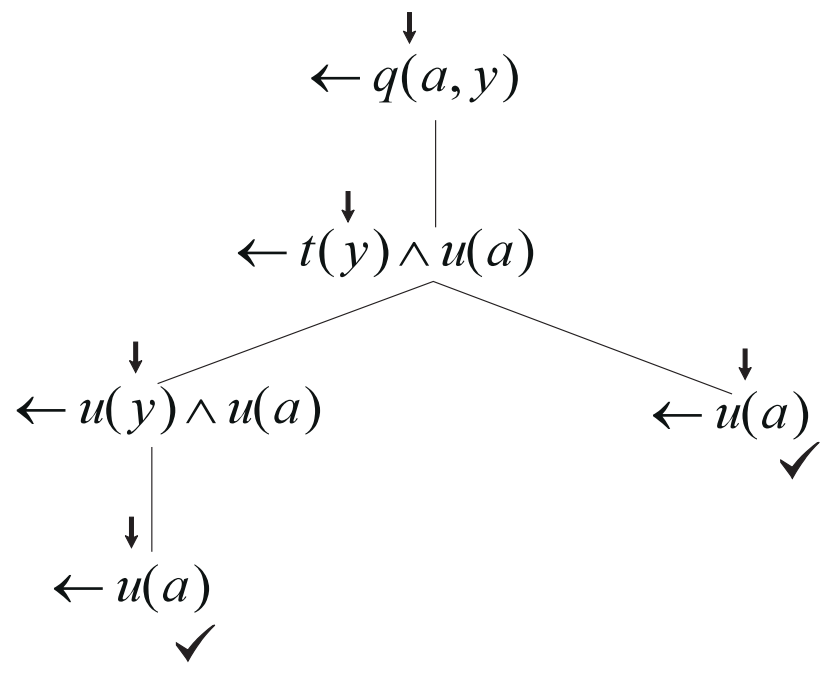

Abbildung 7.3: Restbaum von $f_\epsilon$

**Lemma 7.1.2** *Sei $D$ deduktive Datenbank, $f_\sigma$ ein endlicher unvollständiger SLDNF-Baum von $D \cup \{\leftarrow (L_1 \wedge \ldots \wedge L_n)\sigma\}$ via der Standardberechnungsregel. Sei $f'_\sigma$ der TeilSLDNF-Baum von $f_\sigma$ und $F_\sigma$ die Menge der Restbäume von $f_\sigma$.*

- *Der TeilSLDNF-Baum $f'_\sigma$ ist wieder ein endlicher unvollständiger SLDNF-Baum für $D \cup \{\leftarrow L_1\sigma\}$ via der Standardberechnungsregel.*
- *Jedes $f_{\sigma\theta} \in F_\sigma$, $\theta \in$ überdeckung$(f'_\sigma)$, ist ein endlicher unvollständiger SLDNF-Baum für $D \cup \{\leftarrow (L_2 \wedge \ldots \wedge L_n)\sigma\theta)\}$ via der Standardberechnungsregel.*
- *Für jedes $\theta \in$ überdeckung$(f_\sigma)$ ist ein $f_{\sigma\theta} \in F$.*

**Beweis** Folgt direkt aus dem Aufbau eines SLDNF-Baumes und der Konstruktion von $f'_\sigma$.

Damit erweist sich die Definition des TeilSLDNF-Baumes als komplementär zur Komposition von endlichen unvollständigen SLDNF-Bäumen.

**Definition 7.1.4** *Die Abbildung „beweis" ist eine Funktion von Konjunktionen von Literalen auf Tupel bzw. Mengen von endlichen SLDNF-Widerlegungen bzw. endlichen SLDNF-Bäumen auf folgende Art:*

- *Jeder positive UND-Knoten $W$ der Länge $n$ wird auf eine Folge $(RU_i)$, $0 \leq i \leq n$ von Tupeln von SLDNF-Widerlegungen abgebildet.*
- *Jeder positive ODER-Knoten $A$ wird auf ein Tupel $RU$ von SLDNF-Widerlegungen abgebildet.*
- *Jeder negative ODER-Knoten $A$ wird auf eine Menge von endlichen unvollständigen SLDNF-Bäumen abgebildet.*

- *Jeder negative UND-Knoten A wird auf eine Folge* $(F_i)$ *von Mengen von endlichen unvollständigen SLDNF-Bäumen abgebildet.*

**Definition 7.1.5** *Sei D eine deduktive Datenbank, G eine Konjunktion von Literalen,* $\tau$ *eine beliebige Substitution und* $\mathbf{P}$ *ein SLDNF-Beweis von* $D \cup \{\leftarrow G\tau\}$ *via der Standardberechnungsregel. Der Baum B zu einer Zielklausel* $\leftarrow G$ *zusammen mit den Abbildungen subst, nachfolger und beweis wird induktiv folgendermaßen definiert:*[1]

- *Die Wurzel von B ist* $G \equiv L_1 \wedge \ldots \wedge L_n$ *(und damit ein positiver UND-Knoten). Sei* $R \in \mathbf{P}$ *die SLDNF-Widerlegung von* $D \cup \{\leftarrow G\}$*. Die Folgen* $(U_i) = subst(G)$*,* $(RU_i) = beweis(G)$ *werden folgendermaßen konstruiert:*

  1. $U_0 = \langle\tau\rangle$, $RU_0 = \langle R\rangle$
  2. *Für jedes i mit* $1 \le i \le n$ *wird* $U_i$ *und* $RU_i$ *wie folgt berechnet:*
     (a) *Sei* $\sigma$ *die Substitution aus* $U_{i-1}$ *und* $R'$ *die Widerlegung aus* $RU_{i-1}$.
     (b) $R'$ *wird in eine Teilwiderlegung mit Antwort* $\theta$ *und eine Restwiderlegung* $R'^R$ *zerlegt. Dann berechnen sich* $U_i$ *und* $RU_i$ *zu* $U_i = \langle\sigma\theta\rangle$ *und* $RU_i = \langle R'^R\rangle$

- *Wenn in B ein positiver UND-Knoten* $W \equiv L_1 \wedge \ldots \wedge L_n$ *ist, dann berechnen sich die Abbildungen* $nachfolger(W)$ *sowie subst und beweis für die Nachfolger von W aus den bereits definierten Abbildungen* $(U_i) = subst(W)$*,* $(RU_i) = beweis(W)$*,* $0 \le i \le n$ *folgendermaßen:*

  1. *Für jedes* $1 \le i \le n$ *wird* $A_i$*,* $subst(A_i)$ *und* $beweis(A_i)$ *wie folgt berechnet:*
     (a) *Sei* $\langle\sigma_1, \ldots, \sigma_m\rangle = U_{i-1}$ *und* $\langle R'_1, \ldots, R'_m\rangle = RU_{i-1}$.
     (b) *Falls* $L_i$ *ein Atom ist, dann ist* $A_i \equiv L_i$ *und jedes* $R'_j$*,* $1 \le j \le m$ *wird in eine Teilwiderlegung* $R'^T_j$ *mit Antwort* $\theta_j$ *und eine Restwiderlegung aufgespalten.* $subst(A_i)$ *und* $beweis(A_i)$ *berechnen sich dann zu:*

$$subst(A_i) = \langle\langle\sigma_1, \ldots, \sigma_m\rangle\langle\sigma_1\theta_1, \ldots, \sigma_m\theta_m\rangle\rangle$$

*sowie* $beweis(A_i) = \langle R'^T_1, \ldots, R'^T_m\rangle$*.* $A_i$ *ist dann ein neuer positiver ODER-Knoten.*

[1] Die Begriffe „positiv“ und „negativ“ sowie „UND-Knoten“ und „ODER-Knoten“ werden im folgenden analog zu den entsprechenden Begriffen des Beweisbaumes gebraucht.

(c) *Falls $L_i$ ein negiertes Atom $\neg A'$ ist, dann ist $A_i = A'$ und die Abbildungen berechnen sich zu: $subst(A_i) = \langle\{\sigma_1, \ldots, \sigma_m\}, \emptyset\rangle$ und*

$$beweis(A_i) = F.$$

*Dabei ist $F = \{f_{\sigma_j} \mid 1 \leq j \leq m, f_{\sigma_j} \in \mathbf{P}$ ist endlich gescheiterter SLDNF-Baum für $D \cup \{\leftarrow A_i\sigma_j\}\}$. $A_i$ ist dann ein neuer negativer ODER-Knoten.*

2. *die nachfolger-Abbildung berechnet sich zu*

$$nachfolger(W) = \langle A_1, \ldots, A_n\rangle.$$

- *Wenn in $B$ ein positiver ODER-Knoten $A$ ist, dann berechnen sich die Abbildungen $nachfolger(A)$ sowie subst und beweis für die Nachfolger von $A$ aus den bereits berechneten Abbildungen $subst(A)$ und $beweis(A)$ wie folgt:*

  1. *Falls $A$ ein extensionales Prädikatensymbol besitzt, ist $nachfolger(A) = \langle\rangle$, d.h. der Aufbau stoppt an der Stelle. Ansonsten berechnen sich die Abbildungen $nachfolger(A)$ und subst für die Nachfolger wie folgt:*

     (a) *Sei $subst(A) = \langle\langle\sigma_1, \ldots, \sigma_m\rangle\langle\sigma'_1, \ldots, \sigma_m\rangle\rangle$ und $beweis(A) = \langle R'_1, \ldots, R'_m\rangle$.*

     (b) *Die Abbildung $nachfolger(A)$ wird wie folgt bestimmt: Für alle verschiedenen in dem ersten Schritt von einem $R'_j$, $1 \leq j \leq m$, verwendeten Programmklauseln $A' \leftarrow W$ hat $A$ einen Nachfolger $W\theta$ mit $\theta = mgu(A, A_j)$ [2], d.h. $nachfolger(A) = \{W\theta \mid A' \leftarrow W$ wird in einem $R'_j$, $1 \leq j \leq m$ im ersten Schritt verwendet und $\theta = mgu(A, A'\sigma)\}\}$. Jedes $W\theta$ ist dann ein positiver UND-Knoten.*

     (c) *Die Abbildungen $(U_i) = subst(W')$ und $(RU_i) = beweis(W')$, $1 \leq i \leq n$, $n = länge(W')$ werden für jedes $W'$, wie folgt bestimmt:*

        i. *Für jedes $W\theta \in nachfolger(A)$ wird für jede SLDNF-Widerlegung $R'_j$, in der im ersten Schritt die Regel $A' \leftarrow W$ verwendet wurde, $\sigma_j$ zu $U_0$ und $R''_h$ zu $RU_0$ hinzugefügt. Dabei geht $R''_h$ aus $R'_h$ durch das Weglassen der ersten Zielklausel hervor.*

[2] Dabei muß an die Variablenumbenennung, die beim Aufbau des SLDNF-Baumes verwendet wurde, noch die Bedingung gestellt werden, daß bei verwendeten Programmklauseln abhängig von der Höhe immer die gleiche Variablenumbenennung verwendet wird.

*ii. alle weiteren* $U_i$ *und* $RU_i$, $1 \leq i \leq n$, *werden sukzessive wie folgt bestimmt:*

*A. Sei* $U_{i-1} = \langle \sigma''_1, \ldots, \sigma''_{m'} \rangle$ *und* $RU_{i-1} = \langle R''_1, \ldots, R''_{m'} \rangle$. *Jedes* $R''_d$, $1 \leq d \leq m'$, *wird in eine Teilwiderlegung mit Antwort* $\theta''_d$ *und eine Restwiderlegung* $R''^R_d$ *aufgespalten.*

*B. Dann ist*

$$U_i = \langle \sigma''_1 \theta''_1, \ldots, \sigma''_{m'} \theta''_{m'} \rangle$$

*und*

$$RU_i = \langle R''^R_1, \ldots, R''^R_{m'} \rangle.$$

- *Wenn in* $B$ *ein negativer ODER-Knoten* $A$ *ist, dann berechnen sich die Abbildungen* $nachfolger(A)$ *sowie* $subst$ *und* $beweis$ *für die Nachfolger von* $A$ *aus den bereits berechneten Abbildungen* $subst(A)$ *und* $beweis(A)$ *wie folgt:*

  1. *Falls* $A$ *ein extensionales Prädikat ist, ist* $nachfolger(A) = \emptyset$, *d.h. der Aufbau stoppt an der Stelle. Ansonsten berechnen sich die Abbildungen* $nachfolger(A)$ *und* $subst$ *für die Nachfolger wie folgt:*

     *(a) Sei* $F = beweis(A)$ *und* $\langle S_B, S_E \rangle = subst(A)$.

     *(b) Für alle Programmklauseln* $A' \leftarrow W'$, *die in einem* $f_\sigma \in F$ *im ersten Schritt verwendet wurden, hat* $A$ *einen Nachfolger* $W'\theta$ *mit* $\theta = mgu(A', A)$, *d.h.* $nachfolger(A) = \{W'\theta \mid A' \leftarrow W'$ *wurde in einem* $f_\sigma \in F$ *verwendet und* $\theta = mgu(A, A')\}$. *Alle* $W'\theta$ *sind negative UND-Knoten.*

     *(c) Um für jedes* $W \in nachfolger(A)$ *die Abbildungen* $(S_i) = subst(W)$ *und* $(F_i) = beweis(W)$ *zu bestimmen, berechnen wir zunächst zwei Folgen* $(S'_i)$ *und* $(F'_i)$ *aus denen die Folgen* $(S_i)$ *und* $(F_i)$ *hervorgehen.*

     *I. Zunächst berechnen wir* $F'_0$ *und* $S'_0$ *zu einen gegebenen* $W$: *Von jedem* $f_\sigma \in F$ *erhalten wir durch das Entfernen des Wurzelknotens eine Menge von Bäumen* $f'_\sigma$, *einen für jede im ersten Schritt anwendbare Regel. Wir fügen jeden dieser Bäumen* $f'_\sigma$, *bei denen nun die Wurzel aus der gleichen Programmklausel* $A' \leftarrow W'$ *stammt, aus der auch* $W$ *hervorgegangen ist, zu* $F_0$ *hinzu.*

     *II. Wir setzen* $S'_0 = S_B$.

III. *Ausgehend von* $F_0'$ *und* $S_0'$ *berechnen wir nun sukzessive die restlichen* $F_i'$ *und* $S_i'$ *für* $1 \leq i \leq n$, *wobei* $n = länge(W)$. *Jeden Baum* $f_\sigma$ *aus* $F_{i-1}'$ *spalten wir auf in einen TeilSLDNF-Baum* $f_\sigma'$ *und eine Menge von Restbäumen* $F_{f_\sigma}^R$. *Dann ist*

$$F_i' = \bigcup_{f_\sigma \in F_{i-1}'} F_{f_\sigma}^R$$

*und*

$$S_i' = \bigcup_{f_\sigma \in F_{i-1}'} \{\sigma\theta \mid \theta \in überdeckung(f_\sigma'), \quad f_\sigma' \text{ ist TeilSLDNF-Baum von } f_\sigma\}$$

(d) *Die Folgen* $(S_i)$ *und* $(F_i)$ *bestimmen wir nun folgendermaßen:*

i. $S_0 = S_0'$, $F_0 = F_0'$

ii. *Die restlichen Folgeglieder sind gerade die* $S_j'$ *und* $F_j'$, *für die* $S_j' \neq S_{j-1}'$ *(siehe maximale monoton fallende Teilfolge).*

- *Wenn in* $B$ *ein negativer UND-Knoten* $W = L_1 \wedge \ldots \wedge L_n$ *ist, dann berechnen sich die Abbildungen* $nachfolger(W)$ *sowie subst und beweis für die Nachfolger von* $W$ *aus den bereits berechneten Abbildungen* $subst(W)$ *und* $beweis(W)$ *wie folgt:*

1. *Sei* $(S_i) = subst(W)$ *und* $(F_i) = beweis(W)$ *für* $0 \leq i \leq m$. *Die Nachfolger von* $W$ *gewinnen wir auf folgende Weise:*

(a) *Jedes* $S_i$, $1 \leq i \leq m$ *ist aus einem* $S_j'$ *(siehe negativer UND-Knoten im vorhergehenden Punkt) hervorgegangen, und dafür hat hat* $W$ *einen Nachfolger* $A_i$. $A_i$ *wird aus* $L_j$ *auf folgende Art bestimmt:*

i. *Wenn* $L_j$ *ein Atom ist, dann ist* $A_i = L_j$. $A_i$ *ist dann ein negativer ODER-Knoten.*

ii. *Wenn* $L_j$ *ein negiertes Atom* $\neg A'$ *ist, dann ist* $A_i = A'$. $A_i$ *ist dann ein positiver ODER-Knoten.*

(b) *Die Nachfolger ergeben sich dann zu*

$$nachfolger(W) = \langle A_1, \ldots, A_m \rangle.$$

2. *Für jedes* $A_i$, $1 \leq i \leq m$ *wird* $subst(A_i)$ *und* $beweis(A_i)$ *auf folgende Art gewonnen:*

(a) *Wenn $A_i$ ein negativer ODER-Knoten ist, dann wird jedes $f_\sigma \in F_{i-1}$ aufgespalten in einen TeilSLDNFBaum $f'_\sigma$ und eine Menge von Restbäumen. Dann ist:*

$$beweis(A_i) = \{f'_\sigma \mid f'_\sigma \text{ ist TeilSLDNFBaum von } f_\sigma \in F_{i-1}\}$$

*und*

$$subst(A_i) = \langle S_{i-1}, S_i \rangle.$$

(b) *Wenn $A_i$ ein positiver ODER-Knoten ist, dann ist $subst(A) = \langle U_B, U_B \rangle$, dabei ist $U_B$ ein Tupel genau der $\sigma$, für die $f_\sigma$ ein nur aus der Wurzel bestehender Mißerfolgsbaum ist, d.h. für die eine SLDNF-Widerlegung $R_i \in \mathbf{P}$ für $D \cup \{\leftarrow A_i\sigma\}$ existiert. Somit ist dann $subst(A_i) = \langle \sigma_1, \ldots, \sigma_l \rangle$ mit $\{\sigma_1, \ldots, \sigma_l\} = S_{i-1} \backslash S_i$ und $beweis(A_i) = \langle R_1, \ldots, R_l \rangle$. Dabei ist jedes $R_j$ mit $1 \leq j \leq l$ eine SLDNF-Widerlegung von $D \cup \{\leftarrow A_i\sigma_j\}$.*

Der Konstruktionsprozeß terminiert immer mit extensionalen Prädikaten, da endliche Beweise immer weiter zerlegt werden und im Beweis selbst keine Zyklen auftauchen.

**Lemma 7.1.3** *Sei $D$ eine deduktive Datenbank, $G$ eine Konjunktion von Literalen, $\tau$ eine beliebige Substitution, $\mathbf{P}$ ein unter Verwendung der Standardberechnungsregel durchgeführter SLDNF-Beweis von $D \cup \{\leftarrow G\tau\}$ und $B$ ein aus $\mathbf{P}$ nach Definition 7.1.5 konstruierter Baum zusammen mit den dazugehörigen Abbildungen subst, nachfolger und beweis. Dann gilt für jeden Knoten $k$ in $B$:*

- *Wenn $k$ ein positiver UND-Knoten $L_1 \wedge \ldots \wedge L_n$ ist mit $subst(k) = (U_i)$, $nachfolger(k) = \langle A_1, \ldots, A_n \rangle$ und $beweis(k) = (RU_i)$, $U_i = \langle \sigma_1^i, \ldots, \sigma_m^i \rangle$ und $RU_i = \langle R_1^i, \ldots, R_m^i \rangle$, $0 \leq i \leq n$, dann gilt:*
  - *Jedes $R_j^i$, $1 \leq j \leq m-1$ aus $RU_i$, $0 \leq i \leq n$ ist eine SLDNF-Widerlegung für $D \cup \{\leftarrow (L_{i+1} \wedge \ldots \wedge L_n)\sigma_j^i\}$.*
- *Wenn $k$ ein positiver ODER-Knoten $A$ ist mit $subst(A) = \langle \langle \sigma_1, \ldots, \sigma_m \rangle, \langle \sigma'_1, \ldots, \sigma'_m \rangle \rangle$ und $beweis(A) = \langle R_1, \ldots, R_m \rangle$, dann gilt:*
  - *jedes $R_j$, $1 \leq j \leq m$ ist eine SLDNF-Widerlegung von $D \cup \{\leftarrow A\sigma_j\}$ mit Antwort $\theta_j$, so daß $\sigma'_j = \sigma_j\theta_j$.*
- *Wenn $k$ ein negativer UND-Knoten $L_1 \wedge \ldots \wedge L_n$ ist, mit $nachfolger(k) = \langle A_1, \ldots, A_m \rangle$, $subst(k) = (S_i)$, $0 \leq i \leq m$, $beweis(k) = (F_i)$, $0 \leq i \leq m$, und*

*$L_j$ ist das Literal in $k$, aus dem der Nachfolgeknoten $A_i$, $1 \leq i \leq m$, hervorgegangen ist, dann gilt:*

- *Für jedes $\sigma \in S_i$, $0 \leq i \leq m-1$ existiert ein endlicher unvollständiger SLDNF-Baum $f_\sigma$ in $F_i$ für $D \cup \{\leftarrow (L_j \wedge \ldots \wedge L_n)\sigma\}$*
- *Für jedes $S_i$, $1 \leq i \leq m$, gilt: $S_i = \bigcup_{f_\sigma \in F_{i-1}} \{\sigma\theta \mid \theta \in$ überdeckung$(f'_\sigma)$, $f'_\sigma$ ist TeilSLDNF-Baum von $f_\sigma\}$. Jedes $f'_\sigma$ ist ein endlicher unvollständiger SLDNF-Baum für $D \cup \{\leftarrow L_j\sigma\}$.*

- *Wenn $k$ ein negativer ODER-Knoten $A$ ist mit subst$(A) = \langle S_B, S_E \rangle$ und beweis$(A) = F$, dann gilt:*
  - *für jedes $\sigma \in S_B$ gibt es einen endlichen unvollständigen SLDNF-Baum $f_\sigma \in F$ für $D \cup \{\leftarrow A\sigma\}$.*
  - *$S_E = \bigcup_{\sigma \in S_B} \{\sigma\theta \mid \theta \in$ überdeckung$(f_\sigma)\}$.*

**Beweis** Wir geben hier nur die Beweisidee: Ausgehend von der Wurzel von $B$ gilt:

- In jedem positiven UND-Knoten werden SLDNF-Widerlegungen zerlegt in Teilwiderlegungen und Restwiderlegungen. Jede Teilwiderlegung ist eine Widerlegung für das jeweilige Literal.
- In jedem positiven ODER-Knoten (außer den Blättern) wird eine SLDNF-Widerlegung auf die verwendeten Programmklauseln untersucht. Läßt man dann die erste Zielklausel der Widerlegung weg, ergibt sich wieder eine SLDNF-Widerlegung für den Rumpf der verwendeten Zielklausel.
- In jedem negativen UND-Knoten werden SLDNF-Bäume zerlegt in Teilbäume und eine Menge von Restbäumen. Jeder Teilbaum ist ein endlicher unvollständiger SLDNF-Baum für das jeweilige Literal, jeder Restbaum ein endlicher unvollständiger SLDNF-Baum für den Rest des Knotens.
- In jedem negativen ODER-Knoten (außer den Blättern) werden endliche unvollständige SLDNF-Bäume auf die verwendeten Programmklauseln untersucht. Läßt man die Wurzel des Baumes weg, so erhält man wieder endliche unvollständige SLDNF-Bäume für den Rumpf der jeweiligen verwendeten Zielklausel.

Betrachtet man nun noch die Antwortsubstitutionen, die sich aus den Teilwiderlegungen und TeilSLDNF-Bäumen ergeben, ergibt sich die Behauptung.

**Satz 7.1.1 (Vollständigkeit)** *Sei $D$ eine deduktive Datenbank, $G$ eine Konjunktion von Literalen, $\tau$ eine Substitution,* **P** *ein mit der Standardberechnungsregel bestimmter SLDNF-Beweis von $D \cup \{\leftarrow G\tau\}$ und $B$ ein aus* **P** *nach Definition 7.1.5 konstruierter Baum zusammen mit den dazugehörigen Abbildungen $subst$, $nachfolger$ und $beweis$. Dann ist $B$ ein Standardbeweisbaum für $D \cup \{\leftarrow G\}$ mit Anfangssubstitution $\tau$.*

**Beweis** Beweisidee: Wir zeigen die Behauptung indem wir, ausgehend von der Wurzel, für jeden Nachfolger eines Knotens zeigen, daß er die Bedingungen der Definition der Knoten in einem Beweisbaum erfüllt.

- Die Eigenschaften, die für alle Knoten verlangt werden, folgen nach Konstruktion.
- Die Eigenschaft der Sicherheit folgt aus der Existenz eines SLDNF-Beweises. Die an den jeweiligen Knoten verwendeten Substitutionen finden sich im SLDNF-Beweis an der Stelle wieder, an der das entsprechende negative Literal selektiert wird.
- Da wir nur von erlaubten Programmen ausgehen, und alle Substitutionen Antworten auf SLDNF-Ableitungen sind, sind diese immer Grundsubstitutionen für die ODER-Knoten. Also ist $B$ auch ein Standardbeweisbaum.

## 7.2 Beispiel

Das vorgestellte Verfahren führen wir nun in einem Beispiel Schritt für Schritt nocheinmal vor. Sei $D$ die in Kapitel 4 vorgestellte Datenbank und zusammen mit den normalen Programmklauseln

$$\begin{array}{lcl} ic & \leftarrow & \neg p_1 \\ p_1 & \leftarrow & employee(E) \wedge \neg access(E, menu) \end{array}$$

Der SLDNF-Beweis **P** für $D \cup \{\leftarrow ic\}$ besteht aus folgenden Elementen:

1. drei SLDNF-Widerlegungen $R_1$, $R_2$ und $R_3$:
   (a) $R_1 = \leftarrow ic, \leftarrow \neg p_1, \square$
   (b) $R_2 = \leftarrow access(hans, menu), \leftarrow owner(hans, menu), \square$
   (c) $R_3 = \leftarrow access(peter, menu), \leftarrow manager(peter, E_2) \wedge owner(E_2, menu),$
   $\leftarrow owner(hans, menu), \square$

2. einem endlichen gescheiterten SLDNF-Baum $f_\epsilon$:

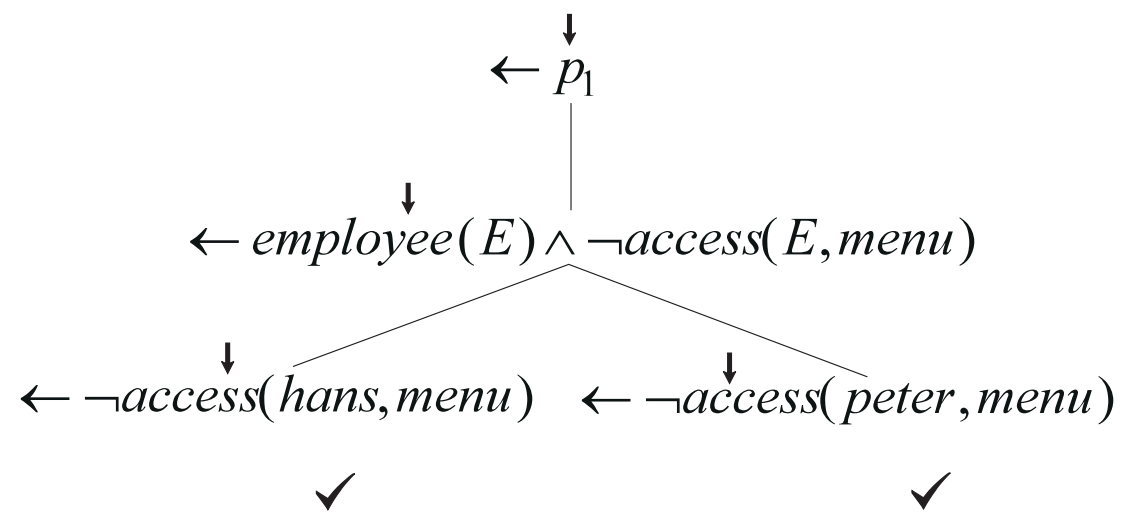

Abbildung 7.4: endlicher gescheiterter SLDNF-Baum $f_\epsilon$ für $D \cup \{\leftarrow p_1\}$

Die Konstruktion des Baumes $B$ mit den Abbildungen $nachfolger$, $subst$ und $beweis$ sieht folgendermaßen aus:

1. Die Wurzel von $B$ ist nach Definition die (nur aus einem Literal bestehende) Konjunktion $ic$.

   - $subst(ic) = (U_i)$, $0 \leq i \leq 1$ mit $U_0 = U_1 = \langle \epsilon \rangle$ und
   - $beweis(ic) = (RU_i)$, $0 \leq i \leq 1$ mit $RU_0 = \leftarrow ic, \leftarrow \neg p_1, \Box$, $RU_1 = \Box$.

   Damit liegt die Wurzel mit ihren Eigenschaften eindeutig fest. Diese Wurzel ist nach Definition ein positiver UND-Knoten.

2. Damit existiert in $B$ ein positiver UND-Knoten, für den die Abbildungen $subst$ und $beweis$ bereits festgelegt wurden. Damit kommen wir nun einen Schritt weiter:

   - $nachfolger(ic) = \langle ic' \rangle$[3]
   - $subst(ic') = \langle \langle \epsilon \rangle, \langle \epsilon \rangle \rangle$
   - $beweis(ic') = \langle \leftarrow ic, \leftarrow \neg p_1, \Box \rangle$. Diese Teilwiderlegung ist identisch zur Ursprungswiderlegung, da auch die Ursprungswiderlegung in der ersten Zielklausel nur ein Literal enthält.

   Damit liegt nun der Nachfolger der Wurzel von $B$ fest. Dieser Nachfolger ist nach Definition ein positiver ODER-Knoten.

[3] In der Definition eines Beweisbaumes haben wir aus Verständnisgründen die Knoten und ihre Beschriftungen identifiziert. Dies führt dann zu Schwierigkeiten, wenn zwei verschiedene Knoten gleich beschriftet sind. Aus diesem Grund heißt der Nachfolgerknoten des positiven UND-Knotens $ic'$, obwohl er genauso wie sein Vorgängerknoten $ic$ heißen müßte.

3. Damit kommen wir zum nächsten Schritt: es wird geprüft, ob $ic$ ein extensionales Prädikatensymbol ist. Dies ist jedoch nicht der Fall, deshalb kommen die weiteren Regeln zum Einsatz. In allen Beweisen, die in $beweis(ic')$ enthalten sind, wurde im ersten Schritt die Programmklausel $ic \leftarrow \neg p_1$ verwendet. Daher ist der einzige Nachfolger von $ic'$ der positive UND-Knoten $\neg p_1$. Die Abbildungen für $p_1$ ergeben sich somit zu:
   - $subst(\neg p_1) = (U_i),\ 0 \leq i \leq 1$ mit $U_0 = U_1 = \langle \epsilon \rangle$, da $p_1$ keine Variablen enthält.
   - $beweis(\neg p_1) = (U_i),\ 0 \leq i \leq 1$ mit $U_0 = \langle \leftarrow \neg p_1, \Box \rangle$ und $U_1 = \langle \Box \rangle$. Dabei ist $U_0$ gerade die zu $ic'$ gehörende Widerlegung, bei der die erste Zielklausel $(\leftarrow ic)$ weggelassen wurde.

4. Die Nachfolger des positiven UND-Knotens $\neg p_1$ werden auf die gleiche Weise bestimmt wie die Nachfolger des positiven UND-Knotens $ic$. In diesem Fall ist der einzige Nachfolger der negative ODER-Knoten $p_1$, d.h.
   - $nachfolger(\neg p_1) = \langle p_1 \rangle$
   - $beweis(p_1) = \{f_\epsilon\}$. Dabei ist $f_\epsilon$ der in Abbildung 7.4 zu sehende endliche gescheiterte SLDNF-Baum von $D \cup \{\leftarrow p_1 \epsilon\}$.
   - $subst(p_1) = \langle \{\epsilon\}, \emptyset \rangle$. Da ein Polaritätswechsel vorliegt, ist auch ein Wechsel in der Betrachtungsweise der Substitutionen nötig: $p_1$ wird nicht auf ein Tupel von Substitutionen, sondern auf eine Menge abgebildet.

5. Die Nachfolger des negativen ODER-Knotens $p_1$ werden nun aus der Menge der SLDNF-Bäume $beweis(p_1)$ bestimmt, da $p_1$ kein extensionales Prädikatensymbol ist. In diesem Fall enthält die Menge $beweis(p_1)$ nur einen Baum $f_\epsilon$. Die Wurzel $\leftarrow p_1 \epsilon$ von $f_\epsilon$ hat nur einen Nachfolger, da es nur die eine anwendbare Programmklausel $p_1 \leftarrow employee(E) \land \neg access(E, menu)$ in $D$ gibt. Daraus ergibt sich als Nachfolger von $p_1$ der folgende negative UND-Knoten:
   - $nachfolger(p_1) = \{employee(E) \land \neg access(E, menu)\}$.
   - Um die Folgen $(F_i) = beweis(employee(E) \land \neg access(E, menu))$ und die Folge $(S_i) = subst(employee(E) \land \neg access(E, menu))$ zu bestimmen, berechnen wir zunächst die Folgen $(F'_i)$ und $(S'_i)$ für $0 \leq i \leq 2$.
     (a) Es ist $F'_0 = \{f^0_\epsilon\}$ und $S'_0 = \epsilon$. Dabei ist $f^0_\epsilon$ der in Abbildung 7.5 zu sehende SLDNF-Baum. Er geht aus $f_\epsilon$ durch das Weglassen des ersten

Knotens hervor.

(b) Weiter ist $F'_1 = \{f^{1a}_{\{E/hans\}}, f^{1b}_{\{E/peter\}}\}$ und $S'_1 = \{\{E/hans\}, \{E/peter\}\}$. Dabei sind die zwei Bäume $f^{1a}_{\{E/hans\}}$ und $f^{1b}_{\{E/peter\}}$, die Restbäume von $f^0_\epsilon$, in Abbildung 7.6 zu sehen.

(c) Die letzten Folgenglieder ergeben sich dann zu $F'_2 = \emptyset$ und $S'_2 = \emptyset$.

Da $S'_i \not\subseteq S'_{i+1}$ [4] für alle $0 \leq i \leq 1$ erfüllt ist, sind alle Literale für den Beweis relevant. Daher ist $(F_i) = (F'_i)$ und $(S_i) = (S'_i)$.

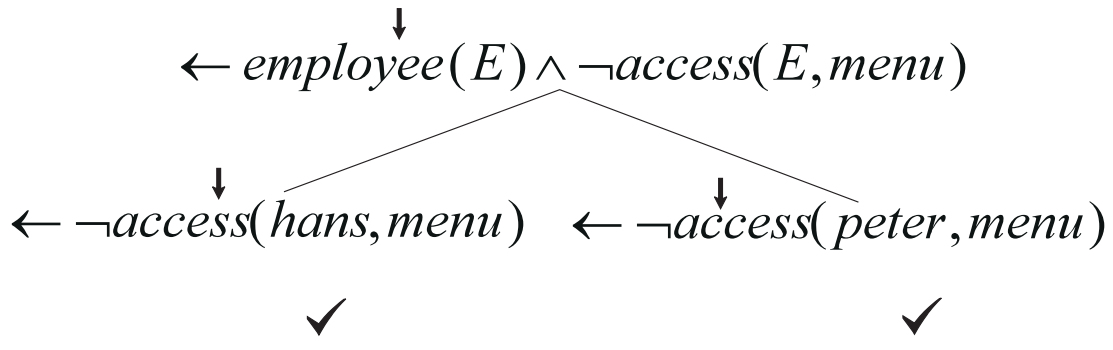

Abbildung 7.5: SLDNF-Baum $f^0_\sigma$

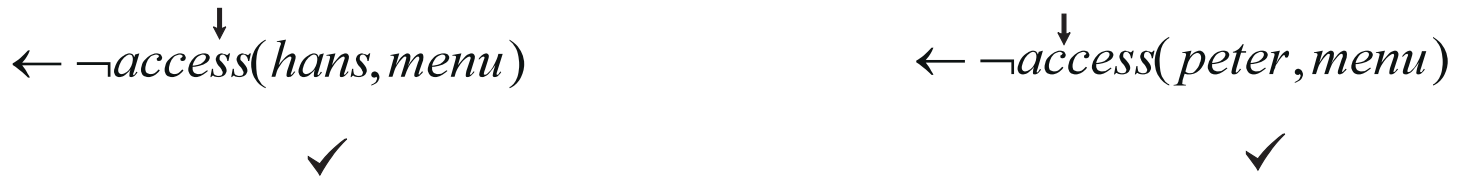

Abbildung 7.6: Restbäume $f^{1a}_{\{E/hans\}}$ und $f^{1b}_{\{E/peter\}}$ von $f^0_\sigma$

6. Die Nachfolger des negativen UND-Knoten $W \equiv employee(E) \wedge \neg access(E, menu)$ werden aus den verwendeten Folgen $(S_i) = subst(W)$ berechnet. Da kein Literal unrelevant ist, ist jedes Literal aus $W$ für einen Nachfolger verantwortlich. Somit ergeben sich die Nachfolger von $W$ zu

   - $nachfolger(W) = \langle employee(E), access(E, menu)\rangle$. Dabei ist der Knoten $employee(E)$ ein negativer und der Knoten $access(E, menu)$ ein positiver ODER-Knoten. Für jeden dieser Knoten sind noch die Abbildungen $subst$ und $beweis$ zu definieren.

     (a) Diese Abbildungen ergeben sich für den Knoten $employee(E)$ zu:

       – $beweis(employee(E)) = \{f'_\epsilon\}$. Dabei ist $f'_\epsilon$ der TeilSLDNFBaum von $f_\epsilon$ und in Abbildung 7.7 zu sehen, sowie

[4] Im allgemeinen ist die Bedingung, die für die Feststellung der Relevanz eines Literals benötigt wird, die Bedingung $S'_i \not\subseteq S'_{i+1}$ (siehe Abschnitt 6.1.2). Beschränkt man sich jedoch, so wie wir, auf erlaubte Programme, so ist die Bedingung äquivalent zur Bedingung $S'_i \neq S'_{i+1}$. Dies folgt aus der Tatsache, daß für erlaubte Programme alle berechneten Antwortsubstitutionen Grundsubstitutionen sind.

– $subst(employee(E)) = \langle \epsilon, \{\{E/hans\}, \{E/peter\}\}\rangle$.

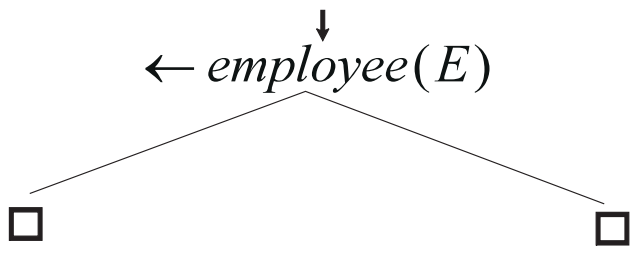

Abbildung 7.7: TeilSLDNFBaum $f'_\epsilon$ von $f_\epsilon$

(b) Für den positiven ODER-Knoten $access(E, menu)$ ergeben sich die Abbildungen zu:

– $beweis(access(E, menu)) = \langle R_2, R_3\rangle$, dabei sind $R_2$ und $R_3$ die zu **P** gehörenden SLDNF-Widerlegungen, sowie

– $subst(access(E, menu)) = \langle\langle\{E/hans\}, \{E/peter\}\rangle , \langle\{E/hans\}, \{E/peter\}\rangle\rangle$.

7. Nun sind der negative ODER-Knoten $employee(E)$ und der positive ODER-Knoten $access(E, menu)$ zu betrachten.

- Das Prädikatensymbol *employee* ist extensional, d.h. der Aufbau stoppt und der Knoten $employee(E)$ hat keine Nachfolger.
- Der Knoten $access(E, menu)$ ist kein extensionales Prädikat. Für jede in einer SLDNF-Widerlegung aus $beweis(access(E, menu))$ im ersten Schritt verwendete Programmklausel hat der Knoten einen Nachfolger. In der Widerlegung $R_2$ wurde die Programmklausel $access(E, F) \leftarrow owner(E, F)$, in der Widerlegung $R_3$ die Programmklausel $access(E, F) \leftarrow manager(E, E_2) \wedge owner(E_2, F)$ angehängt. Damit ergeben sich folgende Abbildungen:

  – $nachfolger(access(E, menu)) = \langle W_1, W_2\rangle$ mit

    ∗ $W_1 \equiv owner(E, menu)$ und

    ∗ $W_2 \equiv manager(E, E_2) \wedge owner(E_2, menu)$.

  Die Beweise und Substitutionstupel für die Knoten $W_1$ und $W_2$ erhalten wir auf folgende Weise:

    ∗ Zu $W_1$:

      · $beweis(W_1) = (R'_i)$ mit $0 \leq i \leq 1$. Dabei ist $R'_0 = \leftarrow owner(hans, menu), \square$, d.h. die Widerlegung $R_2$ ohne die erste

Zielklausel. Die Restwiderlegung von $R'_0$ ergibt sich dann zu $R'_1 = \Box$.

· $subst(W_1) = (U'_i)$ mit $0 \leq i \leq 1$. Dabei ist $U'_0 = U'_1 = \langle\{E/hans\}\rangle$, da keine neuen Variablenbindungen hinzugewonnen werden können.

∗ Bei $W_2$ sind die Beweise und Substitutionstupel etwas umfangreicher, da $W_2$ ein Literal mehr enthält als $W_1$:

· $beweis(W_2) = (R''_i)$ mit $0 \leq i \leq 2$. Dabei ist $R''_0 = \leftarrow manager(peter, E_2) \wedge owner(E_2, menu), \leftarrow owner(hans, menu), \Box$, d.h. die Widerlegung $R_3$ ohne die erste Zielklausel. Die Restwiderlegung von $R''_0$ ergibt sich dann zu $R''_1 = \leftarrow owner(hans, menu), \Box$ und, noch eine Restwiderlegung weiter, ergibt sich $R''_2 = \Box$.

· Die Substitutionen ergeben sich wieder aus den Substitutionen des Vorgängerknotens und der berechneten Antworten der SLDNF-Beweise, die zu $W_2$ gehören. Damit ist dann: $subst(W_1) = (U''_i)$ mit $0 \leq i \leq 2$. Dabei ist $U''_0 = \langle\{E/peter\}\rangle$. In $U''_1$ wird eine Variable mehr gebunden, da das Literal $manager(E, E_2)$ eine neue Variable $E_2$ einführt. Damit ist dann $U''_1 = \langle\{E/peter, E_2/hans\}\rangle$. Da sich weiter nichts ändert ist $U''_2 = U''_1$.

8. Die Nachfolger von $W_1$ und $W_2$ sind nun jeweils Blätter im UND-ODER-Baum, da sie alle extensionale Prädikatensymbole besitzen. Analog zum obigen Vorgehen ergibt sich:

   - Zu $W_1$:
     - $nachfolger(W_1) = \langle owner(E, menu)\rangle$ und
     - $beweis(owner(E, menu)) = \langle(\leftarrow owner(hans, menu), \Box)\rangle$ sowie
     - $subst(owner(E, menu)) = \langle\{E/hans\}\rangle$.
   - $W_2$ hat, nach der Länge der Konjunktion, zwei Nachfolger. Für jeden dieser Nachfolger sind die Abbildungen *subst* und *beweis* separat anzugeben.
     - $nachfolger(W_2) = \langle manager(E, E_2), owner(E_2, menu)\rangle$
     - $beweis(manager(E, E_2)) = \langle(\leftarrow manager(peter, E_2), \Box)\rangle$
     - $subst(manager(E, E_2)) = \langle\langle\{E/peter\}\rangle, \langle\{E/peter, E_2/hans\}\rangle\rangle$

- $beweis(owner(E_2, menu)) = \langle(\leftarrow owner(hans, menu), \Box)\rangle$
- $subst(owner(E, E_2)) = \langle\langle\{E/peter, E_2/hans\}\rangle, \langle\{E/peter, E_2/hans\}\rangle\rangle$

9. Die nun noch nicht behandelten ODER-Knoten haben alle extensionale Prädikatensymbole und deshalb keine Nachfolger.

Stellt man nun die Ergebnisse aller Schritte graphisch dar, so ergibt sich der in Abbildung 7.8 dargestellte Beweisbaum.

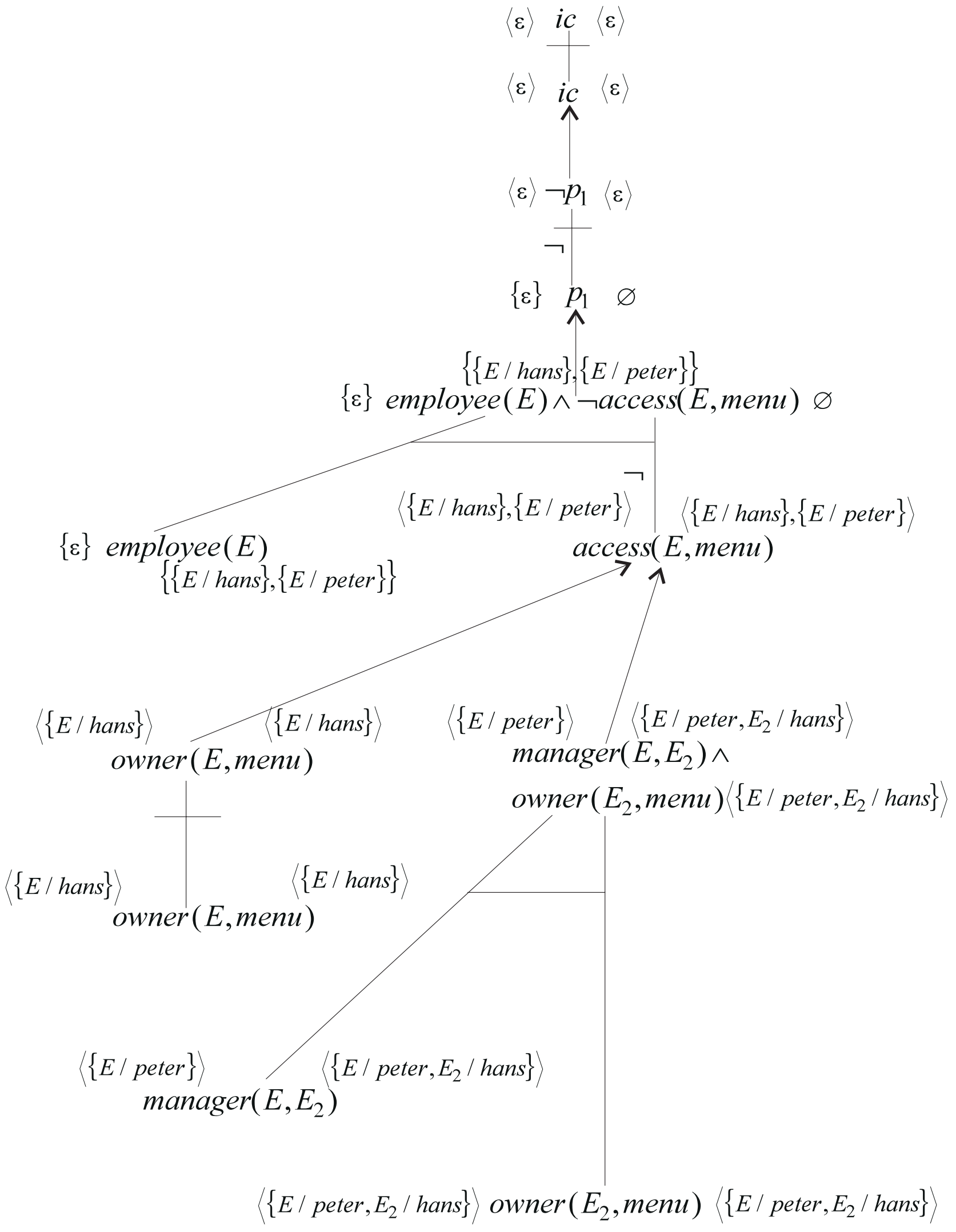

Abbildung 7.8: Resultierender Beweisbaum

# Kapitel 8

# Integritätsüberprüfung durch Beweisbäume

Ein SLDNF-Beweis kann kompakt als Beweisbaum dargestellt werden. Mit Hilfe dieser Repräsentation kann entschieden werden, ob eine gegebene Transaktion $d$ auf einer Datenbank $D$ einen Beweis ungültig macht. Damit ist noch nicht sichergestellt, ob die Integritätsbedingung wirklich verletzt ist [1], aber diese Vorgehensweise hat dennoch große Vorteile:

- Wir bekommen weit schärfere Bedingungen, wann eine Integritätsbedingung möglicherweise verletzt ist.
- Durch geeignete Manipulationen kann ein Beweis, der durch eine Transaktion ungültig wird, in einen neuen, gültigen Beweis transformiert werden. Dabei können unter Umständen große Teile des ursprünglichen Beweises weiterverwendet werden.

Zunächst übertragen wir aber die Ergebnisse auf Beweisbäume, die die Methode von Nicolas (siehe 3.1.1) erzielt hat.

## 8.1 Datenbankänderungen und SLDNF-Beweise

Wir müssen klären, wann eine Datenbankänderung den Beweis einer gegebenen Integritätsbedingung (also einen Beweisbaum) betrifft. Hier unterscheiden wir zwei Fälle:

[1] Dazu muß man *alle* möglichen Beweise kennen. Dies ist aber aus prinzipiellen Gründen nicht möglich.

1. eine *Pflegesituation* und

2. einen *Konflikt*.

Eine *Pflegesituation* liegt vor, wenn der Beweisbaum von einer Datenbankänderung zwar betroffen ist, aber garantiert werden kann, daß der Beweis nach einer geeigneten Modifikation (die ohne Datenbankzugriff auskommt) gültig bleibt. Dies tritt immer dann auf, wenn eine Löschung einen negativen ODER- oder UND-Knoten betrifft.

**Beispiel 8.1.1** *Sei $D$ folgende deduktive Datenbank:*

$$\begin{aligned} p(x) &\leftarrow t(x) \wedge s(x) \\ t(a) &\leftarrow \end{aligned}$$

*$D$ erfüllt die Integritätsbedingung $W \equiv \neg p(a)$. Sei $d = \langle Del, Add \rangle$ mit $Add = \emptyset$ und $Del = \{t(a) \leftarrow\}$. Die resultierende Datenbank $D'$ von $D$ und $d$ erfüllt die Integritätsbedingung $W$ immer noch, trotzdem ist der Beweis hier anders, d.h. der Beweisbaum des SLDNF-Beweises von $D \cup \{\leftarrow W\}$ muß verändert werden. Im Ansatz von Nicolas (siehe 3.1.1) führt eine solche Änderung zu keinen Konsequenzen, da das zu löschende Fakt $t(a) \leftarrow$ nur* negativ *mit der Integritätsbedingung zusammenhängt. Dies sieht zunächst nach einem Nachteil unseres Verfahrens aus, erweist sich aber bei näherer Betrachtung als Vorteil: mit der Beweisänderung, die eine solchen Löschung nach sich zieht, können unter Umständen andere Inkonsistenzen beseitigt werden (siehe Abschnitt 4.4 Punkt 2). Unter der Annahme, daß eine (wirkliche) Integritätsverletzung eher selten auftritt, ist der zusätzliche Aufwand zur Beweisänderung (und zur Erkennung, ob der Beweisbaum gültig bleibt) zu rechtfertigen.*

Nun zu der formalen Definition des Begriffs der *Pflegesituation*:

**Definition 8.1.1 (Pflegesituation)** *Sei $D$ deduktive Datenbank, $\leftarrow W$ eine Zielklausel und $T$ ein Standardbeweisbaum für $D \cup \{\leftarrow W\}$ mit Anfangssubstitution $\epsilon$. Weiter sei $d = \langle Del, Add \rangle$ eine Datenbankänderung von $D$ mit resultierender Datenbank $D'$. Eine* Pflegesituation *von $D$, $d$ und $D'$ tritt auf, wenn eine der folgenden Bedingungen erfüllt ist:*

- *für eine Einheitsklausel $A' \leftarrow \in Del$ gibt es ein negatives ODER-Blatt $A$ in $T$ mit $subst(A) = \langle S_B, S_E \rangle$ und eine Substitution $\sigma \in S_E$, so daß $mgu(A\sigma, A')$ existiert.*[2]

[2] Die Definition ist etwas allgemeiner, als sie sein müßte. Da $A\sigma$ immer ein Grundatom ist ($T$ ist ein erlaubter Beweisbaum), reicht es aus, wenn $A\sigma = A'$ gilt.

- *für eine Programmklausel $A \leftarrow W \in Del$ mit nichtleerem Rumpf $W$ gibt es einen negativen UND-Knoten in $T$, zu dessen Aufbau die Programmklausel $A \leftarrow W$ benutzt wurde.*

Das Auftreten von Pflegesituationen ändert an der Gültigkeit einer Integritätsbedingung prinzipiell nichts, da nur Lösungen weggenommen werden, an denen man nicht interessiert ist. Um es mit Filtern auszudrücken: Es wird jetzt vermehrt an einer bestimmten Stelle herausgefiltert. Da bei einem Polaritätswechsel von positiv nach negativ die Ausgangssubstitutionsmenge immer leer ist, also alles herausgefiltert wurde, ändert sich an der Gültigkeit der Integritätsbedingung nichts.

Der zweite Fall, der bei einer Datenbankänderung auftreten kann, ist der einer *Konfliktsituation*, d.h. eine Datenbankänderung, die die Gültigkeit des Beweises gefährdet.

**Beispiel 8.1.2** *Sei $D$ die deduktive Datenbank aus Beispiel 8.1.1. $D$ erfüllt die Integritätsbedingung $W \equiv \neg p(a)$. Weiterhin sei $d = \langle Del, Add \rangle$ mit $Add = \{s(a)\}$ und $Del = \emptyset$. Die resultierende Datenbank $D'$ von $D$ und $d$ erfüllt die Integritätsbedingung $W$ nicht mehr, da ein negativ im Beweisbaum vorkommendes Atom hinzugefügt wird.*

Die Fälle, die zu einem *Konflikt* führen, erweisen sich als die Bedingungen, die auch in allen anderen Ansätzen bei der Analyse von Datenbankänderungen betrachtet werden:

**Definition 8.1.2 (Konflikt)** *Sei $D$ eine deduktive Datenbank, $\leftarrow W$ eine Zielklausel und $T$ ein Standardbeweisbaum für $D \cup \{\leftarrow W\}$ mit einer beliebigen Anfangssubstitution. Weiter sei $d = \langle Add, Del \rangle$ eine Datenbankänderung auf $D$ mit resultierender Datenbank $D'$. Ein* Konflikt *zwischen $D$ und $d$ tritt auf, wenn eine von folgenden Bedingungen erfüllt ist:*

- *für eine Einheitsklausel $A' \leftarrow \in Del$ gibt es ein positives ODER-Blatt $A$ in $T$ mit $subst(A) = \langle U_B, U_E \rangle$ und eine Substitution $\sigma \in U_E$, so daß $mgu(A\sigma, A')$ existiert.*
- *für eine Einheitsklausel $A' \leftarrow \in Add$ gibt es ein negatives ODER-Blatt $A$ in $T$ mit $subst(A) = \langle S_B, S_E \rangle$ und eine Substitution $\sigma \in S_B$, so daß $mgu(A\sigma, A')$ existiert.*
- *für eine Programmklausel $A' \leftarrow W' \in Del$ gibt es einen positiven UND-Knoten in $T$, zu dessen Aufbau die Programmklausel $A' \leftarrow W'$ benutzt wurde.*
- *für eine Programmklausel $A' \leftarrow W' \in Add$ gibt es einen negativen ODER-Knoten $A$, so daß $mgu(A', A)$ existiert.*

Wir werden nun die Operationen besprechen, die notwendig sind, um einen von einer Datenbankänderung betroffenen Beweisbaum in einen korrekten zu transformieren.

## 8.2 Beweispflege

Bei einer Pflegesituation ist es immer möglich, den Beweisbaum zu restaurieren. Da eine Pflegesituation immer nur negative Knoten betrifft, werden wir zeigen, auf welche Art die Substitutionsmengen zu behandeln sind. Hierzu werden wir zunächst an Beispielen die möglichen Auswirkungen deutlich machen:

**Beispiel 8.2.1** *Sei $A = p(x,a)$ ein negatives ODER-Blatt in einem Standardbeweisbaum mit $subst(A) = \langle S_B, S_E \rangle$. $S_B$ und $S_E$ seien folgendermaßen gegeben:*

- $S_B = \{\{y/b, z/c\}, \{y/a, z/d\}\}$
- $S_E = \{\{x/e, y/b, z/c\}, \{x/f, y/b, z/c\}, \{x/a, y/b, z/c\}\}$

*In einer Datenbankänderung werde das Fakt $p(a,a)$ gelöscht. Dann sind alle Substitutionen zu bestimmen, die nun nicht mehr gültig sind. Sei $\sigma = mgu(p(x,a), p(a,a))$. Es fallen alle Substitutionen $\sigma'$ aus $S_E$ weg, für die gilt: $\sigma \geq \sigma'$, da es für diese Substitutionen nun keine Begründung mehr gibt. Das resultierende $\Delta$ berechnet sich zu:*

$$\Delta = \{\sigma' \mid \sigma' \in S_E \quad \mathit{und} \quad \sigma \geq \sigma'\}.$$

*Das resultierende $S'_E = S_E \setminus \Delta$ ist hier*

$$S'_E = \{\{x/e, y/b, z/c\}, \{x/f, y/b, z/c\}\},$$

*da die Substitution $\{x/a, y/b, z/c\}$ wegfällt.*

Diese Änderungen müssen ab der Stelle, an der sie direkte Auswirkungen haben, durch den Beweisbaum propagiert werden. Hierzu benötigen wir zwei Formen der Beweispflege:

- bottom up: Ausgehend vom Ort der Änderung müssen diese nach oben durch den Beweisbaum propagiert werden.
- top down: Änderungen, die sich an anderer Stelle im Beweisbaum ergeben haben und nach oben propagiert worden sind, müssen an andere Nachfolger des Vorgängerknotens nach unten propagiert werden.

Im folgenden zeigen wir diese beiden Operationen an einem Beispiel.

**Beispiel 8.2.2** *Sei $D$ die Datenbank aus dem Beispiel in Kapitel 4 und $d = \langle Del, Add \rangle$ eine Datenbankänderung auf $D$ mit resultierender Datenbank $D'$. Dabei ist $Del = \{employee(peter)\}$ und $Add = \emptyset$. In Abbildung 8.1 ist ein Ausschnitt* *aus dem Beweisbaum aus Abbildung 4.3 dargestellt für die Integritätsbedingung* $ic \equiv \forall E{:}employee\ access(E, menu)$.

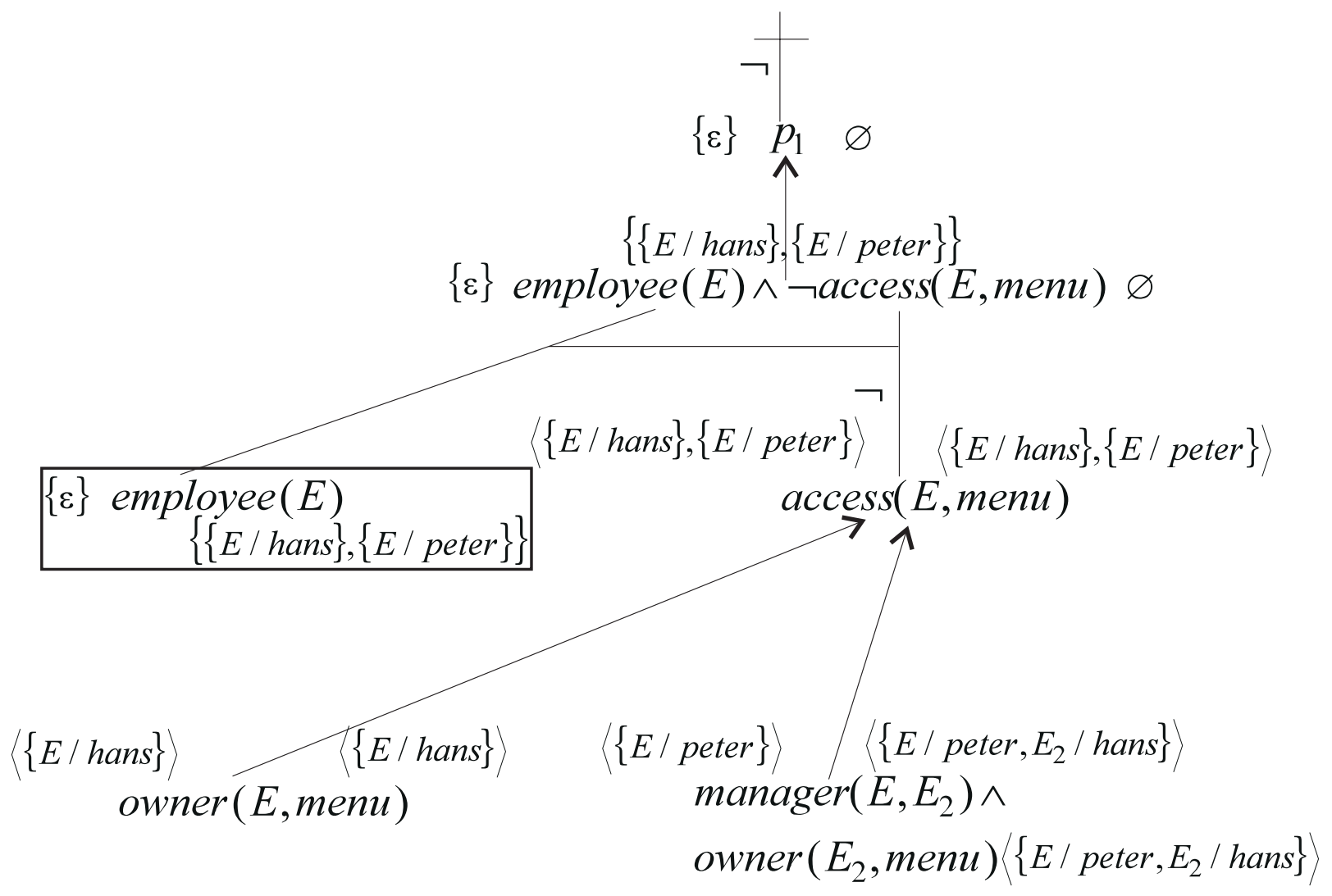

Abbildung 8.1: Ausschnitt aus einem Beweisbaum zur Anfrage $\leftarrow ic$

*Die Datenbankänderung d bedingt eine Pflegesituation bei dem negativen ODER-Blatt* $employee(E)$. *In der zu dem ODER-Knoten* $employee(E)$ *gehörenden Substitutionsmenge* $S_E = \{\{E/hans\}, \{E/peter\}\}$ *sind jetzt Substitutionen enthalten, für die es keinen endlichen unvollständigen SLDNF-Baum für* $D' \cup \{\leftarrow employee(E)\}$ *gibt. Diese Substitutionen müssen aus* $S_E$ *entfernt werden. Diese Löschung wird nach oben weiterpropagiert zum übergeordneten UND-Knoten (bottom-up). Hier setzt sich die Propagierung nach* unten *auf alle rechts von dem employee-Knoten stehenden Knoten fort (top-down). Die Reihenfolge der betrachteten Substitutionsmengen und Tupel wird in Abbildung 8.2 durch die gestrichelten Pfeile markiert, die von dem negativen ODER-Knoten ausgehen, bei dem die Pflegesituation aufgetreten ist.*

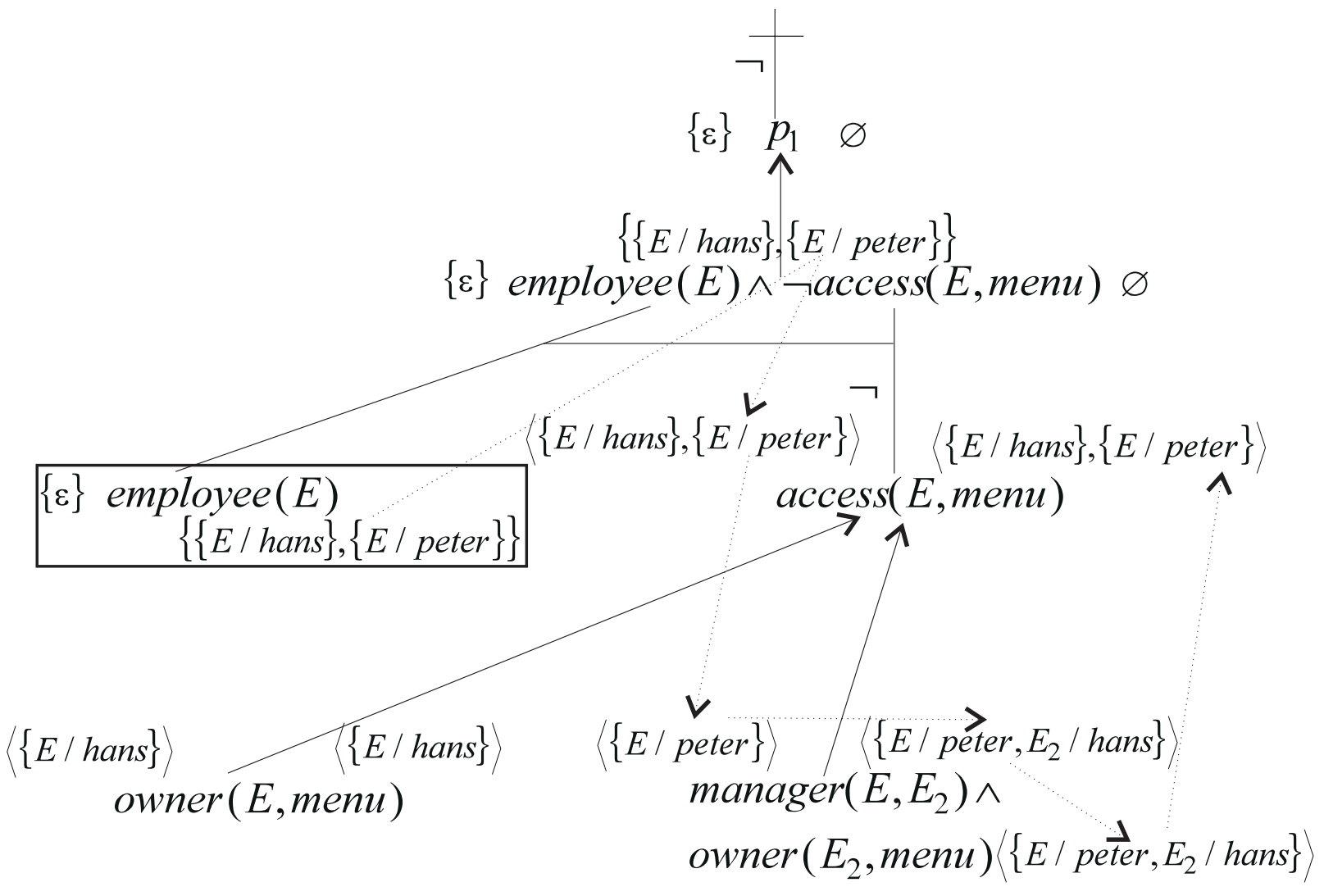

Abbildung 8.2: Zu verändernde Substitutionsmengen und -tupel

*In Abbildung 8.3 ist der Baum nach der Veränderung der Substitutionsmengen zu sehen. Es wurden alle Substitutionen entfernt, für die die oben angegebene Bedingung zutrifft.*

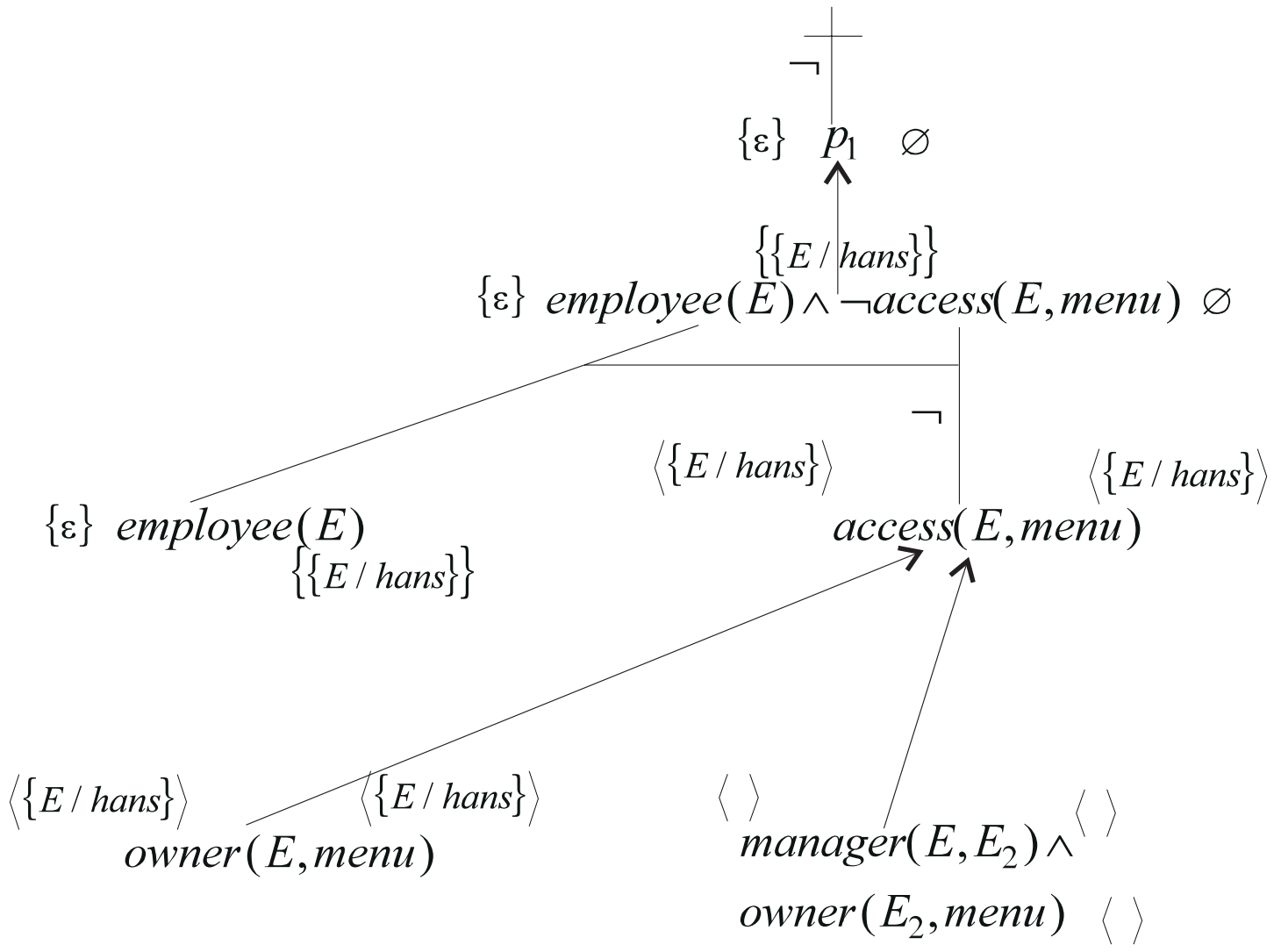

Abbildung 8.3: Gefilterte Substitutionsmengen und -tupel

*Der Teilbaum, der aus der Programmklausel* $access(E, F) \leftarrow manager(E, E_2) \wedge owner(E_2, F)$ *entstand, ist nun überflüssig, da er für keine Substitution mehr zuständig ist. Der entstandene Baum (siehe Abbildung 8.4) ist ein Teil eines Beweisbaumes für die resultierende Datenbank* $D'$.

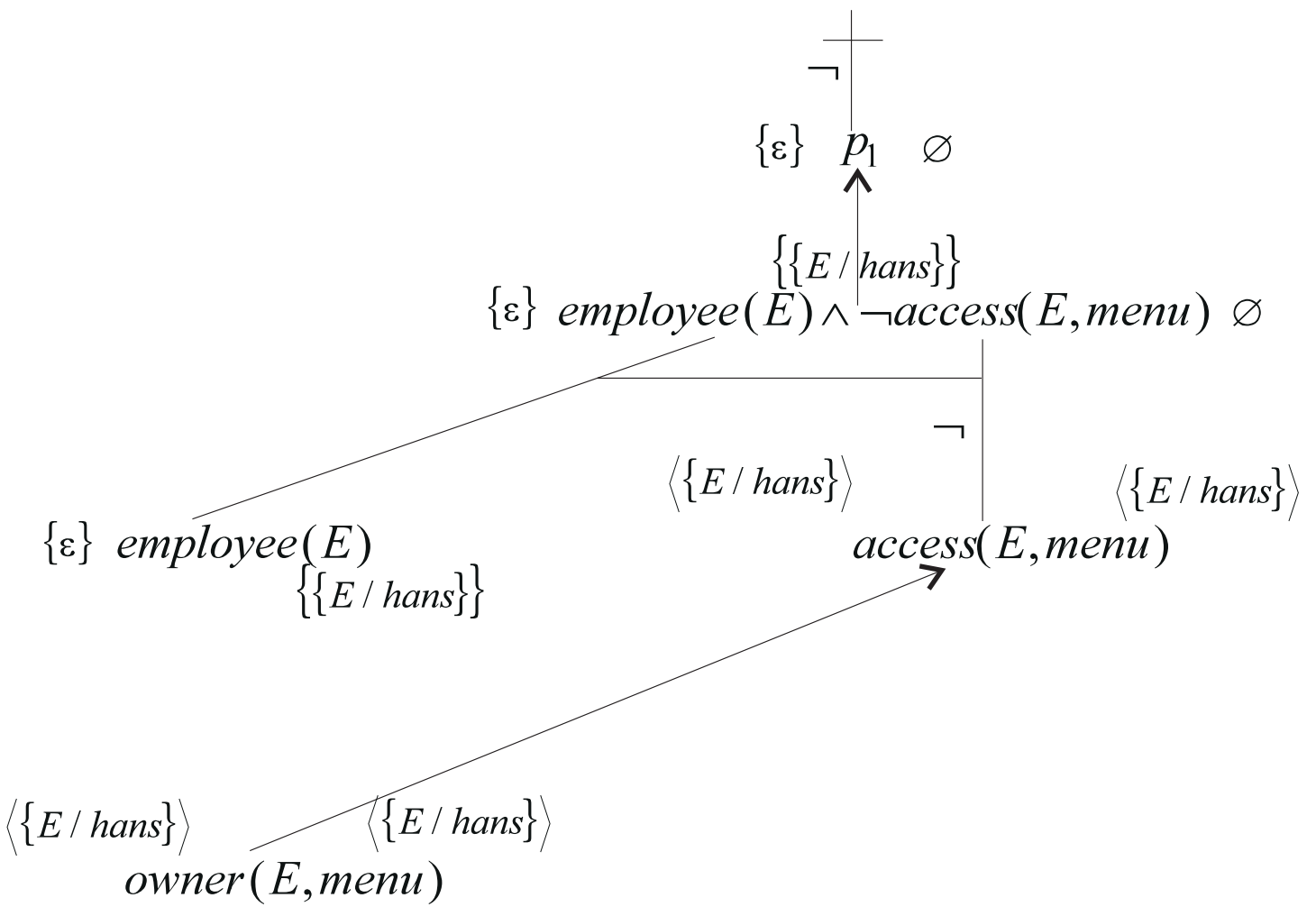

Abbildung 8.4: Teil eines Beweisbaumes für die Datenbank $D'$

Im folgenden beschreiben wir den zur Durchführung der Beweispflege benötigten Algorithmus, der die zur top-down und bottom-up–Propagierung notwendigen Schritte realisiert.

**Definition 8.2.1** *Sei $D$ eine deduktive Datenbank, $d = \langle Del, Add\rangle$ eine Datenbankänderung auf $D$ mit resultierender Datenbank $D'$. Weiter sei $\leftarrow W$ eine Zielklausel und $T$ ein Beweisbaum für $D \cup \{\leftarrow W\}$ mit beliebiger Anfangssubstitution, der aus einem SLDNF-Beweis mit der Standardberechnungsregel nach Definition 7.1.5 erzeugt wurde. Der folgende Algorithmus generiert aus den Abbildungen subst, nachfolger und beweis aus $W$ und $T$ einen neuen Baum.*

*pflege-baum*$(T, Del)$

1. *Finde alle negativen ODER-Blätter $A$ aus $T$, deren Prädikatensymbol mit dem Prädikatensymbol von einem Fakt aus Del in der Transaktion übereinstimmt. Führe start-bottom-up-pflege-neg-ODER-Blatt*$(A, Del)$ *aus.*
2. *Finde alle negativen UND-Blätter $W$, zu deren Aufbau eine Programmklausel aus Del verwendet wurde. Führe start-bottom-up-pflege-neg-UND-Knoten*$(W, Del)$ *aus.*

*start-bottom-up-pflege-neg-ODER-Blatt*$(A, Del)$

1. *Sei die zu* $A$ *gehörenden Substitutionsmengen durch* $subst(A) = \langle S_B, S_E \rangle$ *gegeben.*
2. *Berechne die Substitutionsmenge* $\Delta = \{\sigma \in S_E \mid A\sigma = A', A' \in Del\}$.
3. *Setze* $S_E = S_E \setminus \Delta$.
4. *Sei* $W$ *der Vorgängerknoten von* $A$*, dessen* $i$*-ter Nachfolger* $A$ *ist. Falls* $\Delta = \emptyset$*, ist die Pflege fertig. Ansonsten pflege den Vorgängerknoten* $W$ *mit bottom-up-pflege-neg-UND-Knoten*$(W, \Delta, i)$ *(denn* $W$ *ist ein negativer UND-Knoten, sonst wäre schon* $S_E = \emptyset$*, und damit erst recht* $\Delta$*).*

*start-bottom-up-pflege-neg-UND-Knoten*$(W, Del)$.

1. *Sei die zu* $W$ *gehörende Folge von Substitutionsmengen durch* $subst(W) = (S_j)$*,* $0 \leq j \leq m$*, gegeben und* $A' \leftarrow W' \in Del$ *die Programmklausel aus* $Del$*, aus der* $W$ *hervorgegangen ist. Weiter sei* $A$ *der Vorgänger von* $W$.
2. *Berechne* $\Delta = S_m|_{var(A) \cup dom(S_0)}$.
3. *Lösche* $W$ *und den ganzen darunterliegenden Teilbaum.*
4. *Wenn* $S_m \neq \emptyset$*, dann rufe bottom-up-pflege-neg-ODER-Knoten*$(A, \Delta)$ *auf.*

*bottom-up-pflege-neg-UND-Knoten*$(W, \Delta_i, i)$

1. *Sei die zu* $W$ *gehörende Folge von Substitutionsmengen durch* $subst(W) = (S_j)$*,* $0 \leq j \leq m$*, gegeben. Weiter sei* $A$ *der Vorgänger von* $W$.
2. *Setze* $S_i = S_i \setminus \Delta_i$.
3. *Berechne für alle* $j$ *mit* $i < j \leq m$*:*
   (a) $\Delta_j = \{\sigma \in S_j \mid \sigma' \geq \sigma, \sigma' \in \Delta_{j-1}\}$.
   (b) *Setze* $S_j = S_j \setminus \Delta_j$.
   (c) *Falls* $S_j = S_{j-1}$ *dann lösche* $S_j$ *und den Teilbaum mit Wurzel* $A_j$ *in* $T$*, sonst*
      i. *Falls* $A_j \in nachfolger(W)$ *ein negativer ODER-Knoten und* $\Delta_{j-1} \neq \emptyset$ *ist, dann rufe top-down-pflege-neg-ODER-Knoten*$(A_j, \Delta_{j-1})$ *auf.*
      ii. *Falls* $A_j \in nachfolger(W)$ *ein positiver ODER-Knoten ist, berechne aus* $subst(A_j) = \langle U_B, U_E \rangle$ *und* $\Delta$ *die Menge* $I_j = \{i \mid \sigma_i \in U_B$ *an* $i$*-ter Position und* $\sigma_i \in \Delta_{j-1}\}$*. Wenn* $I_j \neq \emptyset$*, dann rufe top-down-pflege-pos-ODER-Knoten*$(A_j, I)$ *auf.*
4. *Wenn* $\Delta_m \neq \emptyset$*, dann sei* $\Delta' = \Delta_m|_{var(A) \cup dom(S_0)}$ *und rufe bottom-up-pflege-neg-ODER-Knoten*$(A, \Delta')$ *auf.*

*bottom-up-pflege-neg-ODER-Knoten*$(A, \Delta)$

1. *Seien die zu* $A$ *gehörigen Substitutionen durch* $subst(A) = \langle S_B, S_E\rangle$ *gegeben und* $W$ *der Vorgängerknoten von* $A$*, dessen* $i$*-ter Nachfolgeknoten* $A$ *ist.*
2. *Sei* $\Delta' = \{\sigma \in \Delta \mid$ *es gibt kein* $W' \in nachfolger(A)$*, für das* $\sigma \in S_m|_{var(A)\cup dom(S_0)}$ *wobei* $(S_j) = subst(W'),\ 0 \leq j \leq m\}$.
3. *Setze* $S_E = S_E \setminus \Delta'$.
4. *Wenn* $\Delta' \neq \emptyset$*, dann rufe bottom-up-pflege-neg-UND-Knoten*$(W, \Delta', i)$ *auf.*

*top-down-pflege-neg-ODER-Knoten*$(A, \Delta)$

1. *Seien die zu* $A$ *gehörenden Substitutionen durch* $subst(A) = \langle S_B, S_E\rangle$ *gegeben.*
2. *Sei* $\Delta' = \{\sigma' \in S_E \mid \sigma \geq \sigma', \sigma \in \Delta\ \}$.
3. *Setze* $S_B = S_B \setminus \Delta$.
4. *Setze* $S_E = S_E \setminus \Delta'$.
5. *Wenn* $\Delta \neq \emptyset$*, dann führe für alle* $W \in nachfolger(W)$ *top-down-pflege-neg-UND-Knoten*$(W, \Delta)$ *aus.*

*top-down-pflege-pos-ODER-Knoten*$(A, I)$

1. *Seien die zu* $A$ *gehörenden Substitutionen durch* $subst(A) = \langle U_B, U_E\rangle$ *gegeben.*
2. *Entferne für alle* $i \in I$ *die Substitutionen an der* $i$*-ten Position in* $U_B$ *und* $U_E$.
3. *Falls* $A$ *kein Blattknoten ist, bestimme für jeden Nachfolger* $W \in nachfolger(A)$ *die Indexmenge* $I_W$*, die den* $i \in I$ *entsprechen (mit Hilfe der in der Definition des Beweisbaumes vorkommenden Permutation* $\pi$*). Führe top-down-pflege-pos-UND-Knoten*$(W, I_W)$ *aus für alle* $W$*, für die* $I_W \neq \emptyset$.

*top-down-pflege-pos-UND-Knoten*$(W, I)$

1. *Sei die zu* $W$ *gehörende Folge von Substitutionstupeln durch* $subst(W) = (U_j)$*,* $0 \leq j \leq m$*, gegeben.*
2. *Für alle* $0 \leq j \leq m - 1$
   (a) *Falls* $A_{j+1} \in nachfolger(W)$ *ein negativer ODER-Knoten: Bestimme* $\Delta = \{\sigma_i \in U_j \mid i \in I\}$*. Falls* $\Delta \neq \emptyset$*, dann rufe pflege-top-down-neg-ODER-Knoten*$(A_{j+1}, \Delta)$ *auf.*
   (b) *Falls* $A_{j+1} \in nachfolger(W)$ *ein positiver ODER-Knoten ist, dann rufe top-down-pflege-pos-ODER-Knoten*$(A_{j+1}, I)$ *auf.*
3. *Entferne aus allen* $(U_j)$ *alle* $\sigma_i$ *mit* $i \in I$*, die sich also in* $U_j$ *an der* $i$*-ten Position befinden.*

*top-down-pflege-neg-UND-Knoten*$(W, \Delta_0)$

1. *Sei die zu* $W$ *gehörende Folge von Substitutionsmengen durch* $subst(W) = (S_j)$, $0 \leq j \leq m$ *gegeben.*
2. *Für alle* $1 \leq j \leq m$*:*
   (a) *Berechne* $\Delta_j = \{\sigma \in S_j \mid \sigma' \geq \sigma, \sigma' \in \Delta_{j-1}\}$.
   (b) *Setzte* $S_j = S_j \setminus \Delta_j$.
   (c) *Wenn* $S_j = S_{j-1}$, *dann lösche* $S_j$ *und den Teilbaum unter* $W$ *mit der Wurzel* $A_j$, *sonst*
      i. *Falls* $A_j \in nachfolger(W)$ *ein negativer ODER-Knoten und* $\Delta_{j-1} \neq \emptyset$ *ist, dann rufe top-down-pflege-neg-ODER-Knoten*$(A_j, \Delta_{j-1})$ *auf.*
      ii. *Falls* $A_j \in nachfolger(W)$ *ein positiver ODER-Knoten ist, berechne aus* $subst(A_j) = \langle U_B, U_E \rangle$ *und* $\Delta$ *die Menge* $I = \{i \mid \sigma_i \in U_B$ *and i-ter Position und* $\sigma_i \in \Delta_{j-1}\}$. *Wenn* $I \neq \emptyset$, *dann rufe top-down-pflege-pos-ODER-Knoten*$(A_j, I)$ *auf.*

Ein nützlicher Punkt der Beweispflege ist, daß Konfliktsituationen nach einer Beweispflege unter Umständen aufgelöst sind, wie dies in Abschnitt 4.4 Punkt 2 der Fall ist.

## 8.3 Konfliktauflösung

Nachdem wir die Beweispflege betrachtet haben, werden wir im folgenden die zweite Möglichkeit untersuchen, wie eine Beweisänderung einen Beweisbaum betreffen kann. Während eine Pflegesituation für sich allein betrachtet die Gültigkeit eines Beweisbaumes (und damit einer Integritätsbedingung) nicht beeinträchtigt, ist dies bei einem Konflikt anders: Ein Konflikt macht einen Beweisbaum (d.h. *ein* Beweis) *immer* ungültig. Damit ist über die Gültigkeit der Integritätsbedingung nur ausgesagt, daß Beweis, der in dem Beweisbaum repräsentiert wird, nicht mehr möglich ist. Die Konfliktauflösung (d.h. die Operation, die nun notwendig ist) gliedert sich in drei Teile:

1. Die Bereinigung. Hierbei werden fehlende Lösungssubstitutionen im Beweisbaum nach oben propagiert, bis eine Stelle erreicht ist, an der ein neuer Beweisversuch lohnenswert erscheint. Dieser Schritt kann unter Umständen mehrmals ablaufen, wenn nicht direkt ein neuer Beweis gefunden wird.

2. Der Neubeweis. Für fehlende Lösungssubstitutionen wird ein Teil des Beweises neu erstellt, der dann in einen Beweisbaum transformiert wird.

3. Die Eingliederung. Der gefundene Beweisbaum (für fehlende Lösungssubstitutionen) wird in den bereits vorhandenen Baum integriert. Es ergibt sich nun wieder ein Standardbeweisbaum.

Diese drei Schritte werden wir an einem Beispiel verdeutlichen:

**Beispiel 8.3.1** *Sei D die Datenbank aus Kapitel 4 und* $d = \langle Del, Add\rangle$ *mit* $Del = \{manager(peter, hans)\}$ *und* $Add = \{classification(menu, 1)\}$ *eine Datenbankänderung auf D mit resultierender Datenbank* $D'$*. In Abbildung 8.5 ist ein Ausschnitt aus dem Beweisbaum für die Integritätsbedingung* $ic \equiv \forall E{:}employee\ access(E, menu)$ *in Abbildung 4.3 dargestellt.*

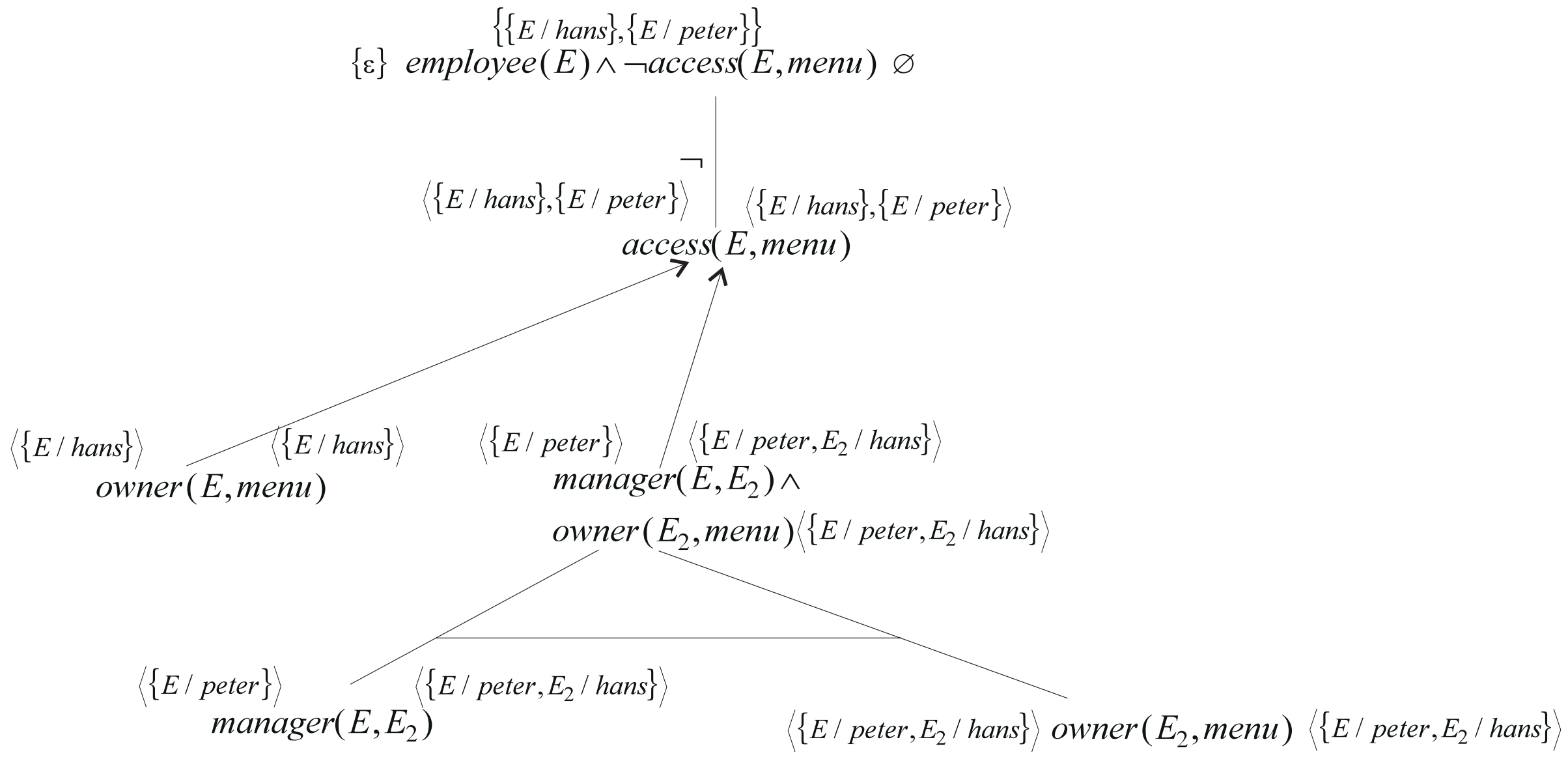

Abbildung 8.5: Ausschnitt aus einem Beweisbaum zur Anfrage $\leftarrow ic$

*Die Datenbankänderung d erzeugt einen Konflikt am positiven ODER-Blatt* $manager(E, E_2)$*, da* $manager(peter, hans) \in Del$ *ist. In dem zum ODER-Knoten* $manager(E, E_2)$ *gehörenden Substitutionstupel* $U_E = \langle\{E/peter, E_2/hans\}\rangle$ *sind nun Substitutionen enthalten, für die in* $D'$ *kein Fakt mehr existiert. Diese Substitutionen werden nun im Beweisbaum nach oben und zur Seite propagiert und somit alle Vorgängersubstitutionen entfernt. Die Propagierung nach oben im Beweisbaum findet dabei auf ähnlich wie bei der Beweispflege statt. Bei allen Knoten links vom Konfliktknoten werden alle Substitutionen entfernt, die allgemeiner als die nun fehlende Substitution sind. Bei allen Knoten rechts vom Konfliktknoten werden alle Substitutionen entfernt, die spezieller sind als die fehlende Substitution. Die Propagierung endet beim ersten Pola-* *ritätswechsel oder bei der Wurzel. Der resultierende Baum ist in Abbildung 8.6 zu sehen. Alle Knoten, bei denen die Substitutionsmengen und -tupel leer werden, können entfernt werden, da sie nichts zum Beweis beitragen.*

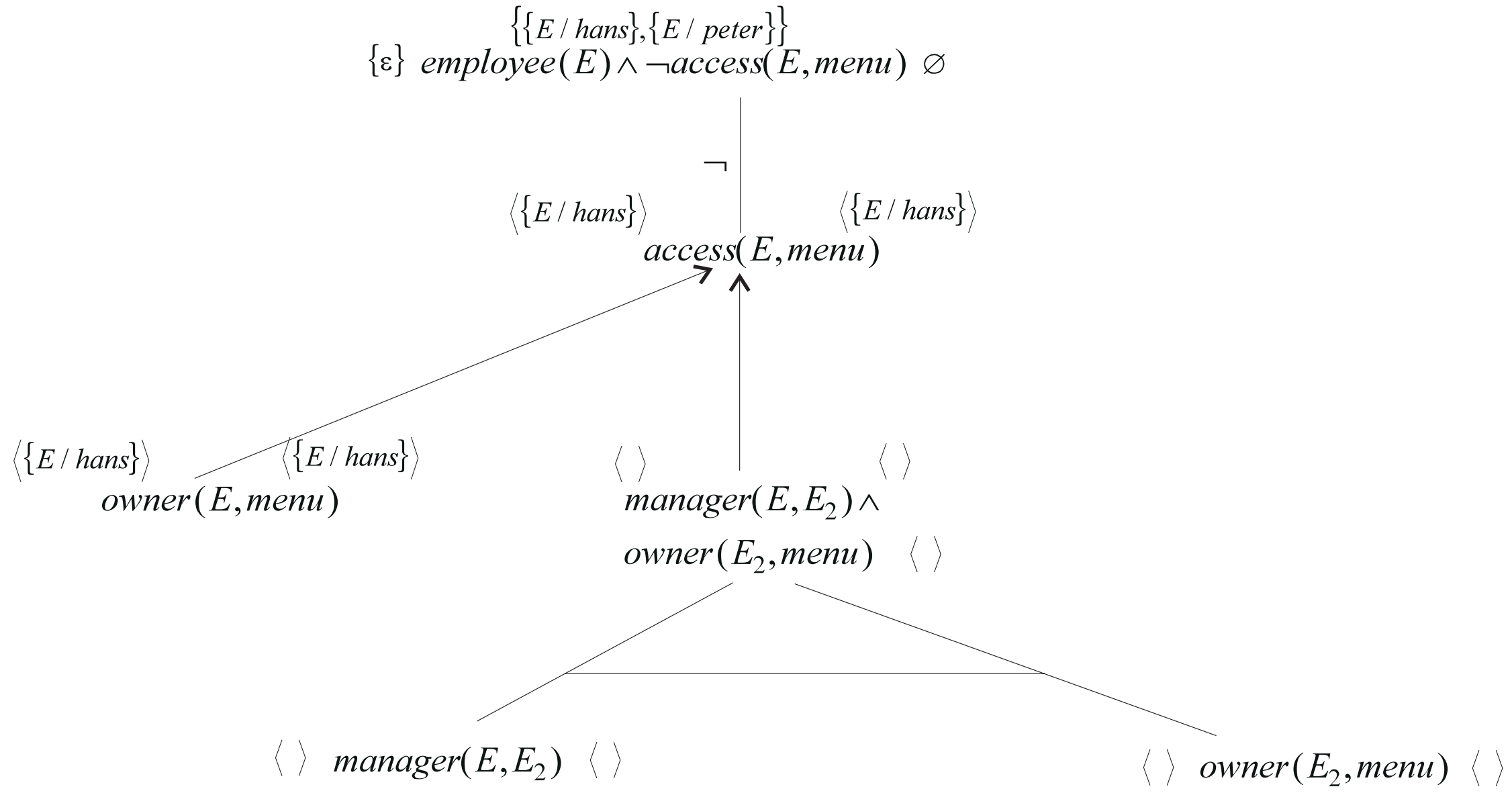

Abbildung 8.6: Teilbaum nach der Konfliktbereinigung

*Nun fehlt im Baum die Begründung für die Substitution* $\{E/peter\}$ *des Knotens* $access(E, menu)$. *Für diesen Knoten versuchen wir einen neuen Beweisbaum zu konstruieren, d.h. wir suchen einen Beweisbaum für* $access(E, menu)$ *mit Anfangssubstitution* $\{E/peter\}$ *in der resultierenden Datenbank* $D'$. *Abbildung 8.7 zeigt diesen Beweisbaum.*

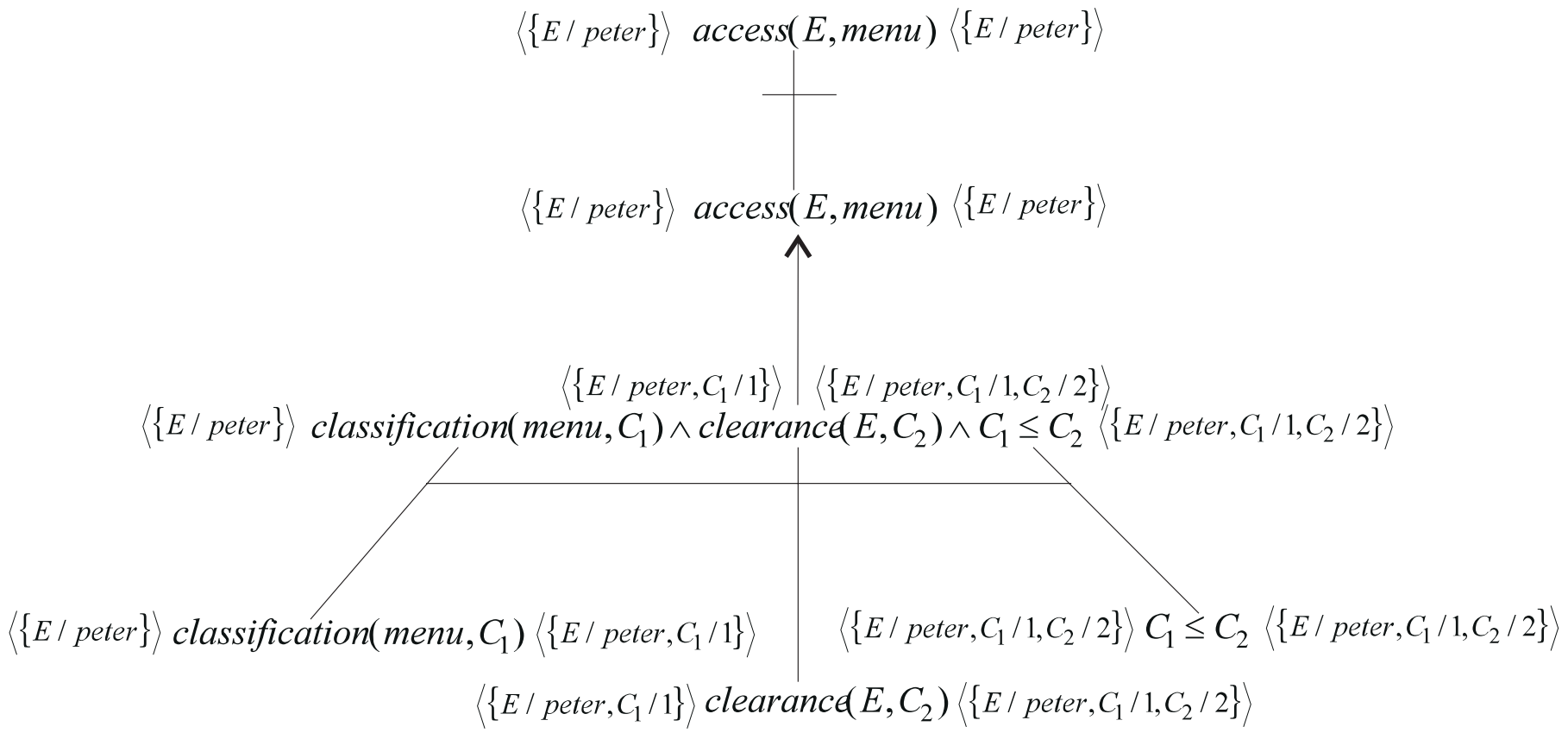

Abbildung 8.7: Neuer Beweisbaum für $\leftarrow acccess(E, menu)\sigma$, $\sigma = \{E/peter\}$

*Dieser neue Beweisbaum muß in den modifizierten integriert werden. In diesem speziellen Fall wird gerade die Wurzel des Beweisbaumes weggelassen, so daß die neue Wurzel nun der* ODER-Knoten $access(E, menu)$ *ist. Dann werden die Substitutionstupel dieses Knotens zu den alten Knoten hinzugefügt. Der darunterliegende Baum wird zu dem alten Beweisbaum unter dem Knoten* $access(E, menu)$ *hinzugefügt. Damit ergibt sich ein neuer Beweisbaum, der in Abbildung 4.5 zu sehen ist.*

### 8.3.1 Bereinigung

Nachdem wir nun die prinzipielle Vorgehensweise an einem Beispiel verdeutlicht haben, geben wir Algorithmen an, die diese Aufgaben erfüllen.

**Definition 8.3.1** *Sei $D$ eine deduktive Datenbank, $d = \langle Del, Add \rangle$ eine Datenbankänderung auf $D$ mit resultierender Datenbank $D'$, $\leftarrow W$ eine Zielklausel und $T$ ein Beweisbaum für $D \cup \{\leftarrow W\}$ mit beliebiger Anfangssubstitution, der aus einem SLDNF-Beweis mit der Standardberechnungsregel nach Definition 7.1.5 erzeugt wurde. Sei $T_p$ der mit $d$ gepflegte Baum $T$. Die im folgenden angegebenen Algorithmen führen die Propagierung einer Änderung nach oben im Beweisbaum durch und geben die Substitutionen zurück, für die im Beweisbaum keine Begründung mehr vorhanden ist. Dabei beginnen wir mit einer Menge $\Delta$ von Substitutionen, von denen wir schon wissen, daß sie keine Lösungen für die Blätter mehr darstellen. Die Argumente sind dabei jeweils die Knoten und eine Menge $\Delta$ von Substitutionen bei negativen Knoten bzw. von Indizes der Substitutionen bei positiven Knoten.*

*bereinige-neg-ODER-Knoten*$(A, \Delta)$

1. *Sei* $subst(A) = \langle S_B, S_E \rangle$.
2. *Sei $W$ der Vorgänger von $A$ und $A$ der $i$-te Nachfolger von $W$.*
3. *Wenn $W$ ein positiver UND-Knoten ist, dann rufe top-down-pflege-neg-ODER-Knoten*$(A, \Delta)$ *auf und gebe* $\langle \neg A, \Delta \rangle$ *zurück, sonst rufe bereinige-neg-UND-Knoten*$(W, i, \Delta)$ *auf.*

*bereinige-neg-UND*$(W, i, \Delta)$

1. *Sei die zu $W$ gehörende Folge von Substitutionsmengen durch* $subst(W) = (S_j)$, $0 \leq j \leq m$ *und die Nachfolger von $W$ durch* $nachfolger(W) = \langle A_1, \ldots, A_m \rangle$ *gegeben. Weiter sei der Vorgänger von $W$ der Knoten $A$.*
2. *Berechne* $\Delta' = \{\sigma \in S_0 \mid \sigma \geq \sigma', \sigma' \in \Delta\}$.
3. *Rufe bereinige-neg-ODER-Knoten*$(A, \Delta')$ *auf.*

*bereinige-pos-ODER-Knoten*$(A, I)$

1. *Seien die zu* $A$ *gehörenden Substitutionen gegeben durch* $subst(A) = \langle U_B, U_E \rangle$ *und* $W$ *der Knoten, dessen* $j$*-ter Nachfolgeknoten* $A$ *ist.*
2. *Falls* $W$ *ein negativer UND-Knoten ist, dann rufe top-down-pflege-pos-ODER-Knoten*$(A, I)$ *auf und gebe* $\langle A, j, \Delta \rangle$ *zurück mit* $\Delta = \{\sigma_i \in U_B \mid i \in I\}$*. Sonst rufe bereinige-pos-UND-Knoten*$(W, j, I)$ *auf.*

*bereinige-pos-UND-Knoten*$(W, i, I)$

1. *Sei die Folge der zu* $W$ *gehörenden Substitutionen gegeben durch* $subst(W) = (U_i)$, $0 \leq j \leq m$ *und* $nachfolger(W) = \langle A_1, \ldots, A_m \rangle$.
2. *Wenn* $W$ *die Wurzel von* $T$ *ist, dann gebe* $\langle$ *Wurzel erreicht mit* $I \rangle$ *zurück. Sonst sei* $A$ *der Vorgänger von* $W$ *mit* $subst(A) = \langle U_B, U_E \rangle$*. Sei* $I' = \{ j \mid j$ *ist die Positionsnummer von* $\sigma_i$ *aus* $U_0$, $i \in I$ *in* $U_B\}$*. Rufe bereinige-pos-ODER-Knoten*$(A, I')$ *auf.*

## 8.4 Vereinigung von zwei Beweisbäumen

Nachdem wir die Konfliktbereinigung eines Beweisbaumes besprochen haben, wenden wir uns dem nächsten noch zu klärenden Punkt zu: Die Integration eines Beweisbaumes in einen Baum, wie ihn die Beweispflege und Konfliktbereinigung erzeugt. Da ein Baum ab dem zu betrachtenden Knoten alle Eigenschaften eines Beweisbaumes erfüllt, wenn die Pflege und Konfliktauflösung durchgeführt wurde, reicht es, die Vereinigung von zwei Beweisbäumen mit gleicher Wurzel zu betrachten. Dabei nehmen wir an, daß jeder Beweis mit der gleichen Berechnungsregel durchgeführt wurde. Im folgenden beschränken wir uns auf die Standardberechnungsregel, jedoch sind die Ergebnisse auch auf andere Berechnungsregeln, die noch näher zu charakterisieren sein werden, übertragbar. Damit reduziert sich die Vereinigung von zwei Beweisbäumen auf die Vereinigung von Knoten gleichen Typs und mit gleichen Beschriftungen. Die Vereinigung zweier Bäume kann damit als sukzessive Vereinigung aller Knoten verstanden werden. Im folgenden verdeutlichen wir die Vereinigung von Knoten gleichen Typs an je einem Beispiel pro Knotentyp.

**Beispiel 8.4.1**

1. *Bei positiven UND-Knoten gestaltet sich die Vereinigung von zwei Knoten einfach: Da jeder Knoten für jedes Literal einen Nachfolger hat und nach Voraussetzung die Reihenfolge der Nachfolger bei beiden Knoten identisch ist, beschränkt sich die*

*Operation auf die Konkatenation der Substitutionstupel an den gleichen Stellen im Knoten. In Abbildung 8.8 sind zwei Knoten in zwei verschiedenen Beweisbäumen dargestellt, deren Vereinigung den dritten Knoten ergibt. Die Eigenschaften eines positiven UND-Knotens werden durch diese Konkatenation nicht beeinträchtigt.*

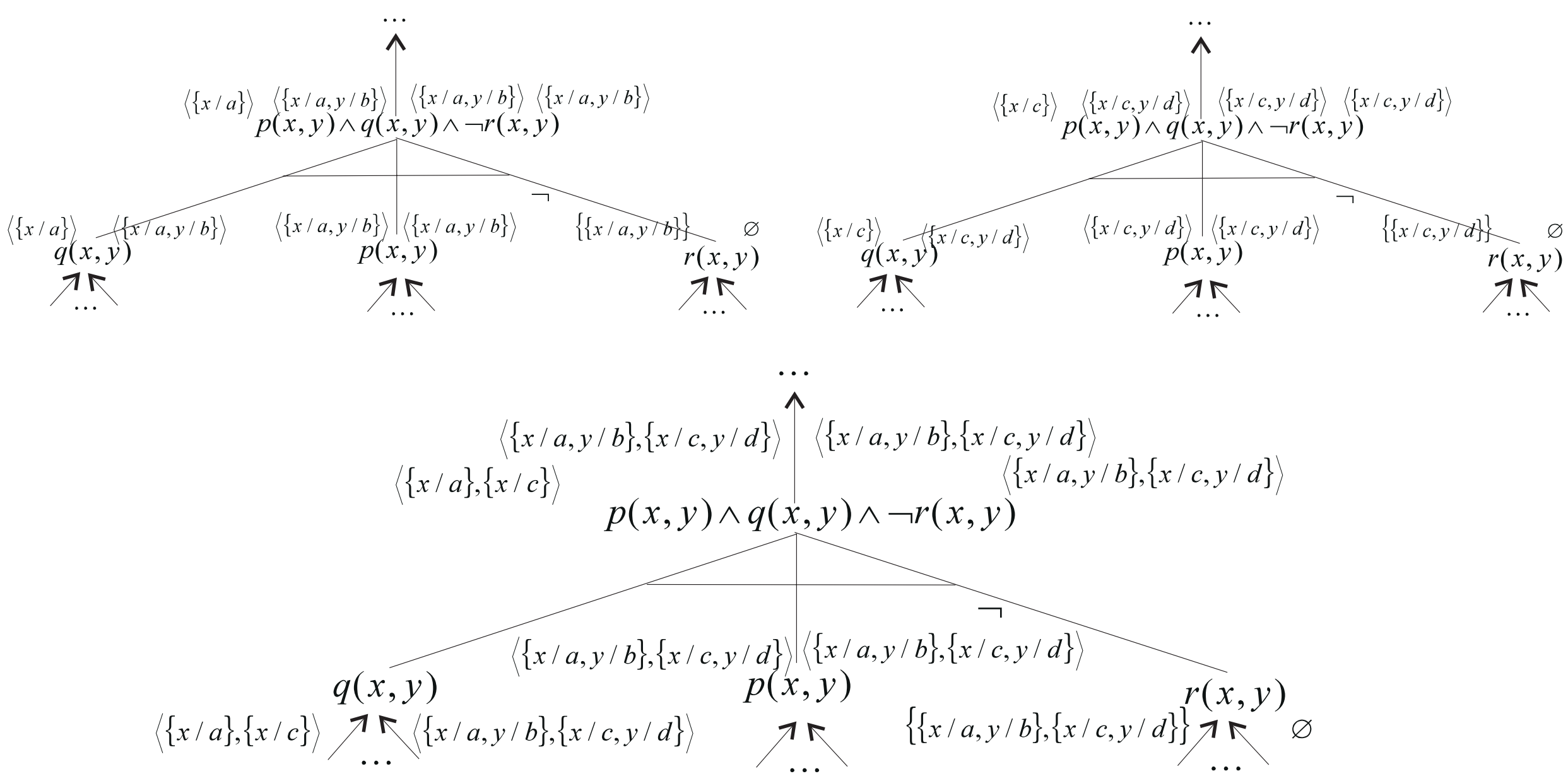

Abbildung 8.8: Zwei positive UND-Knoten und ihre Vereinigung

2. *Bei positiven ODER-Knoten ist die Vereinigung komplizierter. Jeder Knoten kann unterschiedliche Nachfolger haben, abhängig davon, welche Programmklauseln in dem Beweis verwendet wurden. Der resultierende Knoten hat für jede verwendete Programmklausel einen Nachfolger. Die Substitutionstupel ergeben sich als Konkatenation der Substitutionstupel der einzelnen Knoten. Ein Beispiel für eine solche Vereinigung ist in Abbildung 8.9 gegeben. Die Eigenschaften von positiven ODER-Knoten bleiben durch diese Vereinigung erhalten. Auf diese Weise bleiben auch alle Eigenschaften, die die Existenz von SLDNF-Beweisen betreffen, erhalten.*

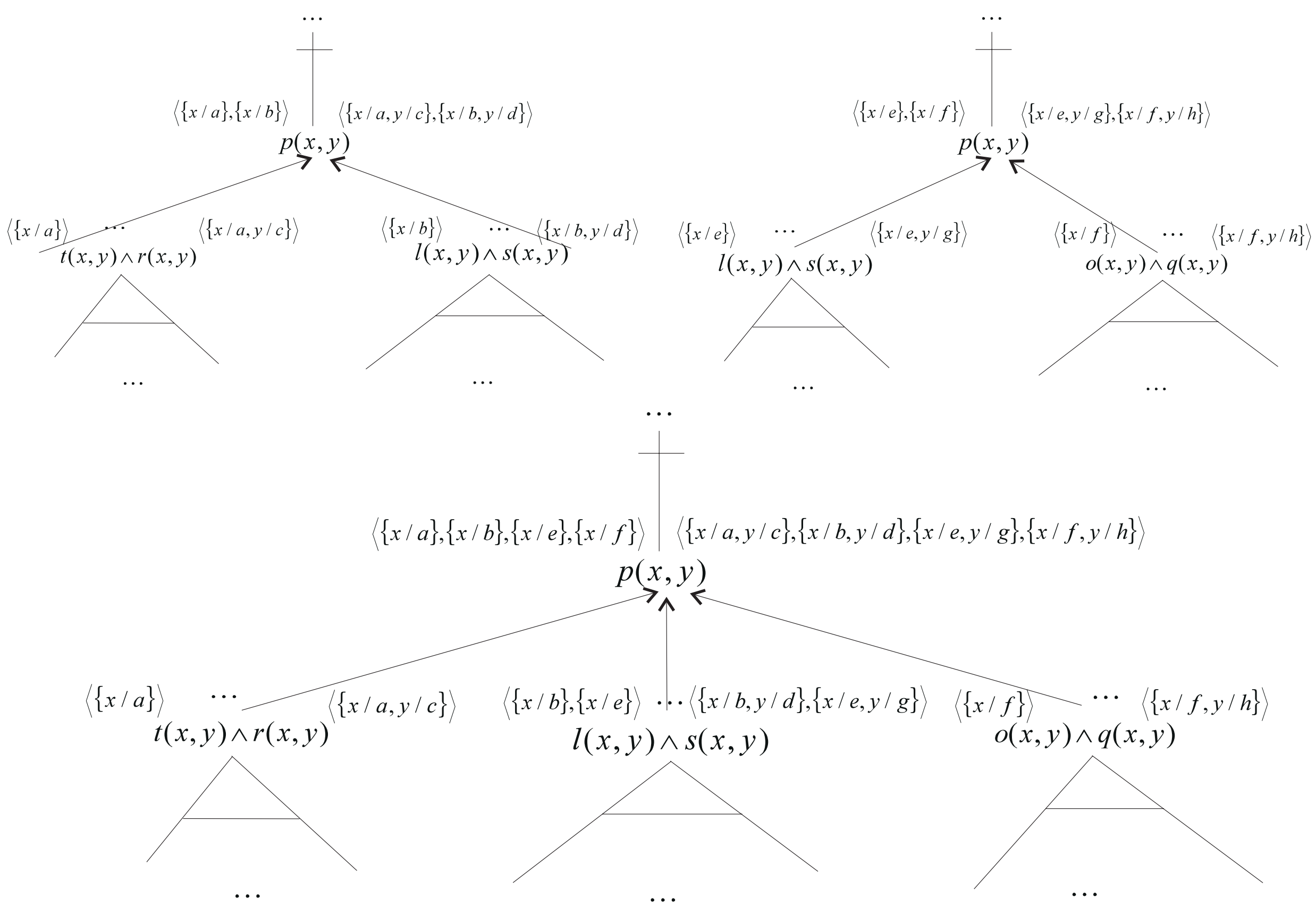

Abbildung 8.9: Zwei positive ODER-Knoten und ihre Vereinigung

3. *Bei negativen UND-Knoten ergibt sich bei der Suche nach einer Vereinigungsoperation folgende Schwierigkeit: Jeder Knoten kann verschiedene Nachfolger haben, trotzdem müssen die Eigenschaften eines negativen UND-Knotens bezüglich der Substitutionsmengen gewahrt bleiben. Der resultierende Knoten hat deswegen für jeden Nachfolger der beiden einzelnen Knoten jeweils einen Nachfolger. Die Substitutionsmengen werden vereinigt, pflanzen sich jedoch auch nach unten in die jeweiligen Teilbäume fort, allerdings nur innerhalb einer Negationsstufe. Damit bleiben ebenfalls alle Eigenschaften eines negativen UND-Knotens in einem Beweisbaum erhalten. Ein Beispiel für diese Art der Vereinigung ist in Abbildung 8.10 zu sehen.*

4. *Bei negativen ODER-Knoten ist die Vereinigung relativ unproblematisch. Da jeder Knoten die gleichen Nachfolger hat, müssen nur die Substitutionsmengen des Knotens vereinigt werden. Ein Beispiel dafür ist in Abbildung 8.11 zu sehen.*

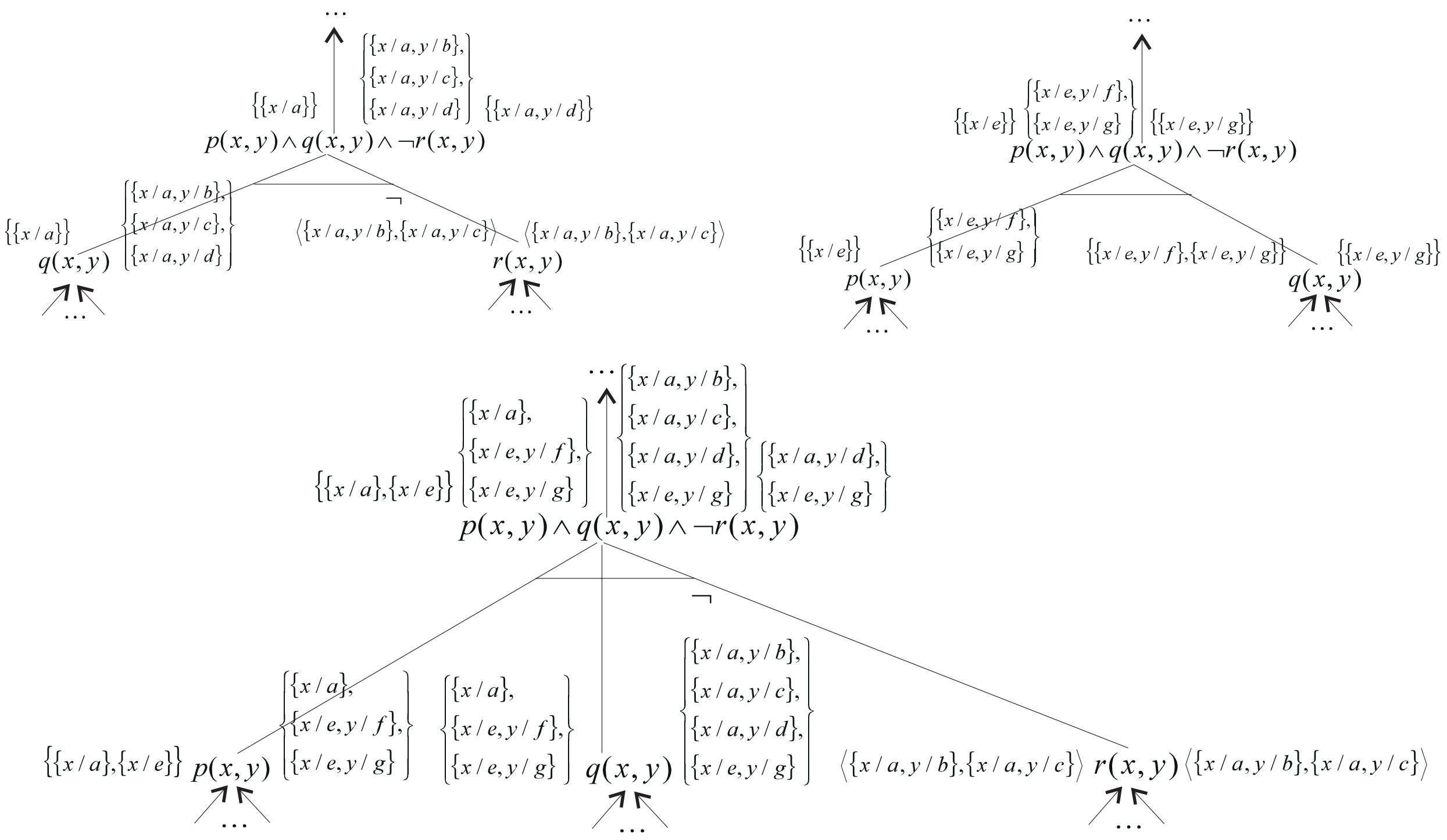

Abbildung 8.10: Zwei negative UND-Knoten und ihre Vereinigung

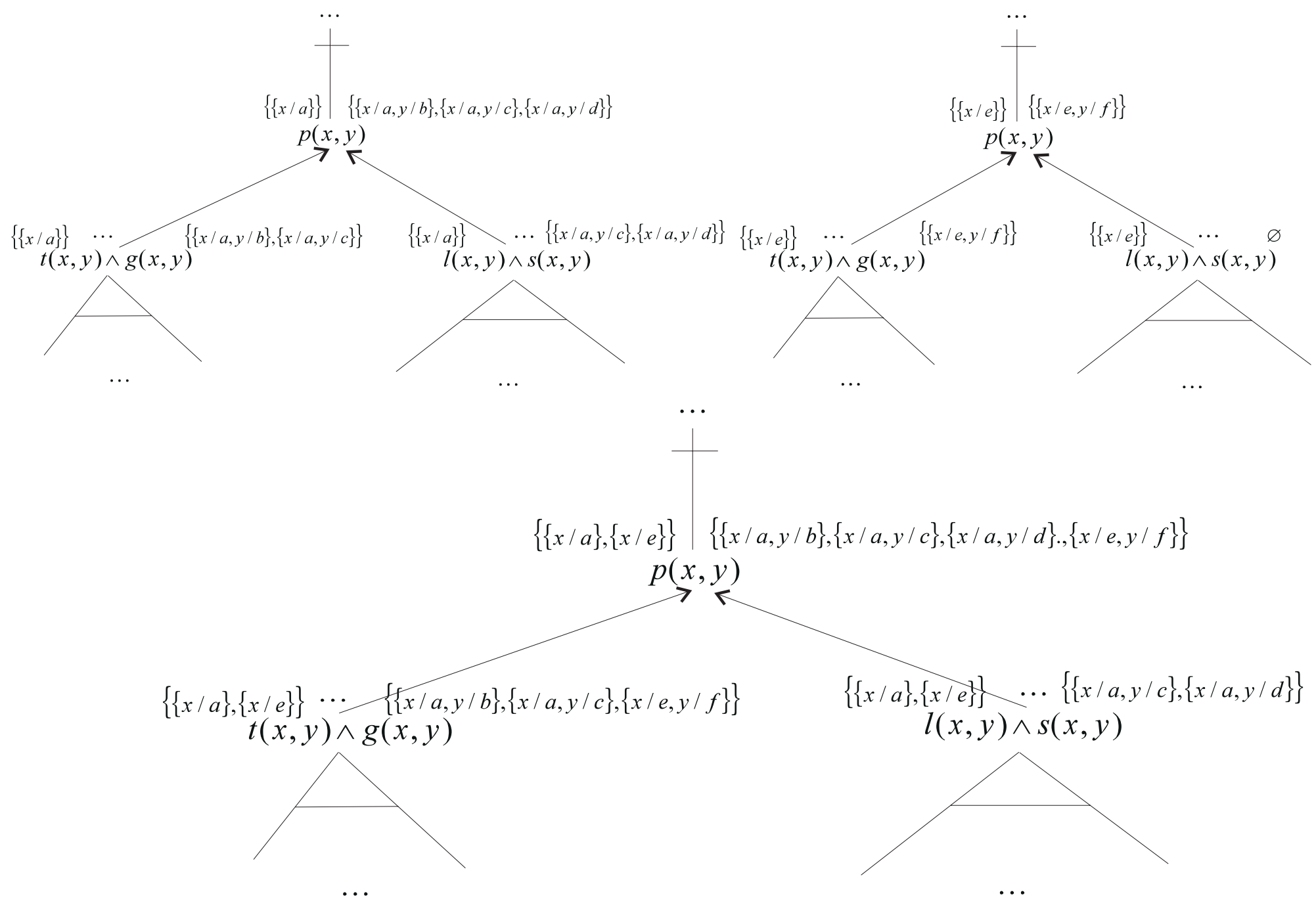

Abbildung 8.11: Zwei negative ODER-Knoten und ihre Vereinigung

Nachdem die grundsätzlichen Elemente der Vereinigung von zwei Knoten an den Beispielen illustriert worden ist, definieren wir nun die genauen Algorithmen, die die beiden Teilbäume sukzessive vereinigen.

**Definition 8.4.1** *Sei $D$ eine deduktive Datenbank, $d = \langle Del, Add \rangle$ eine Datenbankänderung auf $D$ mit resultierender Datenbank $D'$, $\leftarrow W$ eine Zielklausel und $T$ ein Beweisbaum für $D \cup \{\leftarrow W\}$ mit beliebiger Anfangssubstitution, der aus einem SLDNF-Beweis mit der Standardberechnungsregel nach Definition 7.1.5 erzeugt sowie bezüglich $D$ gepflegt und bereinigt wurde. Sei $T'$ ein Beweisbaum bezüglich $D'$, $A$ ein ODER-Knoten aus $T$ und $A'$ ein ODER-Knoten gleicher Polarität und gleicher Beschriftung in $T'$. Die zu $T$ gehörenden Abbildungen heißen subst und $nachfolger$, die zu $T'$ gehörenden Abbildungen heißen subst$'$ und $nachfolger'$. Falls $A$ und $A'$ positive ODER-Knoten sind, führe die Vereinigung durch Aufruf von vereinige-pos-ODER$(A, A')$ durch, sonst durch Aufruf von vereinige-neg-ODER$(A, A')$. Der neue Baum $T$ enthält dann die Vereinigung der beiden Beweisbäume.*

*vereinige-pos-ODER$(A, A')$*

1. *Sei subst$(A) = \langle U_B, U_E \rangle$ und subst$'(A') = \langle U'_B, U'_E \rangle$.*
2. *Setze $U_B = U_B + U'_B$ und $U_E = U_E + U'_E$.*
3. *Die Abbildung $nachfolger(A)$ berechnet folgendermaßen:*
   (a) *Falls $A'$ ein Blattknoten ist dann beende den Algorithmus.*
   (b) *Für jedes $W \in nachfolger(A)$, zu dem es ein $W' \in nachfolger'(A')$ gibt, so daß $W$ und $W'$ von der gleichen Programmklausel abstammen, wird vereinige-pos-und$(W, W')$ aufgerufen.*
   (c) *Für jedes $W' \in nachfolger'(A')$, für das es kein $W \in nachfolger(A)$ gibt, so daß $W'$ und $W$ von der gleichen Programmklausel abstammen, wird $W'$ zu $nachfolger(A)$ hinzugefügt. Außerdem wird der ganze Teilbaum unter $W'$ beibehalten, d.h. subst$(k')$ = subst$'(k')$ und $nachfolger(k') = nachfolger'(k')$ für alle Knoten im Teilbaum, dessen Wurzel $W'$ ist.*

*vereinige-pos-UND$(W, W')$*

1. *Sei subst$(A) = (U_i)$ und subst$'(W') = (U'_j)$.*
2. *Setze $U_i = U_i + U'_i$ für alle $0 \leq i \leq n$, wobei $n$ die Länge von $W$ ist.*
3. *Alle $A_i \in nachfolger(W)$ und alle $A'_i \in nachfolger'(W')$ (für $1 \leq i \leq n$) werden folgendermaßen vereinigt:*
   - *vereinige-pos-ODER$(A_i, A'_i)$, falls $A_i$ ein positiver ODER-Knoten ist, bzw.*
   - *vereinige-neg-ODER$(A_i, A'_i)$, falls $A_i$ ein negativer ODER-Knoten ist.*

*vereinige-neg-ODER(A,A')*

1. *Sei* $subst(A) = \langle S_B, S_E \rangle$ *und* $subst'(A') = \langle S'_B, S'_E \rangle$.
2. *Setze* $S_B = S_B \cup S'_B$ *und* $S_E = S_E \cup S'_E$.
3. *Falls* $A'$ *ein Blattknoten ist, beende den Algorithmus.*
4. *Rufe für jeden Nachfolger* $W$ *von* $A$ *und für jeden Nachfolger* $W'$ *von* $A'$ *(d.h.* $W \in nachfolger(A)$ *und* $W' \in nachfolger(A')$*), zu deren Aufbau die gleiche Programmklausel benutzt wurde, vereinige-neg-UND*$(W, W')$ *auf.*
5. *Für jeden Nachfolger* $W'$ *von* $A'$*, für den es keinen Nachfolger* $W$ *von* $A$ *gibt, der aus der gleichen Programmklausel wie* $W'$ *gebildet wurde, wird der ganze Baum, dessen Wurzel* $W'$ *in* $T'$ *bildet, unter* $A$ *gehängt.* $W'$ *wird zu den Nachfolgern von* $A$ *hinzugefügt. Für alle Knoten* $k'$ *aus diesem Teilbaum ist dann* $subst(k) = subst'(k')$ *und* $nachfolger(k') = nachfolger'(k')$*. An alle Substitutionsmengen für Knoten im angehängten Teilbaum innerhalb der gleichen Polaritätsstufe (d.h. bis zum nächsten Polaritätswechsel) wird die Menge* $S_E$ *hinzugefügt (d.h. vereinigt).*

*vereinige-neg-UND*$(W, W')$

1. *Sei* $W \equiv L_1 \wedge \ldots \wedge L_n$ *und* $W' \equiv L'_1 \wedge \ldots \wedge L'_n$*. Sei* $nachfolger(W) = \langle A_1, \ldots, A_m \rangle$ *und* $nachfolger'(W') = \langle A'_1, \ldots, A'_n \rangle$*. Seien weiterhin* $L_{i_1}, \ldots, L_{i_m}$ *die Literale, aus denen die Atome* $A_1, \ldots, A_m$ *(=* $nachfolger(W)$*) entstanden sind. Genauso seien* $L'_{j_1}, \ldots, L'_{j_n}$ *die Literale, aus denen die Atome* $A'_1, \ldots, A'_n$ *(=* $nachfolger'(W')$*) entstanden sind. Sei* $subst(W) = (S_i)$, $0 \leq i \leq m$*, und* $subst'(W') = (S'_j)$, $0 \leq j \leq n$.
2. *Die Abbildung* $nachfolger(W)$ *berechnet sich folgendermaßen: Die Liste der Indizes* $i_1, \ldots, i_m, j_1, \ldots, j_m$ *wird sortiert und doppelte Vorkommen werden gestrichen. Sei die Ergebnisliste* $k_1, \ldots, k_p$, $p \leq n + m$*. Dann ist* $nachfolger(W) = \langle B_1, \ldots, B_p \rangle$*. Alle* $B_o$ *mit* $1 \leq o \leq p$ *werden folgendermaßen bestimmt:*
   - $B_o = L_{k_o}$*, falls ein nicht negiertes Literal mit Index* $k_o$ *in* $W$ *vorkommt. bzw.* $B_o = L'_{k_o}$ *aus* $W'$*, falls kein nichtnegiertes Literal mit Index* $k_o$ *in* $W$ *existiert.* $B_o$ *ist dann ein negativer ODER-Knoten.*
   - $B_o = B$*, falls in* $W$ *ein negiertes Literal* $L_{k_o} \equiv \neg B$ *vorkommt bzw.* $B_o = B'$*, wenn ein Literal* $L'_{k_o} \equiv \neg B'$ *in* $W'$ *existiert.* $B_o$ *ist dann ein positiver ODER-Knoten.*

   *Der neue Knoten hat also einen Nachfolger für jeden Nachfolger von* $W$ *und* $W'$*. Nun müssen noch die Substitutionsmengen so bestimmt werden, daß sie die Bedingungen für Beweisbäume erfüllen.*

3. *Die neue Folge der zu $W$ gehörenden Substitutionsmengen ($S''_j$) berechnet sich folgendermaßen:*
   - $S''_0 = S_0 \cup S'_0$
   - $S''_o = S_h \cup S'_l$, *dabei sind $h$ und $l$ die größten Indizes, für die $i_h \leq k_o$ und $j_l \leq k_o$ für alle $1 \leq o \leq p$ gilt.*
4. *Für alle $L_{i_h}$ und $L'_{j_l}$, für die $i_h = j_l$ gilt werden die Teilbäume vereinigt, und zwar:*
   (a) *vereinige-pos-ODER($A_i, A'_l$) falls $L_{i_h}$ ein negatives Literal ist, oder*
   (b) *vereinige-neg-ODER($A_i, A'_l$) falls $L_{i_h}$ ein Atom ist.*

   *In diesem Fall ist das gleiche Literal für einen Nachfolger von $W$ und $W'$ verantwortlich.*
5. *Wenn ein Literal $L_{i_h}$ aus $W$ allein für einen Nachfolger verantwortlich ist, es also für $L_{i_h}$ kein $L'_{j_l}$ gibt mit so daß $i_h = j_l$, so müssen die Substitutionsmengen von $W'$ noch zum Teilbaum unter $L_{i_h}$ hinzugefügt werden. Sei $B$ der Nachfolger, für den $L_{i_h}$ verantwortlich ist. Dann wird wird an* allen *Substitutionsmengen von $B$ und der Nachfolger von $B$ innerhalb der gleichen Polaritätsstufe (also insbesondere nur, wenn $L_{i_h}$ ein Atom ist) die Menge $S'_{j_g}$ hinzugefügt. Dabei ist $g$ der größte Index, für den $j_g \leq i_h$ gilt.*
6. *Wenn ein Literal $L'_{j_l}$ aus $W'$ allein für einen Nachfolger verantwortlich ist, es also für $L'_{j_l}$ es kein $L_{i_h}$ mit $j_l = i_h$ gibt, müssen die Substitutionsmengen von $W$ noch zum Teilbaum unter $L'_{j_l}$ hinzugefügt werden. Sei $B$ der Nachfolger, für den $L'_{j_l}$ verantwortlich ist. Dann wird wird an* allen *Substitutionsmengen von $B$ und der Nachfolger von $B$ innerhalb der gleichen Polaritätsstufe (also insbesondere nur, wenn $L'_{j_l}$ ein Atom ist) die Menge $S_{i_g}$ hinzugefügt. Dabei ist $g$ der größte Index, für den $i_g \leq j_l$.*

Mit diesen Verfahren ist die Vereinigung zweier Beweisbäume definiert. Dies erlaubt es jetzt, die Teile eines Beweises zu ersetzen, die im Zuge einer Datenbankänderung ungültig geworden sind. Hierbei suchen wir für diese Teile neue Beweise und machen so den Beweis wieder gültig.

## 8.5 Beweismodifikation

**Definition 8.5.1** *Sei $D$ eine deduktive Datenbank, $d = \langle Del, Add \rangle$ eine Datenbankänderung auf $D$ mit resultierender Datenbank $D'$, $\leftarrow W$ eine Zielklausel und $T$ ein Beweisbaum für $D \cup \{\leftarrow W\}$ mit einer beliebigen Anfangssubstitution, der mit mit der Stand-* *ardberechnungsregel nach Definition 7.1.5 erzeugt wurde. Der folgende Algorithmus löst einen Konflikt auf, der nach erfolgter Beweispflege aufgrund der Datenbankänderung noch vorhanden ist.*

*löse-Konflikt*$(k, p, \Delta)$

1. *Wenn* $k$ *ein positiver ODER-Knoten ist, dann bereinige-pos-ODER*$(k, \Delta)$ *mit Ergebnis neuer Knoten* $k'$ *und Menge* $\Delta'$.
2. *Wenn* $k$ *ein negativer ODER-Knoten ist, dann bereinige-neg-ODER*$(k, \Delta)$ *mit Ergebnis neuer Knoten* $k'$ *und Menge* $\Delta'$.
3. *Wenn* $k$ *ein positiver UND-Knoten ist, dann bereinige-pos-UND*$(k, p, \Delta)$ *mit Ergebnis neuer Knoten* $k'$ *und Menge* $\Delta'$.
4. *Wenn* $k$ *ein negativer UND-Knoten ist, dann bereinige-neg-UND*$(k, p, \Delta)$ *mit Ergebnis neuer Knoten* $k'$ *und Menge* $\Delta'$.
5. *Wenn* $k'$ *die Wurzel von* $T'$ *ist, dann versuche einen Neubeweis von* $k'$. *Wenn der Beweis gelungen ist, dann gebe den neuen Beweisbaum zurück, ansonsten gebe "Integritätsbedingung verletzt" zurück.*
6. *Wenn* $k'$ *ein positiver ODER-Knoten* $A$ *ist, dann*
   (a) *Versuche für jedes für jedes* $\sigma \in \Delta'$ *einen Beweisbaum für* $D' \cup \{\leftarrow A\sigma\}$ *zu generieren. Hierbei betrachten wir das* Literal $A$ *als UND-Knoten mit Anfangssubstitution* $\sigma$. *Sei* $T'$ *einer dieser Beweisbäume. Sei* $A'$ *der Nachfolger der Wurzel in* $T'$. *Rufe für jedes* $A'$ *auf: vereinige-pos-ODER*$(A, A')$.
   (b) *Sei* $\Delta''$ *die Menge aller* $\sigma \in \Delta'$, *für die die Konstruktion eines neuen Beweisbaumes fehlschlägt. Sei* $W$ *der Vorgänger von* $A$. $W$ *ist ein negativer UND-Knoten. Sei* $A$ *der* $j$*-te Nachfolger von* $W$. *Rufe löse-Konflikt*$(W, j, \Delta'')$ *auf.*
7. *Wenn* $k'$ *ein negativer ODER-Knoten* $A$ *ist, dann*
   (a) *Versuche für jedes* $\sigma_i \in \Delta$ *einen Beweisbaum für* $D' \cup \{\leftarrow \neg A\sigma_i\}$ *zu generieren,* $\sigma_i$ *ist dabei die Anfangssubstitution des Beweisbaumes.*
   (b) *Sei* $T'$ *einer dieser Beweisbäume. Sei* $A'$ *der Nachfolger der Wurzel in* $T'$. *Rufe für jedes* $T'$ *und* $A'$ *auf: vereinige-neg-ODER*$(A, A')$.
   (c) *Sei* $\Delta''$ *die Menge aller* $\sigma \in \Delta'$, *für die die Konstruktion eines neuen Beweisbaumes fehlschlägt. Sei* $W$ *der Vorgänger von* $A$. $W$ *ist ein positiver-UND-Knoten mit subst*$(W) = (U_i)$, *dessen* $j-$*ter Nachfolger der Knoten* $A$ *ist. Sei* $I = \{i \mid \sigma \in \Delta'', \sigma$ *befindet sich an der* $i-$*ten Position im Tupel* $U_{j-1}\}$. *Rufe löse-Konflikt*$(W, j, I)$ *auf.*

*Mit den vorangestellten Verfahren sind wir nun in der Lage, unseren eigentlichen Integritätsüberprüfungsalgorithmus zu formulieren:*

*Sei* $\leftarrow W$ *eine Zielklausel,* $D$ *eine deduktive Datenbank,* $d = \langle Del, Add \rangle$ *eine Datenbankänderung von* $D$ *mit resultierender Datenbank* $D'$ *und* $T$ *ein Standardbeweisbaum für* $D \cup \{\leftarrow W\}$*. Der folgende Algorithmus führt die Integritätsüberprüfung durch:*

*überprüfe-baum*$(T, d)$

1. *Rufe pflege-baum*$(T, d)$ *auf, dabei ist der resultierendem Baum* $T_1$.
2. *für alle Konflikt-Knoten* $k$ *aus* $T_1$*:*
   (a) *Wenn* $k$ *ein negativer ODER-Knoten mit subst*$(k) = \langle S_B, S_E \rangle$ *ist, dann sei* $\Delta = \{\sigma \mid \sigma \geq \sigma', k\sigma' = A', A' \leftarrow \in Add\}$.
   (b) *Wenn* $k$ *ein positiver ODER-Knoten mit subst*$(k) = \langle U_B, U_E \rangle$ *ist, dann sei* $\Delta = \{i \mid \sigma_i \in U_E, k\sigma_i = A', A' \leftarrow \in Del$ *und* $\sigma_i$ *befindet sich an der* $i-$*ten Position in* $U_E\}$.
   (c) *Wenn* $k$ *ein negativer UND-Knoten mit subst*$(k) = (S_j)$, $0 \leq j \leq m$, *ist, dann ist* $\Delta = S_0$.
   (d) *Wenn* $k$ *ein positiver UND-Knoten mit subst*$(k) = (U_j)$, $1 \leq j \leq m$, *ist, dann sei* $\Delta = \{1, \ldots, n\}$*. Dabei ist* $n$ *die Anzahl der Substitutionen in* $U_0$.
   (e) *Rufe löse-Konflikt*$(T_1, p, \Delta)$ *auf, welches entweder den neuen resultierendem Baum* $T_1$ *oder eine Meldung zurückgibt. Wenn löse-Konflikt das Ergebnis „Integritätsbedingung verletzt“ zurückgibt, dann gebe „Integrität ist verletzt“ zurück. Beende den Algorithmus.*
3. $T_1$ *ist jetzt ein neuer Beweisbaum für* $W$.

Dieser Algorithmen terminiert genau denn, wenn die darunterliegende Inferenzmaschine terminiert, die neue Beweisbäume sucht. Dies liegt daran, daß die einzelnen Teilalgorithmen bei endlichen Bäumen immer terminieren und die Anzahl der Konfliktknoten nur weniger werden können. Mit diesen Algorithmen können wir folgendes Ergebnis formulieren:

**Satz 8.5.1** *Sei* $\leftarrow W$ *eine Zielklausel,* $D$ *eine deduktive Datenbank,* $d$ *eine Datenbankänderung auf* $D$ *mit resultierender Datenbank* $D'$, $T$ *ein Standardbeweisbaum für* $D \cup \{\leftarrow W\}$*. Wenn der Algorithmus überprüfe-baum*$(T, d)$ *terminiert, dann erfüllt* $D'$ *die Integritätsbedingung* $\leftarrow W$ *genau dann, wenn die Ausgabe ein neuer Beweisbaum* $T'$ *ist.*

**Beweis** Wir geben nur die Beweisidee:

Es reicht, folgendes zu zeigen:

1. Wenn der Algorithmus die Ausgabe “Integritätsbedingung verletzt“ liefert, dann erfüllt die Datenbank $D'$ die Integritätsbedingung nicht. In diesen Fall ist aber ein Neubeweis der Integritätsbedingung versucht worden, und dieser ist gescheitert.

2. Wenn der Algorithmus einen Baum zurückliefert, so ist dieser ein Beweisbaum. Hierfür ist zu zeigen, daß der zurückgelieferte Baum die Eigenschaften eines Beweisbaumes erfüllt. Dies geschieht nach folgendem Schema:

   (a) Die Blätter erfüllen die Eigenschaften eines Beweisbaumes.

   (b) Es gibt keine Konfliktknoten in $T'$.

   (c) Die verlangten Eigenschaften der Substitutionsmengen und –tupel bleiben erhalten, da die Beweispflege an diesen Eigenschaften nichts ändert. Die durch die Bereinigung fehlenden Substitutionen werden durch die Konfliktauflösung und der darauffolgenden Vereinigung mit den neu erzeugten Teilbeweisbäumen wieder hinzugenommen. Weiterhin sind alle Substitutionen in den Mengen und Tupeln voneinander unabhängig, wie in Lemma 6.2.4 gezeigt wurde.

Da die Terminierung des Algorithmus vorausgesetzt wurde, reichen diese beiden Punkte für beide Beweisrichtungen aus.

# Kapitel 9

# Zusammenfassung und Ausblick

## 9.1 Berechnungsregeln

Wir haben eine neue Methode der Beweisdarstellung von Logikprogrammen gegeben und gezeigt, daß diese Darstellung zumindest zu SLDNF-Beweisen, die mit der Standardberechnungsregel erstellt wurden, gleichwertig ist. Die Standardberechnungsregel deckt zwar die typischen Anwendungsfälle ab, ist aber keine genaue Charakterisierung der Berechnungsregeln, für die ein SLDNF-Beweis in einen Beweisbaum transformiert werden kann. Zur weiteren Klärung dieser Frage ist die folgende Beobachtung sehr hilfreich: Die Standardberechnungsregel selektiert alle Literale einer verwendeten Programmklausel, bevor sie ein nicht in der aktuellen verwendeten Programmklausel enthaltenes Literal selektiert. Beweise, die mit Berechnungsregeln solchen Typs erstellt werden, sollten sich für eine Transformation in einen Beweisbaum gut eignen. Die Struktur solcher SLDNF-Beweise kommt der Struktur eines Beweisbaumes sehr entgegen. Gleichzeitig ist aber auch die *Reihenfolge* der Selektion der Literale von Bedeutung, da die Auswertungsreihenfolge in einem Beweisbaum durch die Struktur festgelegt ist. Um eine genaue Charakterisierung zu erhalten, wird man sicherlich die Sichtweise einer Berechnungsregel als Funktion *nur* der aktuellen Zielklausel aufgeben müssen. Es sollten noch weitere Parameter, etwa den aktuellen Pfad oder die *Geschichte* der Ableitung, d.h. die bis dahin im Beweis verwendeten Programmklauseln, betrachtet werden.

## 9.2 Anwendungsmöglichkeiten

Im Vergleich zu einer Menge von SLDNF-Bäumen hat ein Beweisbaum eine sehr übersichtliche Struktur. Die Konstruktion eines solchen Beweisbaumes kann in jeder Höhe eines beliebigen SLDNF-Baumes aus einem SLDNF-Beweis sowohl begonnen als auch abgebrochen werden. Daher eignet sich ein Beweisbaum gut zur statischen Analyse von Beweisen. Eine derartige Analyse ist z.B. für die Fehlersuche in Prologprogrammen nützlich. Vorstellbar sind z.B. Werkzeuge, die einen Teil eines Beweises graphisch als Beweisbaum darstellen. Sie können so eine wesentlich übersichtlichere Form als die übliche Ablaufverfolgung von Prologprogrammen liefern. Es wird dadurch einfach möglich, auch eine tief im Ablauf eines Programmes erzeugte Konstellation direkt zu betrachten. Auf herkömmliche Art und Weise ist dies nur sehr schwierig zu erreichen.

## 9.3 Auswertungsmethoden

Die Bottom-up Anfrageauswertungsmethoden in deduktiven Datenbanken (z.B. magic-sets, naive und seminaive Auswertungsmethoden) betrachten, ebenso wie die Darstellung eines Beweises als Beweisbaum, Substitutionsmengen. Es sollte möglich sein, einen durch eine solche Auswertungsmethode gewonnenen Beweis in einen Beweisbaum zu transformieren. Hier ist aber sicherlich noch Bedarf an einer weiteren Verfeinerung der Methodik.

## 9.4 TMS/JTMS

Die ganze Problematik der inkrementellen Integritätsüberprüfung ist sehr verwandt mit einem anderen, in der Literatur schon sehr gut untersuchten Gebiet: der Erstellung eines TMS (truth maintenance systems) (siehe [Doyle, 1979, de Kleer, 1986]). Hier wie dort wird versucht, die Begründungen für Aussagen zu verwalten. Seltsamerweise ist dieser Zusammenhang in der Literatur nicht zu finden, d.h. es gibt keinerlei Querverweise. Die hier vorgestellte Methode kann als JTMS (justification based truth maintenance system) aufgefaßt werden. Weitere Arbeiten auf diesem Gebiet sind sicher nötig.

## 9.5 Handlungsbedarf

Die Definitionen und Algorithmen, die in dieser Arbeit gegeben wurden, sind im Hinblick auf eine Verwendung in der Praxis sicherlich nicht optimal. So können die Substitutionsmengen an den Knoten eines Beweisbaumes so groß werden, daß eine Realisierung nicht mehr möglich ist. Hier müssen also Verfahren gefunden werden, um die zu speichernden Substitutionsmengen möglichst klein halten bzw. ihre Speicherung überhaupt zu vermeiden. Ein Ansatz hierbei ist z.B., daß Substitutionen an Polaritätswechseln nur für die Variablen des aktuellen Literales wichtig sind. Eventuell reicht auch eine Speicherung der Differenzen von zwei Substitutionsmengen aus.

Weiterhin sind noch viele spezielle Algorithmen auf Beweisbäumen denkbar, die die Integritätsüberprüfung abhängig von der Datenbankänderung effizienter durchführen können. Die könnte z.B. dadurch geschehen, daß bevorzugt innerhalb einer Datenbankänderung nach neuen Lösungen gesucht wird. Dies ist aber ein weites Feld.

Die jetzige Implementation benutzt zum Aufbau des Beweisbaumes einen Metainterpreter. Dies bedeutet einen enormen Effizienzverlust, da auch das *PROLOG*-Backtracking simuliert werden muß. Es sollte möglich sein, eine Programmtransformation durchzuführen, so daß das resultierende Programm zur Auswertung das *PROLOG*-Backtracking benutzt, aber trotzdem alle benötigten Informationen ablegt.

# Anhang A

# Algorithmen

Der Algorithmus zur Übersetzung einer prädikatenlogischen Formel in ein normales Programm wird durch folgende sechs Funktionen realisiert:

1. übersetze-formel(Atom,Formel)
   Eingabe: ein Atom, das als Programmkopf dient und eine beliebige FOL-Formel.
   Rückgabe: Liste von äquivalenten Programmklauseln.

2. neue Regeln(ToDo)
   Eingabe: Liste ToDo von Tupeln, bestehend aus jeweils einem Atom und einer Formel.
   Rückgabe: Liste von Programmklauseln.

3. generiere-Disjunktionen(Atom,Disjunktionen)
   Eingabe: Atom, das als Programmkopf fungiert und eine Liste von Disjunktionen, die jeweils einen Programmrumpf darstellen.
   Rückgabe: Ein Tupel, bestehend aus der Liste von Programmklauseln und einer Liste von Tupeln, die jeweils aus einem Atom und einer Formel bestehen.

4. generiere-rumpf(Konjunktionen)
   Eingabe: eine Liste von konjunktiv miteinander verknüpften Formelelementen.
   Rückgabe: Ein Tupel, bestehend aus einem Klauselrumpf und einer Liste von Tupeln, jeweils bestehend aus einem Atom und einer Formel.

5. hole-disjunktionen(Polarität,Formel)
   Eingabe: Die Polarität der Teilformel, d.h. „−“ für negiert und „+“ für nicht negiert und eine Formel.

Rückgabe: Ein Tupel, bestehend aus einer Liste von disjunktiv miteinander in Beziehung stehenden Elementen und einer Liste von Tupeln, jeweils bestehend aus einem Atom und einer Formel.

6. hole-Konjunktionen(Polarität,Formel)
   Eingabe: Die Polarität der Teilformel, d.h. „−“ für negiert und „+“ für nicht negiert und eine Formel.
   Rückgabe: Ein Tupel, bestehend aus einer Liste von konjunktiv miteinander in Beziehung stehenden Elementen und einer Liste von Tupeln jeweils bestehend aus einem Atom und einer Formel.

**funktion** übersetze-formel(Atom,Formel)
　　$\langle Disj, ToDo_1\rangle$ = hole-disjunktionen(+,Formel)
　　$\langle Programm_1, ToDo_2\rangle$ = generiere-disjunktion(Atom,Disj)
　　$Programm_2$ = neue-regeln($ToDo_1 + ToDo_2$)
　　**gebe-zurück** $Programm_1 + Programm_2$

**funktion** neue-regeln(ToDo)
　　Programmregeln = [ ]
　　**für alle** $\langle Atom, Formel\rangle$ **in** ToDo
　　　　Programmregeln = Programmregeln + übersetze-formel(Atom,Formel)
　　**gebe-zurück** Programmregeln

**funktion** generiere-disjunktion(Atom,Disjunktionen)
　　Programmregeln = [ ]
　　ToDo = [ ]
　　**für alle** Dis **in** Disjunktionen
　　　　$\langle Konj, ToDo_1\rangle$ = hole-Konjunktionen(+, $Dis$)
　　　　$\langle Rumpf, ToDo_2\rangle$ = generiere-Rumpf($Konj$)
　　　　Programmregel = Programmregel +[$Atom \leftarrow Konj$]
　　　　$ToDo = ToDo + ToDo_1 + ToDo_2$
　　**gebe-zurück** $\langle Programmregeln, ToDo\rangle$

**funktion** generiere-rumpf(Konjunktionen)
　　Rumpf = NIL
　　ToDo = [ ]
　　**für alle** Konj **in** Konjunktionen
　　　　**falls** (Konj ein Literal ist) **dann**
　　　　　　$Rumpf = Rumpf \wedge Konj$
　　　　**sonst** ToDo = ToDo + [$\langle p(x_1, \ldots, x_n), Konj\rangle$]
　　　　　　$Rumpf = Rumpf + p(x_1, \ldots, x_n)$
　　　　　　dabei ist $p$ ein neues Prädikatsymbol und die $x_1, \ldots, x_n$ sind alle freien Variablen in Konj
　　**gebe-zurück** $\langle Rumpf, ToDo\rangle$

**funktion** hole-Disjunktionen(Polarität,Formel)

**falls**(Formel = $\neg F_1$) **dann**

**gebe-zurück** hole-Disjunktionen($Polarität^{-1}$,$F_1$)

**sonst falls** (Polarität = + **und** Formel = $F_1 \vee F_2$) **dann**

$\langle Disj_1, ToDo_1\rangle$ = hole-Disjunktionen(+, $F_1$)

$\langle Disj_2, ToDo_2\rangle$ = hole-Disjunktionen(+, $F_2$)

**gebe-zurück** $\langle Disj_1 + Disj_2, ToDo_1 + ToDo_2\rangle$

**sonst falls** (Polarität = − **und** Formel = $F_1 \wedge F_2$) **dann**

$\langle Disj_1, ToDo_1\rangle$ = hole-Disjunktionen(−, $F_1$)

$\langle Disj_2, ToDo_2\rangle$ = hole-Disjunktionen(−, $F_2$)

**gebe-zurück** $\langle Disj_1 + Disj_2, ToDo_1 + ToDo_2\rangle$

**sonst falls** (Polarität = + **und** Formel = $F_1 \leftarrow F_2$) **dann**

$\langle Disj_1, ToDo_1\rangle$ = hole-Disjunktionen(+, $F_1$)

$\langle Disj_2, ToDo_2\rangle$ = hole-Disjunktionen(−, $F_2$)

**gebe-zurück** $\langle Disj_1 + Disj_2, ToDo_1 + ToDo_2\rangle$

**sonst falls** (Polarität = + **und** Formel = $F_1 \rightarrow F_2$) **dann**

$\langle Disj_1, ToDo_1\rangle$ = hole-Disjunktionen(+, $F_2$)

$\langle Disj_2, ToDo_2\rangle$ = hole-Disjunktionen(−, $F_1$)

**gebe-zurück** $\langle Disj_1 + Disj_2, ToDo_1 + ToDo_2\rangle$

**sonst falls** (Polarität = − **und** Formel = $F_1 \leftrightarrow F_2$) **dann**

$\langle Disj_1, ToDo_1\rangle$ = hole-Disjunktionen(−, $F_1 \leftarrow F_2$)

$\langle Disj_2, ToDo_2\rangle$ = hole-Disjunktionen(−, $F_1' \rightarrow F_2'$)

mit $F_i'$ äquivalent zu $F_i$ bis auf Umbenennung aller gebundenen Variablen.

**gebe-zurück** $\langle Disj_1 + Disj_2, ToDo_1 + ToDo_2\rangle$

**sonst falls** (Polarität = − **und** Formel = $\exists\, x_1, \ldots, x_m\; F_1$) **dann**

**gebe-zurück** $\langle[\neg p(x_1, \ldots, x_n)], [\langle p(x_1, \ldots, x_n), F_1\rangle]\rangle$

dabei ist $p$ ein neues Prädikatsymbol und $x_1, \ldots, x_n$ alle freien Variablen in $F_1$

**sonst falls** (Polarität = + **und** Formel = $\exists\, x_1, \ldots, x_m\; F_1$) **dann**

**gebe-zurück** hole-Disjunktionen(+, $F_1$)

**sonst falls** (Formel = $\forall\, x_1, \ldots, x_n\; F_1$) **dann**

**gebe-zurück** hole-Disjunktionen($Polarität^{-1}$, $\exists\, x_1, \ldots, x_n\; \neg F_1$)

**sonst falls** (Polarität = +) **dann**

**gebe-zurück** $\langle[F_1], [\,]\rangle$

**sonst gebe-zurück** $\langle[\neg F_1], [\,]\rangle$

**funktion** hole-Konjunktionen(Polarität,Formel)

**falls**(Formel = $\neg F_1$) **dann**

**gebe-zurück** hole-Konjunktionen($Polarität^{-1}$,$F_1$)

**sonst falls** (Polarität = + **und** Formel = $F_1 \wedge F_2$) **dann**

$\langle Konj_1, ToDo_1\rangle$ = hole-Konjunktionen(+, $F_1$)

$\langle Konj_2, ToDo_2\rangle$ = hole-Konjunktionen(+, $F_2$)

**gebe-zurück** $\langle Konj_1 + Konj_2, ToDo_1 + ToDo_2\rangle$

**sonst falls** (Polarität = − **und** Formel = $F_1 \vee F_2$) **dann**

$\langle Konj_1, ToDo_1\rangle$ = hole-Konjunktionen($-$, $F_1$)
$\langle Konj_2, ToDo_2\rangle$ = hole-Konjunktionen($-$, $F_2$)
**gebe-zurück** $\langle Konj_1 + Konj_2, ToDo_1 + ToDo_2\rangle$
**sonst falls** (Polarität $= -$ **und** Formel $= F_1 \leftarrow F_2$) **dann**
$\langle Konj_1, ToDo_1\rangle$ = hole-Konjunktionen($+$, $F_2$)
$\langle Konj_2, ToDo_2\rangle$ = hole-Konjunktionen($-$, $F_1$)
**gebe-zurück** $\langle Konj_1 + Konj_2, ToDo_1 + ToDo_2\rangle$
**sonst falls** (Polarität $= -$ **und** Formel $= F_1 \rightarrow F_2$) **dann**
$\langle Konj_1, ToDo_1\rangle$ = hole-Konjunktionen($+$, $F_1$)
$\langle Konj_2, ToDo_2\rangle$ = hole-Konjunktionen($-$, $F_2$)
**gebe-zurück** $\langle Konj_1 + Konj_2, ToDo_1 + ToDo_2\rangle$
**sonst falls** (Polarität $= +$ **und** Formel $= F_1 \leftrightarrow F_2$) **dann**
$\langle Konj_1, ToDo_1\rangle$ = hole-Konjunktionen($+$, $F_1 \leftarrow F_2$)
$\langle Konj_2, ToDo_2\rangle$ = hole-Konjunktionen($+$, $F_1' \rightarrow F_2'$)
**gebe-zurück** $\langle Konj_1 + Konj_2, ToDo_1 + ToDo_2\rangle$
**sonst falls** (Polarität $= -$ **und** Formel $= \exists\, x_1, \ldots, x_m\ F_1$) **dann**
**gebe-zurück** $\langle [p(x_1, dots, x_n)], [\langle p(x_1, \ldots, x_n), F_1\rangle]\rangle$
dabei ist $p$ ein neues Prädikatsymbol und $x_1, \ldots, x_n$ alle freien Variablen in $F_1$
**sonst falls** (Polarität $= +$ **und** Formel $= \exists\, x_1, \ldots, x_m\ F_1$) **dann**
**gebe-zurück** hole-Konjunktionen($+$, $F_1$)
**sonst falls** (Formel $= \forall\, x_1, \ldots, x_n\ F_1$) **dann**
**gebe-zurück** hole-Konjunktionen($Polarität^{-1}, \exists\, x_1, \ldots, x_n\ \neg F_1$)
**sonst falls** (Polarität $= +$ **dann**
**gebe-zurück** $\langle [F_1], [\,]\rangle$
**sonst gebe-zurück** $\langle [\neg F_1], [\,]\rangle$

# Literaturverzeichnis

# Danksagung

Neben dem Autor waren noch zahlreiche andere Personen und Institutionen an der Entstehung dieser Arbeit beteiligt. Insbesondere bedanke ich mich

- ... bei Dr. Christoph Lingenfelder. Ihm verdanke ich die Einführung in das Gebiet der Logikprogrammierung und des automatischen Beweisens. Ohne seine Bereitschaft, sehr viel Zeit und Geduld zu opfern, und mir, wenn es nötig war, etwas auch dreimal zu erklären wäre diese Arbeit nie entstanden. Weiterhin verdanke ich Dr. Astrid Schmücker-Schend und Dr. Christoph Lingenfelder eine schöne und interessante Zeit am IWBS der IBM in Heidelberg.
- ... nocheinmal bei Dr. Christoph Lingenfelder, der immer ein Gegenbeispiel für meine schönsten Sätze parat hatte (und hat).
- ... bei der IBM Deutschland Entwicklung GmbH, die den Arbeitsplatz in Böblingen zur Verfügung stellte.
- ... bei dem ESM-Team, als da wären (in weitgehend alphabetischer Reihenfolge): Gabi Bendel, Bernhard Bitzel, Bernhard Berchert, Klaus Deinhart, Armin Hummel, und Dr. Sven Lorenz. Sie alle haben einen nicht unerheblichen Anteil an der Arbeit und bilden zusammen ein hervorragendes Team. Insbesondere Herrn Hummel danke ich für seinen Einsatz, der diese Arbeit bei der IBM erst ermöglichte.
- ... bei meiner Freundin Birgit, die lange Monate auf mich verzichten mußte und, obwohl selbst im Diplomarbeitsstreß, sich immer meine Ideen angehört und meine Ausarbeitung korrekturgelesen hat und die auch sonst einfach ein Schatz ist (Sorry Sven)).
- ... bei meinen Eltern, die mir immer sehr viel Liebe und Geduld entgegenbrachten.
- ... bei den vielen Korrekturlesern, die sich mit viel Arbeit und Mühe durch die Vorabversionen durchgekämpft haben. Insbesondere danke ich (wieder in alphabetischer Reihenfolge): Stefan Bock, Claus George, Folkert Jansen, Knut Meyer und Andreas Schröer. Die Verantwortung für die sicherlich immer noch in großen Mengen vorhandenen Fehler liegt dabei (wie immer) bei mir.
- ... bei den Wissenschaftlern, die mich durch Kommentare und der Zusendung von Literatur unterstützt haben. Insbesondere bei: Hendrik Decker, Stefan Lüttringhaus-Kappel, Ulrike Griefahn und Bern Martens.
- ... bei Prof. Otto Mayer, der die Betreuung dieser Arbeit von Seiten der Universität übernommen hat. Eine Diplomarbeit außerhalb der Universität stellt leider immer noch keine Selbstverständlichkeit dar.